\documentclass[a4paper,fleqn,usenatbib]{mnras}
\usepackage[T1]{fontenc}
\usepackage{ae,aecompl}
\usepackage{graphicx}
\usepackage{amsmath}
\usepackage{amssymb}
\usepackage{natbib}
\usepackage{epstopdf}
\usepackage{caption}
\usepackage{float}
\usepackage{times}
\usepackage{amsmath}
\usepackage{amssymb}
\usepackage{amsfonts}
\usepackage{mathrsfs}
\usepackage{graphicx}
\usepackage{rotating}
\title[Membership probabilities in 70 GES clusters]{The Gaia-ESO Survey: Membership probabilities for stars in 63 open and 7 globular clusters from 3D kinematics}

\author[R. J. Jackson, R. D. Jeffries, N. J. Wright et al.]
{R. J.~Jackson$^1$, R. D.~Jeffries$^1$, N. J.~Wright$^1$, S.~Randich$^2$, G.~Sacco$^2$, A.~Bragaglia$^{3}$, 
\newauthor A. Hourihane$^4$, E.~Tognelli$^{5,6}$, S.~Degl'Innocenti$^{5,6}$, P.G. Prada Moroni $^{5,6}$,
\newauthor  G. Gilmore$^4$, T. Bensby$^7$, E.~Pancino$^{2,8}$, R.~Smiljanic$^9$,  M.~Bergemann$^{10}$, G.~Carraro$^{11}$,    
\newauthor  E. Franciosini$^2$, A.~Gonneau$^4$, P.~Jofr\'e$^{12}$, J.~Lewis$^4$, L.~Magrini$^2$, L.~Morbidelli$^2$,        
\newauthor   L.~Prisinzano$^{13}$, C.~Worley$^4$, S.~Zaggia$^{14}$, G.~Tautvaišiene$^{15}$, M.L.~Gutiérrez Albarrán$^{16}$,    
\newauthor  D.~Montes$^{16}$, F.~Jiménez-Esteban$^{17}$,  
\\
    $^1$ Astrophysics Group, Keele University, Keele, Staffordshire ST5 5BG, United Kingdom\\
    $^2$ INAF - Osservatorio Astrofisico di Arcetri, Largo E. Fermi 5, 50125, Florence, Italy\\
    $^3$ INAF - Osservatorio di Astrofisica e Scienza dello Spazio di Bologna, via Gobetti 93/3, 40129, Bologna, Italy\\
    $^4$ Institute of Astronomy, University of Cambridge, Madingley Road, Cambridge CB3 0HA, United Kingdom\\
    $^5$ Dipartimento di Fisica ‘Enrico Fermi’, Universit\'a di Pisa, Largo Bruno Pontecorvo 3, I-56127 Pisa, Italy\\
    $^6$ INFN, Sezione di Pisa, Largo Bruno Pontecorvo 3, I-56127 Pisa, Italy\\
    $^7$ Lund Observatory, Department of Astronomy and Theoretical Physics, Box 43, SE-221 00 Lund, Sweden\\
    $^8$ Space Science Data Center - Agenzia Spaziale Italiana, via del Politecnico, s.n.c., I-00133, Roma, Italy\\
    $^9$ Nicolaus Copernicus Astronomical Center, Polish Academy of Sciences, ul. Bartycka 18, 00-716, Warsaw, Poland\\
    $^{10}$  Max-Planck Institut f\"{u}r Astronomie, K\"{o}nigstuhl 17, 69117 Heidelberg, Germany\\
    $^{11}$ Dipartimento di Fisica e Astronomia, Universit\`a di Padova, Vicolo dell'Osservatorio 3, 35122 Padova, Italy\\
    $^{12}$ N\'ucleo de Astronom\'{i}a, Facultad de Ingenier\'{i}a, Universidad Diego Portales, Av. Ej\'ercito 441, Santiago, Chile\\
    $^{13}$ INAF - Osservatorio Astronomico di Palermo, Piazza del Parlamento 1, 90134, Palermo, Italy\\
    $^{14}$ INAF - Padova Observatory, Vicolo dell'Osservatorio 5, 35122 Padova Italy\\
    $^{15}$ Institute of Theoretical Physics and Astronomy, Vilnius University, Sauletekio av. 3, 10257 Vilnius, Lithuania.\\
    $^{16}$ Departamento de Física de la Tierra y Astrofís\'ica and IPARCOS UCM, Universidad Complutense de Madrid, E-28040, Madrid, Spain\\
    $^{17}$ Departamento de Astrofísica, Centro de Astrobiología (CSICINTA), Camino Bajo del Castillo s/n, E-28692 Villanueva de la Canada, Madrid, Spain\\
}

\date{}

\pagerange{\pageref{firstpage}--\pageref{lastpage}} \pubyear{2021}

\def\LaTeX{L\kern-.36em\raise.3ex\hbox{a}\kern-.15em
    T\kern-.1667em\lower.7ex\hbox{E}\kern-.125emX}

\begin{document}
\label{firstpage}
\maketitle

\begin{abstract}
Spectroscopy from the final internal data release of the Gaia-ESO Survey (GES) has been combined with {\it Gaia} EDR3 to assign membership probabilities to targets observed towards 63 Galactic open clusters and 7 globular clusters. The membership probabilities are based chiefly on maximum likelihood modelling of the 3D kinematics of the targets, separating them into cluster and field populations.  From 43211 observed targets, 13985 are identified as highly probable cluster members ($P>0.9$), with an average membership probability of 0.993. The addition of GES radial velocities successfully drives down the fraction of false positives and we achieve better levels of discrimination in most clusters over the use of astrometric data alone, especially those at larger distances. Since the membership selection is almost purely kinematic, the union of this catalogue with GES and {\it Gaia} is ideal for investigating the photometric and chemical properties of clusters as a function of stellar mass, age and Galactic position.

\end{abstract}


\section{Introduction}
\label{intro}
The {\it Gaia}-ESO Survey (GES) is a large public survey programme that uses the multi-fibre FLAMES \citep[Fiber Large Array Multi-Element Spectrograph,][]{Pasquini2002a} instrument on the 8-m UT2-{\it Kueyen} telescope of the Very Large Telescope (VLT) to perform chemical and kinematical studies of Galactic stellar populations \citep[see][]{Gilmore2012a, Randich2013a, Pancino2017a}. The GIRAFFE and UVES \citep[Ultraviolet and Visual Echelle Spectrograph,][]{Dekker2000a} spectrographs attached to FLAMES, were used to obtain spectra of about $10^5$ and $10^4$ stars at resolving powers of $\sim 17000$ and $\sim 47000$ respectively. 

About $\sim$40 per cent of the survey time was spent capturing the spectra of stars {\it towards} star clusters and associations at a range of ages (Bragaglia et al. in preparation). In total, GES either observed (or reprocessed similar archival data) and has provided chemical abundances and radial velocities for 85 open clusters and 14 globular clusters as part of the final internal Data Release 6 (hereafter GESiDR6) and these data will be made public via the ESO archive. 
%In this paper we analyse the subset of 68 target clusters that were homogeneously observed using the GIRAFFE HR15N setup (see \S~\ref{2.1}).

Clusters have a special place in astrophysics. They offer large groups of (approximately) coeval stars over a range of masses and evolutionary stages that can be used to test stellar models. At the same time, clusters of different ages can be used to explore or even empirically calibrate the time-dependence of various physical phenomena such as rotation, magnetic activity or the diffusion and mixing of chemical elements. A starting point for such studies is a clean list of cluster members, or at least a list of potential members where any probability of contamination is well understood. An important consideration is to avoid using membership criteria which then bias the phenomena being tested or investigated. For example, one should not use chemical abundances as a membership indicator and then use these members to investigate the dispersion of chemical abundances within a cluster.

The work presented here follows on from \cite{Randich2018a} and \cite{Jackson2020a}, where {\it Gaia} astrometry from {\it Gaia} DR1 and {\it Gaia} DR2, in the form of its first and second data 
releases  \citep[][respectively]{Gaia2016a, Gaia2018a}, was used in conjunction with spectroscopic parameters from earlier GES data releases to define samples of high probability cluster members. \cite{Jackson2020a} (hereafter J20) used data available in GES internal data release 5 (GESiDR5) and {\it Gaia} DR2 \citep{Lindegren2018a} to provide membership probabilities for 32 clusters based on their three-dimensional kinematics. Here, we use temperatures, gravities and radial velocities from GESiDR6, together
with astrometry from {\it Gaia} EDR3 \citep{Lindegren2020a}, to define membership probabilities for sources in 70 clusters from GES. These comprise the full set of 60 open cluster targets in GES that were observed using the GIRAFFE spectrograph with the HR15N order-sorting filter or with UVES (two of which were ``double clusters" -- pairs of clusters that are in the field of view but not kinematically related see \S\ref{2}), to which we added one archival cluster (M67) and 7 globular clusters that were also observed with GIRAFFE$+$HR15N and homogeneously analysed as part of the GES project.

The analysis used to determine membership probability is similar to that described in J20, however the process of selecting a list of {\it potential} cluster members has changed in some key aspects to reflect the broader mix of cluster ages and distances. In general the clusters newly added to this study,  i.e. those reported for the first time in GESiDR6, are older, more distant and contain a higher proportion giant stars that are likely cluster members. For more distant clusters the inclusion of the third dimension of radial velocity ($RV$) from GES in stars as faint as $V \sim 19$ is critical in separating cluster members from contaminants. Overall, our aim is to provide rigorously determined membership lists, with quantitative membership probabilities, that can be used for follow-up investigations.

The paper is organised as follows; \S2 describes the sources of data used for membership analysis, \S3 describes the process of target selection and calculation of membership probability. In \S4 we report the membership probabilities of individual targets for each cluster. The results are discussed in \S5 and a summary and conclusions are provided in \S6. Appendix A reports membership probabilities for a set of supplementary targets with three further appendices (available as supplementary material) showing plots of key results for each of the 70 clusters studied.   

\section{Source data}
\label{2}
GESiDR6 data for our analysis were taken from a library of stacked spectra  for GIRAFFE HR15N and UVES observations of cluster targets and  the GESiDR6 Parameter Catalogue; the latter containing values of effective temperature ($T_{\rm eff}$), gravity $\log g$ and a gravity-sensitive spectral index \citep[$\gamma$, see][]{Damiani2014a} (plus other parameters not used here) derived by the GES Working Groups (WGs) for a large proportion of the cluster targets (see Hourihane et al. in preparation).

\subsection{Cluster targets}
\label{2.1}
Clusters were identified  from the library of  GESiDR6 stacked spectra. The dataset was scanned to identify clusters where multiple targets had been observed using the GIRAFFE 665\,nm (HR15N) order-sorting filter. 68 clusters were identified with unique names, two of which (see below) contained data from two kinematically separate clusters, hence our final list of 70 clusters. We note that the GES project also analysed a further 24 named open clusters. These are generally sparse clusters with few targets (only 6 with $>50$ targets) and were observed with an inhomogeneous set of spectrograph setups that did {\it not} include  HR15N observations; we have not analysed those clusters in this paper.

A small number of targets (21) showed duplicate spectra, one from a GES observation and one from an archival source. In this case the latter was discarded yielding a total of 37930 targets with GIRAFFE HR15N spectra.  Scanning the dataset of UVES 520nm and 580nm spectra associated with these 68 clusters identified UVES spectra for 2508 distinct targets of which 790 also had GIRAFFE HR15N spectra.      

Tables~\ref{table1}~and \ref{table2} show lists of cluster names together with representative values of age, distance modulus and reddening reported in the literature. Also shown are the number of unique targets observed in each cluster with GIRAFFE HR15N and/or UVES setups. Table~\ref{table1} lists properties for the younger clusters ($<$0.6\,Gyr). Data for  sub-clusters NGC\,2451a and NGC\,2451b and for Lambda Ori and Lambda Ori B35 are shown separately since these are known to be spatially and kinematically distinct clusters. Table~\ref{table2} lists the properties of 31 older open clusters with reported ages of 0.6 to 6\,Gyr and 7 globular clusters with reported ages $>$10\,Gyr.

Table~\ref{table3} shows the full list of targets with HR15N and/or UVES spectra that are common to the WG catalogue showing the WG ``cname" together with cluster name, setup and co-ordinates read from the spectrum metadata. The are 41926 entries for 39441 unique targets. Of these, 88 per cent have values of $T_{\rm eff}$ derived by the GES WGs and 67 per cent have WG values of $\log g$.  Table~\ref{tableA2} lists a further 3770 targets common to our list of 70 clusters that were observed with a variety of GIRAFFE setups but not with GIRAFFE+HR15N or UVES. In the interests of homogeneity, these targets were excluded from the main kinematic analysis but their individual membership probabilities are estimated in Appendix A.

\begin{table*}
\caption{Data for younger open clusters. Columns 2--4 show ages, intrinsic distance moduli and reddening from the literature (superscripts refer to references listed below Table\ref{table2}). Columns 5--7 show the numbers of targets observed, the numbers of targets with all data  required for membership analysis (see  \S\ref{2.4}) and the number fitted in the  membership analysis. Columns 8 and 9 show the values of distance modulus and cluster reddening calculated for high probability cluster members (see~\ref{table3}  \S\ref{3.5}).}
\begin{tabular}{lllrrrrrrr}
\hline 
Cluster	 &	Age	&	$(M-m)_0$	&	$E(B-V)$	&	Number	&	Number	&	Number	&	$(M-m)_0^{\rm c}$~~~~	&	$E(B-V)^{\rm c}$	\\
&	(Myr)	&	Literature	&	Literature	&	observed	&	complete	&	fitted	&	members~~~~	&	members\\\hline
NGC 6530 &1$^{65}$&10.48$^{63}$&0.35$^{74}$&                      1972 &  1907 &  1501 &  10.60$\pm$  0.02$\pm$  0.09 &   0.44$\pm$  0.10\\
Trumpler 14 &1--3$^{39}$&12.3$^{39}$&0.4--0.9$^{39}$&             1111 &  1069 &   741 &  12.05$\pm$  0.02$\pm$  0.17 &   0.61$\pm$  0.10\\
Chamaeleon I &2$^{50}$&6.02$^{82}$&$\sim$1$^{50}$&                 708 &   687 &   170 &   6.40$\pm$  0.01$\pm$  0.01 &   0.18$\pm$  0.08\\
Rho Ophiuchus &3$^{26}$&5.4$^{49}$&---&                            311 &   301 &    72 &   5.70$\pm$  0.01$\pm$  0.01 &   0.76$\pm$  0.13\\
NGC 2264 &4$^{80}$&9.4$^{73}$&0.07$^{78}$&                        1876 &  1819 &  1408 &   9.29$\pm$  0.01$\pm$  0.05 &   0.05$\pm$  0.05\\
NGC 2244 &4$^{15}$&11.15$^{15}$&0.48$^{15}$&                       432 &   427 &   375 &  10.90$\pm$  0.03$\pm$  0.10 &   0.49$\pm$  0.09\\
Lambda Ori &6$^{22}$&7.9$^{21}$&0.12$^{20}$&                       608 &   588 &   344 &   8.00$\pm$  0.01$\pm$  0.03 &   0.09$\pm$  0.04\\
Lambda Ori B35 &6$^{22}$&7.9$^{21}$&0.12$^{20}$&                   227 &   215 &   118 &   8.00$\pm$  0.01$\pm$  0.03 &   0.09$\pm$  0.02\\
25 Ori &6$^{25}$&7.63$^{72}$&0.07$^{72}$&                          294 &   284 &   256 &   7.71$\pm$  0.01$\pm$  0.02 &   0.00$\pm$  0.02\\
ASCC 50 &8$^{66}$&9.64$^{46}$&0.23$^{46}$&                        1224 &  1192 &   501 &   9.91$\pm$  0.02$\pm$  0.06 &   0.28$\pm$  0.05\\
Collinder 197 &13$^{19}$&9.38$^{64}$&0.2$^{64}$&                   409 &   395 &   334 &   9.94$\pm$  0.02$\pm$  0.06 &   0.64$\pm$  0.07\\
Gamma Velorum &18$^{43}$&7.72$^{42}$&0.04$^{42}$&                 1262 &  1242 &   497 &   7.73$\pm$  0.01$\pm$  0.02 &   0.04$\pm$  0.03\\
IC 4665 &23$^{67}$&7.69$^{32}$&0.17$^{7}$&                         567 &   562 &   298 &   7.69$\pm$  0.01$\pm$  0.02 &   0.15$\pm$  0.02\\
NGC 2232 &38$^{3}$&7.56$^{32}$&0.03$^{19}$&                       1761 &  1734 &   697 &   7.52$\pm$  0.01$\pm$  0.02 &   0.04$\pm$  0.03\\
NGC 2547 &38$^{67}$&7.97$^{32}$&0.06$^{56}$&                       477 &   472 &   269 &   7.93$\pm$  0.01$\pm$  0.03 &   0.06$\pm$  0.03\\
IC 2602 &44$^{67}$&5.91$^{32}$&0.03$^{37}$&                       1840 &  1817 &   117 &   5.90$\pm$  0.01$\pm$  0.01 &   0.04$\pm$  0.02\\
NGC 2451b &50$^{38}$&7.84$^{38}$&0.01$^{62}$&                     1656 &  1635 &   425 &   7.80$\pm$  0.01$\pm$  0.02 &   0.10$\pm$  0.03\\
NGC 6649 &50$^{81}$&11.$^{81}$&01.37$^{52}$&                       122 &   121 &   116 &  11.69$\pm$  0.03$\pm$  0.14 &   1.43$\pm$  0.05\\
IC 2391 &51$^{67}$&5.9$^{32}$&0.01$^{19}$&                         434 &   426 &    78 &   5.90$\pm$  0.01$\pm$  0.01 &   0.03$\pm$  0.01\\
NGC 2451a &50-80$^{38}$&6.44$^{32}$&0.01$^{62}$&                  1656 &  1637 &   352 &   6.42$\pm$  0.01$\pm$  0.01 &   0.02$\pm$  0.02\\
NGC 6405 &94$^{47}$&8.01$^{47}$&0.14$^{47}$&                       659 &   654 &   373 &   8.31$\pm$  0.01$\pm$  0.03 &   0.14$\pm$  0.04\\
NGC 6067 &120$^{19}$&10.76$^{19}$&0.38$^{19}$&                     532 &   531 &   512 &  11.62$\pm$  0.01$\pm$  0.14 &   0.34$\pm$  0.04\\
NGC 2516 &125$^{51}$&8.09$^{32}$&0.11$^{75}$&                      759 &   745 &   641 &   8.07$\pm$  0.01$\pm$  0.03 &   0.11$\pm$  0.03\\
Blanco 1 &100--150$^{55}$&6.88$^{32}$&0.01$^{19}$&                 463 &   446 &   314 &   6.88$\pm$  0.01$\pm$  0.02 &  -0.01$\pm$  0.03\\
NGC 6709 &150$^{19}$&10.16$^{19}$&0.3$^{19}$&                      684 &   681 &   551 &  10.19$\pm$  0.01$\pm$  0.07 &   0.27$\pm$  0.02\\
NGC 6259 &210$^{54}$&11.61$^{16}$&0.66$^{54}$&                     438 &   423 &   391 &  11.82$\pm$  0.01$\pm$  0.15 &   0.63$\pm$  0.09\\
NGC 6705 &280$^{6}$&11.37$^{43}$&0.42$^{70}$&                     1066 &  1043 &   977 &  11.96$\pm$  0.01$\pm$  0.16 &   0.40$\pm$  0.06\\
Berkeley 30 &300$^{45}$&13.49$^{45}$&0.52$^{45}$&                  226 &   224 &   216 &  13.57$\pm$  0.09$\pm$  0.34 &   0.51$\pm$  0.04\\
NGC 3532 &300$^{18}$&8.46$^{18}$&0.03$^{18}$&                      966 &   952 &   687 &   8.40$\pm$  0.01$\pm$  0.03 &   0.05$\pm$  0.02\\
NGC 6281 &314$^{19}$&8.86$^{19}$&0.15$^{19}$&                      251 &   249 &    63 &   8.62$\pm$  0.01$\pm$  0.03 &   0.18$\pm$  0.02\\
NGC 4815 &560$^{30}$&11.99$^{8}$&0.7$^{8}$&                        126 &   126 &   112 &  12.78$\pm$  0.08$\pm$  0.24 &   0.70$\pm$  0.07\\
NGC 6633 &575$^{79}$&7.99$^{32}$&0.17$^{79}$&                     1595 &   363 &   119 &   7.99$\pm$  0.02$\pm$  0.03 &   0.15$\pm$  0.04\\
\hline
\end{tabular}
\label{table1}
\end{table*}

\begin{table*}
\caption{Data for older open and globular clusters. Columns 2--4 show ages, intrinsic distance moduli and reddening from the literature (superscripts refer to references listed below the Table). Columns 5--7 show the numbers of targets observed, the numbers of targets with all data  required for membership analysis (see \S\ref{2.2}) and the number fitted in the  membership analysis.  Columns 8 and 9 show values of distance modulus and cluster reddening calculated for high probability cluster members. Redenning values were not calculated for the globular clusters (see \S\ref{3.5}) }
\begin{tabular}{lllrrrrrrl}
\hline 
Cluster	 &	Age	&	$(M-m)_0$	&	$E(B-V)$	&	Number	&	Number	&	Number	&	$(M-m)_0^{\rm c}$~~~~	&	$E(B-V)^{\rm c}$	\\
&	(Myr)	&	Literature	&	Literature	&	observed	&	complete	&	fitted	&	members~~~~	&	members\\\hline
Pismis18 &700$^{35}$&11.75$^{60}$&0.5$^{60}$&                      101 &   101 &    90 &  12.27$\pm$  0.03$\pm$  0.19 &   0.65$\pm$  0.03\\
Trumpler 23 &800$^{59}$&11.71$^{12}$&0.58$^{4}$&                    89 &    89 &    83 &  12.20$\pm$  0.03$\pm$  0.18 &   0.68$\pm$  0.04\\
NGC 2355 &900$^{24}$&10.92$^{24}$&0.14$^{24}$&                     208 &   208 &   204 &  11.37$\pm$  0.01$\pm$  0.12 &   0.13$\pm$  0.03\\
NGC 6802 &900$^{76}$&11.28$^{41}$&0.84$^{41}$&                     103 &   103 &    98 &  12.38$\pm$  0.09$\pm$  0.20 &   0.79$\pm$  0.06\\
Ruprecht 134 &1000$^{12}$&12.66$^{12}$&0.5$^{12}$&                 680 &   665 &   602 &  12.04$\pm$  0.05$\pm$  0.17 &   0.46$\pm$  0.08\\
Berkeley 81 &1000$^{53}$&12.39$^{68}$&1.0$^{68}$&                  203 &   203 &   171 &  12.96$\pm$  0.07$\pm$  0.26 &   0.85$\pm$  0.04\\
NGC 6005 &1200$^{60}$&12.16$^{60}$&0.45$^{60}$&                    355 &   353 &   325 &  12.27$\pm$  0.11$\pm$  0.18 &   0.49$\pm$  0.06\\
Pismis 15 &1300$^{11}$&12.31$^{11}$&0.53$^{11}$&                   235 &   235 &   224 &  11.93$\pm$  0.03$\pm$  0.16 &   0.56$\pm$  0.05\\
Trumpler 20 &1400$^{23}$&12.39$^{13}$&0.35$^{13}$&                 552 &   545 &   490 &  12.88$\pm$  0.02$\pm$  0.25 &   0.37$\pm$  0.03\\
Berkeley 44 &1600$^{40}$&12.46$^{41}$&0.98$^{41}$&                  93 &    92 &    83 &  12.49$\pm$  0.05$\pm$  0.21 &   0.90$\pm$  0.07\\
NGC 2141 &1800$^{23}$&13.12$^{23}$&0.4$^{23}$&                     853 &   846 &   801 &  13.37$\pm$  0.02$\pm$  0.31 &   0.35$\pm$  0.04\\
Czernik 24 &2000$^{31}$&13.22$^{31}$&0.54$^{48}$&                  346 &   343 &   302 &  13.18$\pm$  0.07$\pm$  0.28 &   0.65$\pm$  0.03\\
Haffner 10 &2000$^{19}$&12.84$^{19}$&0.55$^{19}$&                  460 &   457 &   428 &  12.94$\pm$  0.05$\pm$  0.25 &   0.52$\pm$  0.05\\
NGC 2158 &2000$^{10}$&13.49$^{19}$&0.36$^{19}$&                    616 &   598 &   571 &  13.20$\pm$  0.03$\pm$  0.29 &   0.48$\pm$  0.04\\
NGC 2420 &2200$^{69}$&11.97$^{71}$&0.05$^{2}$&                     562 &   557 &   520 &  12.03$\pm$  0.01$\pm$  0.17 &   0.04$\pm$  0.03\\
Berkeley 21 &2200$^{83}$&13.98$^{83}$&0.74$^{83}$&                 744 &   738 &   574 &  14.27$\pm$  0.17$\pm$  0.48 &   0.66$\pm$  0.06\\
Berkeley 73 &2300$^{19}$&14.49$^{19}$&0.1$^{19}$&                   76 &    75 &    70 &  14.65$\pm$  0.19$\pm$  0.57 &   0.22$\pm$  0.05\\
Berkeley 22 &2400$^{27}$&13.8$^{27}$&0.72$^{27}$&                  395 &   395 &   352 &  14.16$\pm$  0.12$\pm$  0.45 &   0.58$\pm$  0.07\\
Czernik 30 &2800$^{36}$&14.03$^{36}$&0.24$^{36}$&                  226 &   226 &   193 &  14.49$\pm$  0.12$\pm$  0.53 &   0.31$\pm$  0.03\\
Berkeley 31 &2900$^{17}$&14.4$^{17}$&0.19$^{17}$&                  616 &   614 &   499 &  14.50$\pm$  0.19$\pm$  0.53 &   0.10$\pm$  0.07\\
Berkeley 75 &3000$^{19}$&14.96$^{19}$&0.08$^{19}$&                  75 &    74 &    64 &  14.75$\pm$  0.18$\pm$  0.59 &   0.10$\pm$  0.03\\
Loden 165 &3000$^{9}$&11.39$^{9}$&0.25$^{9}$&                      388 &   387 &   333 &  11.79$\pm$  0.07$\pm$  0.15 &   0.21$\pm$  0.06\\
NGC 6253 &3000$^{5}$&10.9$^{5}$&0.28$^{5}$&                        294 &   235 &   227 &  11.15$\pm$  0.01$\pm$  0.11 &   0.24$\pm$  0.03\\
Messier 67 &3500$^{32}$&9.73$^{32}$&0.04$^{32}$&                   131 &   131 &   130 &   9.67$\pm$  0.01$\pm$  0.06 &   0.05$\pm$  0.01\\
NGC 2425 &3600$^{34}$&12.63$^{34}$&0.28$^{34}$&                    528 &   525 &   481 &  12.69$\pm$  0.02$\pm$  0.23 &   0.35$\pm$  0.04\\
NGC 2243 &3800$^{1}$&12.96$^{1}$&0.05$^{1}$&                       703 &   701 &   614 &  13.29$\pm$  0.02$\pm$  0.30 &   0.04$\pm$  0.04\\
Berkeley 36 &4000$^{57}$&13.84$^{57}$&0.3$^{57}$&                  739 &   737 &   672 &  13.44$\pm$  0.08$\pm$  0.32 &   0.54$\pm$  0.05\\
Trumpler 5 &5000$^{61}$&11.9$^{61}$&0.6$^{61}$&                   1138 &  1132 &  1098 &  12.61$\pm$  0.02$\pm$  0.22 &   0.63$\pm$  0.06\\
Berkeley 32 &5900$^{29}$&12.5$^{77}$&0.15$^{77}$&                  389 &   385 &   348 &  12.72$\pm$  0.02$\pm$  0.23 &   0.16$\pm$  0.04\\
Berkeley 39 &6000$^{44}$&12.9$^{44}$&0.11$^{8}$&                   899 &   897 &   832 &  13.29$\pm$  0.02$\pm$  0.30 &   0.10$\pm$  0.05\\
ESO 92-05 &6000$^{58}$&15.19$^{58}$&0.17$^{58}$&                   212 &   210 &   114 &  16.06$\pm$  0.30$\pm$  1.16 &   0.06$\pm$  0.10\\
NGC 1261 &10200$^{28}$&16.06$^{33}$&0.01$^{33}$&                    80 &    80 &    66 &  17.21$\pm$  0.29$\pm$  2.58 & ---\\
NGC 362 &10400$^{28}$&14.84$^{14}$&0.06$^{14}$&                    148 &   147 &   133 &  15.37$\pm$  0.12$\pm$  0.81 & ---\\
NGC 2808 &10800$^{28}$&14.91$^{33}$&0.22$^{33}$&                   289 &   269 &   191 &  15.39$\pm$  0.04$\pm$  0.82 & ---\\
NGC 1904 &11100$^{28}$&15.56$^{33}$&0.01$^{33}$&                    71 &    71 &    62 &  16.11$\pm$  0.09$\pm$  1.19 & ---\\
NGC 6752 &11800$^{28}$&13.01$^{33}$&0.04$^{33}$&                   353 &   322 &   297 &  13.30$\pm$  0.04$\pm$  0.30 & ---\\
NGC 5927 &12700$^{28}$&14.43$^{33}$&0.45$^{33}$&                   124 &   123 &    88 &  15.39$\pm$  0.18$\pm$  0.82 & ---\\
NGC 104 &13100$^{28}$&13.45$^{14}$&0.05$^{14}$&                    311 &   301 &   296 &  13.41$\pm$  0.04$\pm$  0.32 & ---\\
\hline
 \end{tabular}
\begin{flushleft}
$^{1}$\cite{Anthony-Twarog2005a}
$^{2}$\cite{Anthony-Twarog2006a}
$^{3}$\cite{Binks2021a}
$^{4}$\cite{Bonatto2007a}
$^{5}$\cite{Bragaglia1997b}
$^{6}$\cite{CantatGaudin2014a}
$^{7}$\cite{Cargile2010a}
$^{8}$\cite{Carraro1994a}
$^{9}$\cite{Carraro2001a}
$^{10}$\cite{Carraro2002a}
$^{11}$\cite{Carraro2005a}
$^{12}$\cite{Carraro2006a}
$^{13}$\cite{Carraro2010a}
$^{14}$\cite{Carretta2000a}
$^{15}$\cite{Chen2007a}
$^{16}$\cite{Ciechanowska2006a}
$^{17}$\cite{Cignoni2011a}
$^{18}$\cite{Clem2011a}
$^{19}$\cite{Dias2002a} WEBDA
$^{20}$\cite{Diplas1994a}
$^{21}$\cite{Dolan1999a}
$^{22}$\cite{Dolan2002a}
$^{23}$\cite{Donati2014a}
$^{24}$\cite{Donati2015a}
$^{25}$\cite{Downes2014a}
$^{26}$\cite{Erickson2011a}
$^{27}$\cite{Fabrizio2005a}
$^{28}$\cite{Forbes2010a}
$^{29}$\cite{Friel2010a}
$^{30}$\cite{Friel2014a}
$^{31}$\cite{Froebrich2010a}
$^{32}$\cite{Gaia2018a}
$^{33}$\cite{Harris1996a} GCDB
$^{34}$\cite{Hasegawa2008a}
$^{35}$\cite{Hatzidimitriou2019a}
$^{36}$\cite{Hayes2015a}
$^{37}$\cite{Hill1969a}
$^{38}$\cite{Hunsch2003a}
$^{39}$\cite{Hur2012a}
$^{40}$\cite{Jacobson2016a}
$^{41}$\cite{Janes2011a}
$^{42}$\cite{Jeffries2009a}
$^{43}$\cite{Jeffries2017a}
$^{44}$\cite{Kassis1997a}
$^{45}$\cite{Kharchenko2013a}
$^{46}$\cite{Kharchenko2013a}
$^{47}$\cite{Kilicoglu2016a}
$^{48}$\cite{Kim2005a}
$^{49}$\cite{Loinard2008a}
$^{50}$\cite{Luhman2007a}
$^{51}$\cite{Lyra2006a}
$^{52}$\cite{Madore1975a}
$^{53}$\cite{Magrini2015b}
$^{54}$\cite{Mermilliod2001a}
$^{55}$\cite{Moraux2007a}
$^{56}$\cite{Naylor2006a}
$^{57}$\cite{Ortolani2005a}
$^{58}$\cite{Ortolani2008a}
$^{59}$\cite{Overbeek2017a}
$^{60}$\cite{Piatti1998a}
$^{61}$\cite{Piatti2004a}
$^{62}$\cite{Platais2001a}
$^{63}$\cite{Prisinzano2005a}
$^{64}$\cite{Prisinzano2018a}
$^{65}$\cite{Prisinzano2019a}
$^{66}$~Prisinzano L. - private com.
$^{67}$\cite{Randich2018a}
$^{68}$\cite{Sagar1998a}
$^{69}$\cite{Salaris2004a}
$^{70}$\cite{Santos2005a}
$^{71}$\cite{Sharma2006a}
$^{72}$\cite{Suarez2017a}
$^{73}$\cite{Sung1997a}
$^{74}$\cite{Sung2000a}
$^{75}$\cite{Sung2002a}
$^{76}$\cite{Tang2017a}
$^{77}$\cite{Tosi2007a}
$^{78}$\cite{Turner2012a}
$^{79}$\cite{vanLeeuwen2009a}
$^{80}$\cite{Venuti2018a}
$^{81}$\cite{Walker1987a}
$^{82}$\cite{Whittet1997a}
$^{83}$\cite{Yong2012a}

\end{flushleft}
\label{table2}
\end{table*}

\subsection{Spectral indices}
\label{2.2}

The narrow-band spectral indices $\gamma$ and $\tau$ were calculated from the normalised GIRAFFE spectra using the procedure described in   \cite{Damiani2014a}. The $\tau$ index is used to estimate an approximate value of $T_{\rm eff}$ and the index $\gamma'=\gamma + \tau /6$ is used as a proxy for $\log g$ for targets where it is not reported in the GESiDR6 WG catalogue.

\subsection{Radial velocities}
\label{2.3}
\begin{figure}
  \centering
	\includegraphics[width = 64mm]{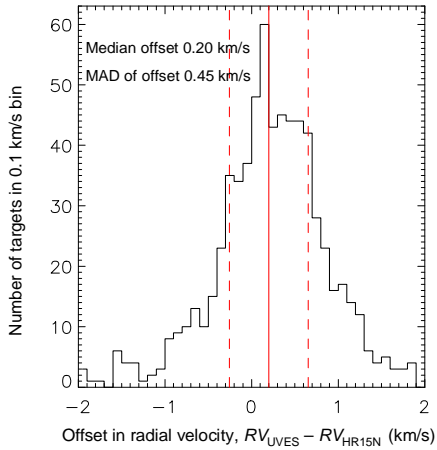}
	\caption{Histogram of the offset between UVES and GIRAFFE HR15N measurements of $RV$ for targets that have repeat measurements. }
	\label{plot_UVES2Giraffe}
\end{figure}

For GIRAFFE spectra, the radial velocities ($RV$) were read from the spectrum {\sc velclass} metadata. For UVES, the mean value  of $RV$ was taken from values reported for the upper and lower CCD spectra \citep{Sacco2014a}. The offset in $RV$ between HR15N and UVES measurements was estimated by comparing $RV$s measured for 790 targets observed using both setups. The results in Fig.~\ref{plot_UVES2Giraffe} show a median offset\footnote{Some of the dispersion or outliers in Fig.~\ref{plot_UVES2Giraffe} could be due to binaries, since the GIRAFFE and UVES observations were not cotemporal.} of 0.2\,km\,s$^{-1}$. Values of  $RV$ measured from UVES spectra were reduced by this value to match the GIRAFFE HR15N $RV$ scale.

For GIRAFFE measurements of $RV$, the expected distribution of measurement precision is non-Gaussian, characterised by the product of a scaling constant $S_{\rm RV}$ and a Student-t distribution  \citep{Jackson2015a}, where $S_{\rm RV}$ is a function of signal to noise ratio ($S/N$), resolving power, $R$ and projected equatorial velocity ($v\sin i$). The scaling constant of uncertainty for short term repeats (e.g. spectra taken consecutively), using the same instrument set up and wavelength calibration, is given by
\begin{equation}
S_{\rm RV,0} = B\frac{(1+([v\sin i]/C)^2)^{3/4}}{(S/N)}\, ,
\end{equation}
where $B$ is an empirically determined parameter that depends on the intrinsic stellar spectrum (largely characterised by the effective temperature) and $C$ is a function of the spectrograph resolving power. For long-term repeats (e.g, spectra taken on different nights), there is an additional contribution 
due to variations in instrument setup and wavelength calibration, $A$, which adds in 
quadrature to give
\begin{equation}
S_{\rm RV} = \sqrt{A^2+ S_{\rm RV,0}^{2}}\, ,
\end{equation}
\noindent{Analysis of 34k short term and 4.6k longer term repeat observations in the GESiDR6 dataset gave the following expressions (in units of km\,s$^{-1}$) for the empirically determined parameters.
\begin{eqnarray}
A & = & {\rm max} (0.26,0.04+13.7/(S/N))\,, \\\nonumber
B & = & 5.85+2.07\tanh{((T_{\rm eff}^d-5000)/500) }, \\\nonumber
C & = & 0.895\,c/\,R\, ,\nonumber
\end{eqnarray}
where $c$ is the speed of light and $T_{\rm eff}^d$ is estimated from the spectral index $\tau$ using the polynomial function given in Section 5 of \citep{Damiani2014a}. These relations are well-calibrated over the temperature range $4000 < T_{\rm eff}^d < 6000$\,K which is sufficient to constrain the value of $B$.}

\begin{figure}
  \centering
	\includegraphics[width = 64mm]{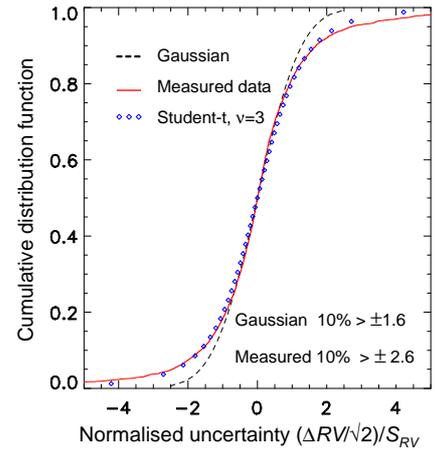}
	\caption{CDF of the normalised value of empirical uncertainty in GIRAFFE HR15N measurements of $RV$ (see \S\ref{2.3}).}
	\label{plot_RV_precision}
\end{figure}

Figure~\ref{plot_RV_precision} shows the cumulative distribution function (CDF) of normalised measurement uncertainty, $(\Delta RV/\sqrt{2})/S_{\rm RV}$, compared to the CDF of a unit Gaussian  distribution. The measured data shows a clear non-Gaussian tail with 10 per cent of targets showing $|\Delta RV/\sqrt{2}|/S_{\rm RV}$>2.6 compared to the $\sim 1$ per cent expected for a unit Gaussian distribution. A Student-t distribution of order $\nu=3$ is a good fit to the empirical CDF of measurement uncertainty.

The measurement uncertainties of UVES $RV$ values are calculated from the difference in $RV$ measured from the upper and lower spectra, $\Delta RV/\sqrt{2}$, in quadrature with a constant term of 0.4\,km\,s$^{-1}$ representing the uncertainty in wavelength calibration \citep{Sacco2014a}. The median uncertainty in UVES and GIRAFFE RV measurements is $<0$.5\,km\,s$^{-1}$.

The reader should be aware that the $RV$ values reported here will differ from the public GES $RV$ catalogue delivered to ESO in two ways (see Hourihane et al. in preparation). (i) Most of our $RV$s come from the GIRAFFE$+$HR15N spectra and these have been shifted by $+0.09$ km\,s$^{-1}$ in the construction of the GES $RV$ catalogue to more closely match the $RV$ values of standard stars. This is a simple shift to all the velocities, that was not applied here, and has no impact on the membership analysis. (ii) In the interests of homogeneity we quote the $RV$ from a single spectrum (the GIRAFFE$+$HR15N value is preferred over UVES) and make no attempt to average the $RV$ values or identify variability for the $\sim 5$ per cent of stars that were observed with both GIRAFFE$+$HR15N, and with UVES or in archival data. 

\subsection{Near infrared and {\it Gaia} data}
\label{2.4}
Target coordinates were cross matched (in a 2~arcsec radius) with the 2MASS \citep{Skrutskie2006a}, VISTA VHSDR6 \citep{McMahon2013} and VVVDR4 \citep{Minniti2010} catalogues to obtain $K_S$  magnitudes on the 2MASS system. $K_S$ data from the VISTA catalogues were used in preference to 2MASS data on 7416 occasions where  $K_S>$13 or the 2MASS quality flag for $K_S$ was not "A". 

%The targets listed in Table~\ref{table3} were cross matched with the {\it Gaia}~EDR3 catalogue to determine their astrometric properties and principle photometric properties ($G$ magnitude and $G_{\rm BP}- G_{\rm RP}$ colour) together with the angular distance, $r$ of the match from the target centre. Targets with valid 2MASS IDs were cross-matched using the gaiaedr3.tmass\_psc\_xsc\_neighbourhood catalogue \citep{Marrese2019a}. Where more than one neighbour was reported, the target with the highest "score" was taken as the valid match. Where the catalogue gave no match for the 2MASS ID, the 2MASS RA and Dec were cross matched to the {\it Gaia}~EDR3 catalogue and we accepted a 1\,arcsec radius for a valid match. Similarly, targets with no valid 2MASS ID were cross-matched by GES RA and Dec to the {\it Gaia}~EDR3 catalogue.  

Targets listed in Table~\ref{table3} were cross matched with the {\it Gaia}~EDR3 catalogue to to determine their astrometric properties and principle photometric properties ($G$ magnitude and $G_{\rm BP}- G_{\rm RP}$ colour) using the {\it Gaia}~ID given in the GESiDR6 parameter catalogue. There were 435 targets missing parallax, proper motion or $G$ magnitude data. These were flagged as incomplete (${\rm Flag}=-1$ in Table~\ref{table3}) and not used in the analysis. In NGC~6633, 1253 erroneous targets observed before the JD2456852\,d were also flagged as $-1$ and excluded from the analysis (see J20 for an explanation).

\subsection{Approximate values of $\gamma$, T$_{\rm eff}$, ~and $K_S$}
\label{2.5}
Values of  $\log g$~or~$\gamma '$, $T_{\rm eff}$~and $K_S$ were required for all targets used in the membership analysis in order to identify giants and estimate target luminosities and masses to model the effects of binarity on cluster membership probability. Approximate values of  $T_{\rm eff}^p$~and $K_S^p$ were calculated where WG values were not available. If both $\log g$ and $\gamma$ were not available in the WG data then a value of $\gamma$ was calculated directly from its spectrum (see \S\ref{2.2}). 

\begin{figure}
  \centering
	\includegraphics[width = 64mm]{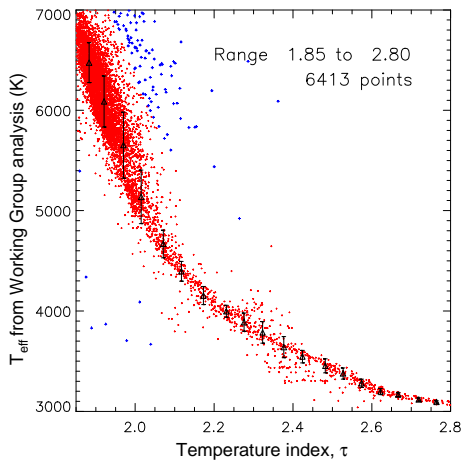}
	\caption{GESiDR6 Working Group $T_{\rm eff}$ versus temperature index $\tau$. Triangles show the median value of $T_{\rm eff}$ in bins of $\tau$. Red points are WG data used to define the median value in each, blue crosses are outliers rejected from the fit.}
	\label{plot_tau2teff}
\end{figure}

\begin{figure}
  \centering
	\includegraphics[width = 64mm]{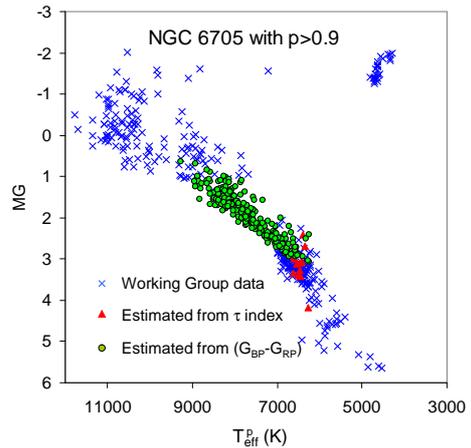}
	\caption{{\it Gaia}~$G$ magnitude versus $T_{\rm eff}^p$ for targets in open cluster NGC\,6705 later identified as cluster members with $P>0.9$. Blue crosses show targets with WG values of $T_{\rm eff}^p$, red triangles show stars where $T_{\rm eff}^p$ is estimated from the spectral index $\tau$, green circles are targets where $T_{\rm eff}^p$ is estimated from their $(G_{\rm BP} - G_{\rm RP})$ colour. }
	\label{plot_NGC6705a}
\end{figure}

Where no WG value of $T_{\rm eff}$ was available an approximate value, $T_{\rm eff}^p$, was found from the temperature index $\tau$ (\S\ref{2.2}) using the calibration curve in Fig.~\ref{plot_tau2teff}, valid in the range 3100 < T$_{\rm eff} < 6700$\,K. If no WG data was available for hotter stars, an approximate value of $T_{\rm eff}^p$ was estimated from the ($G_{\rm BP}- G_{\rm RP}$) colour using Pisa model ($G_{\rm BP}- G_{\rm RP}$)$_0$ to $T_{\rm eff}$ relations \citep{Tognelli2011a}, updated to reflect {\it Gaia}~EDR3 filter characteristics, together with the calculated cluster reddening in Tables~\ref{table1}~and ~\ref{table2} and $R_{\rm Bp-Rp}=1.34$~\citep{Casagrande2018a}. Figure~\ref{plot_NGC6705a} shows results in a typical cluster where the $\tau$ index and ($G_{\rm BP}- G_{\rm RP}$) colour data have been used to supplement WG $T_{\rm eff}$ data to provide a working value of $T_{\rm eff}^p$ in Table~\ref{table3} for$>$99.5 per cent of valid targets.  

$K_S$ data are available from 2MASS and/or VISTA sources for 95 per cent of targets. For the remainder, Pisa model isochrones were used to estimate an approximate value, $K_S^p$ in Table~\ref{table3}, from {\it Gaia}~$G$ magnitude and ($G_{\rm BP}- G_{\rm RP}$) colour using the values of age, $(M-m)_0^{\rm c}$ and $E(B-V)^{\rm c}$ in Tables~\ref{table1} and~\ref{table2}. For this calculation we used $R_G= 2.50$ \citep{Chen2019a}, $R_{Ks}=0.35$ \citep{Yuan2013a} and the relation $(G-K_S)_0$ =  -0.1885 + 2.092(G$_{\rm BP} - G_{\rm RP})_0   -0.1345(G_{\rm BP} - G_{\rm RP})_0^2$ \citep{Busso2018a}.

Tables~\ref{table1}~and~~\ref{table2} show the numbers of targets observed and the number with the complete data required to estimate cluster membership (i.e. {\it Gaia}~data and values of  $T_{\rm eff}^p$, ~and $K_S^p$). 95 per cent of the targets observed have the required data. Tangential velocities (and their uncertainties) were calculated from proper motions for these targets, assuming potential cluster members are at a common distance:
\begin{eqnarray}
V_{\rm RA} & = & 4.74 d_{\rm c}\,{\rm pm_{\rm RA}}\, , \nonumber \\
V_{\rm Dec} & = & 4.74 d_{\rm c}\,{\rm pm_{\rm Dec}}\, ,
\end{eqnarray}
where pm$_{\rm RA}$ and pm$_{\rm Dec}$ are the proper motions in units of mas\,yr$^{-1}$, and $d_c$  is the cluster distance in pc calculated from the distance modulus for high probability cluster members ($(M-m)^c_0$ in Tables~\ref{table1}~and~\ref{table2}). 

\section{Calculated parameters}
\label{3}
\subsection{Selecting potential cluster members}
\label{3.1}
The list of valid targets was screened to remove targets that are highly  unlikely to be cluster members. This process is particularly important for the younger clusters where a large fraction of observed targets are background giants, which can easily be identified from their $\log g$ and parallax. Targets meeting one or more of the following criteria were filtered from the wider sample.
\begin{enumerate}
	\item {\bf Background giants by $\log g$}: Stars in clusters aged $<1$\,Gyr which have $\log g \leq 3.4$,~$4000<T_{\rm  eff}^p/{\rm K}<7000$ {\it and} with a parallax smaller (by at least $2\sigma$) than a value corresponding to the intrinsic distance modulus of the cluster $+2$ mag. These giants are flagged as type 1 in Table~\ref{table3}. 
	\item {\bf Background giants by ${\mathbf \gamma'}$}: Stars where $\log g$ is undefined but the modified index $\gamma' \geq 1.335$ (see \S\ref{2.2}) {\it and} that meet the other conditions of the first criteria. These are flagged type 2 in Table~\ref{table3}.
	\item {\bf Background targets by parallax}: Stars with a {\it Gaia} parallax smaller, by at least 4$\sigma$, than a value corresponding to the intrinsic distance modulus of the cluster $+2$ mag. These are flagged type 3 in Table~\ref{table3}. For two clusters (ASCC 50 and NGC\,6281) it was necessary to reduce the margin of this criteria to cluster distance modulus +1 mag, in order to obtain a satisfactory peak in log likelihood in the membership analysis (see \S\ref{5.1}). 
	\item {\bf Targets with a large velocity offset relative to other cluster members}; Stars with $V_{\rm RA}$, $V_{\rm Dec}$ or $RV$ outside a window of $\pm 75$\,km\,s$^{-1}$ centred on the median velocity of the remaining targets. This rejected targets with bad velocity data whilst retaining almost the entire velocity spectrum of the field population.  
\end{enumerate}

\noindent{Tables~\ref{table1}~and~\ref{table2} show the numbers of targets in each cluster that remain after the screening process and that are used for the membership analysis.}

\subsection{Filtering by quality of the astrometric solution}
\label{3.2}

For the principle analysis in this paper the {\it Gaia} EDR3 data are accepted if there are reported values of $G$ magnitude, parallax and proper motions. An alternative approach examined in J20 is to filter out targets which were observed for only a low number of  visibility periods, ${\rm NPER}<8$ \citep{Arenou2018a}, and those showing a high renormalised unit weighted error, ${\rm RUWE}>1.4$ \citep{Lindegren2018a}. Applying this additional filter reduces the number of targets fitted in the membership analysis by $\sim$9 per cent. The effect of applying this additional filter is discussed in section~\ref{5.2}.

\subsection{Target luminosity and mass}
\label{3.3}
Stellar luminosities ($\log L$) in Table~\ref{table3} were calculated from the $K_S^p$ magnitude using bolometric corrections interpolated from the $T_{\rm eff}^p$ and photometry reported for updated solar-metallicity Pisa model isochrones \citep{Tognelli2011a}. Target masses were used solely to model the distribution of binary velocities in the membership analysis (see J20) and these were estimated from $\log L$ using the same  Pisa model isochrones. The predicted mass at the turn off was adopted as the maximum mass since giants will not be much more massive than this. Note that velocity perturbations due to the modelled binarity in the membership analysis scale only as $M^{1/3}$, so mass uncertainties are of little importance.

\subsection{Probability of cluster membership}
\label{3.4}
Tangential velocities, $RV$s and their associated measurement uncertainties and target masses were used to estimate membership probabilities for individual cluster targets using the maximum likelihood method originally proposed by \cite{Pryor1993a} and later updated by \cite{Cottaar2012a} to include the effect of binarity. Simulated velocity distributions of cluster and field stars are represented by Gaussians whose intrinsic width is then broadened due to the effects of individual uncertainties and, in the case of $RV$, a model for the perturbations caused by a fraction of binary systems following the method described by \cite{Cottaar2012a}. The adopted parameters for the binary distribution come from the binary survey of \citep{Raghavan2010a}: a binary fraction, $f_{\rm B} = 0.46$, a lognormal period distribution with a mean $\log \rm{period}$ = 5.03 (in days) and dispersion 2.28 dex, and a flat mass ratio distribution for $0.1 < q < 1$ . Further detail and a discussion of the effect of these assumptions and of the possibility that binarity perturbs the proper motion distribution is given in J20.

The maximum likelihood calculation used to fit the cluster/field populations for each cluster and assign membership probabilities is detailed in J20. In brief, it was done in two stages. First the mean velocity and dispersion of the background population of field stars was characterised for each velocity component using a series of 1D maximum likelihood analyses. These results were then used in a full 3D analysis to determine expectation values of mean velocity, the intrinsic dispersion for each velocity component and the fraction of targets that are cluster members. The uncertainty in $RV$ is independent of the uncertainty in proper motions allowing the 3D likelihood, $\mathscr{L}_{\rm 3D}$ to be calculated as the product of the likelihood  of fit in $RV$, $\mathscr{L}_{\rm RV}$ and the fit in proper motion space, $\mathscr{L}_{\rm pm}$. Calculation of $\mathscr{L}_{\rm RV}$ takes account of the effects of binarity on measurement uncertainty. Calculation of $\mathscr{L}_{\rm pm}$ takes account of the correlated uncertainty  between  $V_{\rm RA}$ and $V_{\rm Dec}$,  using the covariance matrix element "pmRApmDEcor" in the {\it Gaia} EDR3 dataset to define a weight matrix for each target  to normalise  the uncertainties in proper motion velocities (see Appendix A of \citep{Gaia2016a}. 

A 1D model for each velocity component, consisting of the sum of two Gaussian distributions - one representing the cluster and one the background, was assumed for the majority of clusters. In these cases the 1D likelihood was determined as a function of five free parameters: the intrinsic velocity and dispersion of the cluster and background populations and the overall fraction of the target population that are cluster members.  A more complex model, comprising three Gaussian distributions, one representing the cluster and two representing the distribution of background stars, was used to model 10 of the clusters and gave a significantly higher maximum log likelihood. In 4 cases this was expected since the clusters Gamma Vel \citep{Jeffries2014a}, NGC\,2547 \citep{Sacco2015a} and the cluster pair NGC\,2451a and NGC\,2451b \citep{Hunsch2003a} were previously known to lie close in velocity space to a second velocity grouping.  We found 6 further clusters where the background population is best represented by a the sum of two Gaussian distributions. These are NGC\,6530, NGC\,2264, NGC\,2232, ASCC\,50, NGC\,2244 and NGC\,6405. 

Membership probabilities of individual targets are calculated as the expectation value of being a cluster member evaluated over the full grid of model velocities, velocity dispersions and fractional membership weighted according to the total likelihood summed over all potential cluster members.

An independent 2D calculation of the probability of cluster membership ($P_{\rm 2D}$) was made using just the $V_{\rm RA}$ and $V_{\rm Dec}$. In most cases this was successful in that a clear maximum was found in likelihood space. The 2D calculation was unsuccessful (i.e. no maximum was found) for 6 of the more distant/sparsely populated clusters (ASCC 50, NGC 6005, Haffner\,10, Berkeley\,21, Berkeley\,31 and Loden\,165). Results of the the 2D and 3D analyses are compared in \S\ref{5.2}.

\begin{figure}
  \centering
	\includegraphics[width = 64mm]{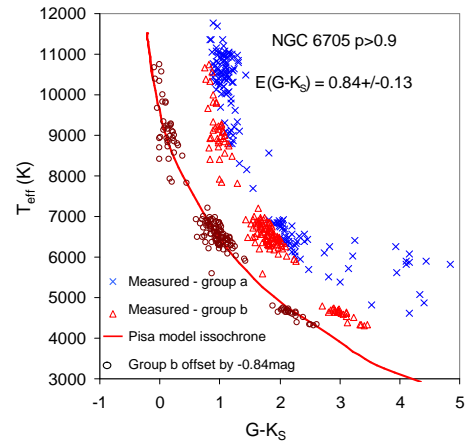}
	\caption{Evaluation  of $E(G-K_S)$ for NGC\,6705. Triangles and crosses show  measured values of $T_{\rm eff}$ as a function of $(G-K_S)$. Group b represents probable single stars and their median offset in colour relative to the Pisa model isochrone is taken to be the reddening -- $E(G-K_s)=0.84$ in this case. Open circles show the effect of subtracting 0.84 from the $G-K_s$ values of group b.}
	\label{plot_NGC6705b}	
\end{figure}

\begin{figure*}
	\begin{minipage}[t]{0.98\textwidth}
	\includegraphics[width = 170mm]{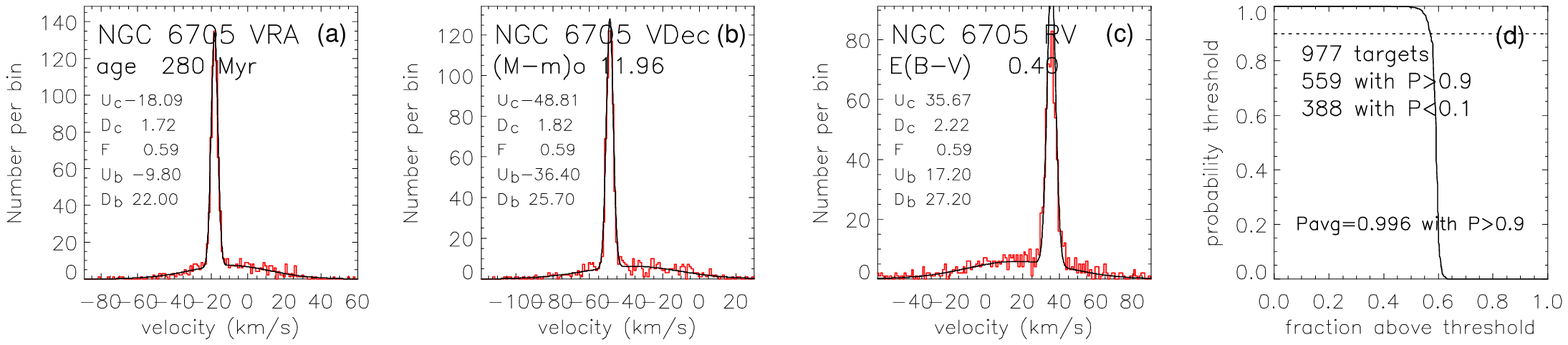}
    \end{minipage}
	\caption{Results of the 3D maximum likelihood analysis of NGC\,6705.  Plots (a), (b) and (c) show histograms of the three velocity components ($V_{\rm RA}$, $V_{\rm Dec}$ and $RV$), together with model probability distributions evaluated at the maximum likelihood values of the fitted parameters, using a median measurement uncertainty. The text shows the best fitting values of cluster velocity ($U_{\rm c}$, in km\,s$^{-1}$), intrinsic dispersion of the cluster velocity ($D_{\rm c}$), the fraction of objects assigned to the cluster population ($F$) and the velocity and dispersion of the (primary) background population ($U_{\rm b}$, $D_{\rm b}$). (d) The fraction of targets with membership probability, $P_{\rm 3D}$ above a threshold level versus the threshold level.}
	\label{plot_results_a}	
\end{figure*}

\begin{figure*}
	\begin{minipage}[t]{0.98\textwidth}
	\includegraphics[width = 170mm]{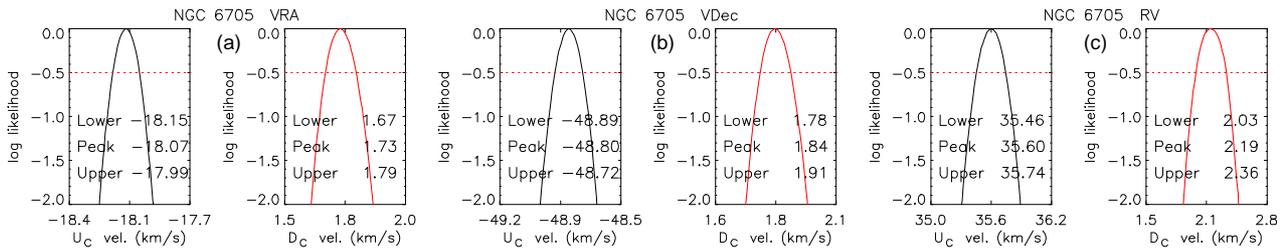}
    \end{minipage}
	\caption{Likelihood of cluster properties for NGC\,6705.   Plots, (a), (b) and (c)  show the variation in log likelihood of the mean cluster velocity, $U_c$ and the intrinsic cluster dispersion $D_c$ for each velocity component. Text on the plots shows the peak and upper and lower $1\sigma$ probability values of the plotted parameter.}
	\label{plot_results_b}	
\end{figure*}

\begin{figure*}
	\begin{minipage}[t]{0.98\textwidth}
	\includegraphics[width = 170mm]{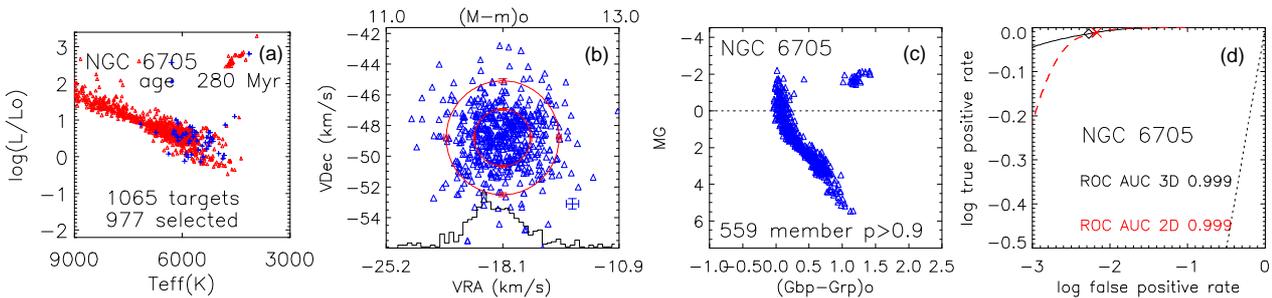}
    \end{minipage}
	\caption{Plots of potential and likely cluster members of NGC\,6705. (a) The HR diagram of valid targets observed in the cluster with stars identified as likely background giants shown as blue crosses. (b) Tangential velocities, $V_{\rm Dec}$ versus $V_{\rm RA}$ for likely ($P_{3D}>0.9$) cluster members where the cross indicates the median measurement uncertainty and the ellipses show one and two times the cluster intrinsic dispersion. The black line on this plot shows a histogram of the relative number of members versus parallax (see scale above the plot).  (c) {\it Gaia}~absolute $G$ magnitude versus $(G_{\rm BP}-G_{\rm RP})_0$ for $P_{\rm 3D}>0.9$ members. (d) The Receiver Operator Characteristic curves for $P_{3D}$~and $P_{\rm 2D}$ membership data (solid black and dashed red curves respectively). 90 per cent probability values are marked on the curve by a black diamond and red cross. The dotted line shows the ROC expected for randomly partitioned data.}
	\label{plot_results_c}	
\end{figure*}

\subsection{Distance modulus and reddening}
\label{3.5}
 Initial likelihood analyses of cluster membership were made using the literature values of distance modulus and reddening in Tables~\ref{table1}~and~\ref{table2}. Taking the initial set of cluster members with $P_{\rm 3D}>0.9$, the analyses were then iterated to determine the values of $(M-m)_0^{\rm c}$ and $E(B-V)^{\rm c}$ shown in the right hand columns of the same tables. $(M-m)_0^{\rm c}$ is estimated from the weighted mean parallax of cluster members taking account of the intrinsic dispersion of the cluster (see \S3.4 of J20). A systematic error equivalent to 0.03\,mas in parallax is shown to account for possible correlated errors in the parallax zero-point \citep{Lindegren2020a}.
 
 A mean $E(B-V)^{\rm c}$ was estimated by comparing the $G-K_s$ of cluster members with Pisa, solar metallicity, model isochrone predictions of $(G-K_s)_0$ at the WG values of $T_{\rm eff}$.  To account for the effects of binaries we used the median value from bluest half of the data, assuming their colours to be representative of single stars. The resulting  $E(G-K_s)$ was converted to $E(B-V)$ using the extinction coefficients from \cite{Yuan2013a} and \cite{Chen2019a}. Fig.~\ref{plot_NGC6705b} shows an example of this process for NGC\,6705. The use of solar metallicity isochrones is considered acceptable for the open clusters which have metallicities in the range $-0.5< [{\rm M/H}]<0.3$. This is not the case for globular clusters which have lower metallicities and in those cases we adopt the literature value of $E(B-V)$ throughout.

$(M-m)_0^{\rm c}$ and $E(B-V)^{\rm c}$ are effectively scaling constants in the likelihood analysis, changing them has {\it no direct effect on the membership probabilities}. Both parameters have a very weak, indirect effect on membership probabilities, since they only affect the estimated mass used to determine the $RV$ offsets of binary stars; but even then, the binary $RV$ offsets scale only as $M^{1/3}$.  

\begin{sidewaystable}
     \centering
\caption{Membership probabilities for GESiDR6 cluster targets. Columns 1 to 5 show the GESiDR6 target names, cluster and filter and co-ordinates. Targets flagged 0 are included in the maximum likelihood analysis (see~\S\ref{3.1}). Column 7 to 9 show the measured $RV$ and scaling constant for empirical uncertainty (see \S~\ref{2.3}), {\it Gaia}~cross-matched ID and $G$ magnitude. Columns 10 and 11 show $T_{\rm eff}^p$ and $K_S^p$ data (which includes estimated values for some targets) used to determine $\log L$ and mass for the  likelihood analysis (see \S~\ref{3.3}). The final columns show the probability that the target is a member of its given cluster using  the full dataset, $P_{\rm 3D}$, or the same probability computed using a dataset filtered to remove targets with suspect {\it Gaia}~data, $P_{\rm QG}$. Targets with a membership probability of -1 were excluded from the membership analysis. A sample of the table is shown here. The full table is available as supplementary material. }
\begin{tabular}{lllllcrrrrrrrr}
\hline
Target & cluster	& setup & RA	 & Dec	 & Flag & $RV$  & {\it Gaia}~ID & $G$     & $T_{\rm eff}^p$  &  $K_S^p$ & $\log L$&\multicolumn{2}{c}{Membership}	\\
cname	& DB name	&	(nm) & (deg) & (deg) &	    &  (km\,s$^{-1}$) & & (mag) & (K)	            & 	 (mag) &         & $P_{\rm 3D}$  & $P_{\rm QG}$	\\\hline
07515457-6047568 &NGC2516   &665& 117.97737& -60.79911& 0&  6.84$\pm$0.55&ID5290899847497012352&15.75&4369&12.87&-0.92& 0.5674& 0.5542\\
07515966-6047220 &NGC2516   &665& 117.99858& -60.78944& 0& 25.90$\pm$0.31&ID5290899881856756224&15.96&4233&12.89&-0.96& 0.9991& 0.9991\\
07520129-6043233 &NGC2516   &665& 118.00538& -60.72314& 0& 14.60$\pm$1.33&ID5290912354441812352&17.42&3767&13.96&-1.46& 0.0000& 0.0000\\
07520389-6050116 &NGC2516   &665& 118.01621& -60.83655& 0& 25.54$\pm$0.31&ID5290898816704856320&15.93&4256&12.87&-0.95& 0.9997& 0.9997\\
07521002-6044245 &NGC2516   &665& 118.04175& -60.74014& 0& 25.63$\pm$0.33&ID5290912113923645952&16.09&4153&13.17&-1.09& 0.9998& 0.9998\\
07523629-6046405 &NGC2516   &665& 118.15121& -60.77792& 0& 31.33$\pm$0.79&ID5290899396521946240&17.44&3883&13.98&-1.45& 0.0000& 0.0000\\
\hline
\multicolumn{12}{l}{Values of $P_{\rm 3D}$ and  $P_{\rm QG}$ are {\it not} reported for Loden\,165 since the maximum likelihood calculation failed for this cluster (see \S ~\ref{5.1}).}

\end{tabular}
\label{table3}
\end{sidewaystable}

\begin{table*}
\caption{Results of the membership analysis for younger open clusters. Columns 2 to 7 show the mean and rms values of central velocity and intrinsic cluster dispersion used to determine the probability of cluster
membership (see \S\ref{results_b}). Column~8 shows the fraction of the targets analysed  that are expected cluster members. Columns 9 and 10 show the number of targets with membership probability $P_{\rm 3D}>0.9$ and $P_{\rm QG}>0.9$ respectively.}

\begin{tabular}{lrrrrrrrrr}
\hline
Cluster	&	\multicolumn{3}{c}{Cluster central velocity (km\,s$^{-1}$)}	&	\multicolumn{3}{c}{Intrinsic dispersion of cluster (km\,s$^{-1}$)}	&	Fraction	&
\multicolumn{2}{c}{Members $P$$>$0.9}\\
	&	$U_{\rm RA}$	&	$U_{\rm Dec}$ & $U_{\rm RV}$& $D_{\rm RA}$	&	$D_{\rm Dec}$ &	$D_{\rm RV}$	&	members	&	$P_{\rm 3D}$		&	$P_{\rm QG}$	\\\hline
NGC 6530            &   8.48$\pm$   0.17& -12.82$\pm$   0.12&   0.15$\pm$   0.16&   3.32$\pm$   0.15&   2.23$\pm$   0.12&   2.48$\pm$   0.16&   0.34$\pm$   0.01& 452& 420\\
Trumpler 14         & -82.15$\pm$   0.26&  29.71$\pm$   0.26&  -7.68$\pm$   0.55&   4.38$\pm$   0.24&   4.59$\pm$   0.21&   8.08$\pm$   0.70&   0.53$\pm$   0.02& 320& 302\\
Chamaeleon I        & -20.24$\pm$   0.10&   0.35$\pm$   0.13&  15.61$\pm$   0.14&   0.91$\pm$   0.09&   1.21$\pm$   0.09&   0.90$\pm$   0.13&   0.55$\pm$   0.04&  90&  62\\
Rho Ophiuchus       &  -4.22$\pm$   0.12& -17.20$\pm$   0.14&  -6.41$\pm$   0.30&   0.76$\pm$   0.09&   0.92$\pm$   0.10&   1.36$\pm$   0.33&   0.63$\pm$   0.06&  45&  39\\
NGC 2264            &  -6.58$\pm$   0.08& -12.64$\pm$   0.05&  20.31$\pm$   0.13&   1.64$\pm$   0.06&   0.94$\pm$   0.04&   2.39$\pm$   0.12&   0.38$\pm$   0.01& 502& 424\\
NGC 2244            & -13.05$\pm$   0.14&   1.05$\pm$   0.21&  30.64$\pm$   0.29&   1.25$\pm$   0.14&   2.10$\pm$   0.20&   2.47$\pm$   0.28&   0.35$\pm$   0.03& 120& 110\\
Lambda Ori          &   1.98$\pm$   0.08&  -3.40$\pm$   0.11&  26.80$\pm$   0.13&   1.09$\pm$   0.07&   1.47$\pm$   0.09&   1.27$\pm$   0.12&   0.60$\pm$   0.03& 200& 164\\
Lambda Ori B35      &   4.16$\pm$   0.16&  -4.60$\pm$   0.07&  27.86$\pm$   0.19&   1.11$\pm$   0.12&   0.48$\pm$   0.06&   0.79$\pm$   0.25&   0.42$\pm$   0.05&  49&  38\\
25 Ori              &   2.24$\pm$   0.05&  -0.19$\pm$   0.06&  20.65$\pm$   0.09&   0.58$\pm$   0.04&   0.75$\pm$   0.05&   0.64$\pm$   0.09&   0.68$\pm$   0.03& 170& 138\\
ASCC 50             & -25.04$\pm$   0.14&  17.84$\pm$   0.12&  21.70$\pm$   0.11&   1.77$\pm$   0.16&   1.52$\pm$   0.10&   1.01$\pm$   0.13&   0.41$\pm$   0.02& 194& 173\\
Collinder 197       & -26.47$\pm$   0.14&  18.37$\pm$   0.17&  20.82$\pm$   0.13&   1.37$\pm$   0.12&   1.63$\pm$   0.14&   0.73$\pm$   0.15&   0.37$\pm$   0.03& 110&  95\\
Gamma Velorum       & -10.65$\pm$   0.05&  15.57$\pm$   0.09&  18.24$\pm$   0.15&   0.77$\pm$   0.05&   1.31$\pm$   0.07&   1.64$\pm$   0.15&   0.45$\pm$   0.02& 209& 181\\
IC 4665             &  -1.52$\pm$   0.08& -13.92$\pm$   0.07& -13.66$\pm$   0.14&   0.41$\pm$   0.09&   0.32$\pm$   0.10&   0.22$\pm$   0.17&   0.11$\pm$   0.02&  33&  29\\
NGC 2232            &  -7.17$\pm$   0.04&  -2.63$\pm$   0.05&  25.38$\pm$   0.09&   0.35$\pm$   0.03&   0.38$\pm$   0.04&   0.18$\pm$   0.15&   0.13$\pm$   0.01&  82&  78\\
NGC 2547            & -15.67$\pm$   0.04&   7.88$\pm$   0.04&  12.77$\pm$   0.08&   0.50$\pm$   0.04&   0.49$\pm$   0.04&   0.54$\pm$   0.07&   0.62$\pm$   0.03& 165& 138\\
IC 2602             & -12.68$\pm$   0.12&   7.69$\pm$   0.12&  17.61$\pm$   0.12&   0.82$\pm$   0.11&   0.82$\pm$   0.11&   0.37$\pm$   0.22&   0.53$\pm$   0.05&  61&  45\\
NGC 2451b           & -16.71$\pm$   0.12&   8.24$\pm$   0.07&  14.93$\pm$   0.12&   0.95$\pm$   0.09&   0.54$\pm$   0.06&   0.44$\pm$   0.19&   0.16$\pm$   0.02&  63&  51\\
NGC 6649            &   0.22$\pm$   0.28&  -1.42$\pm$   0.23& -14.38$\pm$   1.11&   1.96$\pm$   0.24&   1.60$\pm$   0.22&   6.46$\pm$   1.01&   0.62$\pm$   0.05&  65&  64\\
IC 2391             & -17.70$\pm$   0.09&  16.68$\pm$   0.09&  14.80$\pm$   0.17&   0.58$\pm$   0.06&   0.58$\pm$   0.09&   0.39$\pm$   0.25&   0.61$\pm$   0.05&  48&  38\\
NGC 2451a           & -19.20$\pm$   0.17&  13.88$\pm$   0.08&  23.37$\pm$   0.09&   1.07$\pm$   0.12&   0.50$\pm$   0.07&   0.16$\pm$   0.15&   0.13$\pm$   0.02&  40&  33\\
NGC 6405            &  -2.91$\pm$   0.09& -12.91$\pm$   0.11&  -8.74$\pm$   0.14&   0.71$\pm$   0.08&   0.87$\pm$   0.09&   0.71$\pm$   0.15&   0.19$\pm$   0.02&  60&  54\\
NGC 6067            & -19.37$\pm$   0.10& -25.71$\pm$   0.11& -37.96$\pm$   0.23&   1.27$\pm$   0.08&   1.39$\pm$   0.09&   1.84$\pm$   0.24&   0.39$\pm$   0.02& 181& 179\\
NGC 2516            &  -9.05$\pm$   0.04&  21.92$\pm$   0.04&  23.92$\pm$   0.05&   0.87$\pm$   0.03&   0.89$\pm$   0.03&   0.75$\pm$   0.06&   0.75$\pm$   0.02& 476& 420\\
Blanco 1            &  21.04$\pm$   0.05&   2.96$\pm$   0.04&   5.98$\pm$   0.07&   0.48$\pm$   0.04&   0.48$\pm$   0.05&   0.23$\pm$   0.10&   0.43$\pm$   0.03& 128& 123\\
NGC 6709            &   7.61$\pm$   0.08& -18.27$\pm$   0.12&  -9.24$\pm$   0.13&   0.61$\pm$   0.07&   0.94$\pm$   0.09&   0.54$\pm$   0.15&   0.15$\pm$   0.01&  71&  64\\
NGC 6259            & -11.53$\pm$   0.16& -31.66$\pm$   0.13& -32.74$\pm$   0.41&   1.74$\pm$   0.13&   1.45$\pm$   0.11&   3.26$\pm$   0.49&   0.38$\pm$   0.03& 136& 125\\
NGC 6705            & -18.07$\pm$   0.08& -48.80$\pm$   0.09&  35.60$\pm$   0.15&   1.74$\pm$   0.06&   1.85$\pm$   0.06&   2.20$\pm$   0.15&   0.59$\pm$   0.02& 559& 526\\
Berkeley 30         &  -4.26$\pm$   0.30&  -7.26$\pm$   0.37&  47.71$\pm$   0.47&   0.43$\pm$   0.38&   1.13$\pm$   0.43&   2.07$\pm$   0.64&   0.34$\pm$   0.04&  60&  61\\
NGC 3532            & -23.61$\pm$   0.05&  11.90$\pm$   0.05&   5.46$\pm$   0.06&   1.04$\pm$   0.04&   1.02$\pm$   0.04&   0.88$\pm$   0.05&   0.73$\pm$   0.02& 490& 419\\
NGC 6281            &  -4.60$\pm$   0.31& -10.43$\pm$   0.16&  -4.41$\pm$   0.16&   1.48$\pm$   0.27&   0.73$\pm$   0.16&   0.57$\pm$   0.16&   0.44$\pm$   0.06&  26&  25\\
NGC 4815            & -99.15$\pm$   0.34& -15.76$\pm$   0.28& -27.35$\pm$   0.75&   2.20$\pm$   0.29&   1.61$\pm$   0.27&   4.55$\pm$   0.66&   0.50$\pm$   0.05&  55&  54\\
NGC 6633            &   2.13$\pm$   0.22&  -3.54$\pm$   0.19& -28.16$\pm$   0.23&   1.05$\pm$   0.18&   0.89$\pm$   0.15&   0.71$\pm$   0.28&   0.21$\pm$   0.04&  21&  14\\
\hline
 \end{tabular}
 \label{table4}
\end{table*}

\begin{table*}
\caption{Results of the membership analysis for older open clusters and for globular clusters. Columns 2 to 7 show the mean and rms values of central velocity and intrinsic cluster dispersion used to determine the probability of cluster membership.(see \S\ref{results_b}). Column~8 shows the fraction of the targets analysed  that are expected cluster members. Columns 9 and 10 show the number of targets with membership probability $P_{\rm 3D}>0.9$ and $P_{\rm QG}>0.9$ respectively.}

\begin{tabular}{lrrrrrrrrr}
\hline
Cluster	&	\multicolumn{3}{c}{Cluster central velocity (km\,s$^{-1}$)}	&	\multicolumn{3}{c}{Intrinsic dispersion of cluster (km\,s$^{-1}$)}	&	Fraction	&
\multicolumn{2}{c}{Members $P$$>$0.9}\\
	&	$U_{\rm RA}$	&	$U_{\rm Dec}$ & $U_{\rm RV}$& $D_{\rm RA}$	&	$D_{\rm Dec}$ &	$D_{\rm RV}$	&	members	&	$P_{\rm 3D}$		&	$P_{\rm QG}$	\\\hline
Pismis 18           & -76.36$\pm$   0.18& -30.46$\pm$   0.24& -28.13$\pm$   0.47&   0.72$\pm$   0.15&   0.97$\pm$   0.21&   1.23$\pm$   0.63&   0.31$\pm$   0.05&  23&  23\\
Trumpler 23         & -55.29$\pm$   0.20& -62.04$\pm$   0.23& -61.48$\pm$   0.31&   1.04$\pm$   0.20&   1.30$\pm$   0.18&   1.26$\pm$   0.35&   0.47$\pm$   0.05&  39&  37\\
NGC 2355            & -34.21$\pm$   0.10&  -9.45$\pm$   0.10&  36.37$\pm$   0.09&   1.11$\pm$   0.08&   1.08$\pm$   0.07&   0.59$\pm$   0.09&   0.69$\pm$   0.03& 141& 139\\
NGC 6802            & -40.13$\pm$   0.19& -91.74$\pm$   0.22&  13.47$\pm$   0.41&   1.04$\pm$   0.17&   1.27$\pm$   0.20&   1.92$\pm$   0.45&   0.56$\pm$   0.05&  55&  55\\
Ruprecht 134        & -19.39$\pm$   0.15& -29.41$\pm$   0.19& -40.53$\pm$   0.20&   1.31$\pm$   0.21&   1.71$\pm$   0.17&   0.97$\pm$   0.28&   0.18$\pm$   0.02& 102&  93\\
Berkeley 81         & -21.76$\pm$   0.28& -35.99$\pm$   0.26&  47.92$\pm$   0.22&   0.70$\pm$   0.40&   1.20$\pm$   0.34&   0.69$\pm$   0.44&   0.34$\pm$   0.04&  56&  56\\
NGC 6005            & -54.88$\pm$   0.21& -51.34$\pm$   0.28& -24.42$\pm$   0.22&   1.12$\pm$   0.22&   1.59$\pm$   0.32&   1.04$\pm$   0.25&   0.19$\pm$   0.02&  55&  51\\
Pismis 15           & -60.65$\pm$   0.22&  38.68$\pm$   0.19&  35.13$\pm$   0.16&   1.03$\pm$   0.22&   0.86$\pm$   0.19&   0.46$\pm$   0.22&   0.19$\pm$   0.03&  36&  36\\
Trumpler 20         &-126.92$\pm$   0.13&   2.97$\pm$   0.12& -39.81$\pm$   0.14&   1.43$\pm$   0.13&   1.26$\pm$   0.13&   1.29$\pm$   0.15&   0.38$\pm$   0.02& 172& 167\\
Berkeley 44         &  -0.54$\pm$   0.24& -42.91$\pm$   0.18&  -8.54$\pm$   0.23&   1.19$\pm$   0.24&   0.75$\pm$   0.20&   0.84$\pm$   0.26&   0.51$\pm$   0.05&  40&  41\\
NGC 2141            &  -1.91$\pm$   0.10& -16.85$\pm$   0.09&  26.39$\pm$   0.06&   1.38$\pm$   0.10&   1.48$\pm$   0.09&   0.93$\pm$   0.07&   0.76$\pm$   0.01& 598& 595\\
Czernik 24          &   5.03$\pm$   0.26& -55.63$\pm$   0.20&  22.22$\pm$   0.35&   0.98$\pm$   0.31&   1.02$\pm$   0.22&   2.15$\pm$   0.39&   0.27$\pm$   0.03&  77&  77\\
Haffner 10          & -22.73$\pm$   0.32&  28.42$\pm$   0.33&  88.00$\pm$   0.09&   4.45$\pm$   0.33&   4.65$\pm$   0.40&   0.82$\pm$   0.10&   0.62$\pm$   0.03& 247& 239\\
NGC 2158            &  -3.98$\pm$   0.14& -41.18$\pm$   0.12&  28.68$\pm$   0.15&   1.93$\pm$   0.14&   1.75$\pm$   0.11&   2.05$\pm$   0.14&   0.67$\pm$   0.02& 361& 346\\
NGC 2420            & -14.81$\pm$   0.05& -24.61$\pm$   0.06&  74.62$\pm$   0.05&   0.75$\pm$   0.05&   0.92$\pm$   0.05&   0.59$\pm$   0.05&   0.75$\pm$   0.02& 388& 384\\
Berkeley 21         &  17.54$\pm$   0.32& -35.10$\pm$   0.32&   1.57$\pm$   0.38&   0.81$\pm$   0.54&   1.84$\pm$   0.48&   2.67$\pm$   0.77&   0.34$\pm$   0.02& 178& 178\\
Berkeley 73         &   7.13$\pm$   0.53&  43.14$\pm$   0.54&  97.33$\pm$   0.41&   1.16$\pm$   0.74&   0.92$\pm$   0.79&   1.57$\pm$   0.51&   0.66$\pm$   0.06&  41&  41\\
Berkeley 22         &  19.92$\pm$   0.42& -13.04$\pm$   0.32&  94.60$\pm$   0.19&   1.87$\pm$   0.54&   1.00$\pm$   0.50&   1.12$\pm$   0.36&   0.52$\pm$   0.03& 170& 170\\
Czernik 30          & -23.80$\pm$   0.35&  -2.91$\pm$   0.38&  81.99$\pm$   0.15&   1.20$\pm$   0.41&   1.69$\pm$   0.44&   0.58$\pm$   0.25&   0.38$\pm$   0.04&  69&  69\\
Berkeley 31         &   5.29$\pm$   0.62& -34.50$\pm$   0.52&  56.97$\pm$   0.16&   3.57$\pm$   1.39&   3.68$\pm$   0.69&   1.16$\pm$   0.22&   0.34$\pm$   0.02& 136& 136\\
Berkeley 75         &  -9.74$\pm$   0.31&  47.97$\pm$   0.43& 124.98$\pm$   0.13&   0.36$\pm$   0.34&   0.54$\pm$   0.49&   0.21$\pm$   0.20&   0.71$\pm$   0.06&  43&  43\\
Loden 165$^{**}$    & -76.13$\pm$   0.70&  34.09$\pm$   0.61&  -0.84$\pm$   2.06&   3.00$\pm$   0.00&   3.00$\pm$   0.00&  16.68$\pm$   1.59&   0.27$\pm$   0.03&  77&  70\\
NGC 6253            & -36.70$\pm$   0.09& -42.71$\pm$   0.10& -28.93$\pm$   0.11&   1.05$\pm$   0.07&   1.17$\pm$   0.09&   0.90$\pm$   0.11&   0.72$\pm$   0.03& 162& 154\\
Messier 67          & -44.77$\pm$   0.08& -11.78$\pm$   0.08&  34.00$\pm$   0.11&   0.88$\pm$   0.06&   0.85$\pm$   0.07&   0.84$\pm$   0.12&   0.92$\pm$   0.02& 118& 106\\
NGC 2425            & -58.42$\pm$   0.11&  33.15$\pm$   0.12& 103.68$\pm$   0.08&   1.08$\pm$   0.10&   1.15$\pm$   0.12&   0.58$\pm$   0.09&   0.31$\pm$   0.02& 141& 138\\
NGC 2243            & -27.19$\pm$   0.08& 118.54$\pm$   0.09&  59.79$\pm$   0.05&   1.30$\pm$   0.08&   1.23$\pm$   0.08&   0.59$\pm$   0.07&   0.88$\pm$   0.01& 539& 538\\
Berkeley 36         & -39.65$\pm$   0.16&  21.98$\pm$   0.14&  62.84$\pm$   0.12&   1.20$\pm$   0.17&   0.89$\pm$   0.16&   1.17$\pm$   0.14&   0.34$\pm$   0.02& 212& 212\\
Trumpler 5          &  -9.71$\pm$   0.08&   4.31$\pm$   0.08&  51.60$\pm$   0.09&   1.72$\pm$   0.08&   1.61$\pm$   0.07&   1.86$\pm$   0.09&   0.76$\pm$   0.01& 814& 808\\
Berkeley 32         &  -5.93$\pm$   0.14& -26.58$\pm$   0.12& 105.89$\pm$   0.07&   1.42$\pm$   0.13&   1.18$\pm$   0.10&   0.60$\pm$   0.09&   0.70$\pm$   0.03& 240& 238\\
Berkeley 39         & -37.29$\pm$   0.09& -35.18$\pm$   0.08&  58.85$\pm$   0.05&   1.29$\pm$   0.10&   1.22$\pm$   0.09&   0.72$\pm$   0.06&   0.62$\pm$   0.02& 499& 507\\
ESO 92-05           &-231.28$\pm$   1.00& 190.66$\pm$   0.98&  61.47$\pm$   0.19&   2.07$\pm$   1.56&   2.70$\pm$   1.54&   0.42$\pm$   0.41&   0.80$\pm$   0.04&  89&  89\\
NGC 1261            & 208.60$\pm$   1.15&-268.51$\pm$   1.69&  72.38$\pm$   0.44&   7.85$\pm$   1.05&  11.62$\pm$   1.68&   3.01$\pm$   0.35&   0.96$\pm$   0.02&  63&  59\\
NGC 362             & 375.88$\pm$   0.76&-142.68$\pm$   0.66& 222.91$\pm$   0.47&   7.72$\pm$   0.64&   6.33$\pm$   0.56&   4.14$\pm$   0.41&   0.88$\pm$   0.04& 104&  91\\
NGC 2808            &  56.73$\pm$   0.75&  15.98$\pm$   0.94& 103.66$\pm$   0.76&   7.84$\pm$   0.61&   9.42$\pm$   0.88&   7.22$\pm$   0.58&   0.73$\pm$   0.04& 112& 122\\
NGC 1904            & 194.63$\pm$   0.90&-127.13$\pm$   0.89& 205.85$\pm$   0.40&   6.64$\pm$   0.68&   6.37$\pm$   0.72&   2.58$\pm$   0.34&   0.94$\pm$   0.03&  59&  58\\
NGC 6752            & -68.96$\pm$   0.37& -86.93$\pm$   0.36& -26.13$\pm$   0.32&   5.98$\pm$   0.28&   5.92$\pm$   0.27&   4.49$\pm$   0.24&   0.96$\pm$   0.01& 284& 238\\
NGC 5927            &-287.38$\pm$   1.05&-180.65$\pm$   0.96&-103.29$\pm$   0.53&   8.74$\pm$   0.89&   7.82$\pm$   0.75&   4.13$\pm$   0.40&   0.89$\pm$   0.03&  80&  71\\
NGC 104             & 120.10$\pm$   0.61& -59.03$\pm$   0.62& -18.07$\pm$   0.51&  10.36$\pm$   0.47&  10.46$\pm$   0.50&   7.49$\pm$   0.38&   0.98$\pm$   0.01& 286& 255\\
\hline
\multicolumn{9}{l}{** Components of the intrinsic cluster dispersion are artificially constrained in maximum likelihood fit for Loden\,165 (see \S~\ref{5.1}).}
 \end{tabular}
 \label{table5}
\end{table*}
\section{Results}
\label{results}

\subsection{Measured data and model fits}
\label{results_a}
As an example of the results, Fig.~\ref{plot_results_a} shows the outcomes of the maximum likelihood analysis for the relatively distant ($d_{\rm c}=2.5$\,kpc) open cluster NGC\,6705. 
Figures~\ref{plot_results_a}a--c show histograms of each velocity component in 1\,km\,s$^{-1}$ bins, with a black curve indicating a "best fit"  probability distribution based on the maximum likelihood values of cluster central velocity and intrinsic dispersion and a median value of observational uncertainty. The observational uncertainty in $RV$ reflects the combined effects of the non-Gaussian distribution of measurement uncertainty in $RV$ (see \S\ref{2.3}) and the $RV$ offsets calculated for a  proportion of randomly oriented binary systems (see \S\ref{3.4}).

Figure~\ref{plot_results_a}d shows the fraction of targets with $P_{\rm 3D}$ above a threshold level versus that threshold level. Text on the plot indicates the proportion of targets that are considered likely cluster members with $P_{\rm 3D}$ above a threshold value of 0.9. The sharpness of the ``step" indicates how well the analysis discriminates between members and background stars. For example in NGC\,6705, 57 per cent of targets are identified as likely ($P_{\rm 3D}>0.9$) members, 39 per cent as probable background stars ($P_{\rm 3D}<0.1$), with only 4 per cent having intermediate values and effectively unresolved membership. 

Similar figures for each cluster are shown in Appendix~\ref{appb} (available on-line only) in order of the cluster age shown in Tables~\ref{table1}~and~\ref{table2}.

\subsection{Cluster velocity and dispersion}
\label{results_b}
Figure~\ref{plot_results_b} shows log likelihood as a function of the three central velocities and intrinsic velocity dispersions for NGC\,6705. The dotted lines mark the 1$\sigma$ level (a log likelihood increment of -0.5 relative to the maximum value). Text on the plots show the maximum likelihood value of each parameter and values at the upper and lower 1$\sigma$ levels. The range explored for each parameter is set to be greater than $\pm 5 \sigma$ from the maximum likelihood value of that parameter, but with a minimum value of zero for the velocity dispersion. Similar figures for each cluster are shown in Appendix~C in age order (available on-line only).
 
NGC\,6705 is a well-defined cluster with almost 1000 targets, over 50 per cent of which are probable cluster members. These produce clearly identified likelihood peaks for each of the cluster velocity parameters. For more distant clusters and those with small numbers of likely cluster members these curves can be broader and in a few cases become irregular. ESO\,92-05, shows a secondary peak in the likelihood distribution of $RV$ dispersion, although  the dominant peak is readily identified. For Loden\,165, the 3D analysis fails to find any peak at all in the likelihood distribution of $D_{\rm c}$, invalidating calculations of membership probability for this cluster (see \S\ref{5.1}).

The likelihood curves were used to determine the mean and rms values of cluster central velocity and intrinsic dispersion shown in Tables~\ref{table4}~and~\ref{table5}. These values are the result of fitting the measured velocities in a fixed co-ordinate system to determine the probability of cluster membership. They do not necessarily characterise the true shape of the cluster in velocity space. This might require a more general model of the cluster that allows for free rotation of the cluster axes, bulk rotation of the cluster and, depending on the size of the cluster, modelling of ``perspective expansion" \citep{vanLeeuwen2009a, Kuhn2019a},  none of which are crucial to the cluster membership calculations. In addition, the assumed unresolved binary fraction will affect the inferred intrinsic velocity dispersion in $RV$, but probably has a negligible influence on the tangential velocity dispersions (see \S 5.4 of J20).

\subsection{Target membership probabilities}
\label{result_table3}
Membership probabilities, $P_{\rm 3D}$, for individual targets in each cluster are reported in Table~\ref{table3}, which contains 41926 entries for 39441 unique targets of which 12110 are $P_{\rm 3D} > 0.9$ cluster members. Targets with a membership probability of -1 were not included in the maximum likelihood analysis. Membership probabilities, $P_{\rm QG}$, are also given for the slightly smaller sample of potential cluster members, where targets with potentially unreliable {\it Gaia} data were excluded from the analysis (see \S~\ref{3.2}). 

Targets identified as cluster members were used to calculate the distance moduli and reddening values -- $(M-m)^{\rm c}_o$, and \mbox{$E(B-V)^c$} -- shown in Tables~\ref{table1}~and~\ref{table2} (see \S\ref{3.5}). The numbers of targets in each cluster with $P_{\rm 3D} > 0.9$  and $P_{\rm QG} > 0.9$ are also reported in Tables~\ref{table4}~and~\ref{table5}.

Membership probabilities of supplementary targets observed with alternate GIRAFFE setups are discussed in Appendix A. These targets are not included in the main cluster kinematic analyses and their membership probabilities are calculated as the expectation value of their being members of the cluster as defined only by the Giraffe+HR15N and UVES targets. From a list of 3770 additional targets in Table~\ref{tableA2}, 1875 are identified as $P_{\rm 3D} > 0.9$ cluster members.

\subsection{Properties of potential and likely cluster members}
\label{results_c}
Figure~\ref{plot_results_c} shows four plots of targets and likely cluster members in NGC~6705. Similar figures for all clusters, in age order, are provided in Appendix~\ref{appc} (available on-line only).
\begin{enumerate}
	\item Figure~~\ref{plot_results_c}a shows a Hertzsprung-Russell (HR) diagram for all the targets used in the analysis (those flagged $\ge 0$ in Table~\ref{table3}). Targets identified as background giants or distant background stars (flagged > 0) are identified separately as blue crosses. 
	
	\item Figure~~\ref{plot_results_c}b shows the tangential velocity components $V_{\rm RA}$ versus $V_{\rm Dec}$ for ($P_{\rm 3D} > 0.9$) cluster members. Ellipses represent one and two times the intrinsic cluster dispersion  with error bars indicating their uncertainties. This plot also shows a histogram of the number of members versus distance modulus over a $\pm$1 magnitude range relative to the cluster centre (see scale above the plot). The histogram shows a relatively broad peak centered on 11.9 mag, indicating that at this distance, parallax measurements are of limited value in determining the probability of cluster membership.  
	
	\item Figure~~\ref{plot_results_c}c shows an absolute $G$ magnitude versus $(G_{\rm Bp}-G_{\rm Rp})_0$ colour magnitude diagram for cluster members using distance $(M-m)^{\rm c} _0$ and $E(B-V)^{\rm c}$ from Tables~\ref{table1}~and~\ref{table2}. 
	
	\item Figure~~\ref{plot_results_c}d shows the ``Receiver Operator Characteristic" (ROC) curve for NGC\,6705. This shows the true positive rate (TPR) versus the false positive rate (FPR) for increasing threshold probabilities of cluster membership. A random discriminator would have equal fractions of true and false positives for any probability threshold (shown as a dotted line in Fig.~\ref{results_c}), whereas a perfect test would show a true positive rate of 1 above arbitrarily low probability thresholds. This provides a graphical illustration of the degree to which our calculated membership probabilities are an effective discriminator of cluster stars from field stars. ROC curves are shown for both $P_{\rm 3D}$ and $P_{\rm 2D}$ (see \S \ref{3.4}), although in the case of NGC~6705 the area under the ROC curve was almost identical. The results for $P_{\rm 3D}$ are in general better and provide more discrimination than $P_{\rm 2D}$ for more distant or sparsely populated clusters (see \S\ref{5.2}). 
	
	%Text on the plots shows the area under the ROC curve (AUC). For NGC\,6807 the AUCs show both $P_{\rm 3D}$ and $P_{\rm 2D}$ to be excellent discriminators. Results for more distant or sparsely populated clusters are are more variable (see \S\ref{5.2})
\end{enumerate}

\section{Discussion}
\label{5}
\begin{figure}
  \centering
	\includegraphics[width = 85mm]{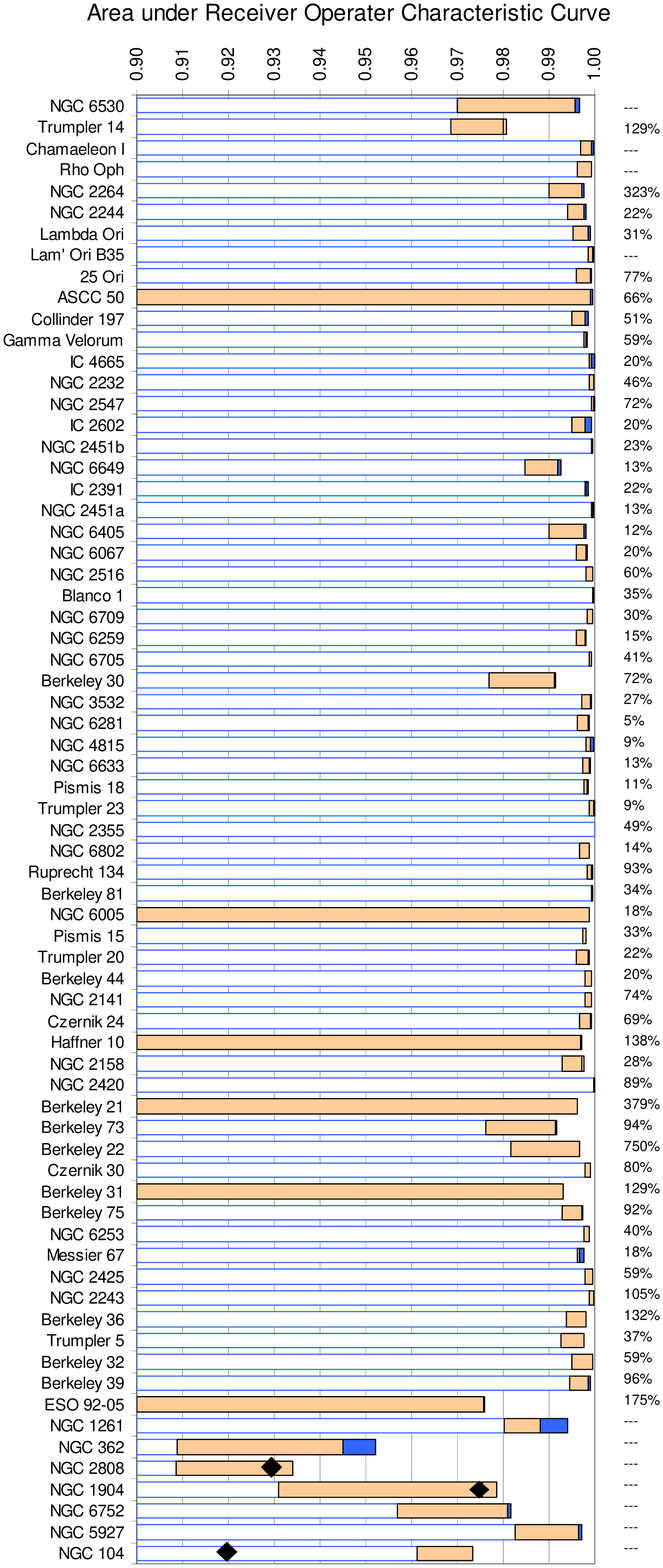}
	\caption{Stacked bar chart showing the area under the Receiver Operator Characteristic curves (ROC AUC) shown in Appendix C for {\it Gaia}-ESO clusters. The right edge of the unfilled bar indicates the ROC AUC for $P_{\rm 2D}$ data, the right edge of the light brown bar shows the ROC "area under the curve" (AUC) statistic for $P_{\rm 3D}$. The right edge of the blue bar gives the ROC AUC for based on $P_{\rm QG}$, as defined in \S \ref{5.2}, Where the AUC is lower than derived from $P_{\rm 3D}$ its value is marked with a diamond. Percentages on the right hand side of the plot show the ratio of the number {\it Gaia}-ESO cluster members compared to the number identified purely from {\it Gaia} DR2 data by \protect\cite{CantatGaudin2018a} (see \S\ref{5.4}).}
	\label{plot_ROC}
\end{figure}

The main objective of this paper is to determine, on an individual basis, the probability that targets observed towards open and globular clusters as part of GES are members of that cluster. It is not  intended, or possible, to provide a complete list of cluster members because the GES target sample was not spatially complete.

For most clusters the GES selected targets come from contiguous central areas of the clusters and will not have probed regions that are more distant from the centre. In addition, during the process of allocating targets to fibres in a discrete set of FLAMES fields, it is inevitable that there will be potential targets that could not be observed. On the other hand, the target selection was usually unbiased with respect to position in the CMD (between bright and faint limits of approximately $11<V<19$) by virtue of selecting targets from a CMD region that comfortably encompasses the likely locus of cluster members including equal-mass binary systems (Bragaglia et al. in preparation). The exceptions were (i) in almost all clusters, the relatively few UVES fibers in each field were preferentially allocated to (brighter) objects that had previous evidence of cluster membership; (ii) in the case of the globular clusters, both UVES and GIRAFFE fibers were preferentially allocated to targets with previous evidence for cluster membership.  

Table~\ref{table3} provides the main catalogue of members with measured, GIRAFFE+HR15N and/or UVES spectra and Table~\ref{tableA2} is a similar catalogue for supplementary target members observed with other GIRAFFE setups. These can be used to measure stellar properties and chemical abundances for a quantifiably reliable sample of cluster members. The accuracy of the discrimination achieved by the kinematic analyses was discussed in J20; tests made using independent membership criteria in young clusters showed that the derived membership probabilities give a reliable indication of both cluster membership {\it and} the residual contamination in any high probability sample of cluster members.

\subsection{Selecting potential cluster members}
\label{5.1}
Whilst the method used to assess membership probability was uniformly applied, the quantity and resolution of target data varies widely between clusters. This is best seen from the histograms in Appendix~\ref{appb}  indicating the number of potential cluster members, the level of contamination from likely background stars and the scatter in velocities due to measurement uncertainties. These determine whether the maximum likelihood analysis is able to resolve a group of member stars in velocity space that are distinct from the velocity distribution of the general background population. The criteria for success are that the plots in Appendix~C of the three components of cluster central velocity $U_i$, and intrinsic dispersion $D_i$, show clearly defined peaks in log likelihood. If peaks are not found in one or more of $U_i$ and $D_i$ then the method breaks down and will identify increased numbers of false positives with incorrect membership probabilities.

The discrimination achieved by the maximum likelihood analysis depends in part on the ratio of true cluster members to the numbers in the background population. For this reason we filtered the initial target list to remove stars that are considered extremely unlikely to be cluster members by discarding  background giants and/or distant background stars to produce the list of potential cluster members ( see \S\ref{3.1}). This yielded satisfactory solutions for 67 of the 70 clusters -- i.e they show clear peaks in log likelihood for $U_i$ and $D_i$. However 3 clusters initially failed to show a peak in one or more $D_i$ components indicating further filtering was required.

For two clusters (ASCC\,50 and NGC\,6281) the criterion used to filter out distant background stars was tightened to reject stars with parallax smaller by at least 4$\sigma$ than a value corresponding to \mbox{$(M-m)_0^{\rm c}+1$}\ mag (compared to  the standard criterion of \mbox{$(M-m)_0^{\rm c}+2$}\ mag). This change produced the required peak in log likelihood of $D_i$ for both clusters and relatively precise estimates of distance modulus ($9.92\pm 0.06$ and $8.62\pm 0.03$ respectively), such that the margin of 1\,mag used was still significantly above the required $4\sigma$ level. 

Loden\,165 however, is a poorly defined cluster with no identifiable peak in the $RV$ histogram (see Appendix~\ref{appb}). A solution was computed by capping the intrinsic dispersions, $D_{\rm RA}$ and $D_{\rm Dec}$ at a maximum value of 3.0\,km\,s$^{-1}$. This produced a {\it false solution}, the results of which are shown graphically in Appendix~\ref{appc}. From the CMD it can be seen that  high probability ($P_{\rm 3D}>0.9$) cluster members do not fall around a common isochrone suggesting the sample includes an excess numbers of false positives. No satisfactory solution was found for Loden\,165, hence no membership probabilities are  reported in Table~\ref{table3}.

\subsection{The effects of selection criteria on discrimination}
\label{5.2}
Figure~\ref{plot_ROC} summarises the area under the curve (AUC, also known as the c-statistic) of the ROC for each cluster in age order (excluding Loden\,165). The AUC is equivalent to the probability that our classification will give a randomly chosen member a higher probability of membership than a randomly chosen non-member and can be used as a figure of merit for judging how successful a binary classifier is at discriminating between the two possibilities \citep[e.g.,][]{Hosmer2000}. The discrimination achieved between members and background targets depends on the data available for a particular cluster and the method of analysis. In this paper we report results of 3D analyses, $P_{\rm 3D}$ using both proper motion and $RV$ data, which always provide better discrimination than a 2D analysis of proper motion data alone, $P_{\rm 2D}$. 

Figure~\ref{plot_ROC} compares the AUC for both 2D and 3D analyses.  For about half of the clusters -- those that are nearby and hence have cluster members that form a distinct and narrow peak in proper motion space -- the 2D membership AUC is already very high $(>0.99)$ and the addition of the third dimension of $RV$ makes only a small improvement to the AUC. However for many of the remaining clusters, particularly those that are more distant or for which the proper motion coincides with the peak of the proper motion of field stars, the use of $RV$ data improves the AUC, and hence the level of discrimination, significantly. In fact, for five open clusters (ASCC 50, NGC\,6005, Haffner 10, Berkeley 21 and Berkeley 31), the 2D analysis failed to converge at all. i.e. A 3D analysis was mandatory to kinematically identify cluster members.

\begin{figure}
  \centering
	\includegraphics[width = 80mm]{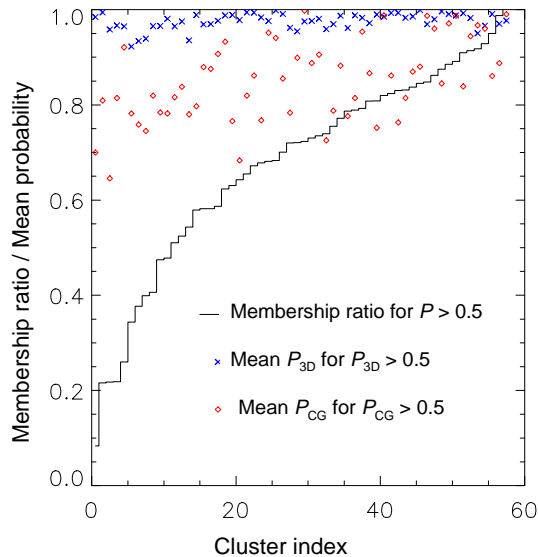}
	\caption{Comparison of cluster membership for targets in Table~\ref{table3} of this paper with results reported in \protect\cite{CantatGaudin2018a} for 58 open clusters common to both datasets. Clusters are ordered by increasing ratio of common targets identified as likely cluster members ($P_{\rm 3D}>0.5$) in Cantat-Gaudin et al. to those identified in this paper (see \S\ref{5.4}), which is shown as a black line. The blue and red points are the average membership probability for targets with $P>0.5$ in each cluster from this paper and from Cantat-Gaudin et al. respectively.}
	\label{plot_cantat_gaudin}
\end{figure}

The seven globular clusters (at the bottom of Fig.~\ref{plot_ROC})  have lower AUC values than  most of the open clusters. This is the result of the target selection process rather than any difference in cluster kinematics or the method of analysis. For the open clusters $\sim 50$ per cent or more of the targets are likely background stars with $P_{\rm 3D} < 0.1$. Thus the distribution of background stars is well defined by the maximum likelihood analysis allowing  members of well populated clusters to be discriminated for the background population. In contrast, for the the globular clusters, the target samples were already pre-selected to be likely cluster members; they contain, on average, only 5 per cent of probable background stars. This amounts to only a few measured stars which is insufficient to fully characterise background populations that are well separated from the distribution of cluster members in velocity space (see Appendix~\ref{appb}). For these clusters it is not possible to clearly identify {\it non}-members. Instead, individual $P_{\rm 3D}$ values must be used to assess the {\it probability} that outliers in kinematic space are cluster members. 

Membership probabilities were also calculated for a restricted list of targets that was filtered to remove those with {\it potentially} problematic {\it Gaia} data (see \S~\ref{3.2}). The membership probabilities in this case are referred to as $P_{\rm QG}$. For individual stars present in both lists $P_{\rm 3D}$ and $P_{\rm QG}$ are almost always very similar: $<1$ per cent of targets have a difference $|\Delta P|>0.01$.
Filtering targets with suspect {\it Gaia} data also has little effect on discrimination. 
%For about half the clusters it produced a marginal increase in AUC (see Fig.~\ref{plot_ROC}) but the difference is almost always small compared to the effects of moving from a 2D to a 3D analysis. 
For the open clusters it produced either no change or a marginal increase ($<$0.2 per cent) in AUC. Globular clusters, which contain a higher proportion of targets with suspect {\it Gaia} data and small numbers of non-members, show larger changes in AUC, both positive and negative (see Fig.~\ref{plot_ROC}). 
For most applications $P_{\rm 3D}$ is preferred since it identifies about 7~per cent more $P>0.9$ cluster members and any target selected as a likely member on the basis of its kinematic properties almost certainly has good proper motion data (see J20 for a detailed discussion). However, for particular applications the reader may wish to avoid stars with suspect {\it Gaia} data in which case $P_{\rm QG}$ is available in Table~$\ref{table3}$. 

\subsection{Comparison with previous kinematic analysis}
\label{5.3}
J20 reported membership probabilities for 32 open clusters using GESiDR5 data cross-matched with {\it Gaia}~DR2. The method of kinematic analysis used here is similar to that used in J20 although the method of selecting potential cluster members has been updated to reflect the greater mix of cluster ages and distances in GESiDR6 data. We compare results between J20 and the present analysis by counting the number of targets that are identified as cluster members $P_{\rm 3D} > 0.9$ in the present analysis but have a membership probability  $P_{\rm 3D} < 0.8$ in the J20 analysis. There are  a total of 288 GES targets in the 32 clusters that show this change in membership probability. Of these, 121 are targets with the same {\it Gaia}~source ID in both analyses with the remainder showing different {\it Gaia}~source IDs.

NGC\,6530 has 29 new members compared with the J20 analysis (and with the same {\it Gaia} counterpart) and shows a significant improvement in the quality of fit of the model distribution to the measured $V_{\rm RA}$ data (See Appendix B and the equivalent plot in J20) due, most probably, to the increase in the number of targets selected as potential cluster members. In J20, targets with no GES  $T_{\rm eff}$ were excluded from the analysis whereas in the current analysis $T_{\rm eff}^p$ is estimated where no GES value is available (see \S\ref{2.5}). Other clusters show smaller numbers of mismatched membership for targets with the same {\it Gaia}~source ID, most of which are associated with relatively large changes ($>3\sigma$) in proper motion between {\it Gaia} DR2 and EDR3. For example NGC\,2264 has 14 new members, 12 with a $>3\sigma$ change in proper motion.

There are 167 targets that are newly assigned membership in this paper but have {\it different} {\it Gaia}~source IDs in J20. In J20 GES targets were identified with the nearest source in the 2MASS catalogue. In the present paper we adopt the {\it Gaia} ID given in the GESiDR6 catalogue (see Hourihane et al. in preparation). This provides a better, but still not a perfect, cross-match to {\it Gaia}~source IDs for a target list that includes non-2MASS sources in several of the more distant clusters. Any mismatched targets are unlikely to have been ascribed cluster membership so this difficulty is a source of incompleteness rather than contamination.

%There are 164 targets that are newly assigned membership in this paper but have {\it different} {\it Gaia}~source IDs in J20. i.e. the GES target was matched to a different {\it Gaia} source with different astrometric properties. In J20 GES targets were identified with the nearest source in the 2MASS catalogue. This condition was changed in the present paper to reflect the number of more distant clusters where 2MASS targets are more densely packed and where a number of non-2MASS targets were included in the target list. Valid 2MASS targets were cross-matched with {\it Gaia} ERD3 using their 2MASS source ID (see \S2.4), with the remaining targets being linked to the nearest {\it Gaia}~source. This provides a better, but still  imperfect, cross-match to {\it Gaia}~source IDs for a target list that includes non-2MASS sources in several of the more distant clusters. For 3 per cent of targets two or more neighbours are found within 1\,arcsec. In this case the {\it Gaia}~ target with the highest "score" or lowest angular distance are selected \citep[see \S2.4 and][]{Marrese2019a}. In 75 per cent of these cases this is also the neighbour with the brightest $G$ magnitude. From this we estimate that the cross-match between the GES target and the {\it Gaia} source ID may be uncertain for $\sim$1 per cent of GES targets, the error rate being highest for fainter targets in the more distant clusters. Any mismatched targets are unlikely to have been ascribed cluster membership so this difficulty is a source of incompleteness rather than contamination.

\subsection{Comparison with other work}
\label{5.4}
\cite{CantatGaudin2018a} applied an unsupervised membership assignment algorithm to {\it Gaia} DR2 data to  characterise Milky Way star clusters, producing membership lists and cluster parameters for 1229 clusters. Cross referencing target RA and Dec in Table~\ref{table3} with the results of Cantat-Gaudin et al. identified matches for 58 of the 63 open clusters considered in this paper; the exceptions are the young clusters Chamaeleon I, Lambda Ori B35, NGC\,6530 and Rho Ophiuchus, and Loden\,165 (where we have no membership data). For each cluster we compare the number of stars with membership probability $P_{\rm CG} \ge 0.5$ in Table~1 of \cite{CantatGaudin2018a} with the number of $P_{\rm 3D} \ge 0.5$ targets in Table~\ref{table3}; the ratio giving a rough indication of the relative completeness of our list of GES$+$EDR3 cluster members compared to a membership list derived solely from {\it Gaia} DR2 data. Results for each cluster are shown on the RHS of the plot in in Fig.~\ref{plot_ROC}. The median value over the 58 matched clusters is 40 per cent. i.e. Cantat-Gaudin et al. have identified more stars as members. However within this number there are significant variations according to cluster age and distance. The analysis based solely on {\it Gaia} DR2 data fails to identify any members in 3 of the 6 youngest clusters (aged $<$4\,Myr in Table~\ref{table1}) and in two other cases, Trumpler\,14 and NGC\,2264,  Cantat-Gaudin et al. report fewer members than are identified here. The median ratio increases to 70 per cent  for more distant clusters ($>$2\,kpc) nine of which show higher numbers of targets identified from the {\it Gaia}-ESO data.

Next we compare the number of targets common to both datasets that are likely cluster members: with $P \ge 0.5$ in both datasets. 
%Note that this is an upper limit to the true value since Cantat-Gaudin et al. do not report probability values if they were lower than 0.1. 
However, the values specific to each cluster vary a lot. The results are shown in Fig.~\ref{plot_cantat_gaudin} as a solid line, with the clusters ordered by increasing values of the membership ratio. The clusters where the ratio is $<$0.5 tend to be the more distant clusters where the addition of $RV$ data becomes more important and where clusters tend to be projected against higher levels of field star contamination. Also shown in Fig.~\ref{plot_cantat_gaudin} are the average values of  membership probability for likely cluster members (those with $P_{\rm 3D}>0.5$ or $P_{\rm CG}>0.5$) from Table~\ref{table3} and from \cite{CantatGaudin2018a}. The mean values of membership probability for these samples across all clusters is 98 per cent for the data in Table~\ref{table3} compared to 85 per cent for data from Cantat-Gaudin et al. indicating that the the number of {\it false} positives (i.e. the level of contamination expected) is, on average, 6 times lower using Table~3. This ratio turns out to be quite independent of the chosen cut-off level used to define likely cluster members. 

The differing numbers of members and levels of discrimination between these two datasets likely reflects how samples of potential cluster members were selected for analysis and the available kinematic data. \cite{CantatGaudin2018a} placed no restrictions on potential targets, apart from a cut off in $G$ magnitude for their analysis of {\it Gaia} astrometric data.  For many clusters the sky positions of the Cantat-Gaudin et al. members extend well beyond the confines of the GES survey, particularly for nearby clusters.  Conversely the analysis reported here considers a restricted list of potential cluster members selected as targets for GES observations with some further screening to remove targets considered highly unlikely to be cluster members (see \S\ref{3.1}). This selection process together with use of GES $RV$ data alongside improved {\it Gaia} EDR3 astrometric data achieves a higher level of discrimination and much lower fraction of false positives, albeit from a smaller parent sample.

%There are three exceptions (IC 2391, ...) to this general trend where the analysis of \cite{CantatGaudin2018a} yields a slightly higher mean probability. These clusters are nearby and have distinctive proper motions such that selection via tangential motion and parallax (which we have not used) is highly effective and the addition of $RV$ data is of little importance.

\section{Summary}
\label{summary}
As a part of a the {\it Gaia}-ESO survey (GES) 39441 targets towards 63 open clusters and 7 globular clusters were observed using the GIRAFFE medium resolution spectrograph (with the HR15N setup) and/or the UVES high resolution spectrograph. Parameters from these spectra, reported in the final GES data release, have been combined with {\it Gaia}~EDR3, 2MASS and VISTA data to determine the  probability of individual targets being cluster members.

After an initial filtering process, a maximum likelihood technique was used to determine membership probabilities for each of these stars based solely on their 3D kinematics.  Reliable solutions were obtained for 69 of the 70 clusters  with a high discrimination achieved between members and non-members of the individual clusters: the areas under their Receiver Operator Characteristic curves are $\geq 0.975$  for all open clusters. Of the potential cluster members that go into the kinematic analysis for  69 clusters, 47 per cent were identified as likely members with an average probability  of $P_{\rm 3D} =0.993$, 48 per cent were flagged as non-members with an average of $P_{\rm 3D}<0.005$ and only 5 per cent were of uncertain status ($0.1<P_{\rm 3D}<0.9)$. 
The inclusion of the GES spectroscopic data and radial velocities is of significant benefit to membership discrimination and especially in the reduction of false positives. This is particularly apparent in more distant clusters and those with significant contamination in the proper motion plane by field stars.

Results are presented in the form of a main catalogue of compiled data that includes the membership probability for 26351 potential cluster members observed using GIRAFFE+HR15N setup and/or UVES. Membership probabilities are shown for two cases: $P_{\rm 3D}$ calculated for the the full list of potential cluster members; and $P_{\rm QG}$ calculated for a restricted list filtered to remove $\sim 9$ per cent of targets with potentially suspect {\it Gaia}~data. Membership probabilities are also given for a further 3193 potential cluster members that were observed using alternate GIRAFFE setups and were excluded from the main kinematic analysis.

Since the membership criteria are almost purely kinematic, and independent of stellar photometry and chemistry, then the union of this catalogue with GES, {\it Gaia} EDR3 or other datasets is ideal for studying the behaviour of photometric, chemical or  other physical properties as a function of mass, age or Galactic position, without compromising investigations by using those same properties as membership criteria. Examples include testing stellar evolutionary models using HR and colour-magnitude diagrams \citep[e.g,][]{Randich2018a}. following the evolution of magnetic activity, rotation and light element depletion \citep[e.g.,][]{Jeffries2017a, Gutierrez2020a}, studying chemical inhomogeneity within clusters \citep[e.g.][]{Spina2015a, Spina2018a} or investigating chemical abundance gradients with age and Galactocentric radius \citep[e.g.][]{Spina2017a, Magrini2018a}.

\section{acknowledgements}
RJJ, RDJ and NJW wish to thank the UK Science and Technology Facilities Council
for financial support. 

Based on data products from observations made
with ESO Telescopes at the La Silla Paranal Observatory under programme
ID 188.B-3002. These data products have been processed by the Cambridge
Astronomy Survey Unit (CASU) at the Institute of Astronomy, University
of Cambridge, and by the FLAMES/UVES reduction team at
INAF/Osservatorio Astrofisico di Arcetri. These data have been obtained
from the {\it Gaia}-ESO Survey Data Archive, prepared and hosted by the Wide
Field Astronomy Unit, Institute for Astronomy, University of Edinburgh,
which is funded by the UK Science and Technology Facilities Council.
This work was partly supported by the European Union FP7 programme
through ERC grant number 320360 and by the Leverhulme Trust through
grant RPG-2012-541. We acknowledge the support from INAF and Ministero
dell' Istruzione, dell' Universit\`a' e della Ricerca (MIUR) in the
form of the grant "Premiale VLT 2012". T.B. was funded by grant No. 2018-04857 from The Swedish Research Council. The results presented here
benefit from discussions held during the {\it Gaia}-ESO workshops and
conferences supported by the ESF (European Science Foundation) through
the GREAT Research Network Programme.

This work has made use of data from the European Space Agency (ESA) mission
{\it Gaia}  (\url{https://www.cosmos.esa.int/gaia}), processed by the {\it Gaia}
Data Processing and Analysis Consortium (DPAC,
\url{https://www.cosmos.esa.int/web/gaia/dpac}\\ \url{/consortium}). Funding for the DPAC
has been provided by national institutions, in particular the institutions
participating in the {\it Gaia} Multilateral Agreement.

This paper also uses data products from: the VISTA Hemisphere Survey, ESO programme ID:
179.A-2010; the VISTA Variables in the Via Lactea Survey, ESO programme ID: 179.B-2002 and the the Two Micron All Sky Survey, which is a joint project of the University of Massachusetts and the Infrared Processing and Analysis Center/California Institute of Technology, funded by the National Aeronautics and Space Administration and the National Science Foundation.

\section{Data availability statement}
The reduced stacked spectra underlying this article can be obtained via the European Southern Observatory (ESO) archive and is identified as the "Phase 3 release" of Gaia-ESO Survey DR4. Raw data can also be obtained from the ESO archive. The full catalogue of stellar parameters derived from these spectra by the various GES working groups will be deposited in this same archive shortly. All the other data used in this paper ({\it Gaia}, 2MASS, VISTA) are also available in public archives.

\bibliographystyle{aa} 
\bibliography{references} 

\appendix
\section{Supplementary targets}
In this section we evaluate the membership probability of supplementary targets observed towards the clusters listed in Tables \ref{table1} \&~ \ref{table2} using alternative GIRAFFE setups. These are targets with a measured $RV$ reported in the GESiDR6 Parameter Catalogue but which were {\it not} observed with the GIRAFFE HR15N setup or with UVES. These targets could not pass through all the same filtering steps described in $\S$~3.1 since they did not have the $\gamma$ and $\tau$ spectroscopic indices.  A total of 3770 additional targets were identified in  30 clusters , the remaining 40 clusters showed no additional targets. Selection criteria for these additional targets was varied. In some cases targets were chosen on the same basis as GIRAFFE+HR15N targets but in other clusters the targets had previous evidence of cluster membership. It is for these reasons that we chose to treat these targets separately and the membership probabilities may be less accurate than for the main sample.

Potential cluster members had their luminosity and mass estimated as in \S \ref{3.3}, the only difference in the target data being the source of the measured $RV$ and its uncertainty. For the main kinematic analyses the $RV$ of targets observed using GIRAFFE was taken from the spectrum {\sc velclass} metadata and its uncertainty estimated from an empirical analysis of repeat observations (see \S2.3). For these additional targets the measured $RV$ and uncertainty were taken from the GESiDR6 Parameter Catalogue in which the reported $RV$ was averaged over observations made using a variety of GIRAFFE setups. A small (0.09\,km\,s$^{-1}$) correction was added to the catalogue $RV$ to match the GIRAFFE HR15N $RV$ scale. 

\begin{table}
\caption{Numbers of supplementary targets with $RV$ reported in the GESiDR6 Parameter Catalogue that were {\it not} observed with the GIRAFFE HR15N setup or with UVES.}

\begin{tabular}{lrrr}
\hline
Cluster	&	Number of & Number & Members\\
        &   targets   & fitted & $P_{\rm 3D}$$>$0.9\\
        \hline
Trumpler 14    &    41&    24&    10\\
NGC 2232       &    61&    25&     0\\
NGC 6649       &    25&    22&    11\\
NGC 6405       &    38&    33&    20\\
NGC 6067       &    82&    59&    29\\
NGC 6709       &    46&    45&    13\\
NGC 6259       &    28&    23&     6\\
NGC 6705       &    11&    10&     9\\
Berkeley 30    &    55&    34&    11\\
NGC 3532       &   172&   148&   122\\
NGC 6281       &    69&    49&    30\\
NGC 4815       &    50&    45&    23\\
NGC 6633       &    68&    20&     8\\
Pismis 18      &    34&    30&    10\\
Trumpler 23    &    59&    54&    27\\
NGC 6802       &    47&    46&    39\\
Berkeley 81    &    64&    44&     9\\
NGC 6005       &   198&   171&    44\\
Pismis 15      &    92&    83&    25\\
Trumpler 20    &   699&   645&   296\\
Haffner 10     &    76&    53&     7\\
NGC 6253       &   352&   309&   115\\
NGC 2243       &     6&     6&     6\\
Berkeley 32    &    49&    47&    41\\
NGC 1261       &   179&   144&   126\\
NGC 362        &   255&   210&   133\\
NGC 2808       &   144&   113&    56\\
NGC 1904       &   154&   115&    85\\
NGC 6752       &   353&   335&   322\\
NGC 104        &   263&   251&   242\\
\hline
 \end{tabular}
 \label{tableA1}
\end{table}

\begin{table*}
\caption{Membership probabilities of supplementary targets. Columns 1 to 4 show GESiDR6 target name, cluster and co-ordinates. Targets flagged 0 are identified as potential cluster members. Column 6 and 7 show the  {\it Gaia}~cross-matched ID and $G$ magnitude. Columns 8 and 9 show $T_{\rm eff}^p$ and $K_S^p$ data (which includes estimated values for some targets) used to determine $\log L$ and mass. The final column show the probability that the target is a member of its given cluster. Targets with a membership probability of -1 could not be fitted as cluster members. A sample of the table is shown here. The full table is available as supplementary material.}
\begin{tabular}{llllclrrrrr}
\hline
Target & cluster & RA	 & Dec	 & Flag  & {\it Gaia}~ID & $G$   & $T_{\rm eff}^p$  & $K_S^p$ & $\log L$  & Membership   \\
cname  & DB name & (deg) & (deg) &	     &               & (mag) & (K)	             & 	(mag) &            & $P_{\rm  3D}$\\
\hline
18331347-1023200 &NGC6649   & 278.30613& -10.38889& 0&ID4155024758859315456&16.07& 8825&12.53& 1.73& 0.9199\\
18331394-1029175 &NGC6649   & 278.30808& -10.48819& 0&ID4155014176059792384&16.42& 9139&13.07& 1.56& 0.9917\\
18331844-1024428 &NGC6649   & 278.32683& -10.41189& 0&ID4155017848247221888&15.96&10051&12.60& 1.88& 0.5234\\
18331952-1026068 &NGC6649   & 278.33133& -10.43522& 0&ID4155017749472616704&16.23& 8192&12.59& 1.61& 0.0000\\
18332318-1023026 &NGC6649   & 278.34658& -10.38406& 0&ID4155024071664517632&16.20& 9506&12.85& 1.70& 0.9898\\
\hline

\end{tabular}
\label{tableA2}
\end{table*}
Membership probabilities for these additional targets were calculated as the expectation value of being a cluster member evaluated over the grid of model velocities, dispersions and fractional membership weighted according to the total likelihood summed over potential cluster members that were observed with Giraffe+HR15N and/or UVES -- thus giving the probability that an additional target is a member of the cluster as defined by main GIRAFFE+HR15N and UVES dataset.

This approach is considered more robust to systematic differences in the $RV$ measured using different GIRAFFE setups and/or variations in the method of target selection. Any systematic offset in $RV$ for an individual target will tend to reduce $P_{\rm 3D}$ Similarly if an additional target is selected from a dataset previously identified as cluster members then $P_{\rm 3D}$ should give a conservative estimate of membership probability based on the kinematics of the Giraffe+HR15N and UVES stars. These additional targets thus have no effect on the best fit cluster parameters in Tables~\ref{table4} and ~\ref{table5}.

Table \ref{tableA1} shows the numbers of supplementary targets, potential cluster members and $P_{\rm 3D}>0.9$ cluster members in each cluster. There are a total of 1875 additional $P_{\rm 3D}>0.9$ cluster members, 911 in open clusters and 964 in globular clusters. Membership probabilities for individual additional targets are given in Table~\ref{tableA2}.

\section{Measured data and model fits}
Plots showing histograms of target velocities and the results of the maximum likelihood analysis for each of the 70 clusters in age order are shown in Appendix B which is available as supplementary material to this paper. A representative set of the four plots is shown for  typical cluster in Fig.~\ref{plot_results_a}. 
\label{appb}
\section{Cluster velocity and dispersion}
Plots showing the variation in log likelihood of cluster properties as a function of velocity are shown for each of the 70 clusters in age order are shown in Appendix C which is available as supplementary material to this paper. A representative set of the six plots for a typical cluster is shown in Fig.~\ref{plot_results_b}.
\label{appc}
\section{plots of cluster members}
\label{appc}
Plots showing properties of  potential and likely cluster members for each of the 70 clusters in age order are shown in Appendix D which is available as supplementary material to this paper. A set of the four plots is shown for a typical cluster in Fig.~\ref{plot_results_c}. 
%%%%%%%%%%%%%%%%%%%%%%%%%%%%%%%%%%%%
\clearpage
\newpage
%%%%%%%%%%%%%%%%%%%%%%%%%%%%%%%%%%%%
\begin{figure*}
\begin{minipage}[t]{0.98\textwidth}
\centering
\includegraphics[width = 145mm]{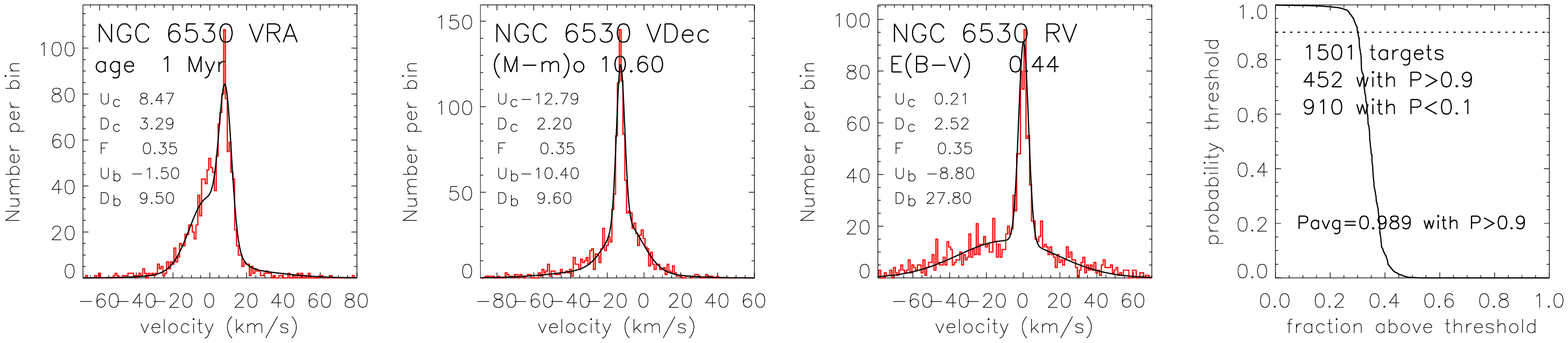}\\
\end{minipage}
\label{figB:1}
\end{figure*}
%%%%%%%%%%%%%%%%%%%%%%%%%%%%%%%%%%%%
\begin{figure*}
\begin{minipage}[t]{0.98\textwidth}
\centering
\includegraphics[width = 145mm]{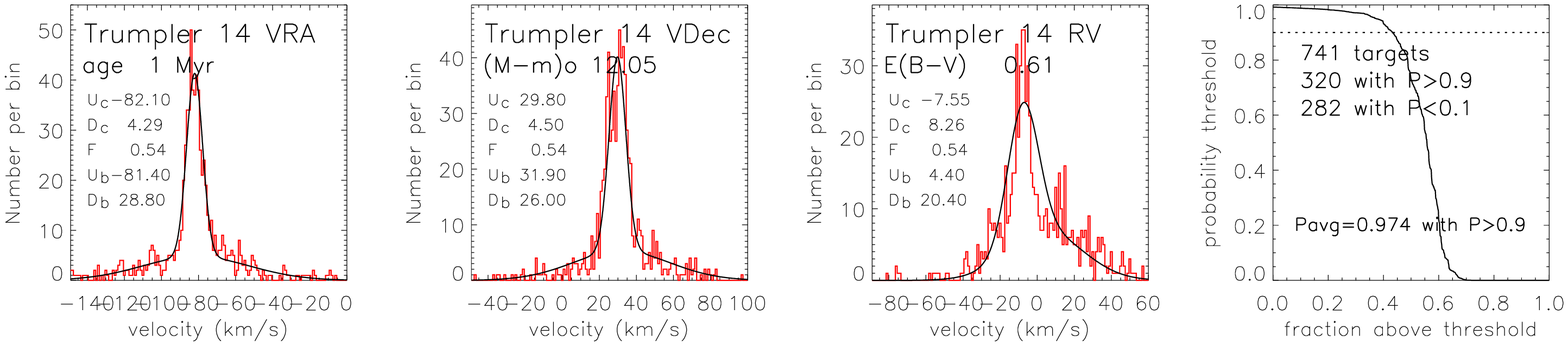}\\
\end{minipage}
\label{figB:2}
\end{figure*}
%%%%%%%%%%%%%%%%%%%%%%%%%%%%%%%%%%%%
\begin{figure*}
\begin{minipage}[t]{0.98\textwidth}
\centering
\includegraphics[width = 145mm]{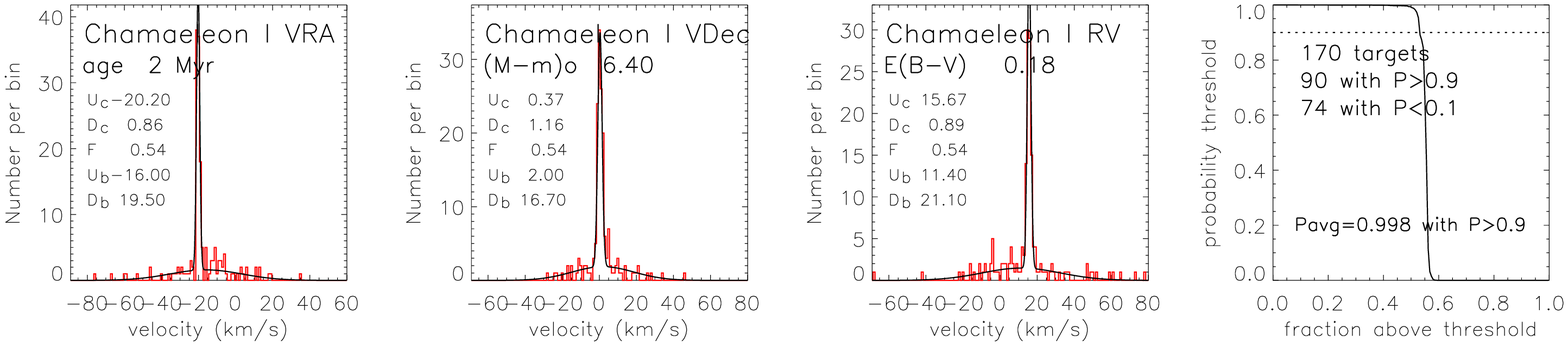}\\
\end{minipage}
\label{figB:3}
\end{figure*}
%%%%%%%%%%%%%%%%%%%%%%%%%%%%%%%%%%%%
\begin{figure*}
\begin{minipage}[t]{0.98\textwidth}
\centering
\includegraphics[width = 145mm]{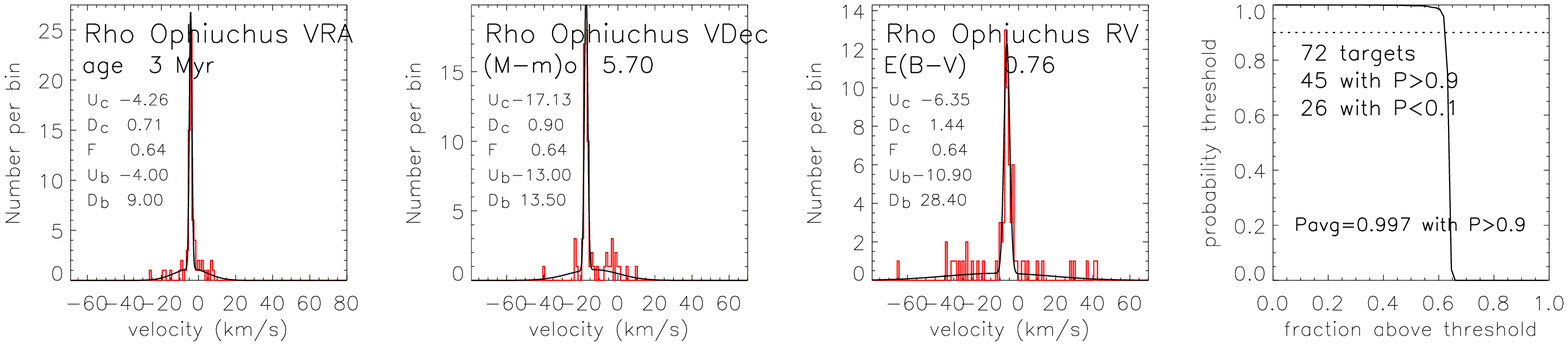}\\
\end{minipage}
\label{figB:4}
\end{figure*}
%%%%%%%%%%%%%%%%%%%%%%%%%%%%%%%%%%%%
\begin{figure*}
\begin{minipage}[t]{0.98\textwidth}
\centering
\includegraphics[width = 145mm]{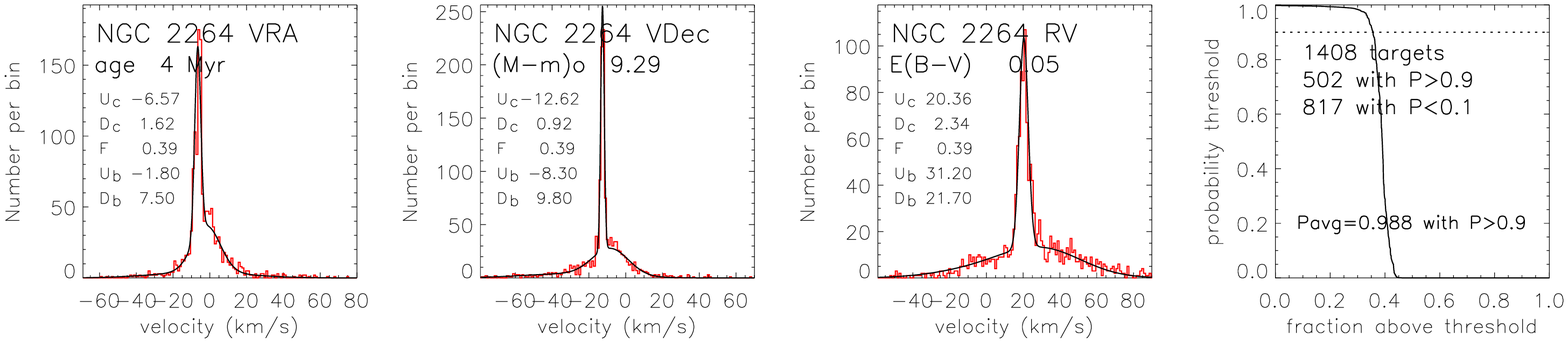}\\
\end{minipage}
\label{figB:5}
\end{figure*}
%%%%%%%%%%%%%%%%%%%%%%%%%%%%%%%%%%%%
\begin{figure*}
\begin{minipage}[t]{0.98\textwidth}
\centering
\includegraphics[width = 145mm]{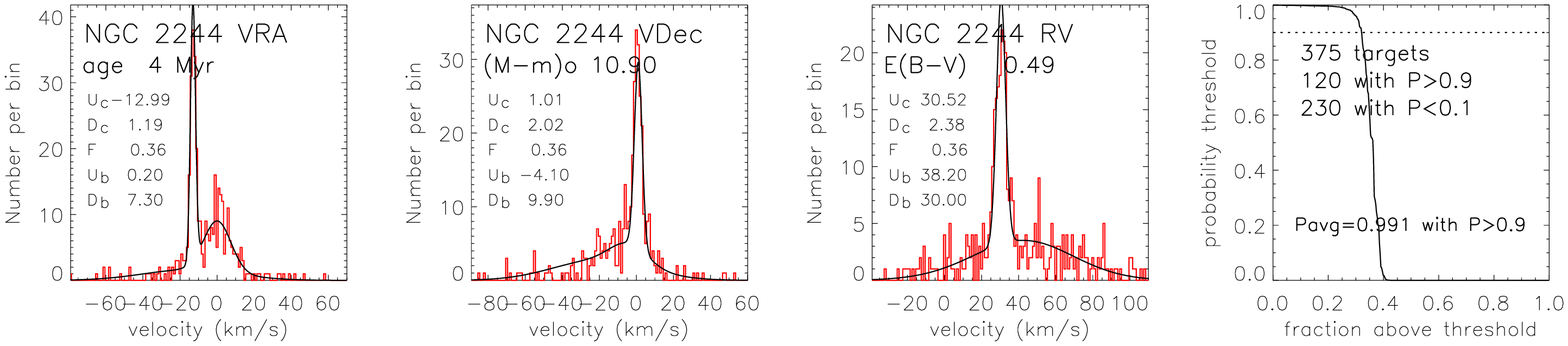}\\
\end{minipage}
\label{figB:6}
\end{figure*}
%%%%%%%%%%%%%%%%%%%%%%%%%%%%%%%%%%%%
\clearpage
\newpage
%%%%%%%%%%%%%%%%%%%%%%%%%%%%%%%%%%%%
\begin{figure*}
\begin{minipage}[t]{0.98\textwidth}
\centering
\includegraphics[width = 145mm]{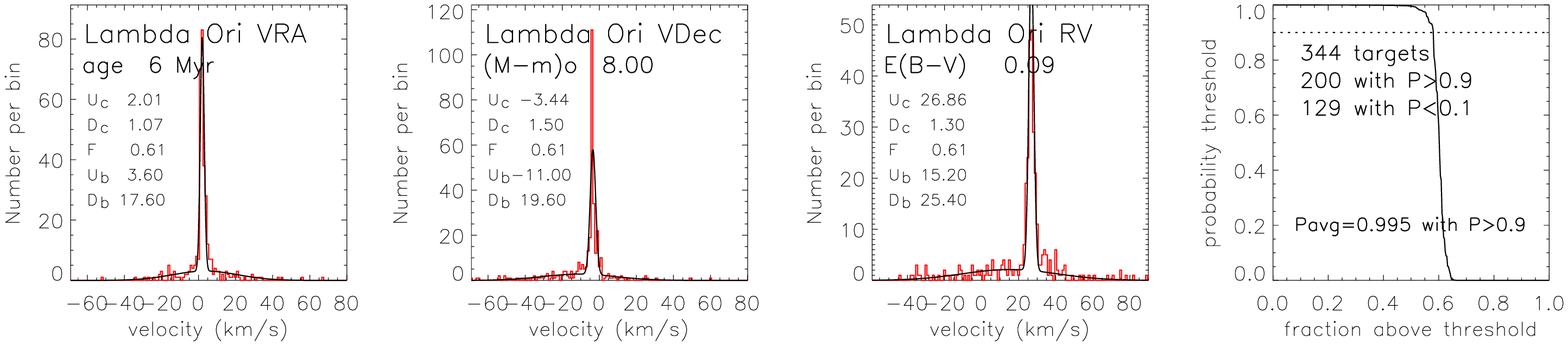}\\
\end{minipage}
\label{figB:7}
\end{figure*}
%%%%%%%%%%%%%%%%%%%%%%%%%%%%%%%%%%%%
\begin{figure*}
\begin{minipage}[t]{0.98\textwidth}
\centering
\includegraphics[width = 145mm]{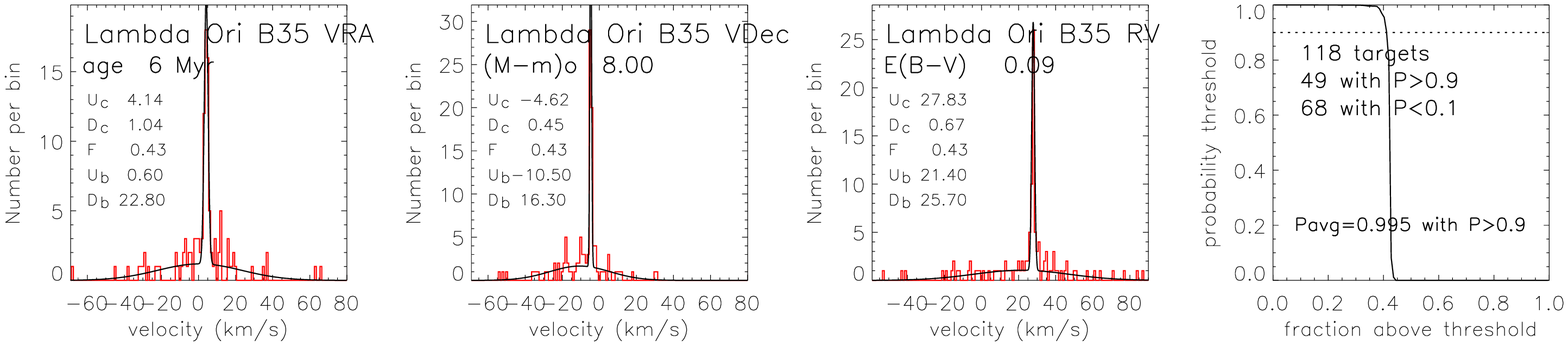}\\
\end{minipage}
\label{figB:8}
\end{figure*}
%%%%%%%%%%%%%%%%%%%%%%%%%%%%%%%%%%%%
\begin{figure*}
\begin{minipage}[t]{0.98\textwidth}
\centering
\includegraphics[width = 145mm]{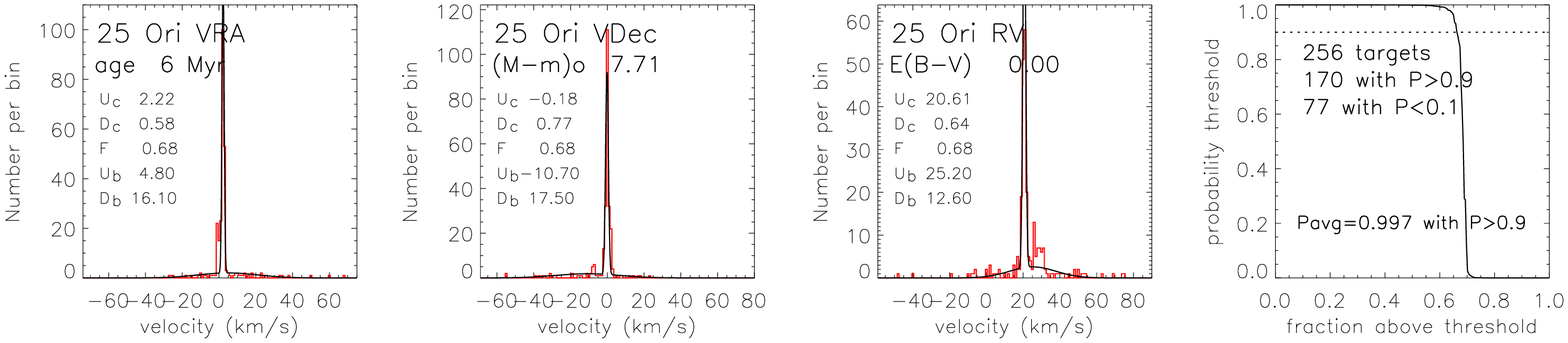}\\
\end{minipage}
\label{figB:9}
\end{figure*}
%%%%%%%%%%%%%%%%%%%%%%%%%%%%%%%%%%%%
\begin{figure*}
\begin{minipage}[t]{0.98\textwidth}
\centering
\includegraphics[width = 145mm]{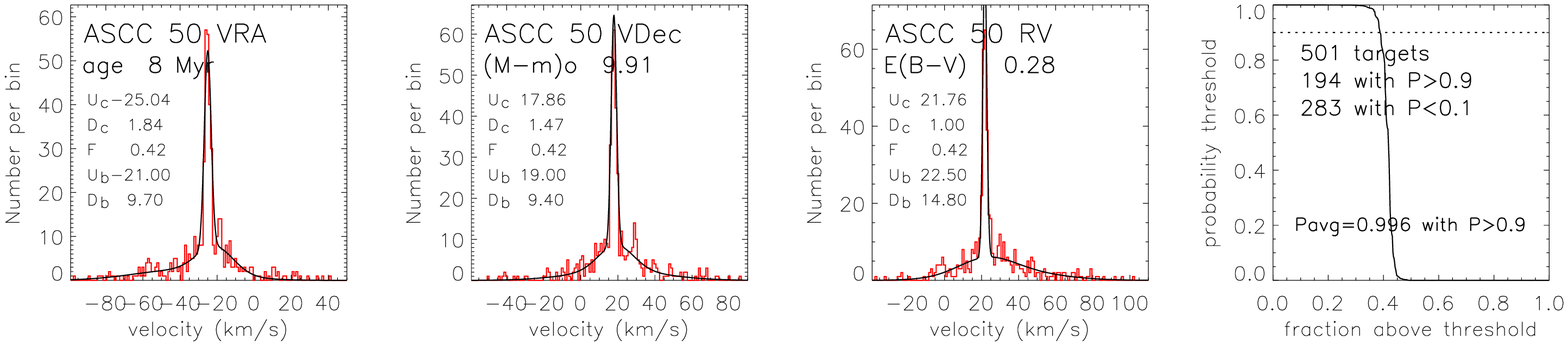}\\
\end{minipage}
\label{figB:10}
\end{figure*}
%%%%%%%%%%%%%%%%%%%%%%%%%%%%%%%%%%%%
\begin{figure*}
\begin{minipage}[t]{0.98\textwidth}
\centering
\includegraphics[width = 145mm]{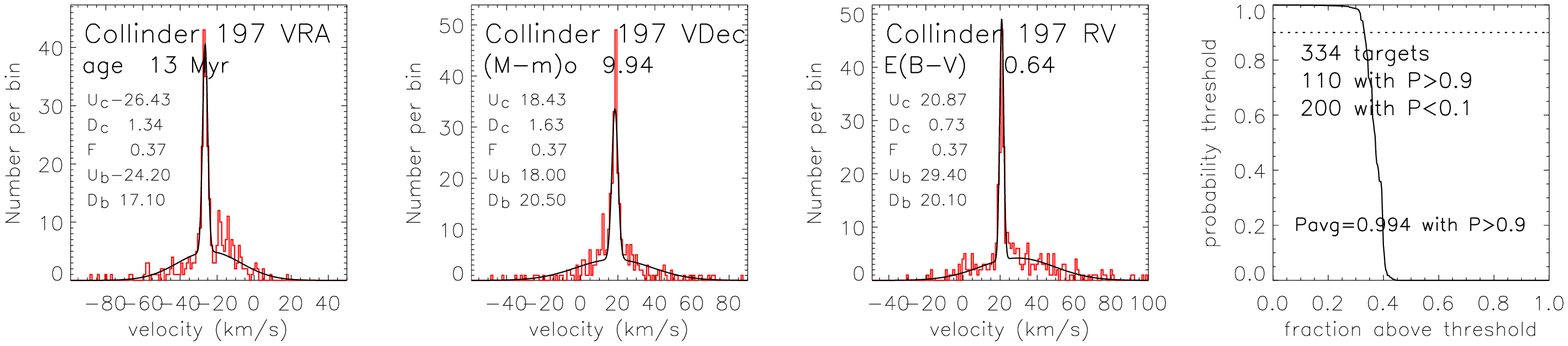}\\
\end{minipage}
\label{figB:11}
\end{figure*}
%%%%%%%%%%%%%%%%%%%%%%%%%%%%%%%%%%%%
\begin{figure*}
\begin{minipage}[t]{0.98\textwidth}
\centering
\includegraphics[width = 145mm]{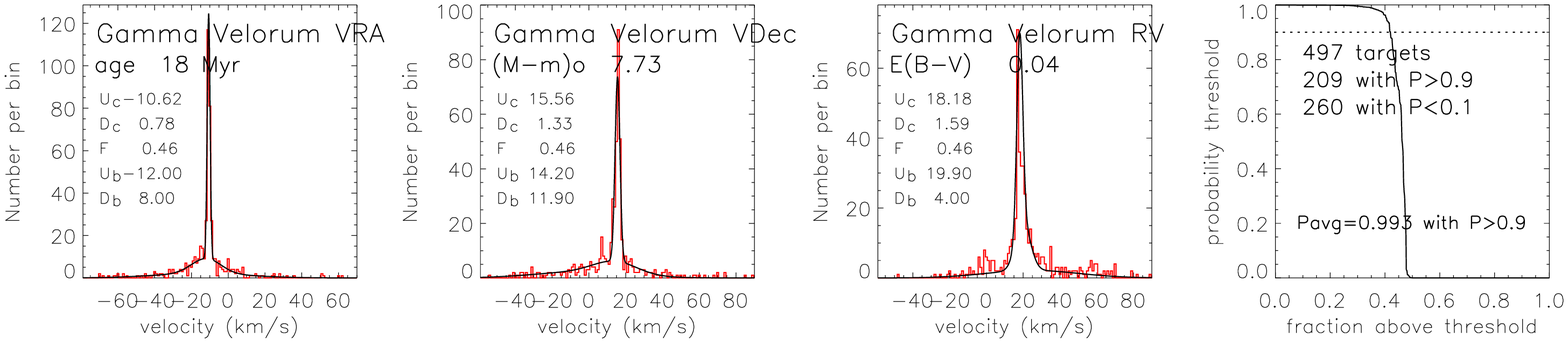}\\
\end{minipage}
\label{figB:12}
\end{figure*}
%%%%%%%%%%%%%%%%%%%%%%%%%%%%%%%%%%%%
\clearpage
\newpage
%%%%%%%%%%%%%%%%%%%%%%%%%%%%%%%%%%%%
\begin{figure*}
\begin{minipage}[t]{0.98\textwidth}
\centering
\includegraphics[width = 145mm]{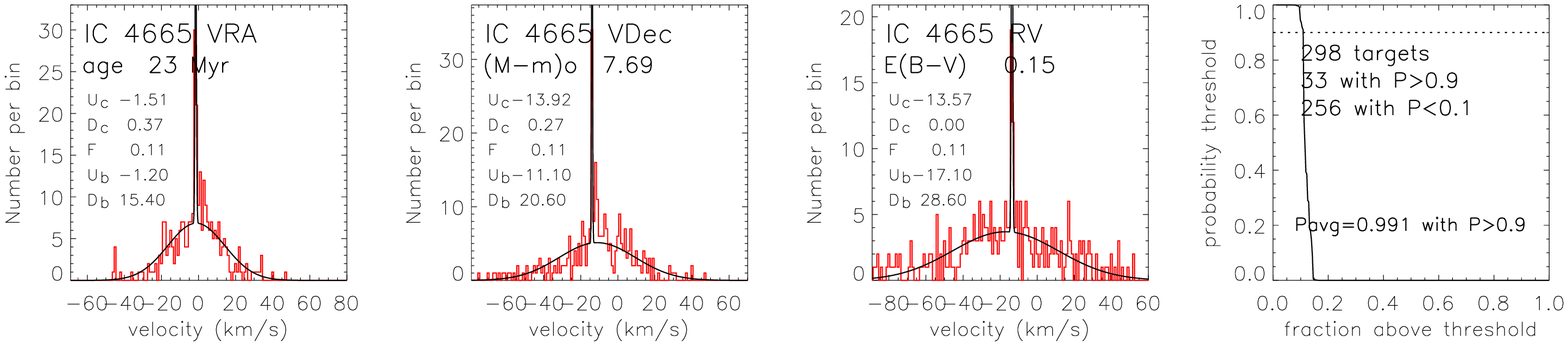}\\
\end{minipage}
\label{figB:13}
\end{figure*}
%%%%%%%%%%%%%%%%%%%%%%%%%%%%%%%%%%%%
\begin{figure*}
\begin{minipage}[t]{0.98\textwidth}
\centering
\includegraphics[width = 145mm]{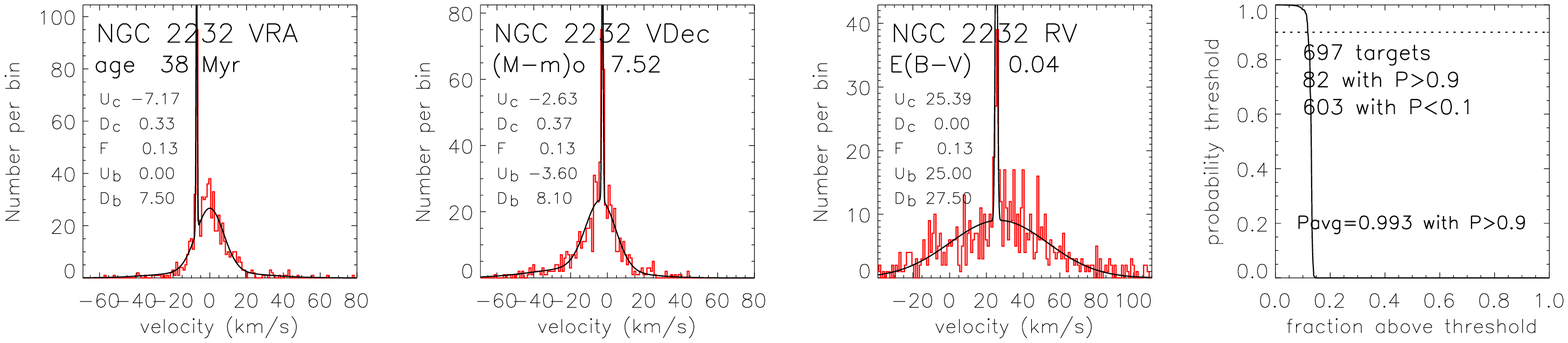}\\
\end{minipage}
\label{figB:14}
\end{figure*}
%%%%%%%%%%%%%%%%%%%%%%%%%%%%%%%%%%%%
\begin{figure*}
\begin{minipage}[t]{0.98\textwidth}
\centering
\includegraphics[width = 145mm]{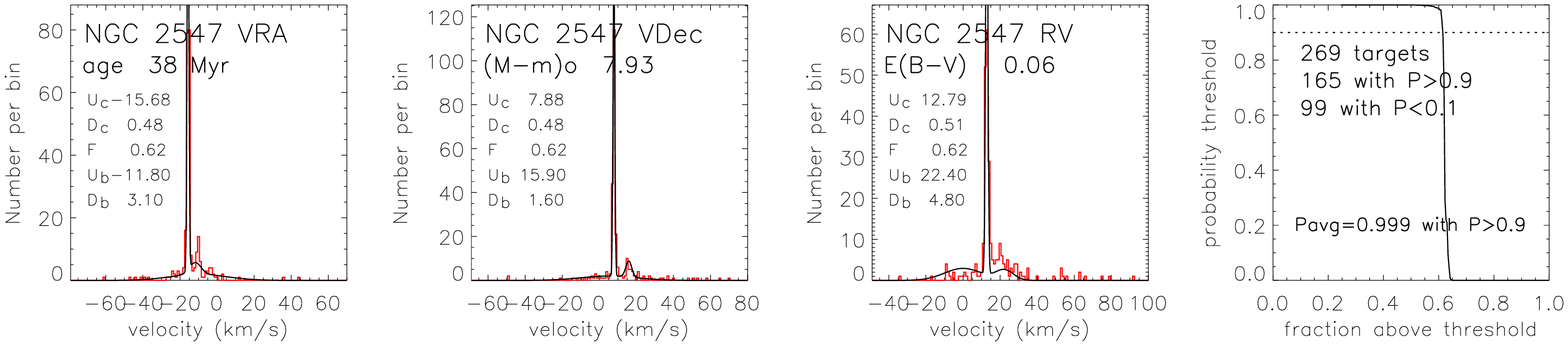}\\
\end{minipage}
\label{figB:15}
\end{figure*}
%%%%%%%%%%%%%%%%%%%%%%%%%%%%%%%%%%%%
\begin{figure*}
\begin{minipage}[t]{0.98\textwidth}
\centering
\includegraphics[width = 145mm]{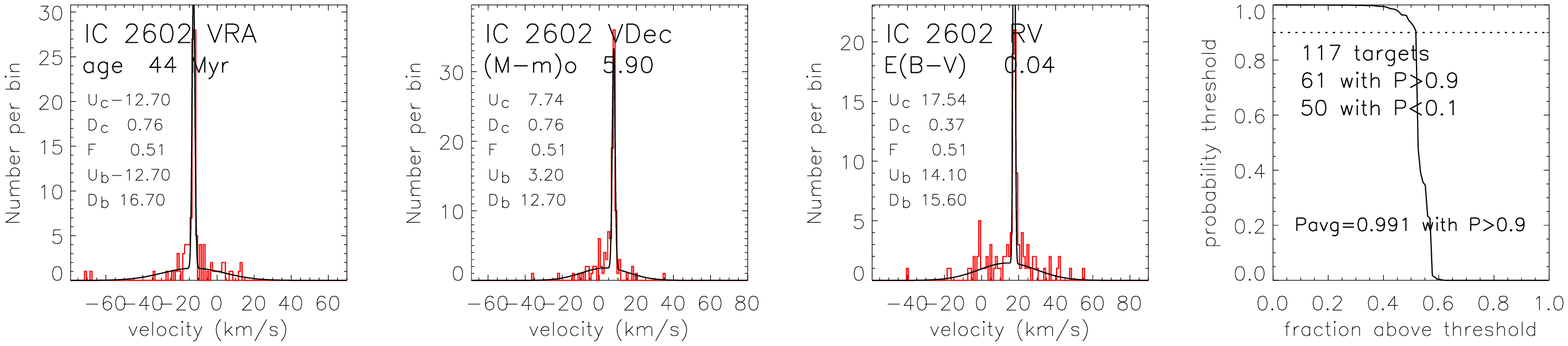}\\
\end{minipage}
\label{figB:16}
\end{figure*}
%%%%%%%%%%%%%%%%%%%%%%%%%%%%%%%%%%%%
\begin{figure*}
\begin{minipage}[t]{0.98\textwidth}
\centering
\includegraphics[width = 145mm]{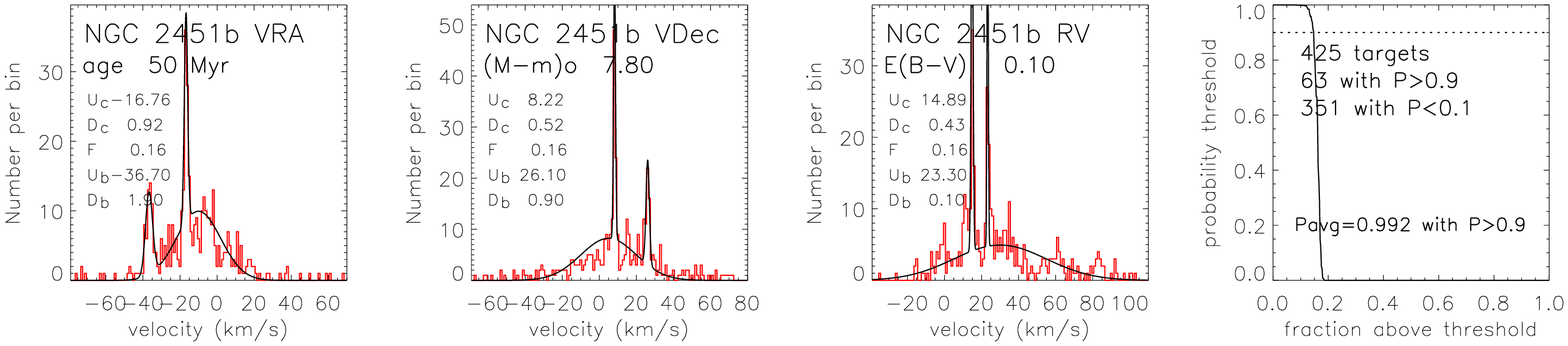}\\
\end{minipage}
\label{figB:17}
\end{figure*}
%%%%%%%%%%%%%%%%%%%%%%%%%%%%%%%%%%%%
\begin{figure*}
\begin{minipage}[t]{0.98\textwidth}
\centering
\includegraphics[width = 145mm]{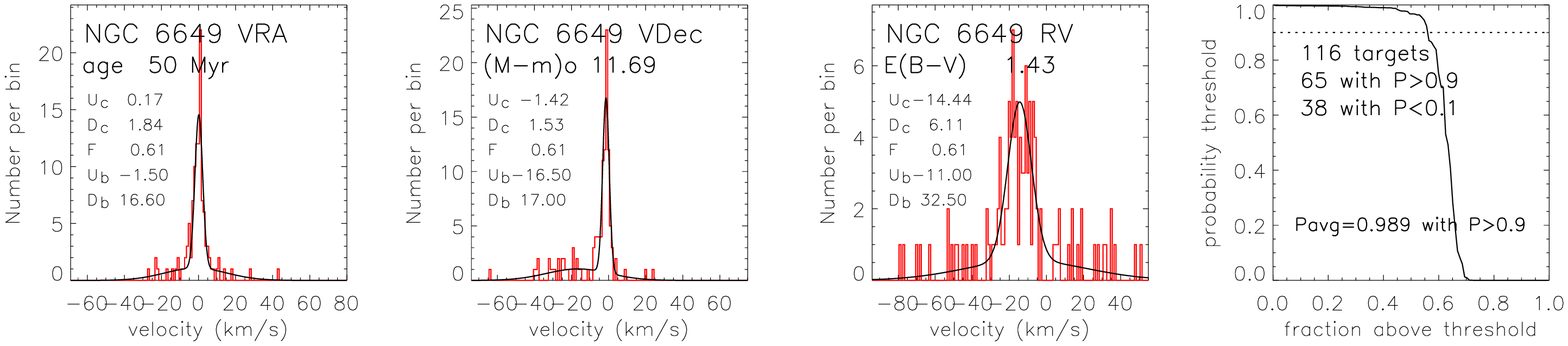}\\
\end{minipage}
\label{figB:18}
\end{figure*}
%%%%%%%%%%%%%%%%%%%%%%%%%%%%%%%%%%%%
\clearpage
\newpage
%%%%%%%%%%%%%%%%%%%%%%%%%%%%%%%%%%%%
\begin{figure*}
\begin{minipage}[t]{0.98\textwidth}
\centering
\includegraphics[width = 145mm]{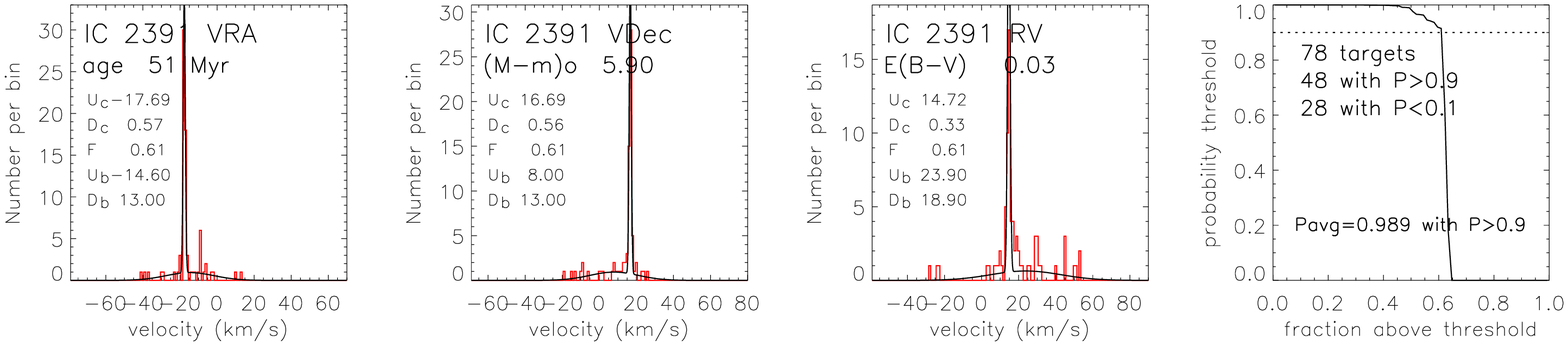}\\
\end{minipage}
\label{figB:19}
\end{figure*}
%%%%%%%%%%%%%%%%%%%%%%%%%%%%%%%%%%%%
\begin{figure*}
\begin{minipage}[t]{0.98\textwidth}
\centering
\includegraphics[width = 145mm]{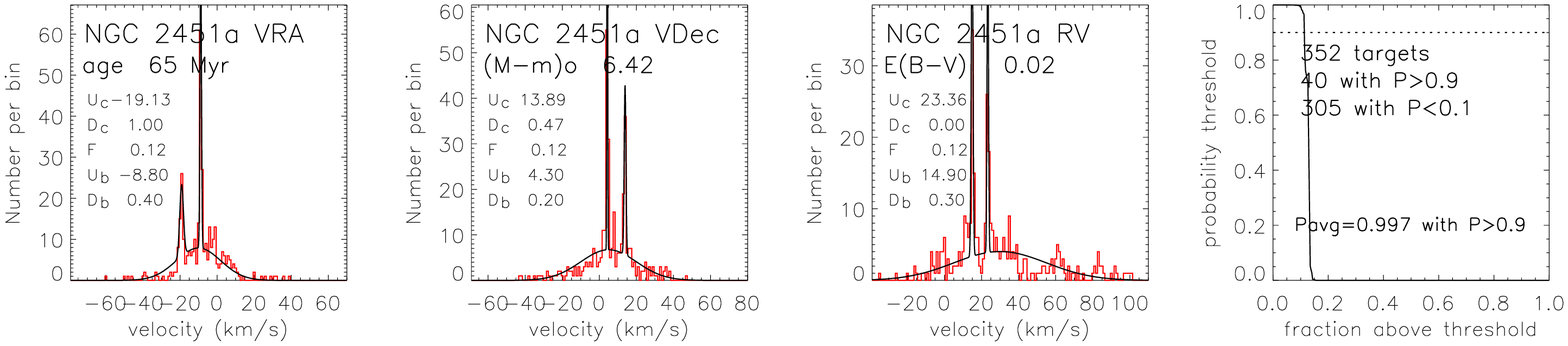}\\
\end{minipage}
\label{figB:20}
\end{figure*}
%%%%%%%%%%%%%%%%%%%%%%%%%%%%%%%%%%%%
\begin{figure*}
\begin{minipage}[t]{0.98\textwidth}
\centering
\includegraphics[width = 145mm]{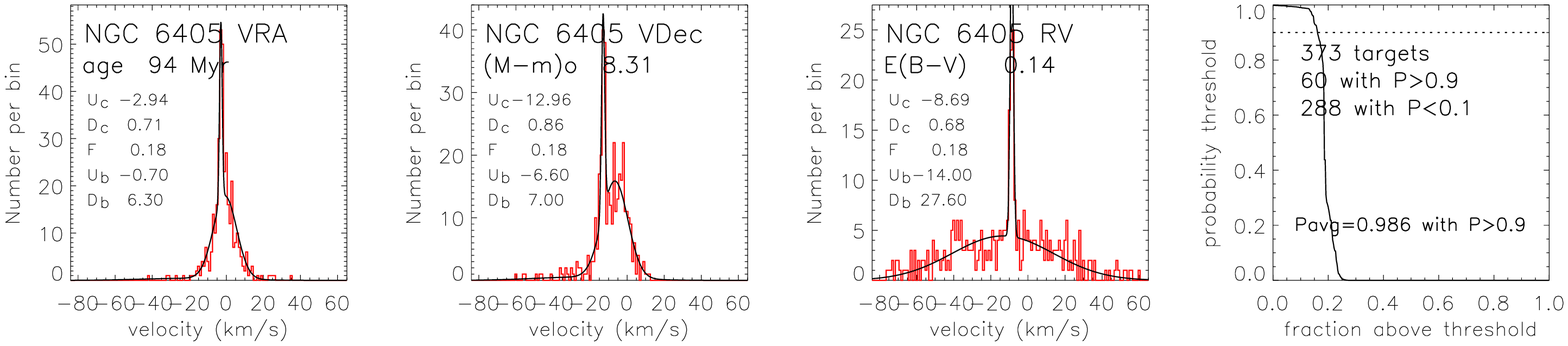}\\
\end{minipage}
\label{figB:21}
\end{figure*}
%%%%%%%%%%%%%%%%%%%%%%%%%%%%%%%%%%%%
\begin{figure*}
\begin{minipage}[t]{0.98\textwidth}
\centering
\includegraphics[width = 145mm]{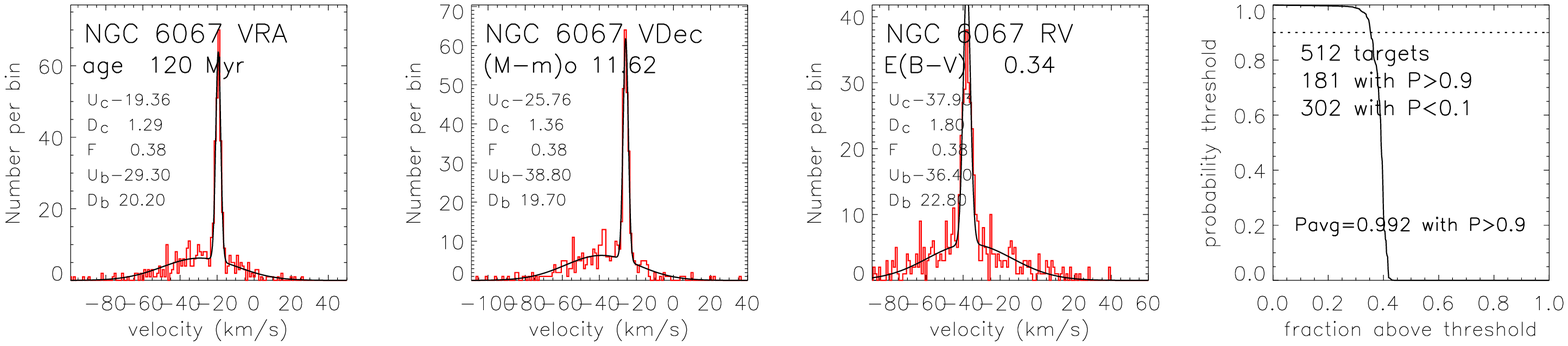}\\
\end{minipage}
\label{figB:22}
\end{figure*}
%%%%%%%%%%%%%%%%%%%%%%%%%%%%%%%%%%%%
\begin{figure*}
\begin{minipage}[t]{0.98\textwidth}
\centering
\includegraphics[width = 145mm]{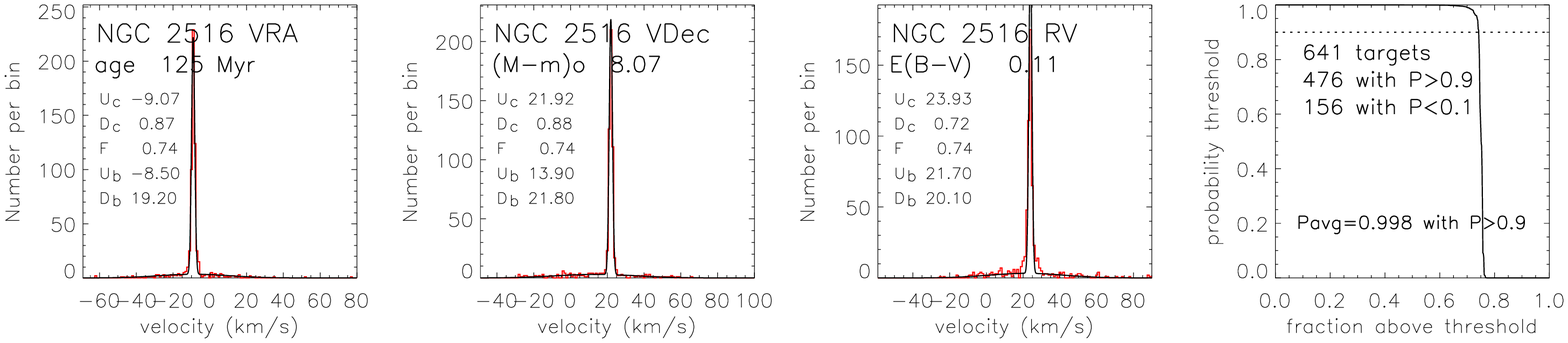}\\
\end{minipage}
\label{figB:23}
\end{figure*}
%%%%%%%%%%%%%%%%%%%%%%%%%%%%%%%%%%%%
\begin{figure*}
\begin{minipage}[t]{0.98\textwidth}
\centering
\includegraphics[width = 145mm]{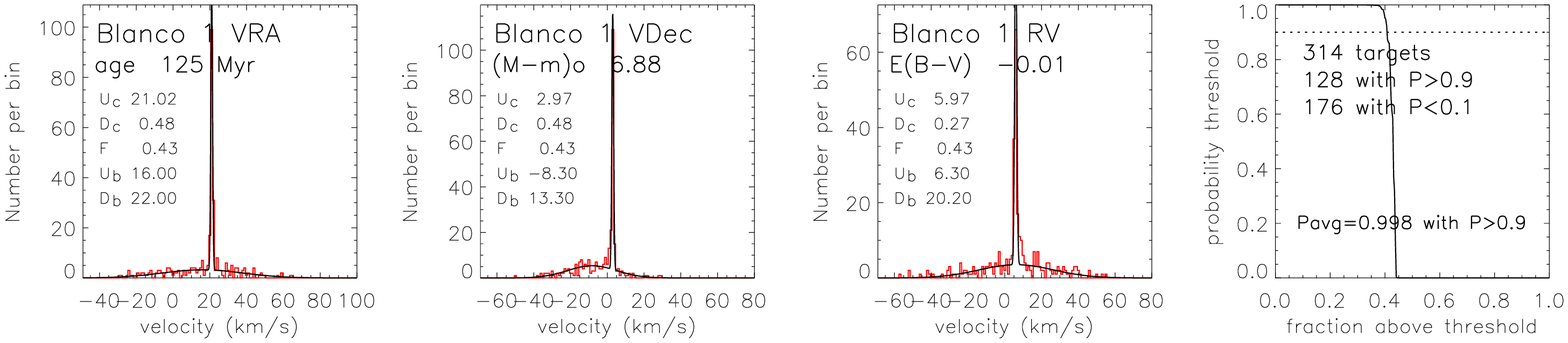}\\
\end{minipage}
\label{figB:24}
\end{figure*}
%%%%%%%%%%%%%%%%%%%%%%%%%%%%%%%%%%%%
\clearpage
\newpage
%%%%%%%%%%%%%%%%%%%%%%%%%%%%%%%%%%%%
\begin{figure*}
\begin{minipage}[t]{0.98\textwidth}
\centering
\includegraphics[width = 145mm]{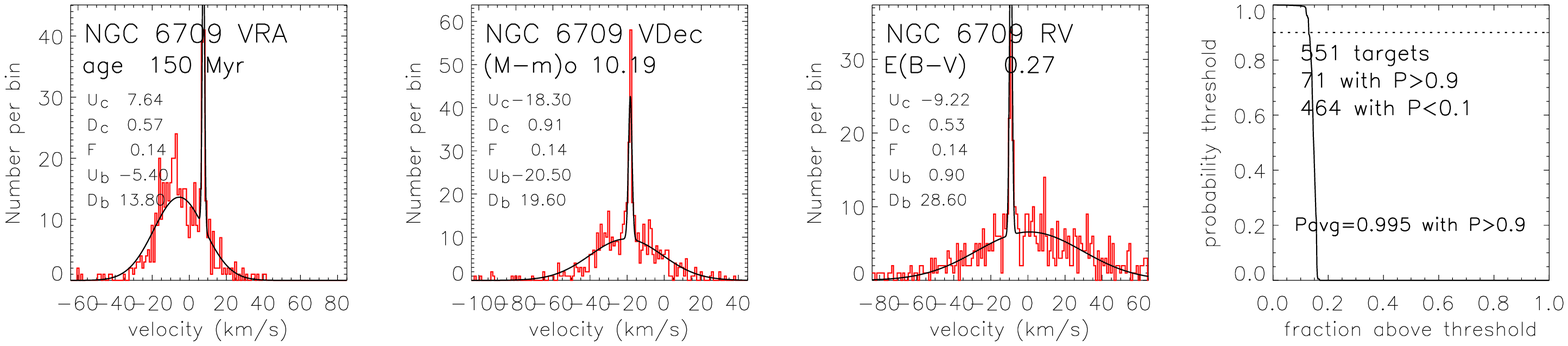}\\
\end{minipage}
\label{figB:25}
\end{figure*}
%%%%%%%%%%%%%%%%%%%%%%%%%%%%%%%%%%%%
\begin{figure*}
\begin{minipage}[t]{0.98\textwidth}
\centering
\includegraphics[width = 145mm]{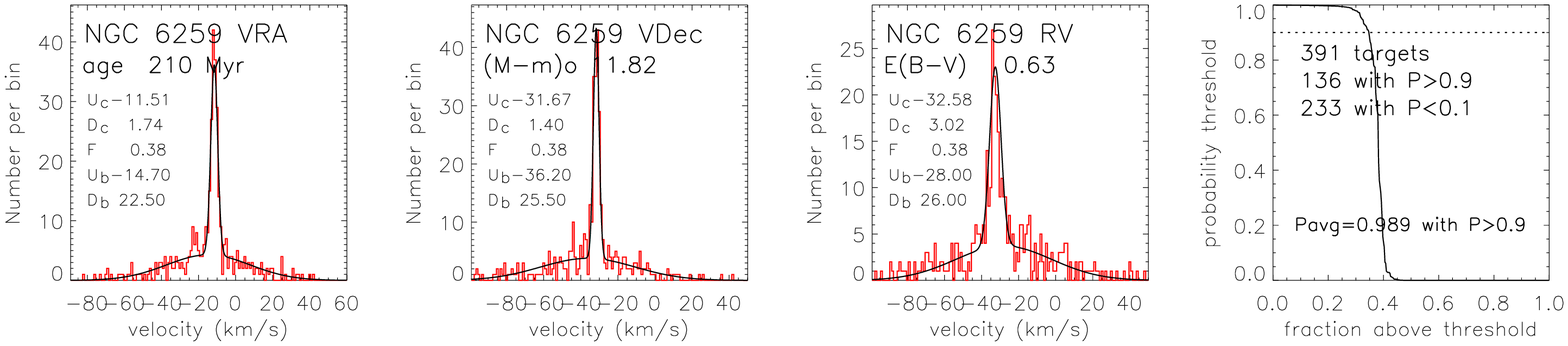}\\
\end{minipage}
\label{figB:26}
\end{figure*}
%%%%%%%%%%%%%%%%%%%%%%%%%%%%%%%%%%%%
\begin{figure*}
\begin{minipage}[t]{0.98\textwidth}
\centering
\includegraphics[width = 145mm]{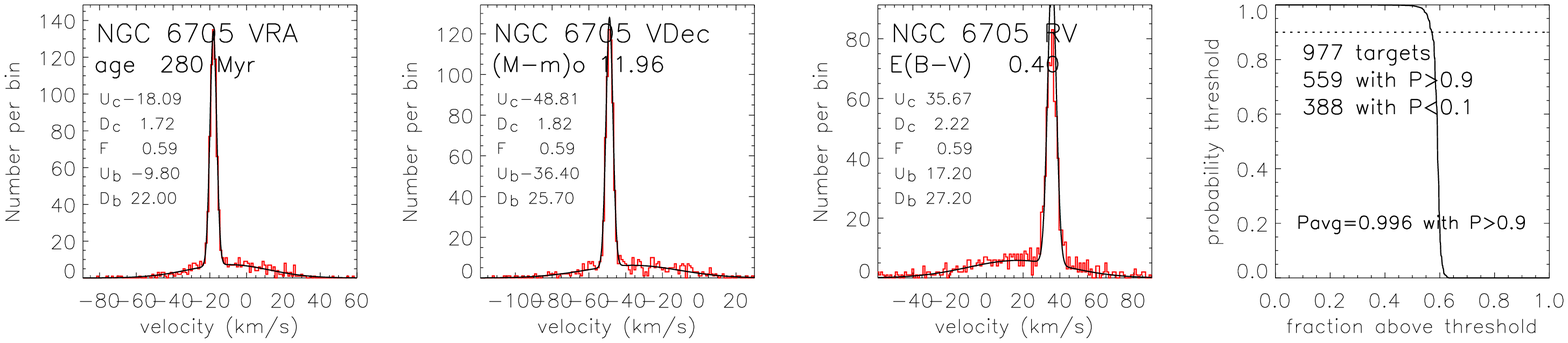}\\
\end{minipage}
\label{figB:27}
\end{figure*}
%%%%%%%%%%%%%%%%%%%%%%%%%%%%%%%%%%%%
\begin{figure*}
\begin{minipage}[t]{0.98\textwidth}
\centering
\includegraphics[width = 145mm]{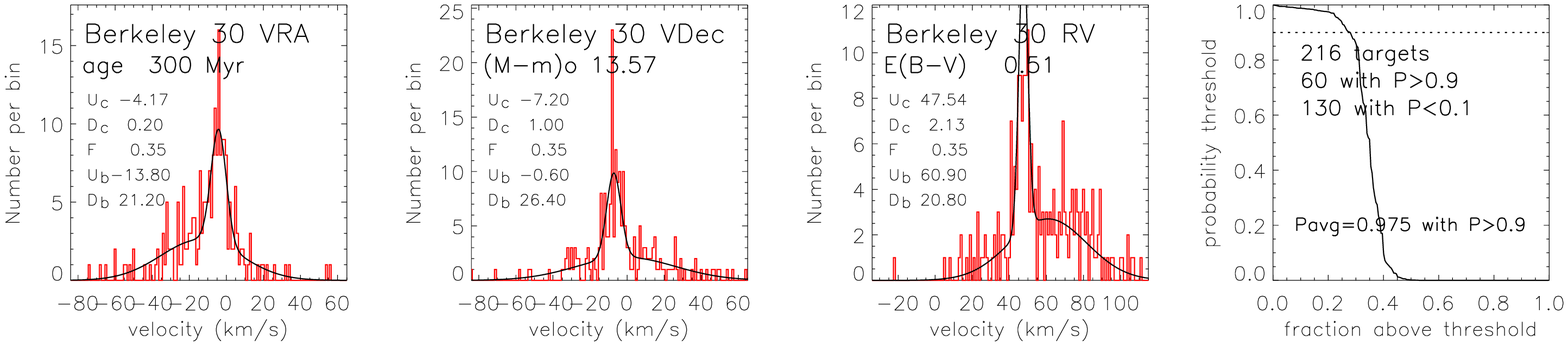}\\
\end{minipage}
\label{figB:28}
\end{figure*}
%%%%%%%%%%%%%%%%%%%%%%%%%%%%%%%%%%%%
\begin{figure*}
\begin{minipage}[t]{0.98\textwidth}
\centering
\includegraphics[width = 145mm]{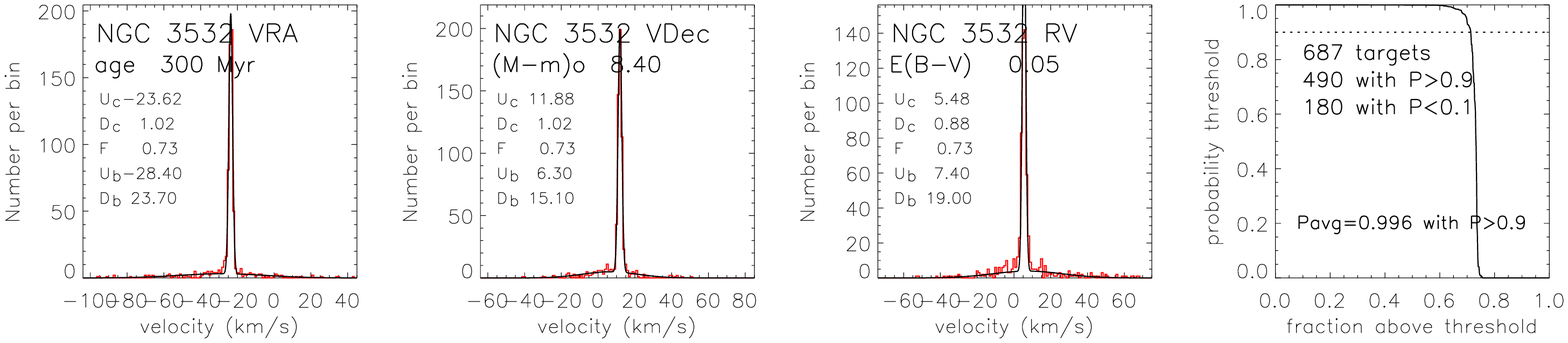}\\
\end{minipage}
\label{figB:29}
\end{figure*}
%%%%%%%%%%%%%%%%%%%%%%%%%%%%%%%%%%%%
\begin{figure*}
\begin{minipage}[t]{0.98\textwidth}
\centering
\includegraphics[width = 145mm]{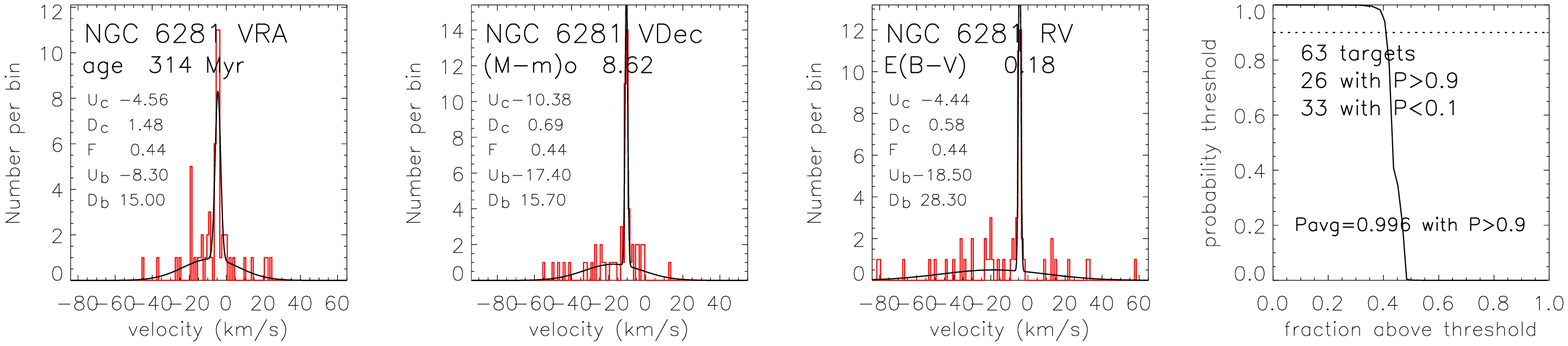}\\
\end{minipage}
\label{figB:30}
\end{figure*}
%%%%%%%%%%%%%%%%%%%%%%%%%%%%%%%%%%%%
\clearpage
\newpage
%%%%%%%%%%%%%%%%%%%%%%%%%%%%%%%%%%%%
\begin{figure*}
\begin{minipage}[t]{0.98\textwidth}
\centering
\includegraphics[width = 145mm]{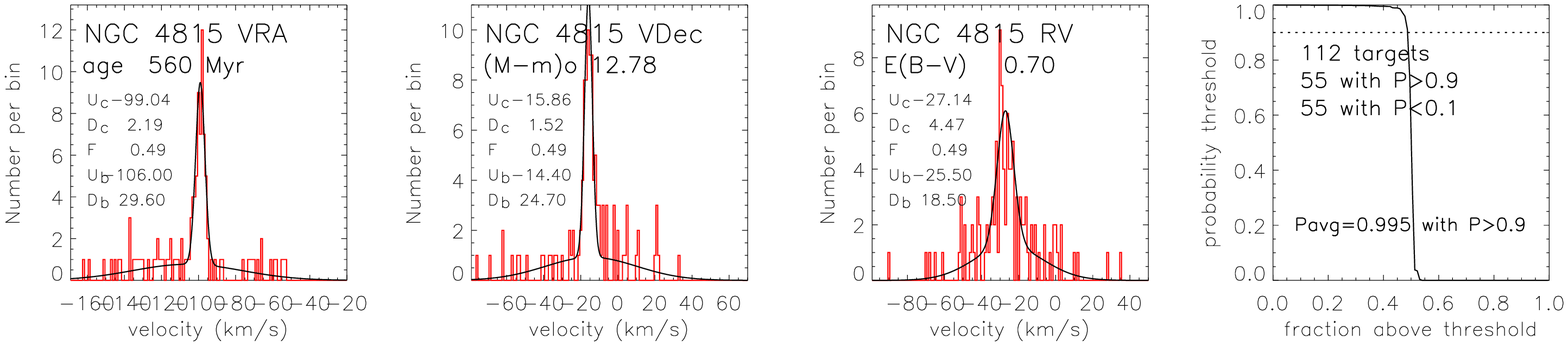}\\
\end{minipage}
\label{figB:31}
\end{figure*}
%%%%%%%%%%%%%%%%%%%%%%%%%%%%%%%%%%%%
\begin{figure*}
\begin{minipage}[t]{0.98\textwidth}
\centering
\includegraphics[width = 145mm]{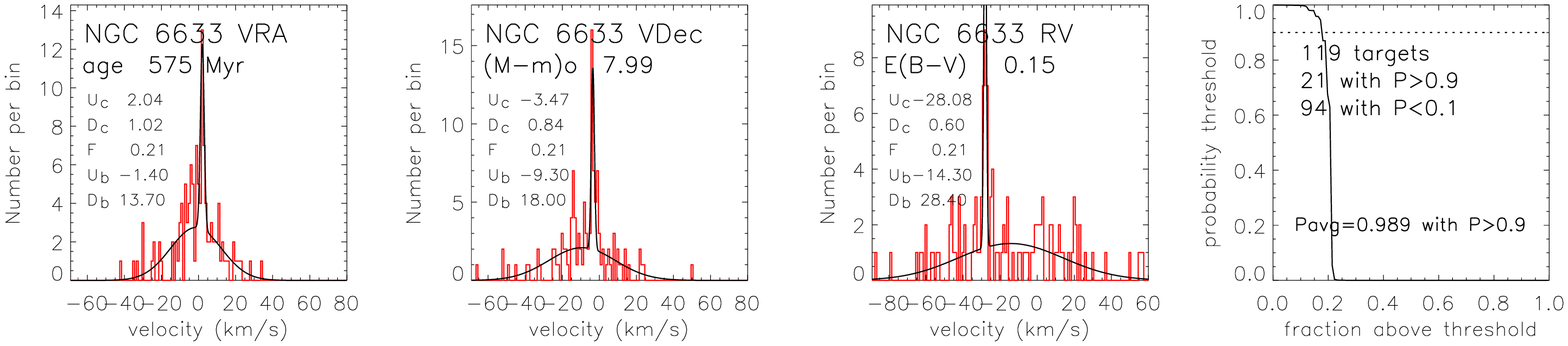}\\
\end{minipage}
\label{figB:32}
\end{figure*}
%%%%%%%%%%%%%%%%%%%%%%%%%%%%%%%%%%%%
\begin{figure*}
\begin{minipage}[t]{0.98\textwidth}
\centering
\includegraphics[width = 145mm]{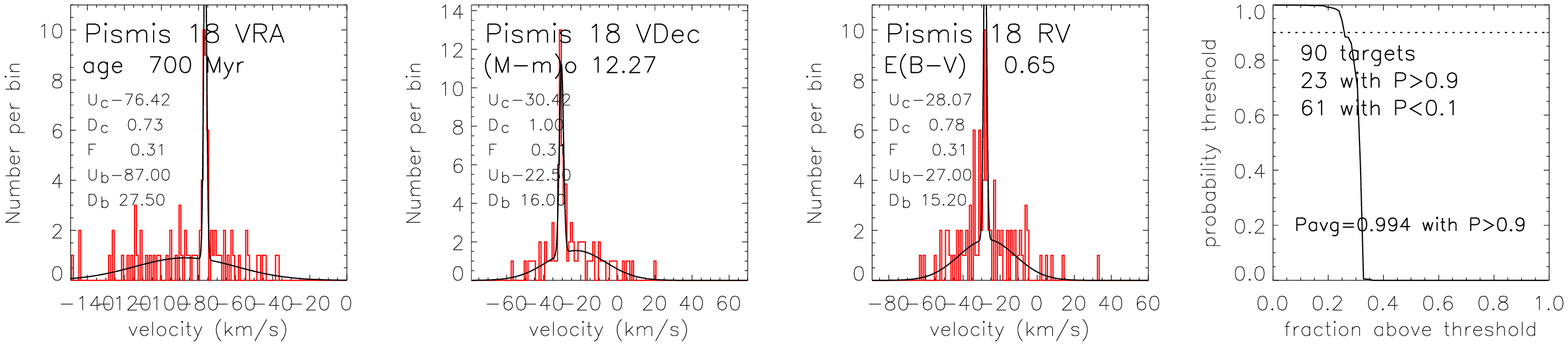}\\
\end{minipage}
\label{figB:33}
\end{figure*}
%%%%%%%%%%%%%%%%%%%%%%%%%%%%%%%%%%%%
\begin{figure*}
\begin{minipage}[t]{0.98\textwidth}
\centering
\includegraphics[width = 145mm]{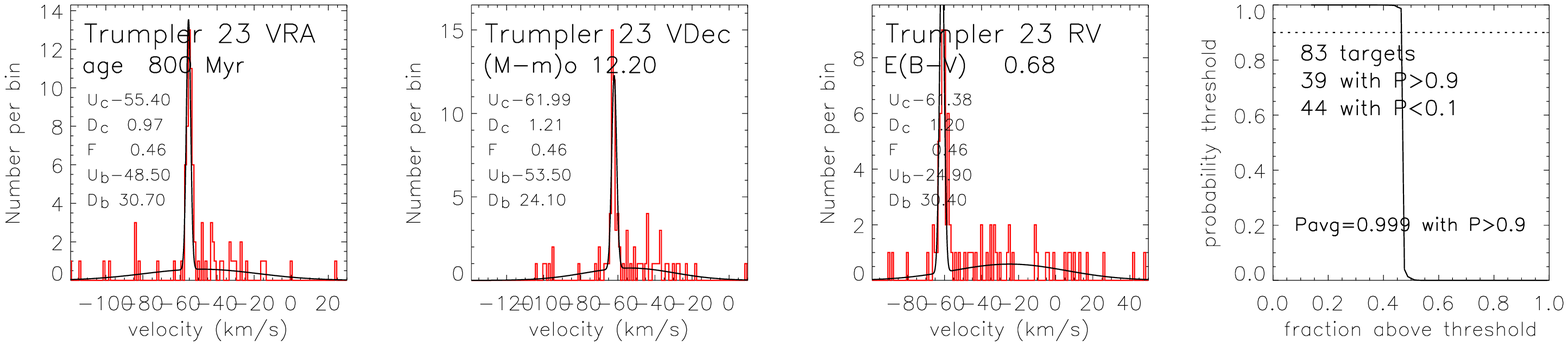}\\
\end{minipage}
\label{figB:34}
\end{figure*}
%%%%%%%%%%%%%%%%%%%%%%%%%%%%%%%%%%%%
\begin{figure*}
\begin{minipage}[t]{0.98\textwidth}
\centering
\includegraphics[width = 145mm]{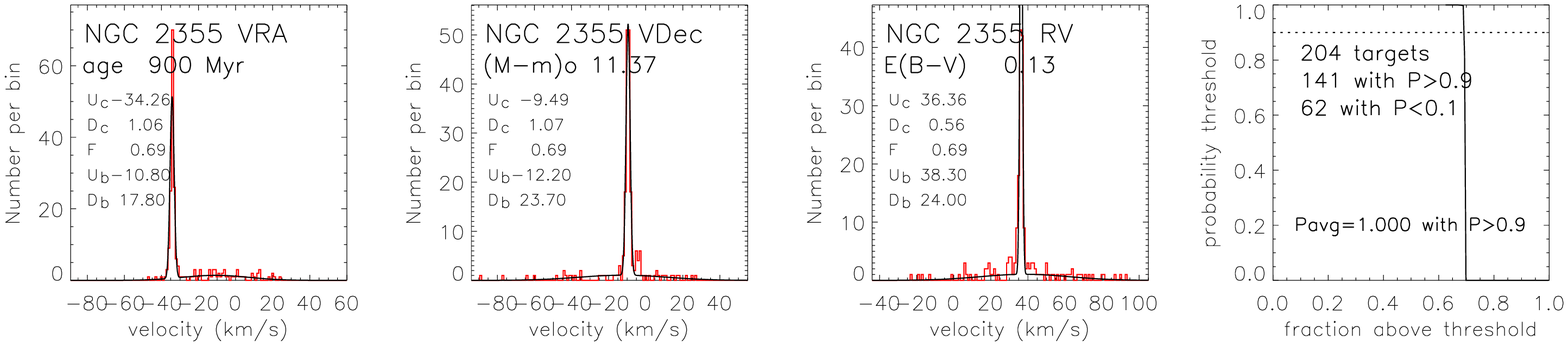}\\
\end{minipage}
\label{figB:35}
\end{figure*}
%%%%%%%%%%%%%%%%%%%%%%%%%%%%%%%%%%%%
\begin{figure*}
\begin{minipage}[t]{0.98\textwidth}
\centering
\includegraphics[width = 145mm]{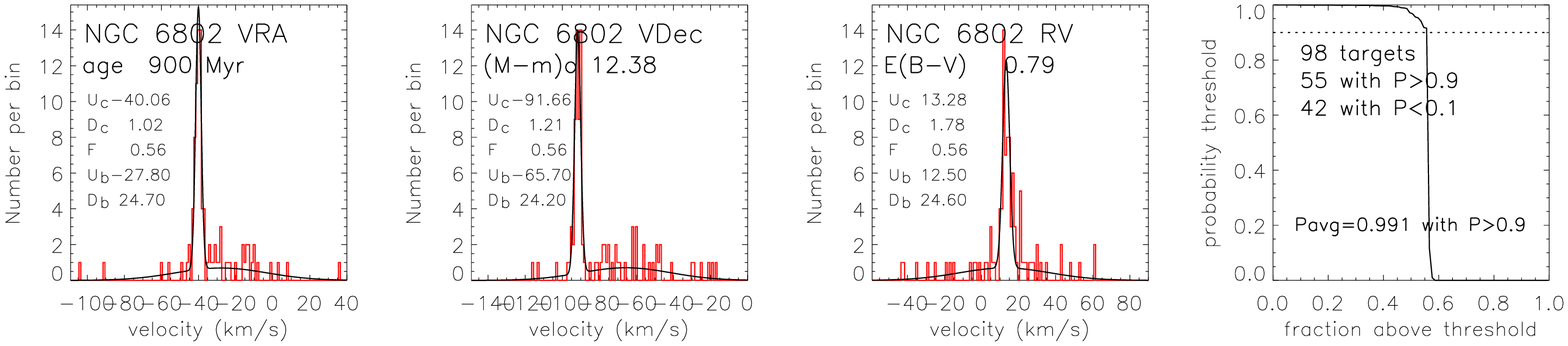}\\
\end{minipage}
\label{figB:36}
\end{figure*}
%%%%%%%%%%%%%%%%%%%%%%%%%%%%%%%%%%%%
\clearpage
\newpage
%%%%%%%%%%%%%%%%%%%%%%%%%%%%%%%%%%%%
\begin{figure*}
\begin{minipage}[t]{0.98\textwidth}
\centering
\includegraphics[width = 145mm]{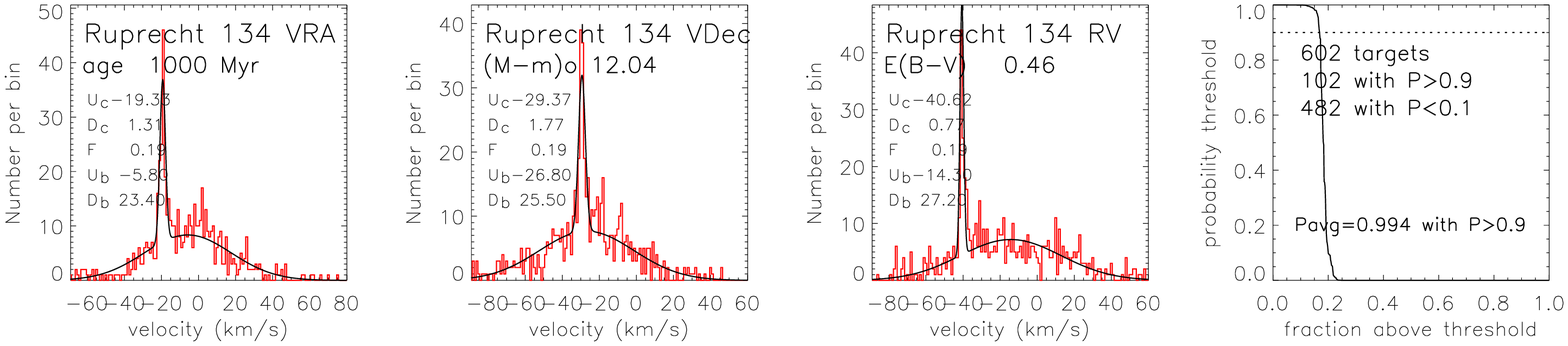}\\
\end{minipage}
\label{figB:37}
\end{figure*}
%%%%%%%%%%%%%%%%%%%%%%%%%%%%%%%%%%%%
\begin{figure*}
\begin{minipage}[t]{0.98\textwidth}
\centering
\includegraphics[width = 145mm]{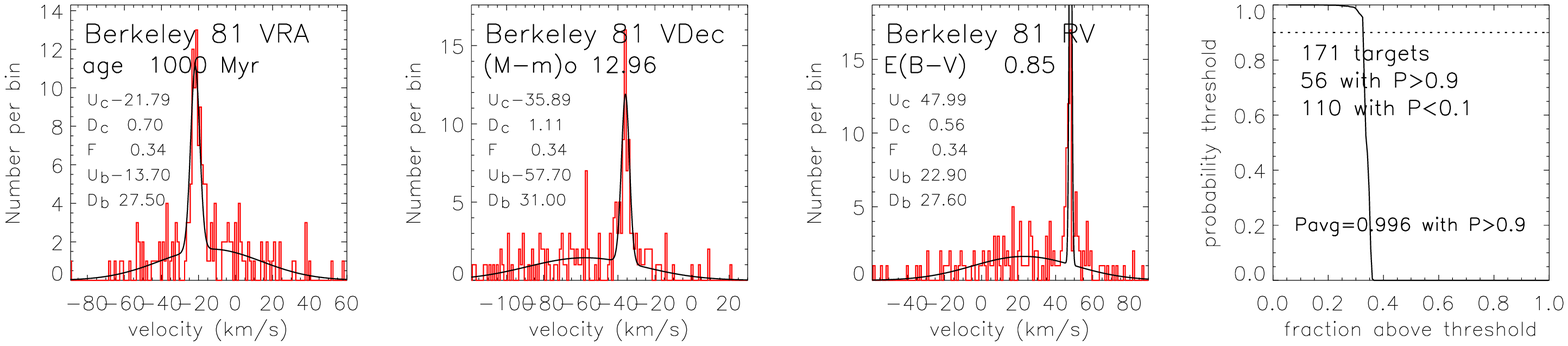}\\
\end{minipage}
\label{figB:38}
\end{figure*}
%%%%%%%%%%%%%%%%%%%%%%%%%%%%%%%%%%%%
\begin{figure*}
\begin{minipage}[t]{0.98\textwidth}
\centering
\includegraphics[width = 145mm]{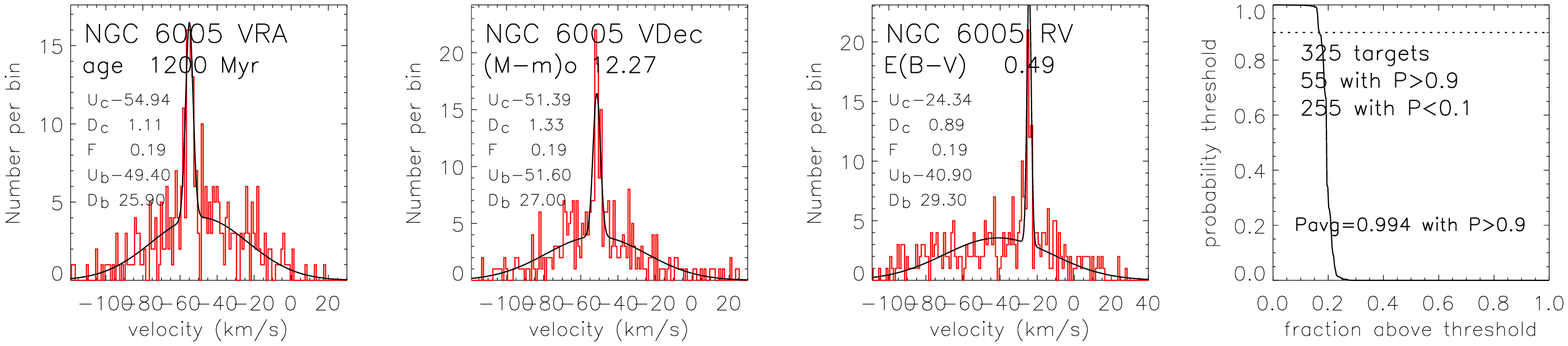}\\
\end{minipage}
\label{figB:39}
\end{figure*}
%%%%%%%%%%%%%%%%%%%%%%%%%%%%%%%%%%%%
\begin{figure*}
\begin{minipage}[t]{0.98\textwidth}
\centering
\includegraphics[width = 145mm]{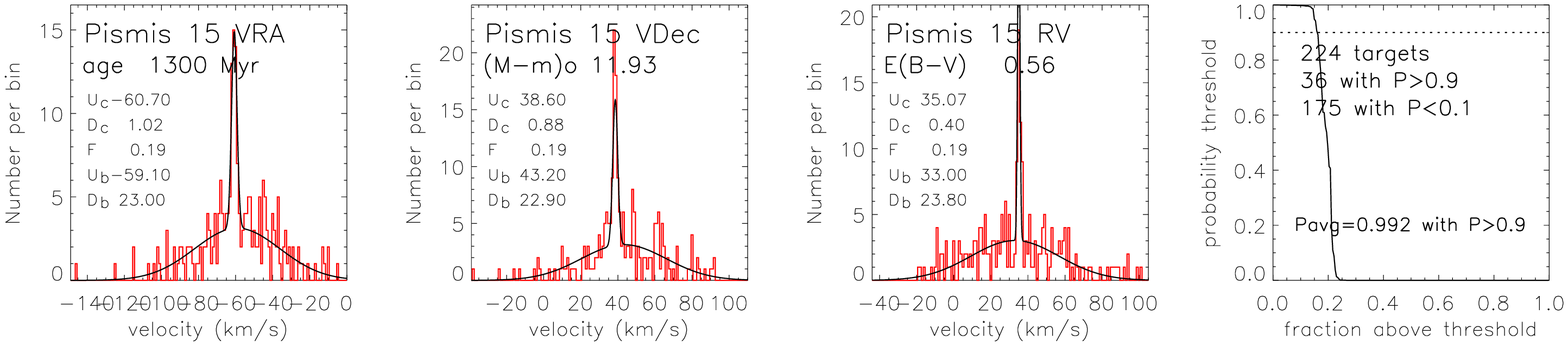}\\
\end{minipage}
\label{figB:40}
\end{figure*}
%%%%%%%%%%%%%%%%%%%%%%%%%%%%%%%%%%%%
\begin{figure*}
\begin{minipage}[t]{0.98\textwidth}
\centering
\includegraphics[width = 145mm]{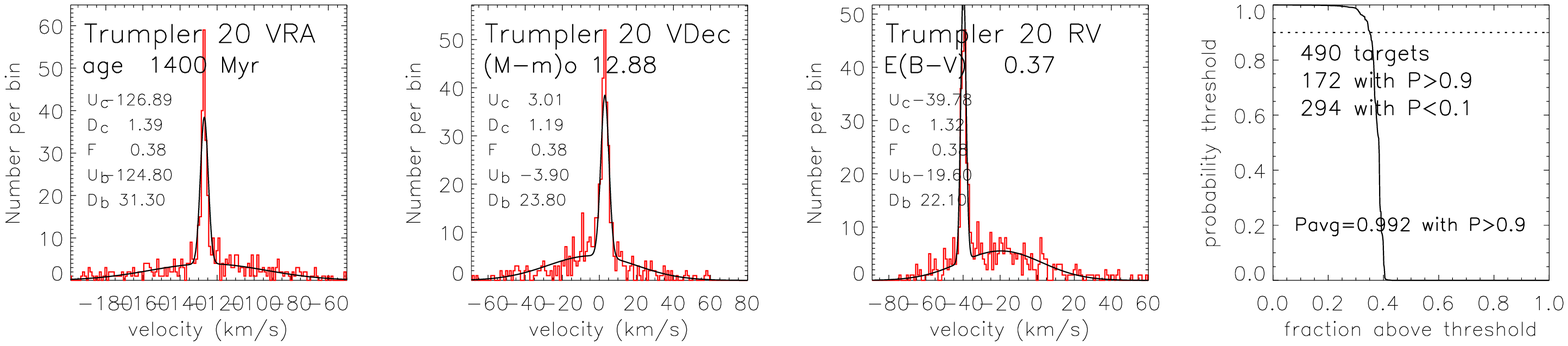}\\
\end{minipage}
\label{figB:41}
\end{figure*}
%%%%%%%%%%%%%%%%%%%%%%%%%%%%%%%%%%%%
\begin{figure*}
\begin{minipage}[t]{0.98\textwidth}
\centering
\includegraphics[width = 145mm]{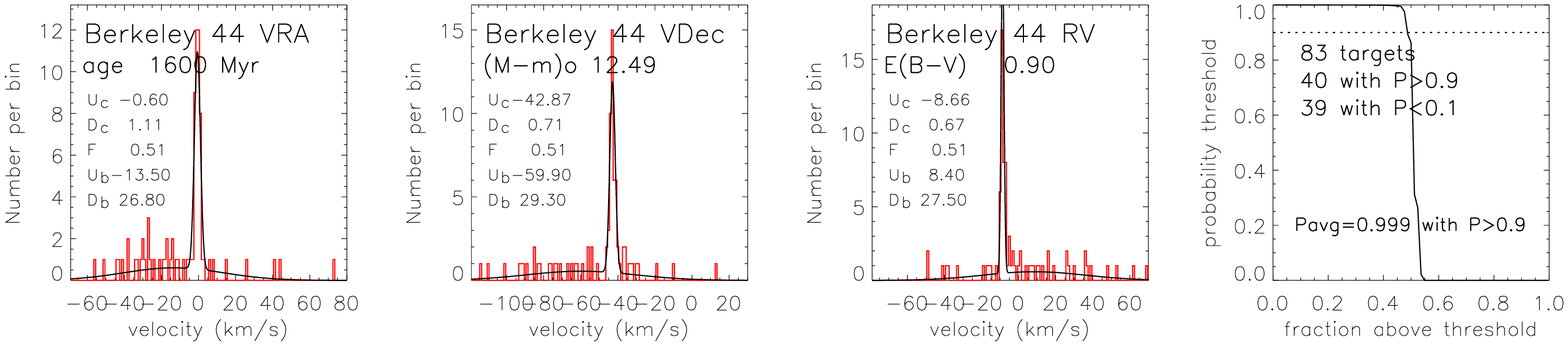}\\
\end{minipage}
\label{figB:42}
\end{figure*}
%%%%%%%%%%%%%%%%%%%%%%%%%%%%%%%%%%%%
\begin{figure*}
\begin{minipage}[t]{0.98\textwidth}
\centering
\includegraphics[width = 145mm]{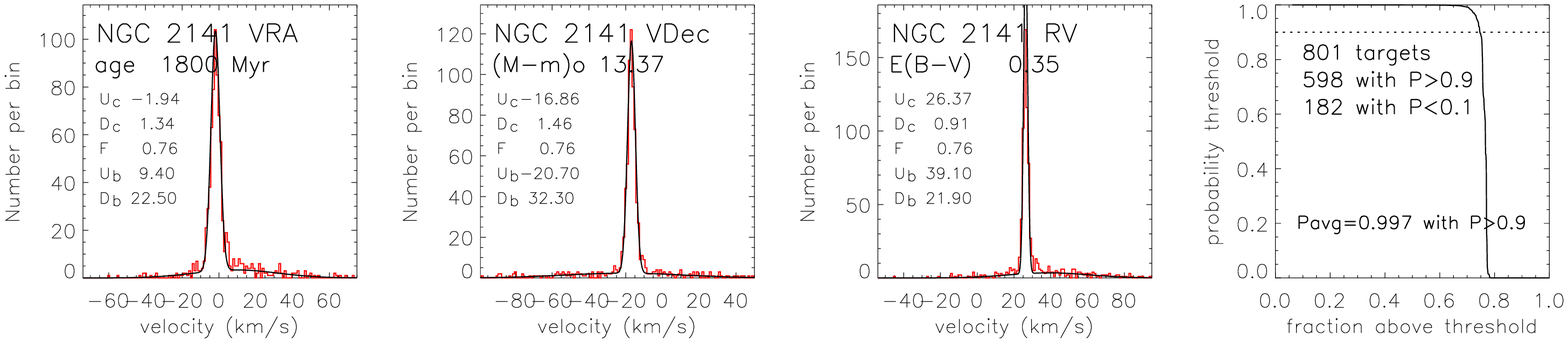}\\
\end{minipage}
\label{figB:43}
\end{figure*}
%%%%%%%%%%%%%%%%%%%%%%%%%%%%%%%%%%%%
\begin{figure*}
\begin{minipage}[t]{0.98\textwidth}
\centering
\includegraphics[width = 145mm]{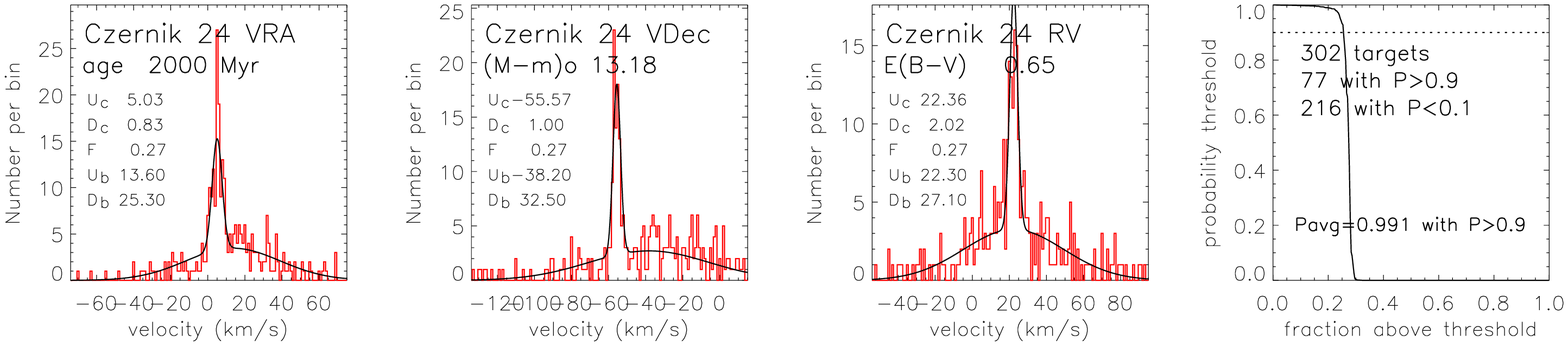}\\
\end{minipage}
\label{figB:44}
\end{figure*}
%%%%%%%%%%%%%%%%%%%%%%%%%%%%%%%%%%%%
\begin{figure*}
\begin{minipage}[t]{0.98\textwidth}
\centering
\includegraphics[width = 145mm]{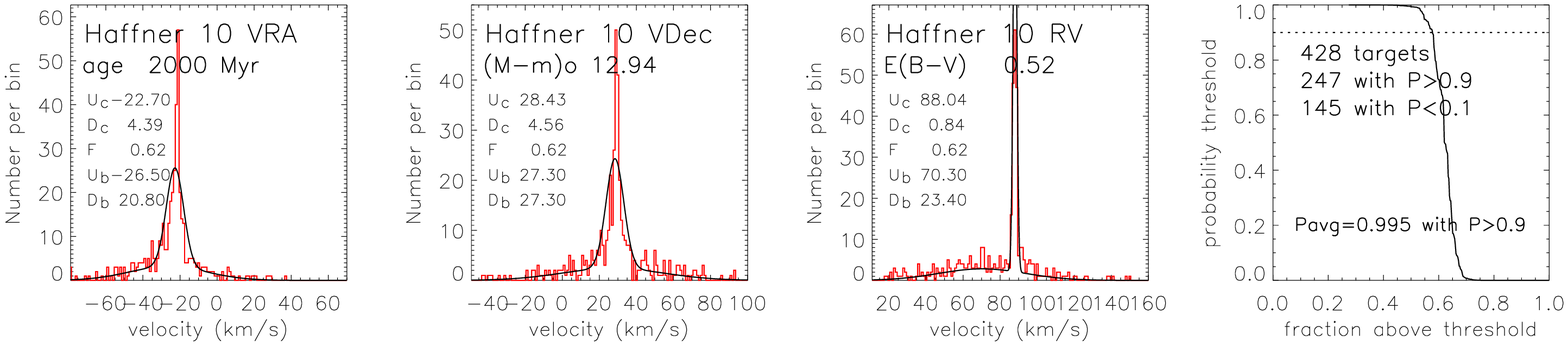}\\
\end{minipage}
\label{figB:45}
\end{figure*}
%%%%%%%%%%%%%%%%%%%%%%%%%%%%%%%%%%%%
\begin{figure*}
\begin{minipage}[t]{0.98\textwidth}
\centering
\includegraphics[width = 145mm]{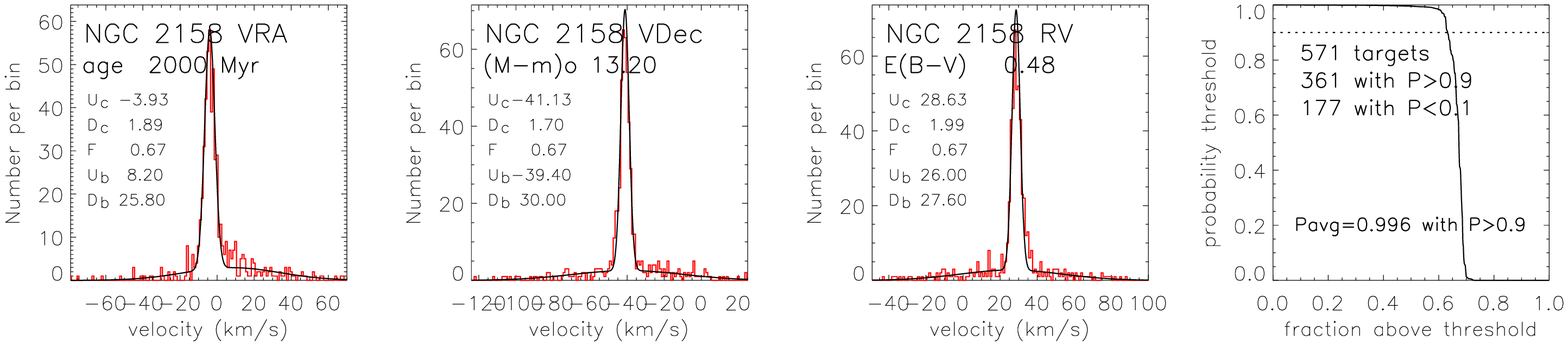}\\
\end{minipage}
\label{figB:46}
\end{figure*}
%%%%%%%%%%%%%%%%%%%%%%%%%%%%%%%%%%%%
\begin{figure*}
\begin{minipage}[t]{0.98\textwidth}
\centering
\includegraphics[width = 145mm]{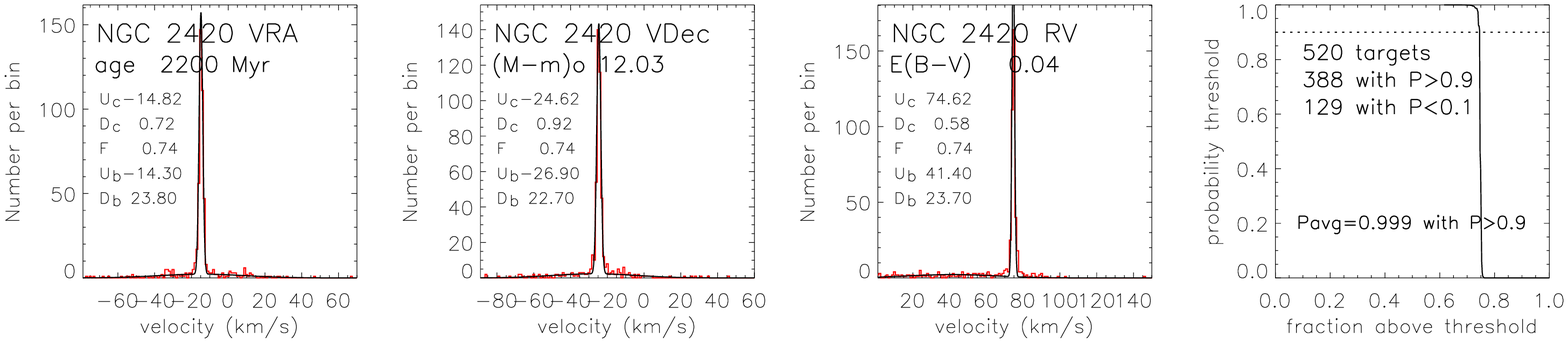}\\
\end{minipage}
\label{figB:47}
\end{figure*}
%%%%%%%%%%%%%%%%%%%%%%%%%%%%%%%%%%%%
\begin{figure*}
\begin{minipage}[t]{0.98\textwidth}
\centering
\includegraphics[width = 145mm]{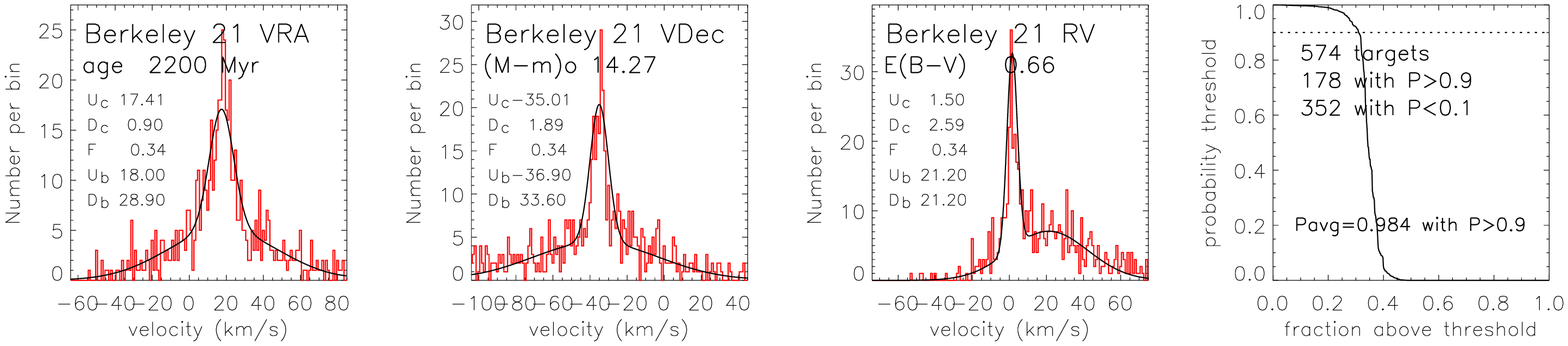}\\
\end{minipage}
\label{figB:48}
\end{figure*}
%%%%%%%%%%%%%%%%%%%%%%%%%%%%%%%%%%%%
\clearpage
\newpage
%%%%%%%%%%%%%%%%%%%%%%%%%%%%%%%%%%%%
\begin{figure*}
\begin{minipage}[t]{0.98\textwidth}
\centering
\includegraphics[width = 145mm]{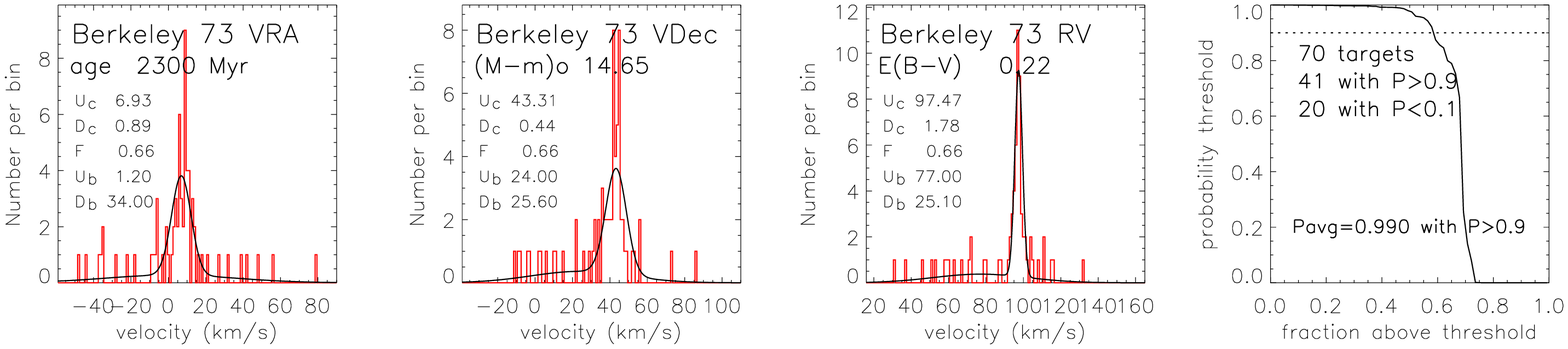}\\
\end{minipage}
\label{figB:49}
\end{figure*}
%%%%%%%%%%%%%%%%%%%%%%%%%%%%%%%%%%%%
\begin{figure*}
\begin{minipage}[t]{0.98\textwidth}
\centering
\includegraphics[width = 145mm]{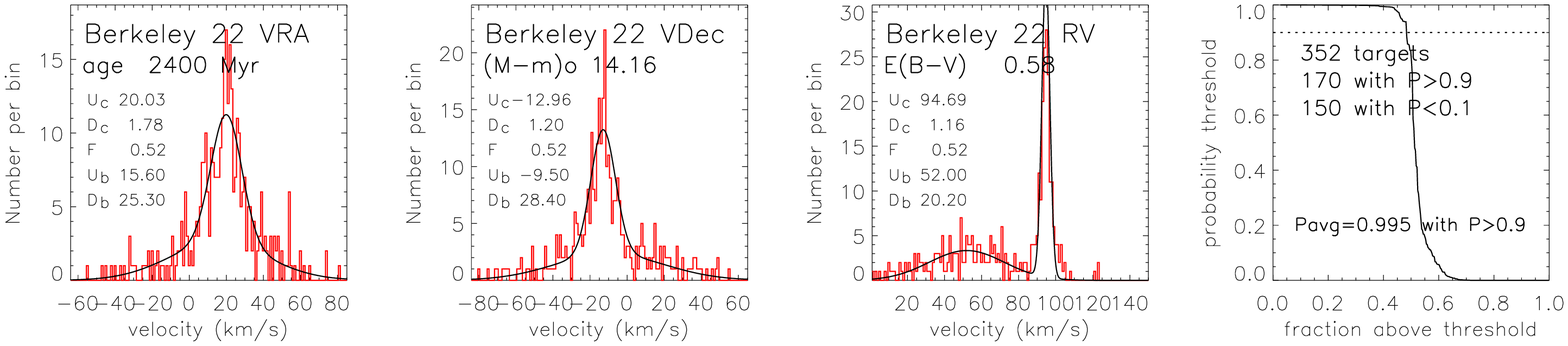}\\
\end{minipage}
\label{figB:50}
\end{figure*}
%%%%%%%%%%%%%%%%%%%%%%%%%%%%%%%%%%%%
\begin{figure*}
\begin{minipage}[t]{0.98\textwidth}
\centering
\includegraphics[width = 145mm]{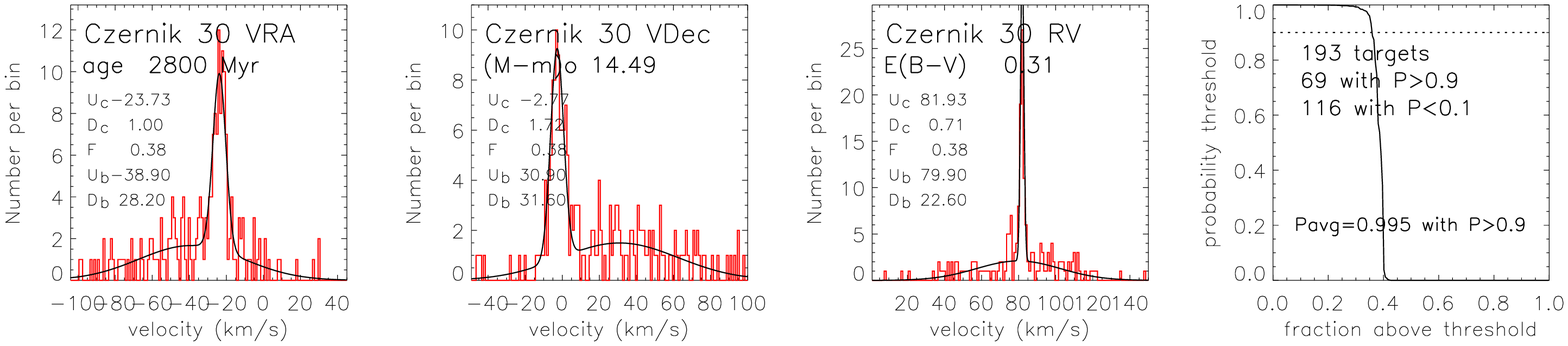}\\
\end{minipage}
\label{figB:51}
\end{figure*}
%%%%%%%%%%%%%%%%%%%%%%%%%%%%%%%%%%%%
\begin{figure*}
\begin{minipage}[t]{0.98\textwidth}
\centering
\includegraphics[width = 145mm]{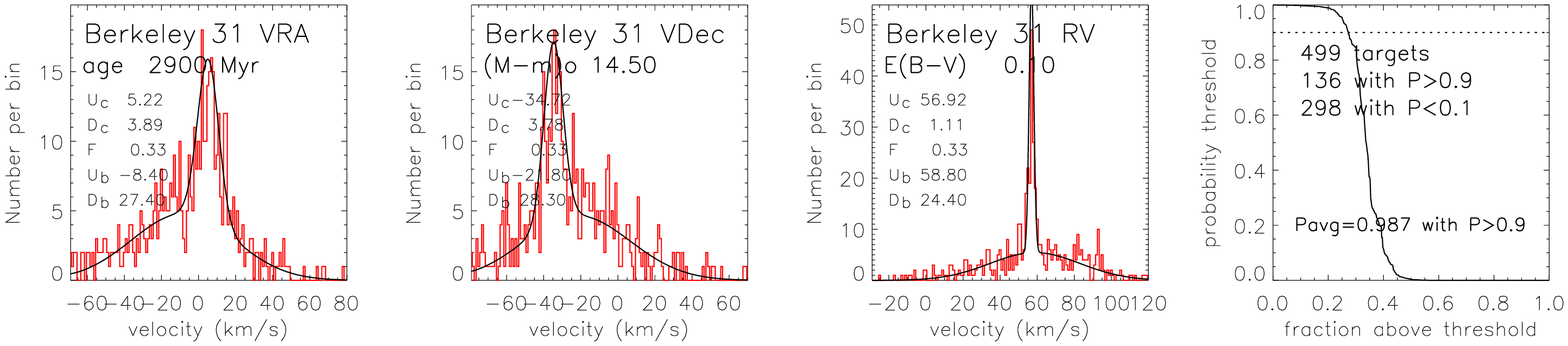}\\
\end{minipage}
\label{figB:52}
\end{figure*}
%%%%%%%%%%%%%%%%%%%%%%%%%%%%%%%%%%%%
\begin{figure*}
\begin{minipage}[t]{0.98\textwidth}
\centering
\includegraphics[width = 145mm]{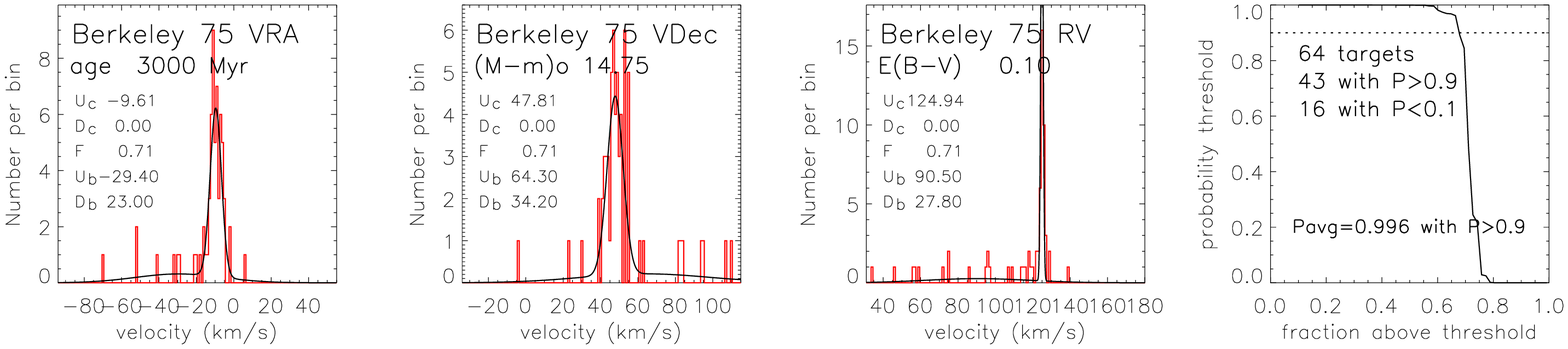}\\
\end{minipage}
\label{figB:53}
\end{figure*}
%%%%%%%%%%%%%%%%%%%%%%%%%%%%%%%%%%%%
\begin{figure*}
\begin{minipage}[t]{0.98\textwidth}
\centering
\includegraphics[width = 145mm]{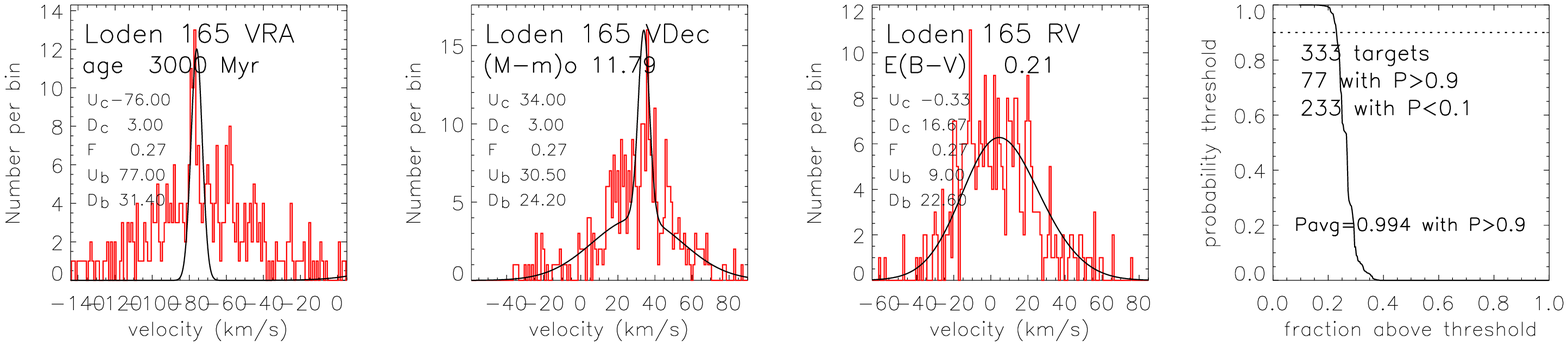}\\
\end{minipage}
\label{figB:54}
\end{figure*}
%%%%%%%%%%%%%%%%%%%%%%%%%%%%%%%%%%%%
\clearpage
\newpage
%%%%%%%%%%%%%%%%%%%%%%%%%%%%%%%%%%%%
\begin{figure*}
\begin{minipage}[t]{0.98\textwidth}
\centering
\includegraphics[width = 145mm]{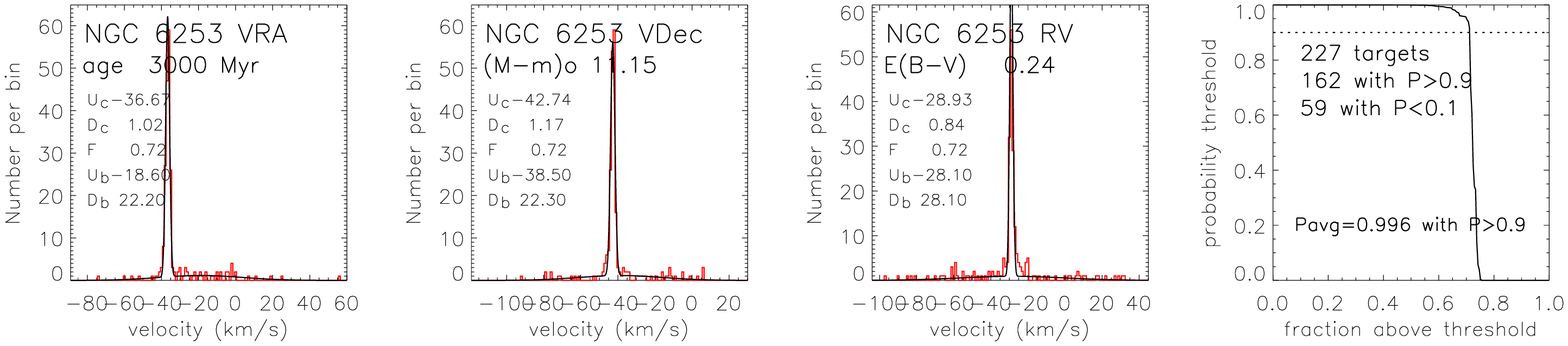}\\
\end{minipage}
\label{figB:55}
\end{figure*}
%%%%%%%%%%%%%%%%%%%%%%%%%%%%%%%%%%%%
\begin{figure*}
\begin{minipage}[t]{0.98\textwidth}
\centering
\includegraphics[width = 145mm]{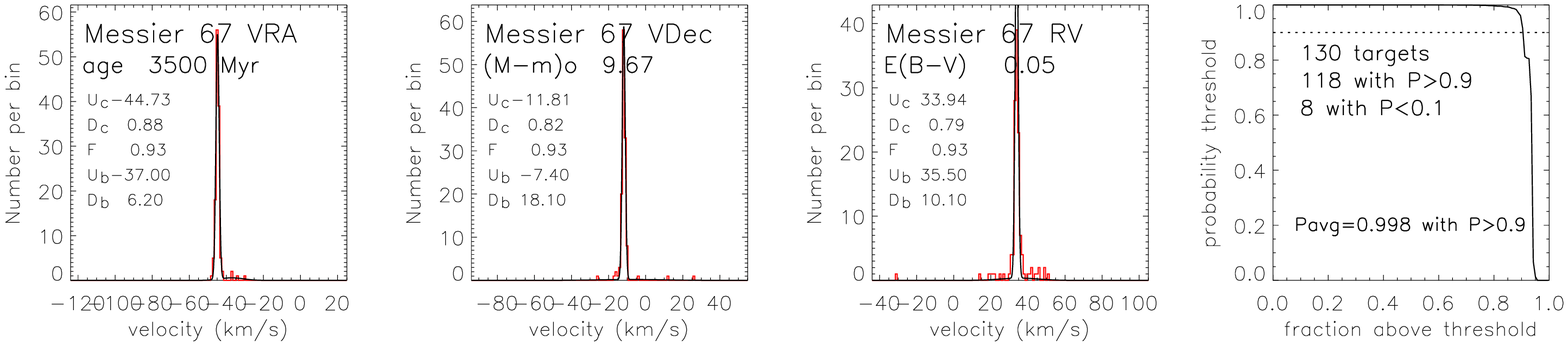}\\
\end{minipage}
\label{figB:56}
\end{figure*}
%%%%%%%%%%%%%%%%%%%%%%%%%%%%%%%%%%%%
\begin{figure*}
\begin{minipage}[t]{0.98\textwidth}
\centering
\includegraphics[width = 145mm]{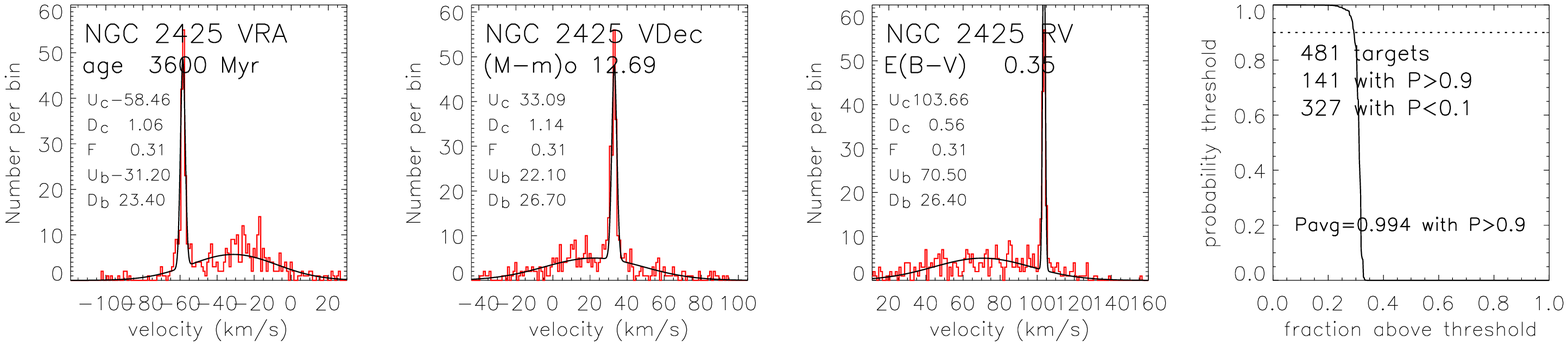}\\
\end{minipage}
\label{figB:57}
\end{figure*}
%%%%%%%%%%%%%%%%%%%%%%%%%%%%%%%%%%%%
\begin{figure*}
\begin{minipage}[t]{0.98\textwidth}
\centering
\includegraphics[width = 145mm]{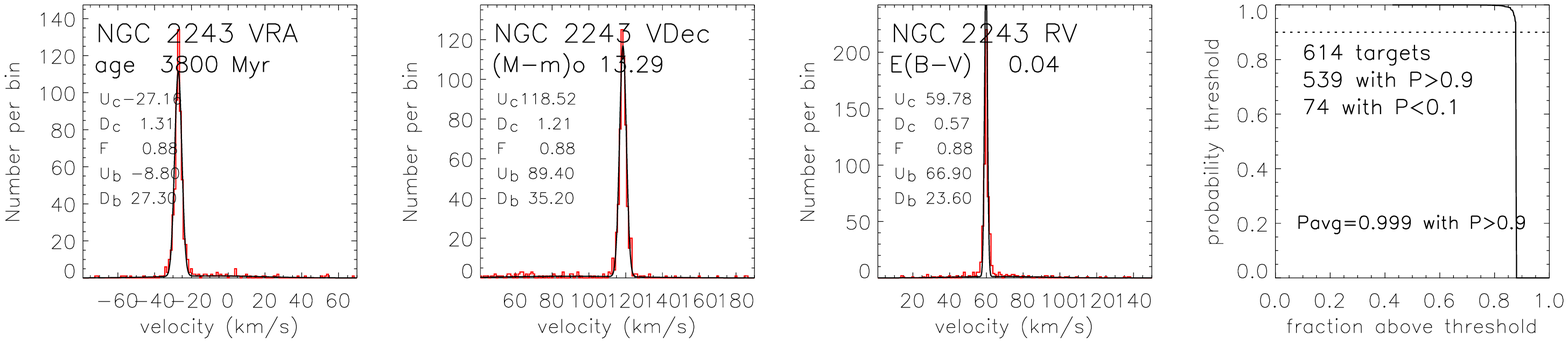}\\
\end{minipage}
\label{figB:58}
\end{figure*}
%%%%%%%%%%%%%%%%%%%%%%%%%%%%%%%%%%%%
\begin{figure*}
\begin{minipage}[t]{0.98\textwidth}
\centering
\includegraphics[width = 145mm]{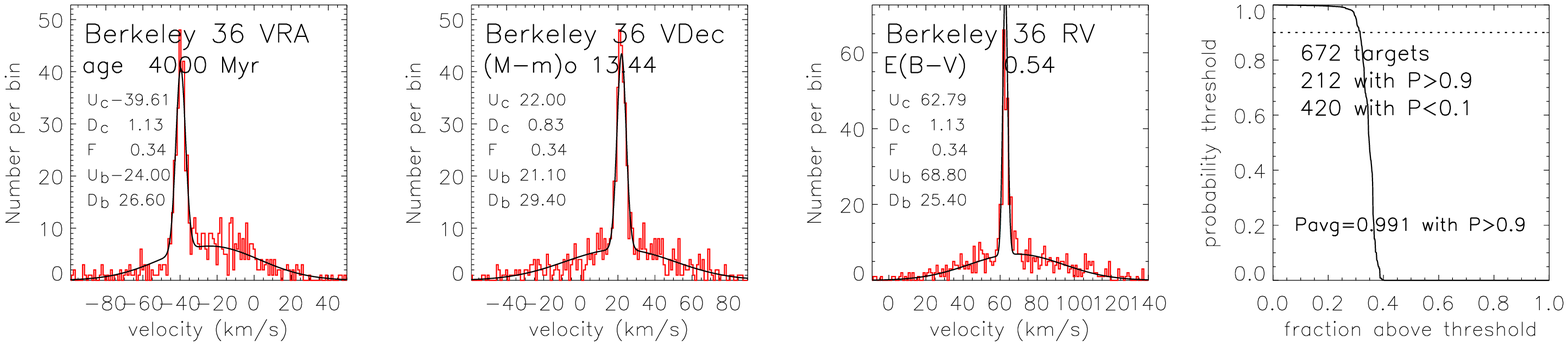}\\
\end{minipage}
\label{figB:59}
\end{figure*}
%%%%%%%%%%%%%%%%%%%%%%%%%%%%%%%%%%%%
\begin{figure*}
\begin{minipage}[t]{0.98\textwidth}
\centering
\includegraphics[width = 145mm]{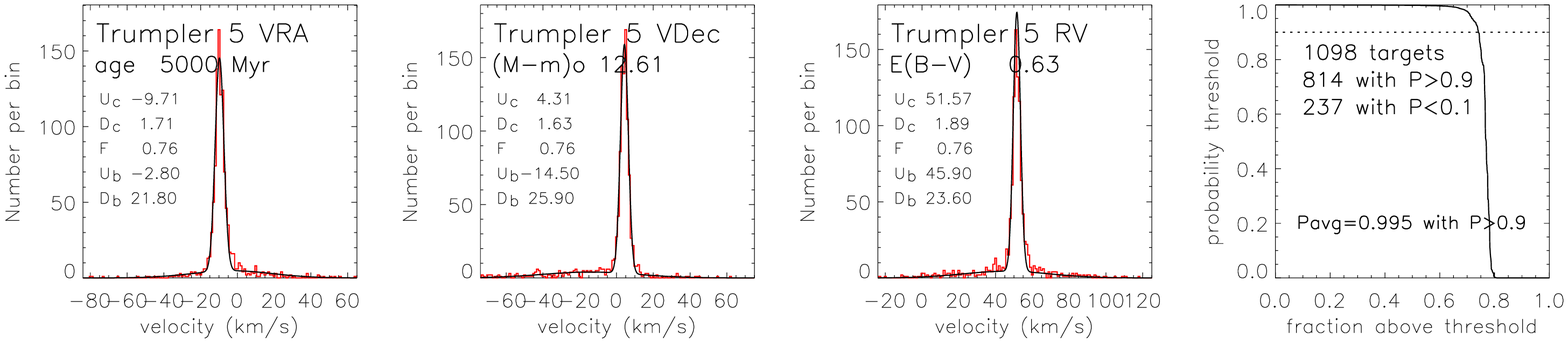}\\
\end{minipage}
\label{figB:60}
\end{figure*}
%%%%%%%%%%%%%%%%%%%%%%%%%%%%%%%%%%%%
\clearpage
\newpage
%%%%%%%%%%%%%%%%%%%%%%%%%%%%%%%%%%%%
\begin{figure*}
\begin{minipage}[t]{0.98\textwidth}
\centering
\includegraphics[width = 145mm]{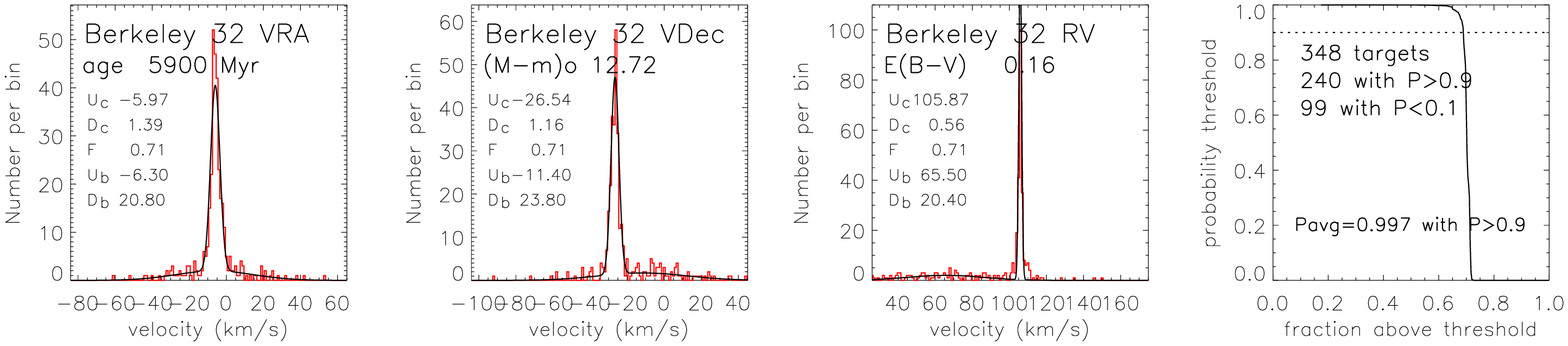}\\
\end{minipage}
\label{figB:61}
\end{figure*}
%%%%%%%%%%%%%%%%%%%%%%%%%%%%%%%%%%%%
\begin{figure*}
\begin{minipage}[t]{0.98\textwidth}
\centering
\includegraphics[width = 145mm]{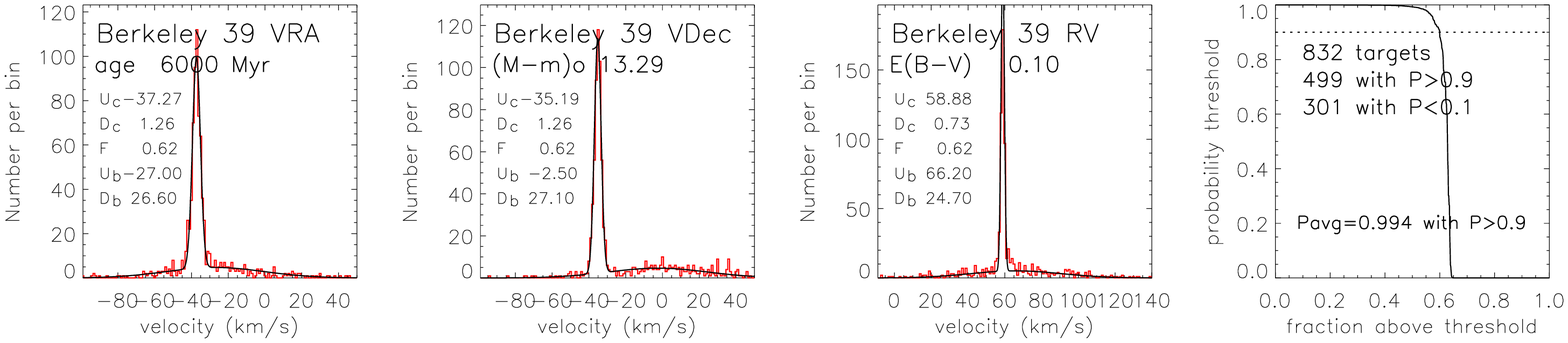}\\
\end{minipage}
\label{figB:62}
\end{figure*}
%%%%%%%%%%%%%%%%%%%%%%%%%%%%%%%%%%%%
\begin{figure*}
\begin{minipage}[t]{0.98\textwidth}
\centering
\includegraphics[width = 145mm]{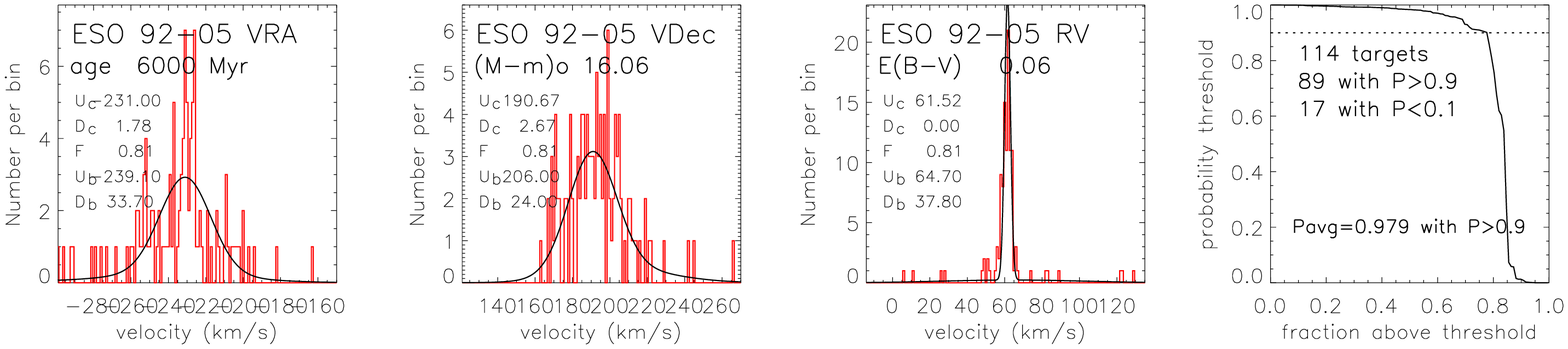}\\
\end{minipage}
\label{figB:63}
\end{figure*}
%%%%%%%%%%%%%%%%%%%%%%%%%%%%%%%%%%%%
\begin{figure*}
\begin{minipage}[t]{0.98\textwidth}
\centering
\includegraphics[width = 145mm]{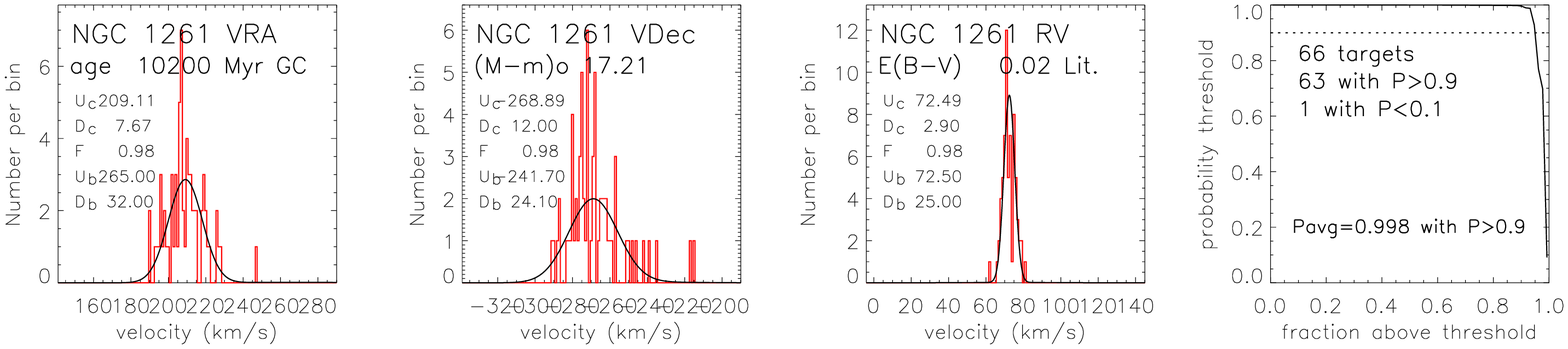}\\
\end{minipage}
\label{figB:64}
\end{figure*}
%%%%%%%%%%%%%%%%%%%%%%%%%%%%%%%%%%%%
\begin{figure*}
\begin{minipage}[t]{0.98\textwidth}
\centering
\includegraphics[width = 145mm]{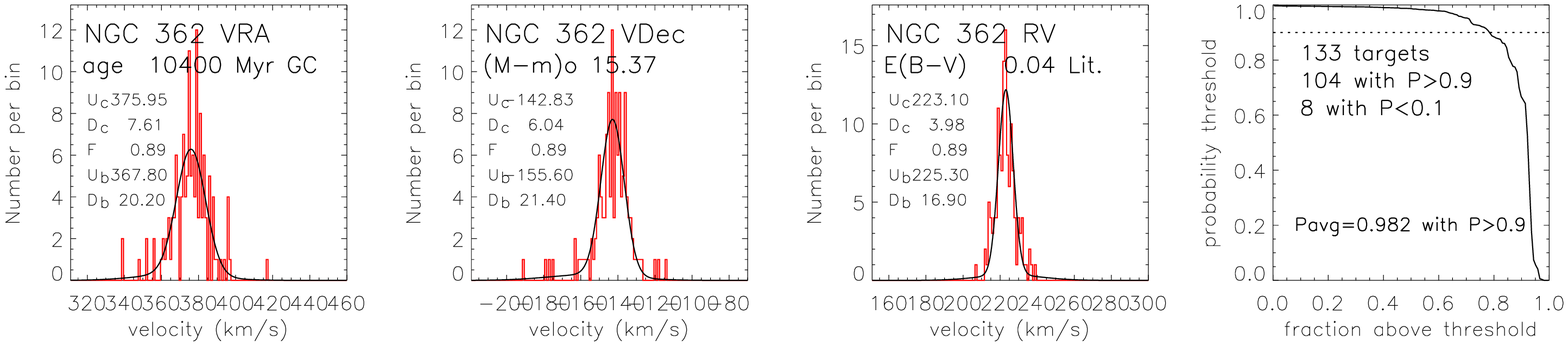}\\
\end{minipage}
\label{figB:65}
\end{figure*}
%%%%%%%%%%%%%%%%%%%%%%%%%%%%%%%%%%%%
\begin{figure*}
\begin{minipage}[t]{0.98\textwidth}
\centering
\includegraphics[width = 145mm]{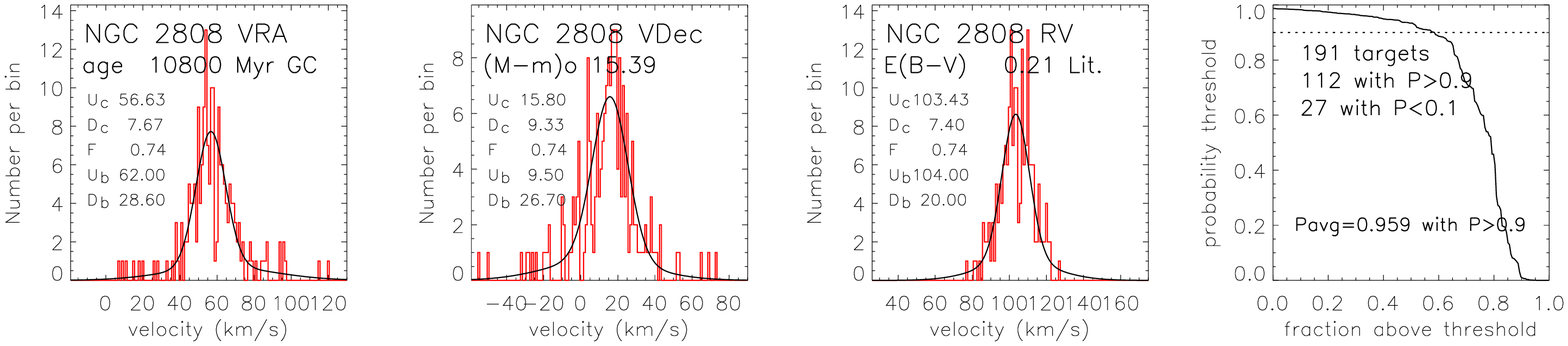}\\
\end{minipage}
\label{figB:66}
\end{figure*}
%%%%%%%%%%%%%%%%%%%%%%%%%%%%%%%%%%%%
\clearpage
\newpage
%%%%%%%%%%%%%%%%%%%%%%%%%%%%%%%%%%%%
\begin{figure*}
\begin{minipage}[t]{0.98\textwidth}
\centering
\includegraphics[width = 145mm]{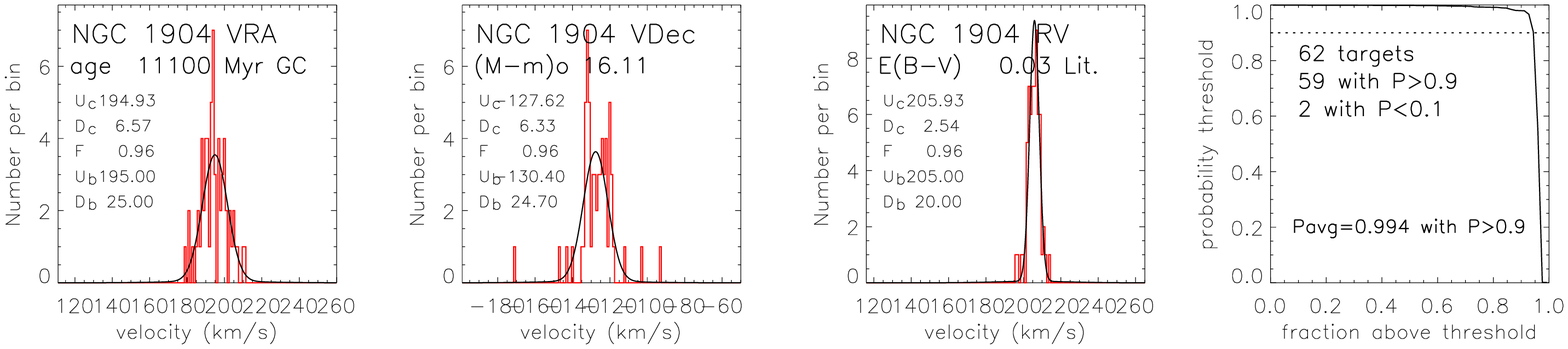}\\
\end{minipage}
\label{figB:67}
\end{figure*}
%%%%%%%%%%%%%%%%%%%%%%%%%%%%%%%%%%%%
\begin{figure*}
\begin{minipage}[t]{0.98\textwidth}
\centering
\includegraphics[width = 145mm]{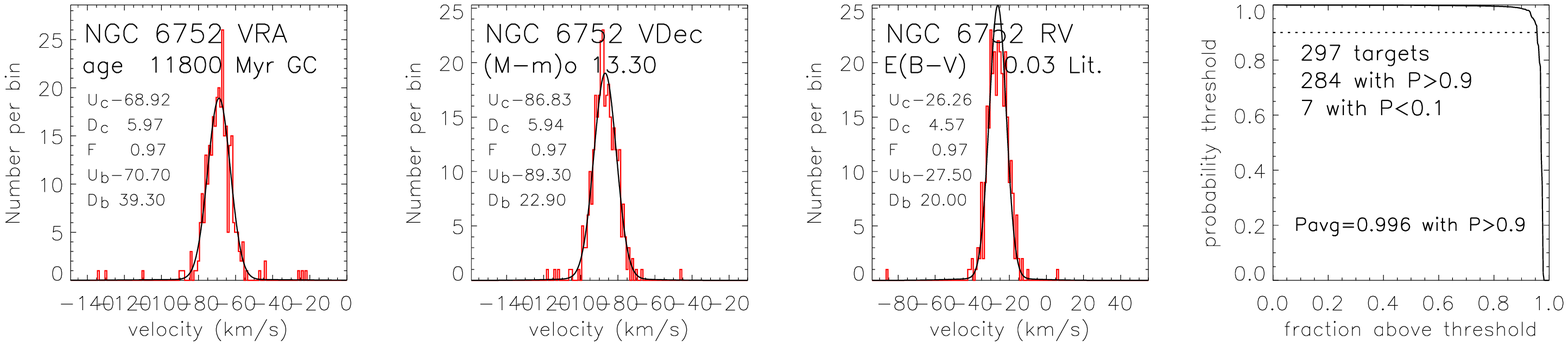}\\
\end{minipage}
\label{figB:68}
\end{figure*}
%%%%%%%%%%%%%%%%%%%%%%%%%%%%%%%%%%%%
\begin{figure*}
\begin{minipage}[t]{0.98\textwidth}
\centering
\includegraphics[width = 145mm]{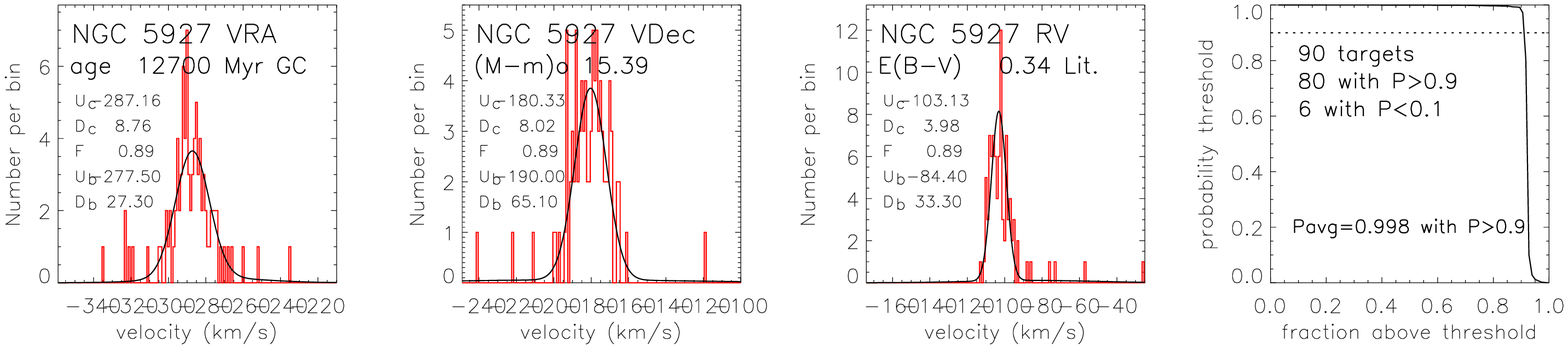}\\
\end{minipage}
\label{figB:69}
\end{figure*}
%%%%%%%%%%%%%%%%%%%%%%%%%%%%%%%%%%%%
\begin{figure*}
\begin{minipage}[t]{0.98\textwidth}
\centering
\includegraphics[width = 145mm]{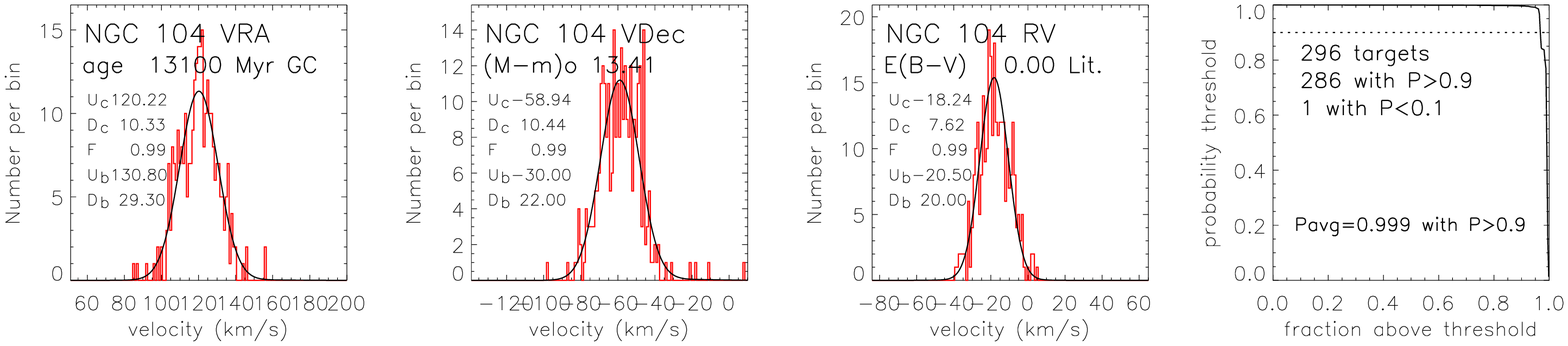}\\
\end{minipage}
\label{figB:70}
\end{figure*}
%%%%%%%%%%%%%%%%%%%%%%%%%%%%%%%%%%%%
%%%%%%%%%%%%%%%%%%%%%%%%%%%%%%%%%%%%
%%%%%%%%%%%%%%%%%%%%%%%%%%%%%%%%%%%%
%%%%%%%%%%%%%%%%%%%%%%%%%%%%%%%%%%%%
%%%%%%%%%%%%%%%%%%%%%%%%%%%%%%%%%%%%
%%%%%%%%%%%%%%%%%%%%%%%%%%%%%%%%%%%%
%%%%%%%%%%%%%%%%%%%%%%%%%%%%%%%%%%%%%
\clearpage
\newpage
%%%%%%%%%%%%%%%%%%%%%%%%%%%%%%%%%%%%
\begin{figure*}
\begin{minipage}[t]{0.98\textwidth}
\centering
\includegraphics[width = 145mm]{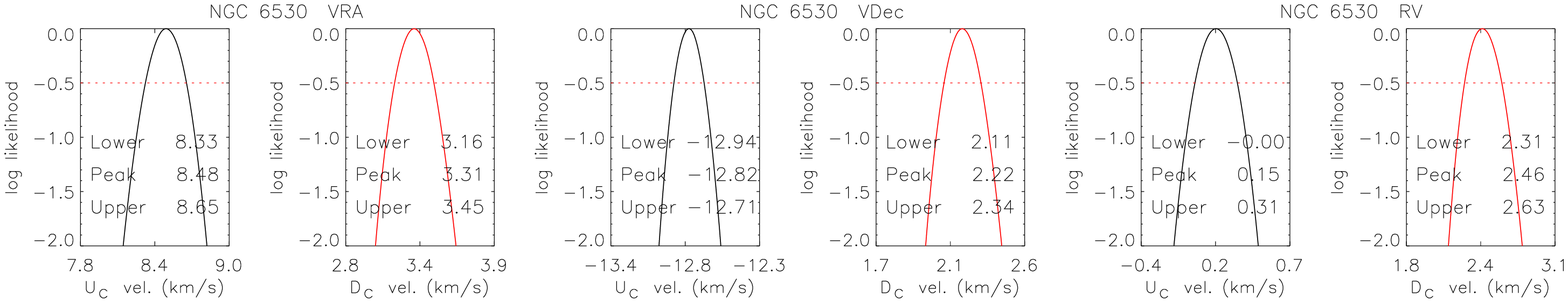}\\
\end{minipage}
\label{figB:1}
\end{figure*}
%%%%%%%%%%%%%%%%%%%%%%%%%%%%%%%%%%%%
\begin{figure*}
\begin{minipage}[t]{0.98\textwidth}
\centering
\includegraphics[width = 145mm]{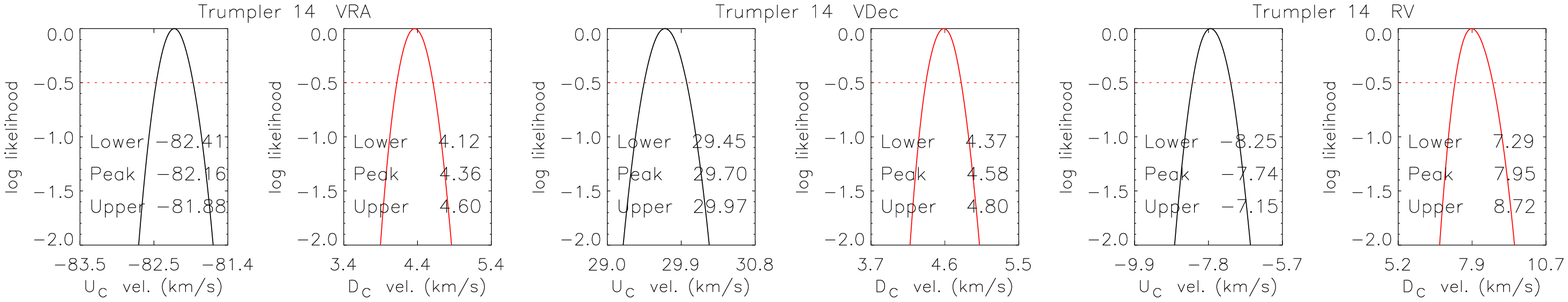}\\
\end{minipage}
\label{figB:2}
\end{figure*}
%%%%%%%%%%%%%%%%%%%%%%%%%%%%%%%%%%%%
\begin{figure*}
\begin{minipage}[t]{0.98\textwidth}
\centering
\includegraphics[width = 145mm]{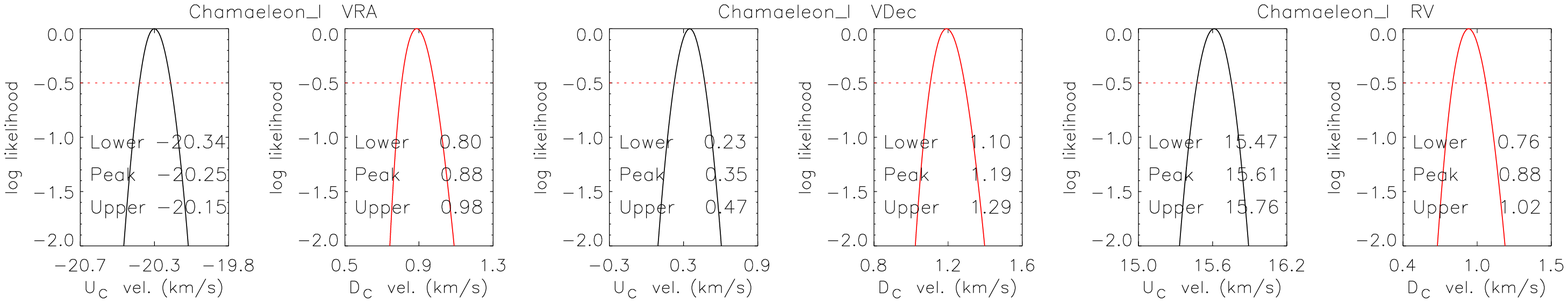}\\
\end{minipage}
\label{figB:3}
\end{figure*}
%%%%%%%%%%%%%%%%%%%%%%%%%%%%%%%%%%%%
\begin{figure*}
\begin{minipage}[t]{0.98\textwidth}
\centering
\includegraphics[width = 145mm]{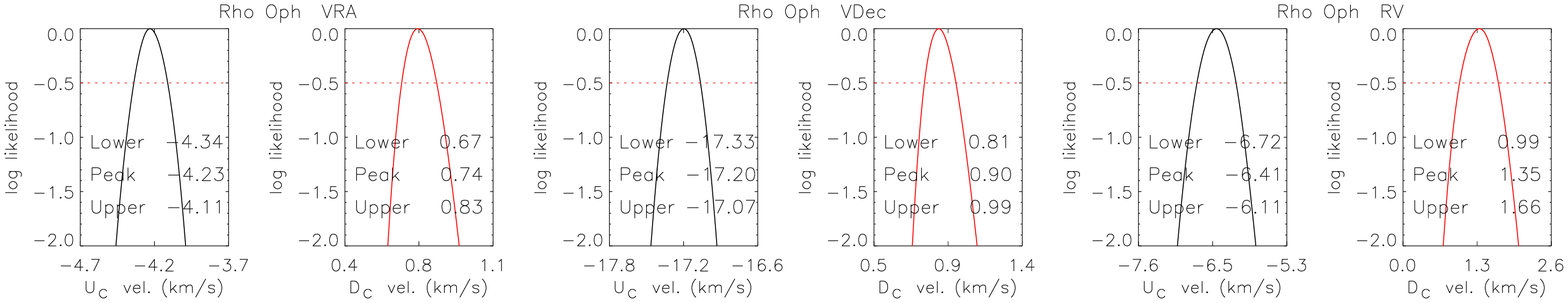}\\
\end{minipage}
\label{figB:4}
\end{figure*}
%%%%%%%%%%%%%%%%%%%%%%%%%%%%%%%%%%%%
\begin{figure*}
\begin{minipage}[t]{0.98\textwidth}
\centering
\includegraphics[width = 145mm]{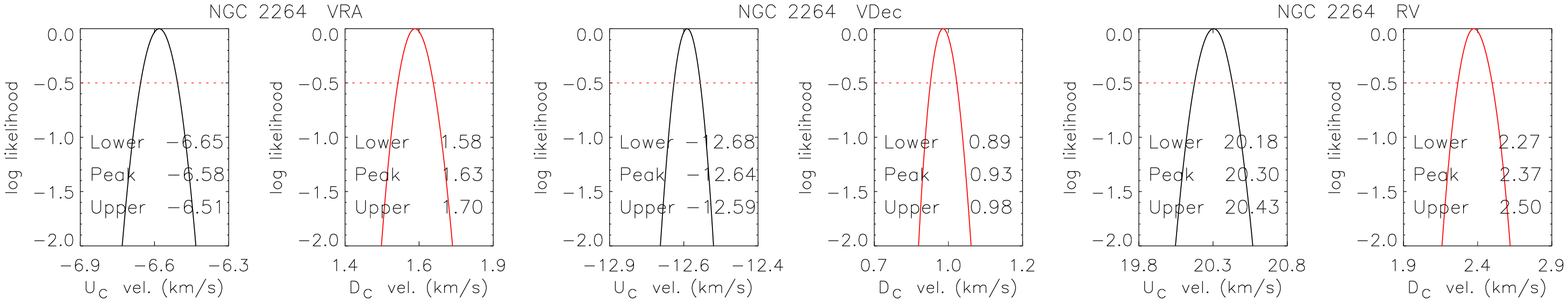}\\
\end{minipage}
\label{figB:5}
\end{figure*}
%%%%%%%%%%%%%%%%%%%%%%%%%%%%%%%%%%%%
\begin{figure*}
\begin{minipage}[t]{0.98\textwidth}
\centering
\includegraphics[width = 145mm]{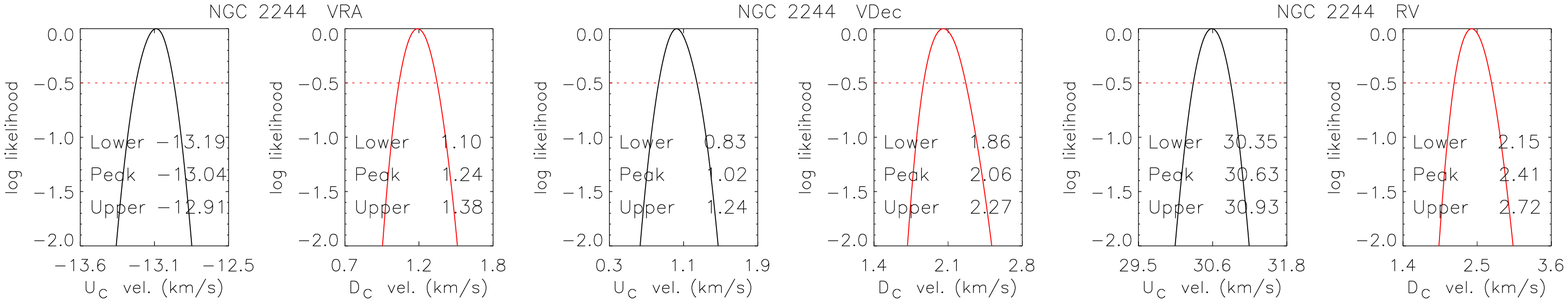}\\
\end{minipage}
\label{figB:6}
\end{figure*}
%%%%%%%%%%%%%%%%%%%%%%%%%%%%%%%%%%%%
\clearpage
\newpage
%%%%%%%%%%%%%%%%%%%%%%%%%%%%%%%%%%%%
\begin{figure*}
\begin{minipage}[t]{0.98\textwidth}
\centering
\includegraphics[width = 145mm]{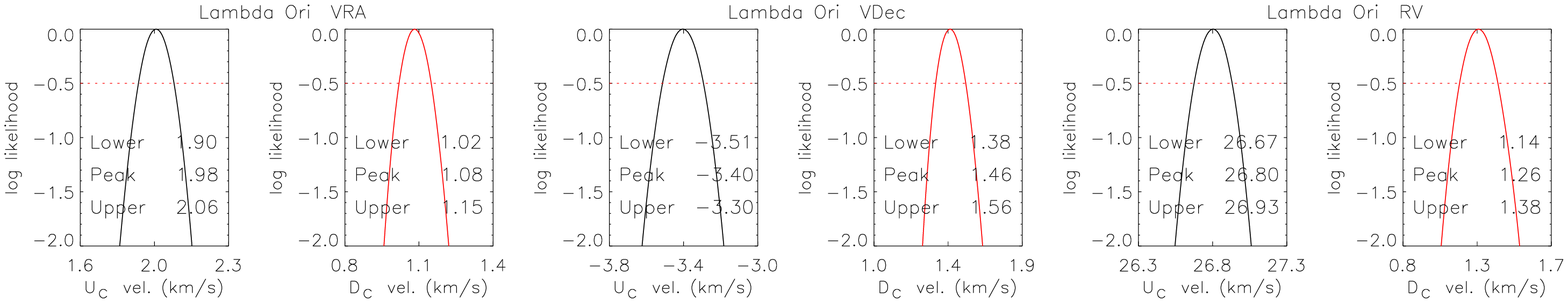}\\
\end{minipage}
\label{figB:7}
\end{figure*}
%%%%%%%%%%%%%%%%%%%%%%%%%%%%%%%%%%%%
\begin{figure*}
\begin{minipage}[t]{0.98\textwidth}
\centering
\includegraphics[width = 145mm]{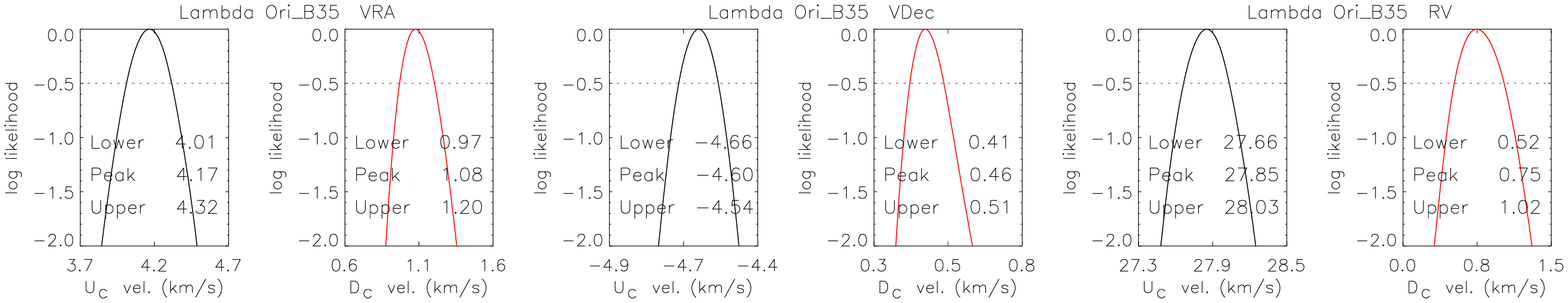}\\
\end{minipage}
\label{figB:8}
\end{figure*}
%%%%%%%%%%%%%%%%%%%%%%%%%%%%%%%%%%%%
\begin{figure*}
\begin{minipage}[t]{0.98\textwidth}
\centering
\includegraphics[width = 145mm]{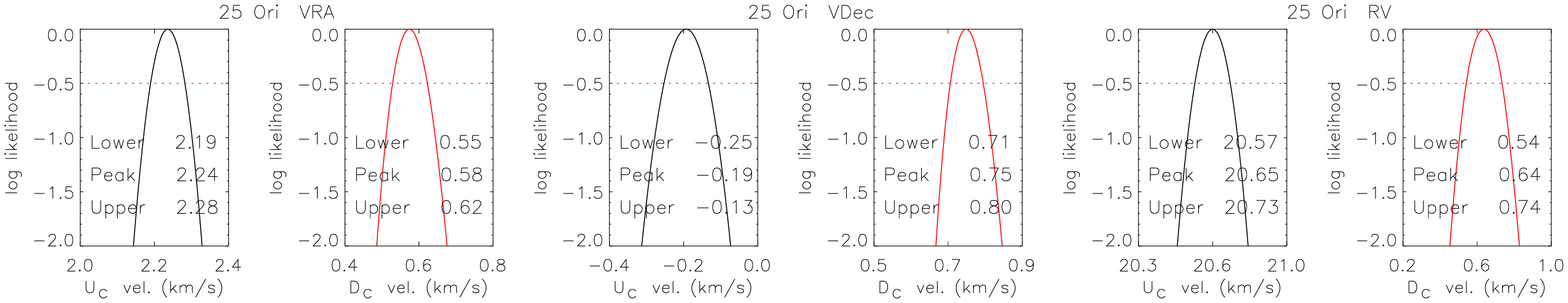}\\
\end{minipage}
\label{figB:9}
\end{figure*}
%%%%%%%%%%%%%%%%%%%%%%%%%%%%%%%%%%%%
\begin{figure*}
\begin{minipage}[t]{0.98\textwidth}
\centering
\includegraphics[width = 145mm]{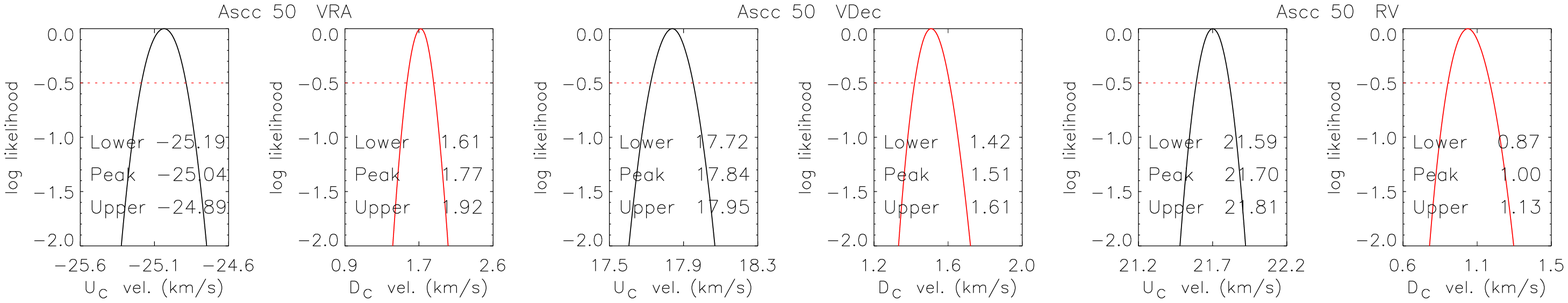}\\
\end{minipage}
\label{figB:10}
\end{figure*}
%%%%%%%%%%%%%%%%%%%%%%%%%%%%%%%%%%%%
\begin{figure*}
\begin{minipage}[t]{0.98\textwidth}
\centering
\includegraphics[width = 145mm]{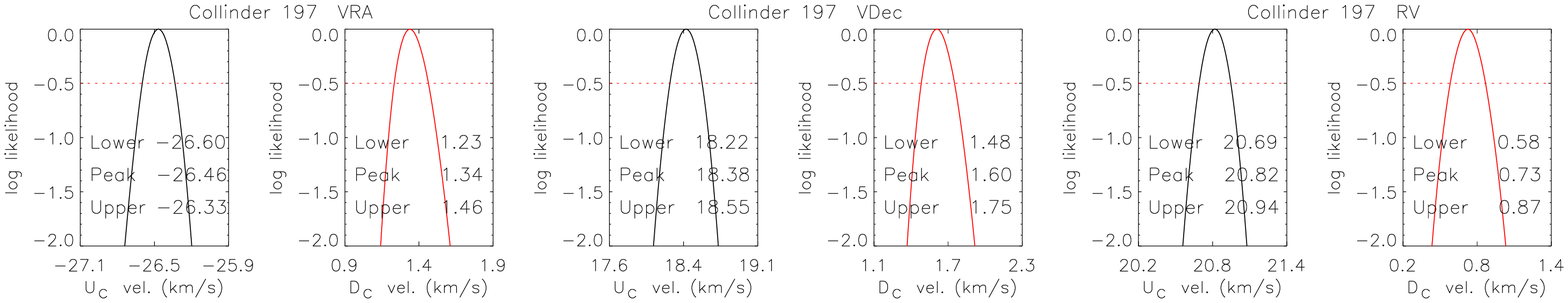}\\
\end{minipage}
\label{figB:11}
\end{figure*}
%%%%%%%%%%%%%%%%%%%%%%%%%%%%%%%%%%%%
\begin{figure*}
\begin{minipage}[t]{0.98\textwidth}
\centering
\includegraphics[width = 145mm]{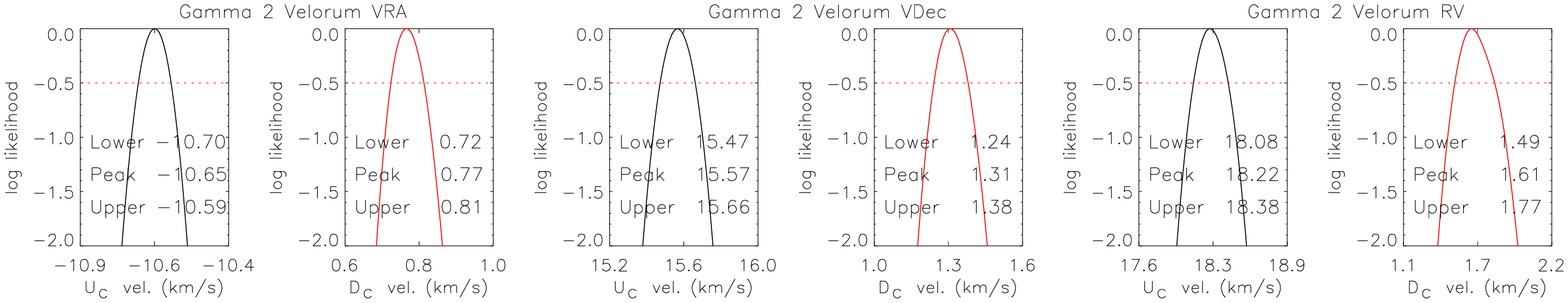}\\
\end{minipage}
\label{figB:12}
\end{figure*}
%%%%%%%%%%%%%%%%%%%%%%%%%%%%%%%%%%%%
\clearpage
\newpage
%%%%%%%%%%%%%%%%%%%%%%%%%%%%%%%%%%%%
\begin{figure*}
\begin{minipage}[t]{0.98\textwidth}
\centering
\includegraphics[width = 145mm]{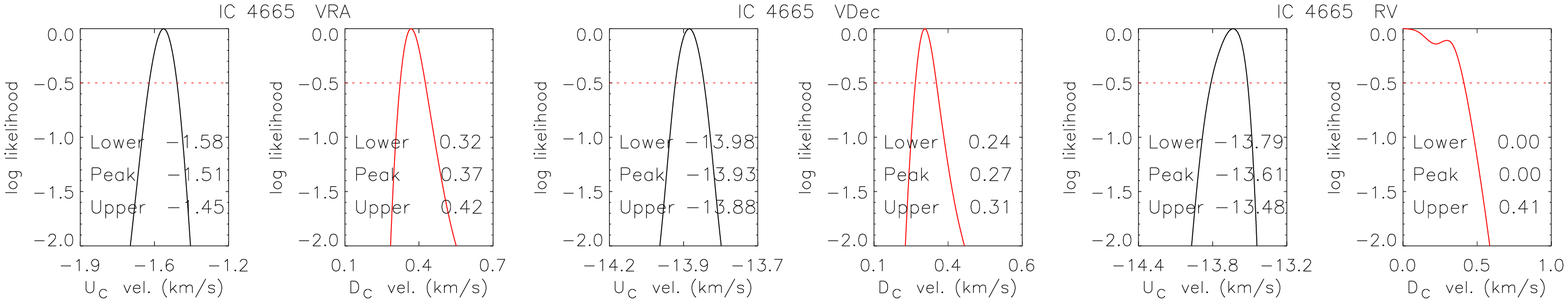}\\
\end{minipage}
\label{figB:13}
\end{figure*}
%%%%%%%%%%%%%%%%%%%%%%%%%%%%%%%%%%%%
\begin{figure*}
\begin{minipage}[t]{0.98\textwidth}
\centering
\includegraphics[width = 145mm]{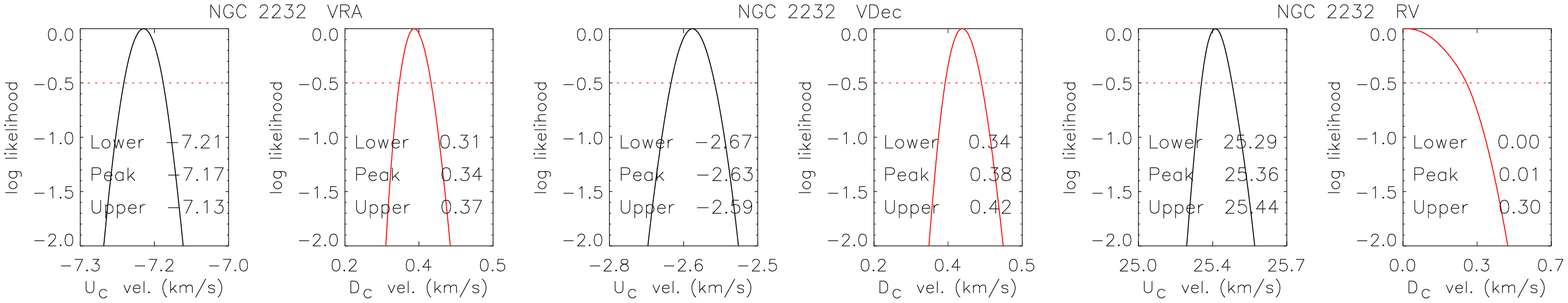}\\
\end{minipage}
\label{figB:14}
\end{figure*}
%%%%%%%%%%%%%%%%%%%%%%%%%%%%%%%%%%%%
\begin{figure*}
\begin{minipage}[t]{0.98\textwidth}
\centering
\includegraphics[width = 145mm]{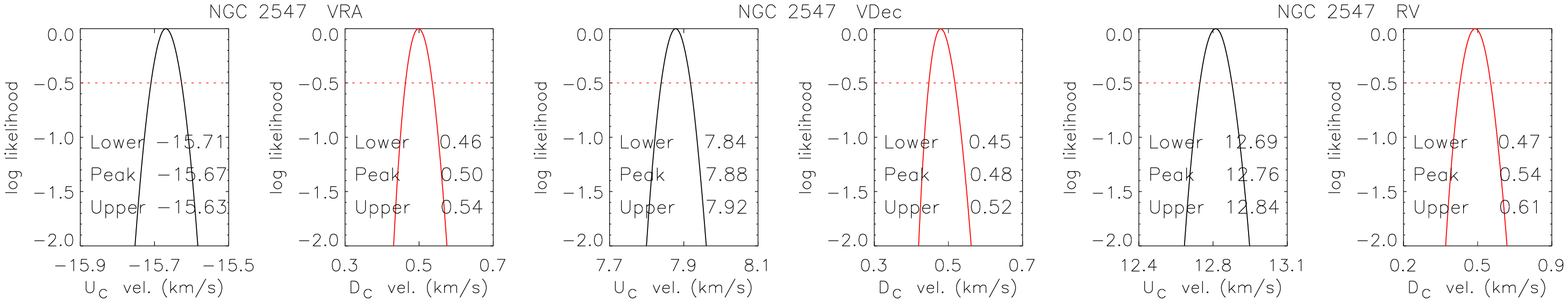}\\
\end{minipage}
\label{figB:15}
\end{figure*}
%%%%%%%%%%%%%%%%%%%%%%%%%%%%%%%%%%%%
\begin{figure*}
\begin{minipage}[t]{0.98\textwidth}
\centering
\includegraphics[width = 145mm]{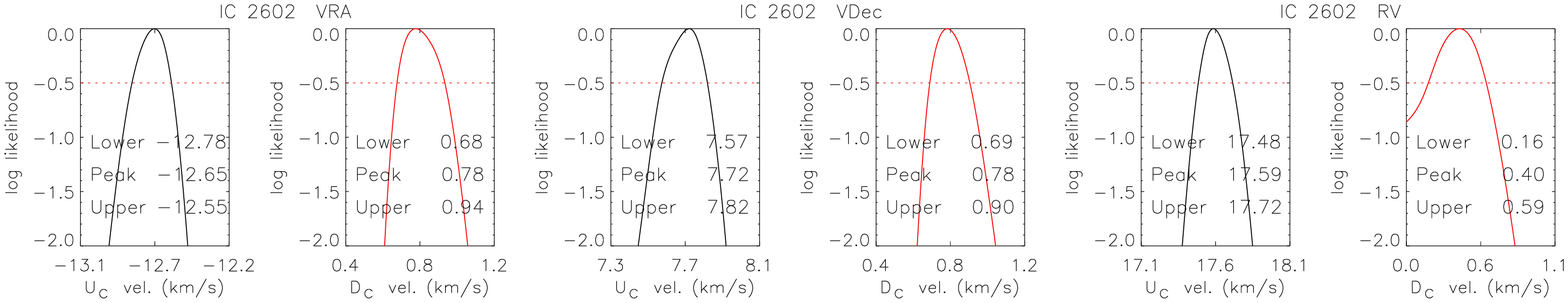}\\
\end{minipage}
\label{figB:16}
\end{figure*}
%%%%%%%%%%%%%%%%%%%%%%%%%%%%%%%%%%%%
\begin{figure*}
\begin{minipage}[t]{0.98\textwidth}
\centering
\includegraphics[width = 145mm]{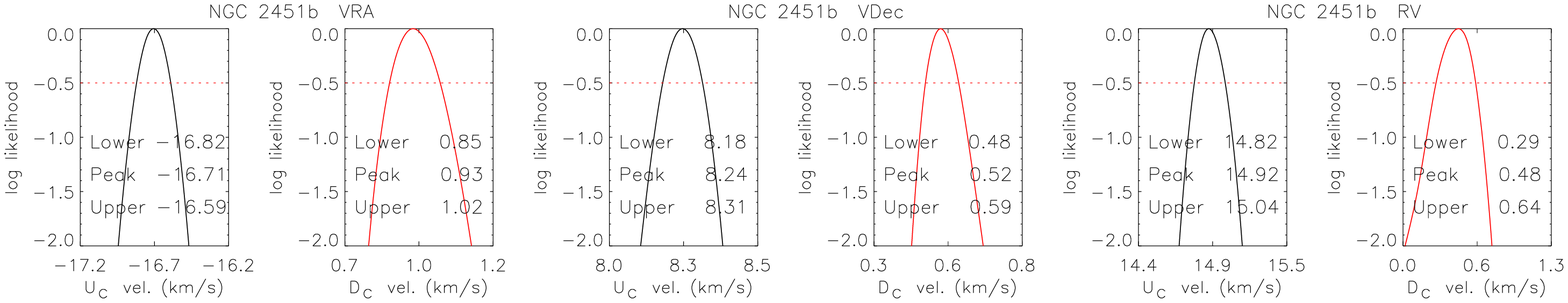}\\
\end{minipage}
\label{figB:17}
\end{figure*}
%%%%%%%%%%%%%%%%%%%%%%%%%%%%%%%%%%%%
\begin{figure*}
\begin{minipage}[t]{0.98\textwidth}
\centering
\includegraphics[width = 145mm]{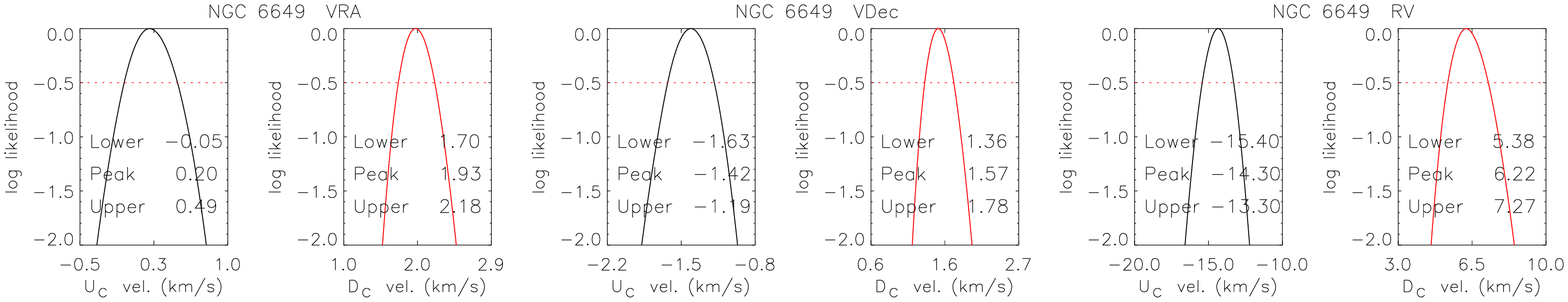}\\
\end{minipage}
\label{figB:18}
\end{figure*}
%%%%%%%%%%%%%%%%%%%%%%%%%%%%%%%%%%%%
\clearpage
\newpage
%%%%%%%%%%%%%%%%%%%%%%%%%%%%%%%%%%%%
\begin{figure*}
\begin{minipage}[t]{0.98\textwidth}
\centering
\includegraphics[width = 145mm]{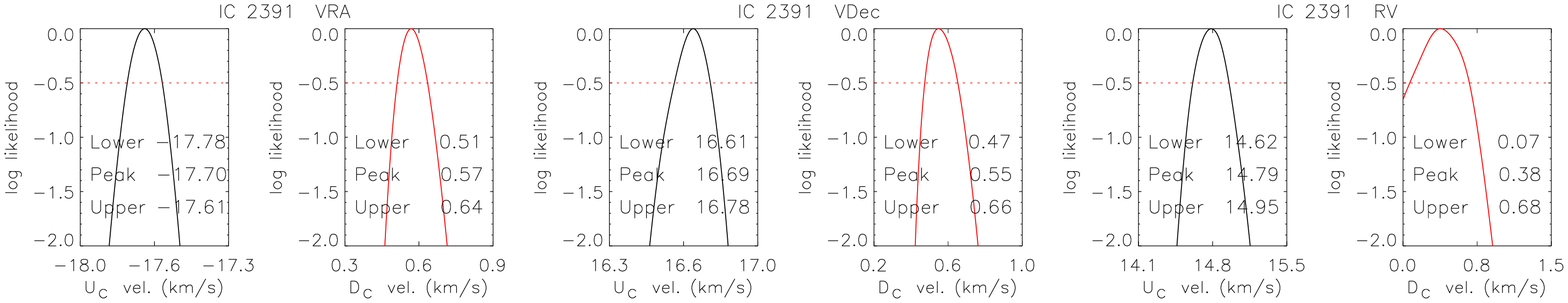}\\
\end{minipage}
\label{figB:19}
\end{figure*}
%%%%%%%%%%%%%%%%%%%%%%%%%%%%%%%%%%%%
\begin{figure*}
\begin{minipage}[t]{0.98\textwidth}
\centering
\includegraphics[width = 145mm]{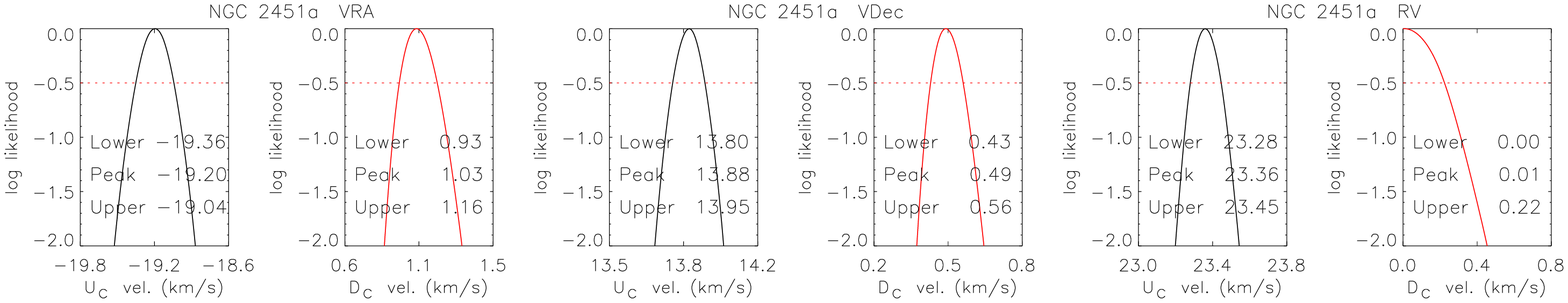}\\
\end{minipage}
\label{figB:20}
\end{figure*}
%%%%%%%%%%%%%%%%%%%%%%%%%%%%%%%%%%%%
\begin{figure*}
\begin{minipage}[t]{0.98\textwidth}
\centering
\includegraphics[width = 145mm]{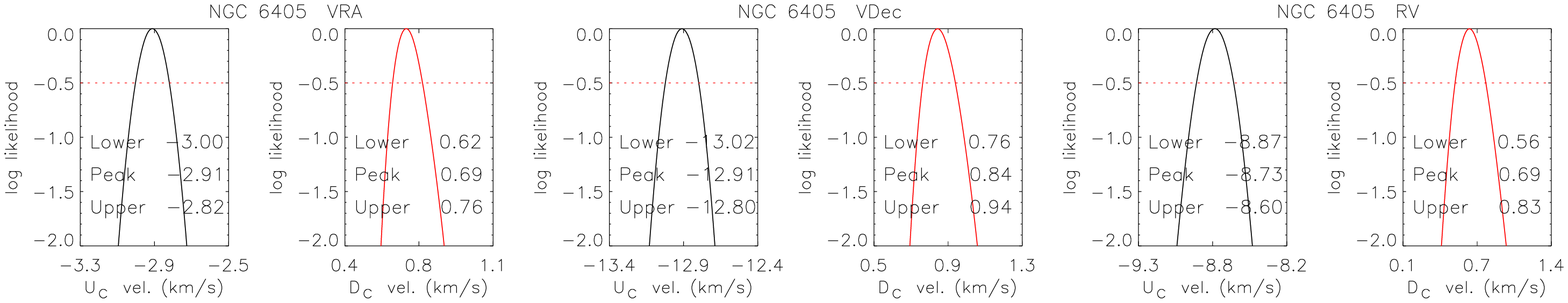}\\
\end{minipage}
\label{figB:21}
\end{figure*}
%%%%%%%%%%%%%%%%%%%%%%%%%%%%%%%%%%%%
\begin{figure*}
\begin{minipage}[t]{0.98\textwidth}
\centering
\includegraphics[width = 145mm]{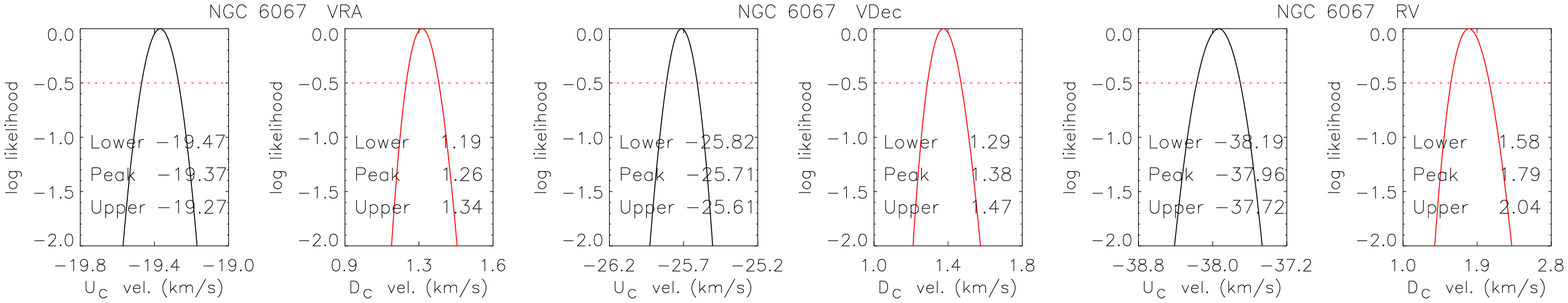}\\
\end{minipage}
\label{figB:22}
\end{figure*}
%%%%%%%%%%%%%%%%%%%%%%%%%%%%%%%%%%%%
\begin{figure*}
\begin{minipage}[t]{0.98\textwidth}
\centering
\includegraphics[width = 145mm]{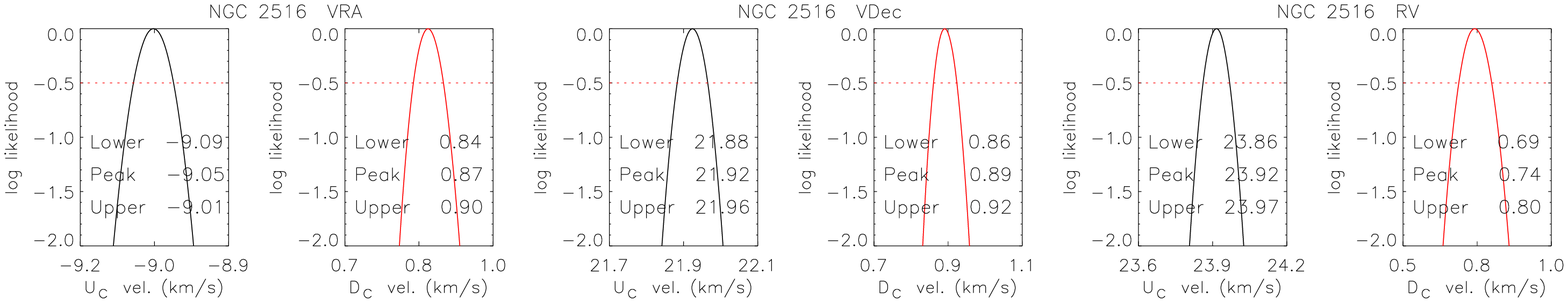}\\
\end{minipage}
\label{figB:23}
\end{figure*}
%%%%%%%%%%%%%%%%%%%%%%%%%%%%%%%%%%%%
\begin{figure*}
\begin{minipage}[t]{0.98\textwidth}
\centering
\includegraphics[width = 145mm]{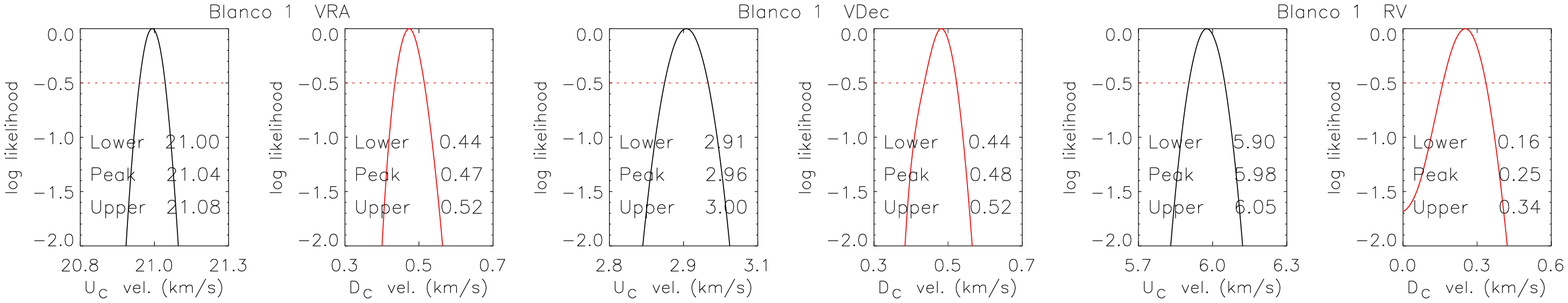}\\
\end{minipage}
\label{figB:24}
\end{figure*}
%%%%%%%%%%%%%%%%%%%%%%%%%%%%%%%%%%%%
\clearpage
\newpage
%%%%%%%%%%%%%%%%%%%%%%%%%%%%%%%%%%%%
\begin{figure*}
\begin{minipage}[t]{0.98\textwidth}
\centering
\includegraphics[width = 145mm]{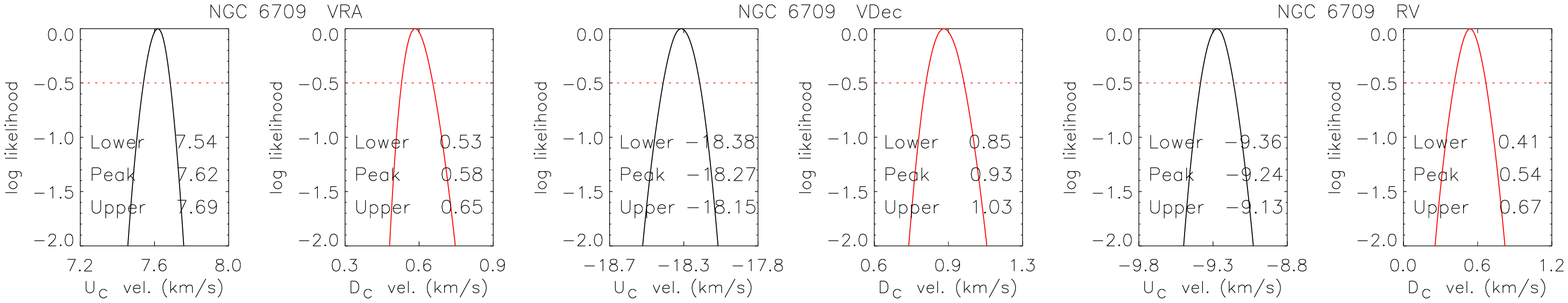}\\
\end{minipage}
\label{figB:25}
\end{figure*}
%%%%%%%%%%%%%%%%%%%%%%%%%%%%%%%%%%%%
\begin{figure*}
\begin{minipage}[t]{0.98\textwidth}
\centering
\includegraphics[width = 145mm]{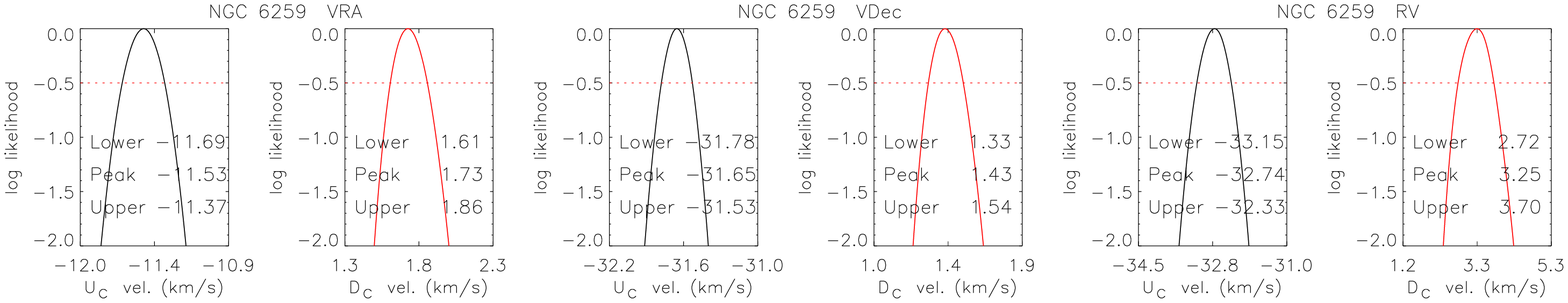}\\
\end{minipage}
\label{figB:26}
\end{figure*}
%%%%%%%%%%%%%%%%%%%%%%%%%%%%%%%%%%%%
\begin{figure*}
\begin{minipage}[t]{0.98\textwidth}
\centering
\includegraphics[width = 145mm]{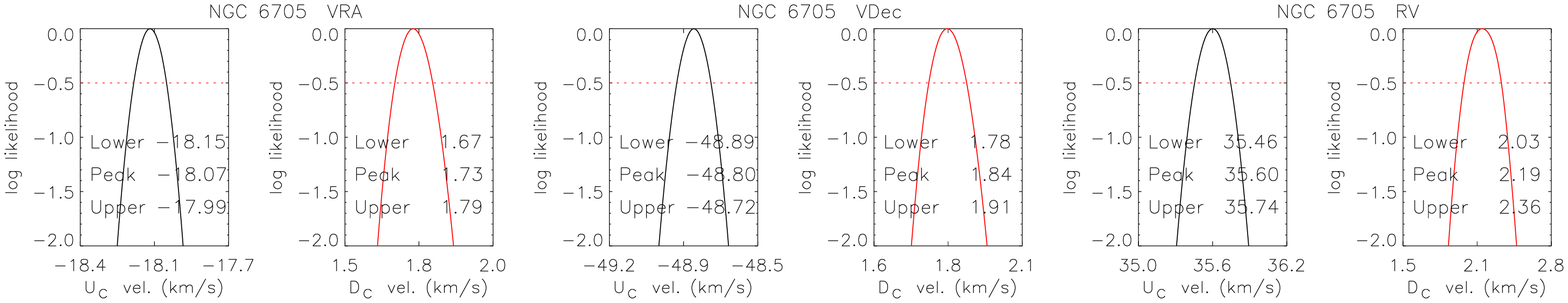}\\
\end{minipage}
\label{figB:27}
\end{figure*}
%%%%%%%%%%%%%%%%%%%%%%%%%%%%%%%%%%%%
\begin{figure*}
\begin{minipage}[t]{0.98\textwidth}
\centering
\includegraphics[width = 145mm]{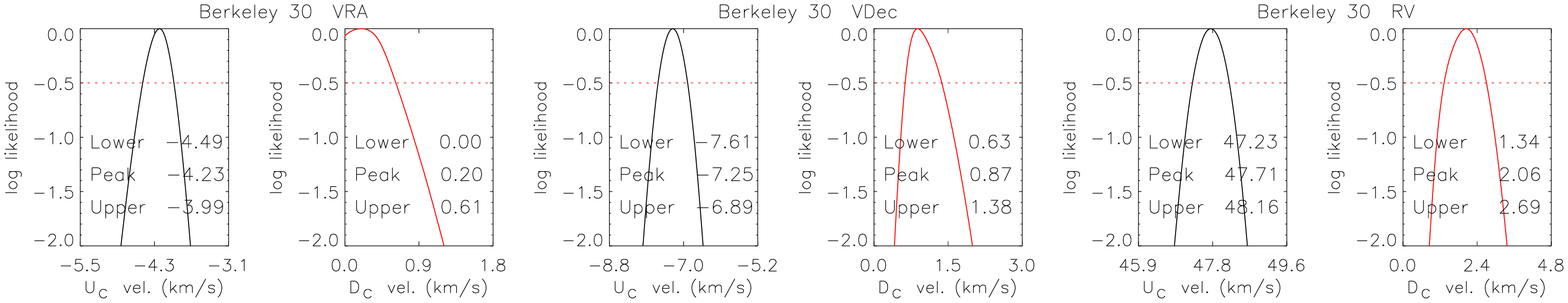}\\
\end{minipage}
\label{figB:28}
\end{figure*}
%%%%%%%%%%%%%%%%%%%%%%%%%%%%%%%%%%%%
\begin{figure*}
\begin{minipage}[t]{0.98\textwidth}
\centering
\includegraphics[width = 145mm]{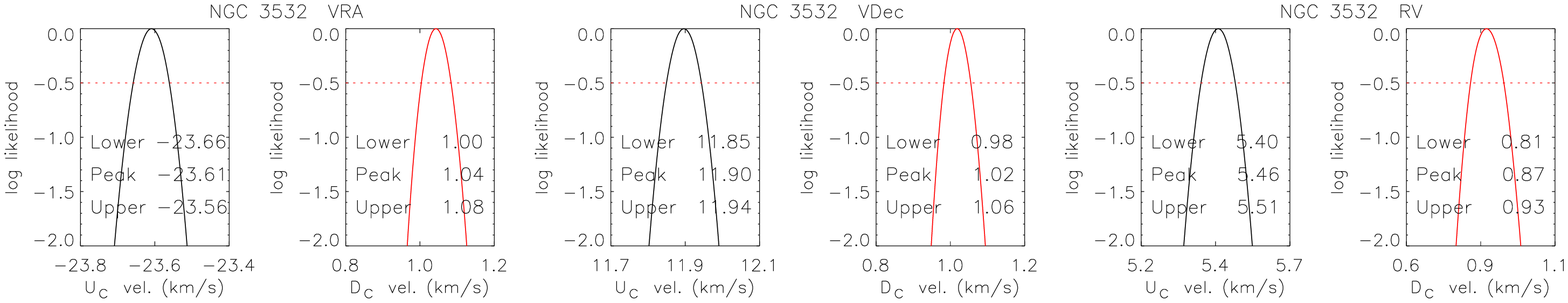}\\
\end{minipage}
\label{figB:29}
\end{figure*}
%%%%%%%%%%%%%%%%%%%%%%%%%%%%%%%%%%%%
\begin{figure*}
\begin{minipage}[t]{0.98\textwidth}
\centering
\includegraphics[width = 145mm]{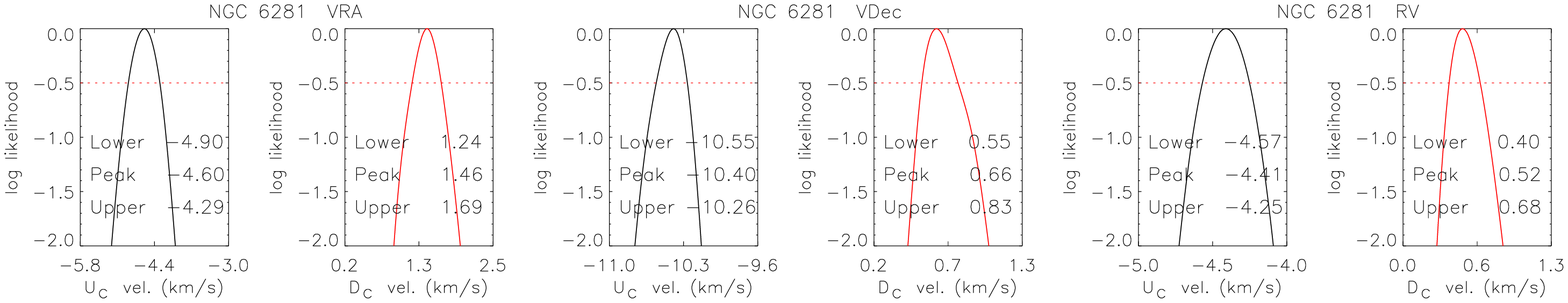}\\
\end{minipage}
\label{figB:30}
\end{figure*}
%%%%%%%%%%%%%%%%%%%%%%%%%%%%%%%%%%%%
\clearpage
\newpage
%%%%%%%%%%%%%%%%%%%%%%%%%%%%%%%%%%%%
\begin{figure*}
\begin{minipage}[t]{0.98\textwidth}
\centering
\includegraphics[width = 145mm]{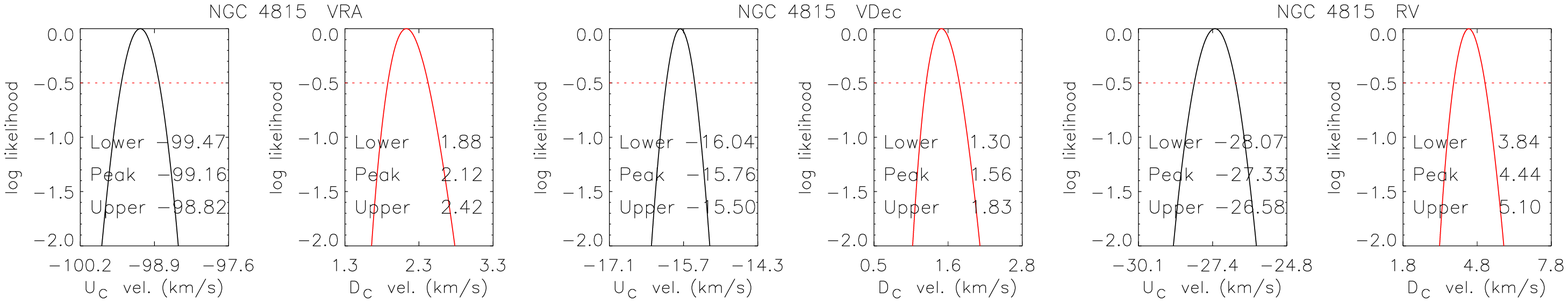}\\
\end{minipage}
\label{figB:31}
\end{figure*}
%%%%%%%%%%%%%%%%%%%%%%%%%%%%%%%%%%%%
\begin{figure*}
\begin{minipage}[t]{0.98\textwidth}
\centering
\includegraphics[width = 145mm]{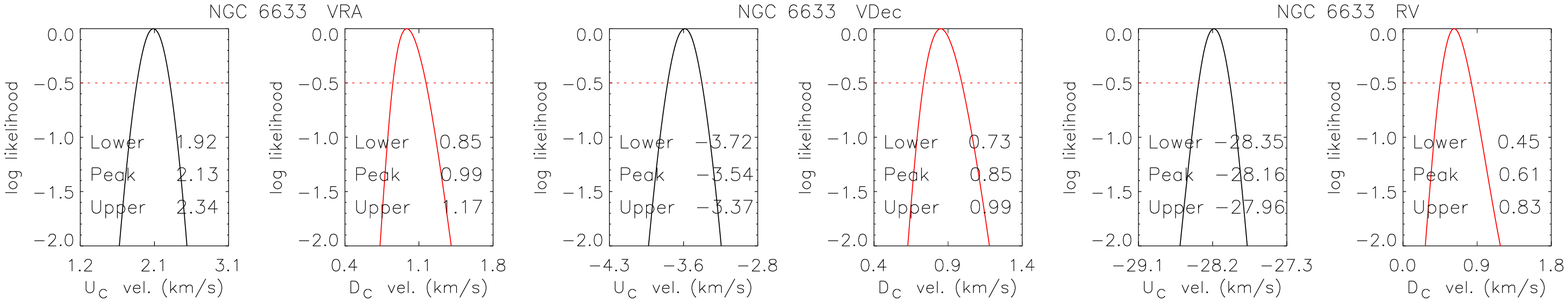}\\
\end{minipage}
\label{figB:32}
\end{figure*}
%%%%%%%%%%%%%%%%%%%%%%%%%%%%%%%%%%%%
\begin{figure*}
\begin{minipage}[t]{0.98\textwidth}
\centering
\includegraphics[width = 145mm]{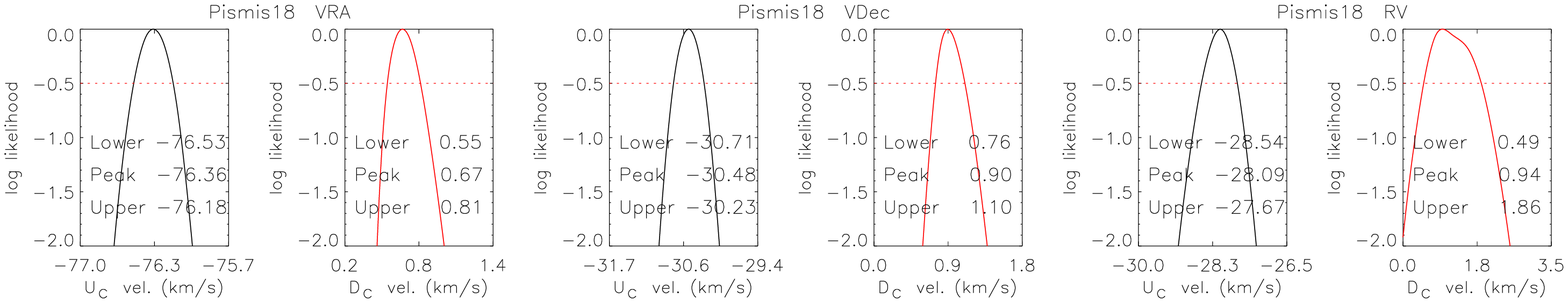}\\
\end{minipage}
\label{figB:33}
\end{figure*}
%%%%%%%%%%%%%%%%%%%%%%%%%%%%%%%%%%%%
\begin{figure*}
\begin{minipage}[t]{0.98\textwidth}
\centering
\includegraphics[width = 145mm]{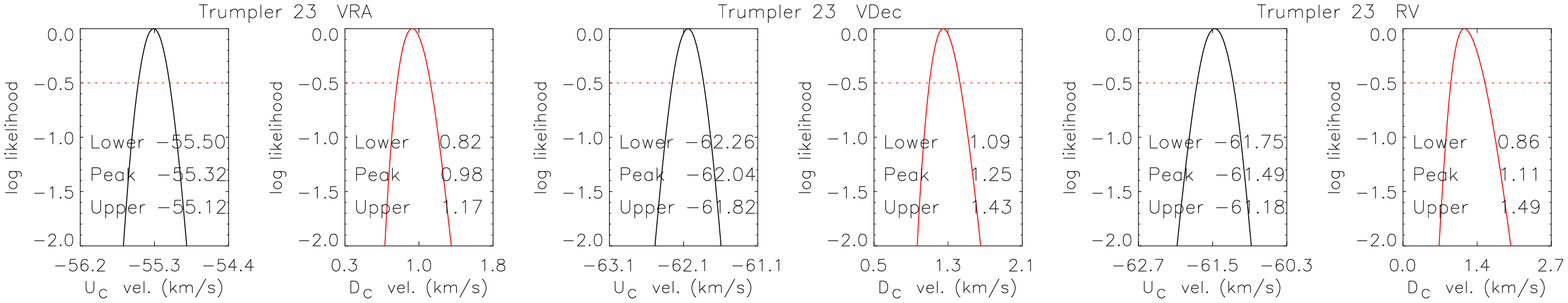}\\
\end{minipage}
\label{figB:34}
\end{figure*}
%%%%%%%%%%%%%%%%%%%%%%%%%%%%%%%%%%%%
\begin{figure*}
\begin{minipage}[t]{0.98\textwidth}
\centering
\includegraphics[width = 145mm]{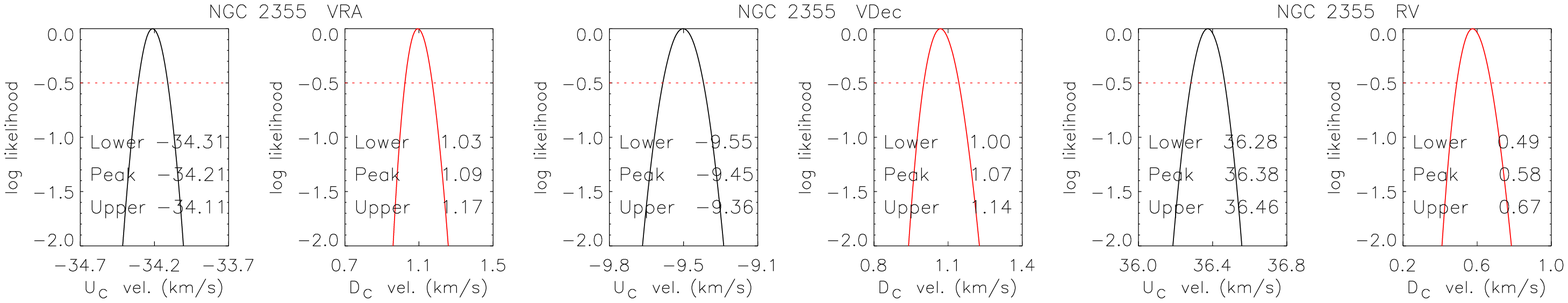}\\
\end{minipage}
\label{figB:35}
\end{figure*}
%%%%%%%%%%%%%%%%%%%%%%%%%%%%%%%%%%%%
\begin{figure*}
\begin{minipage}[t]{0.98\textwidth}
\centering
\includegraphics[width = 145mm]{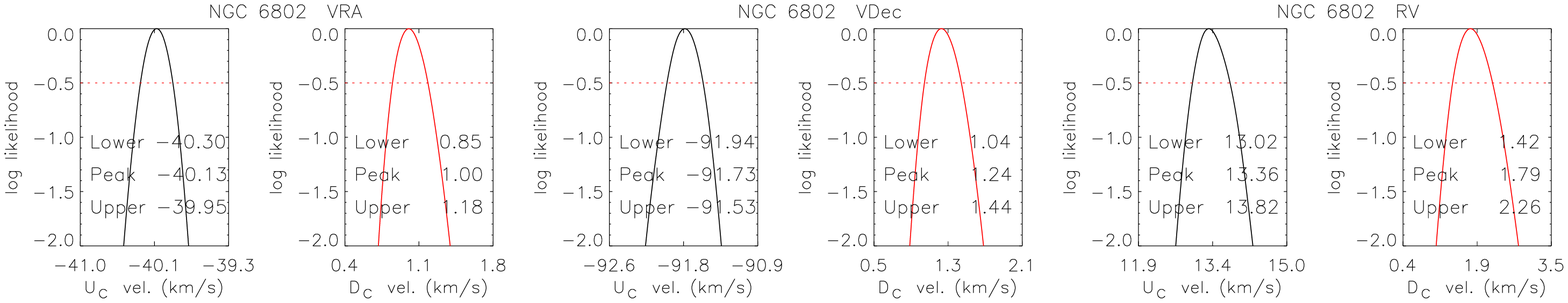}\\
\end{minipage}
\label{figB:36}
\end{figure*}
%%%%%%%%%%%%%%%%%%%%%%%%%%%%%%%%%%%%
\clearpage
\newpage
%%%%%%%%%%%%%%%%%%%%%%%%%%%%%%%%%%%%
\begin{figure*}
\begin{minipage}[t]{0.98\textwidth}
\centering
\includegraphics[width = 145mm]{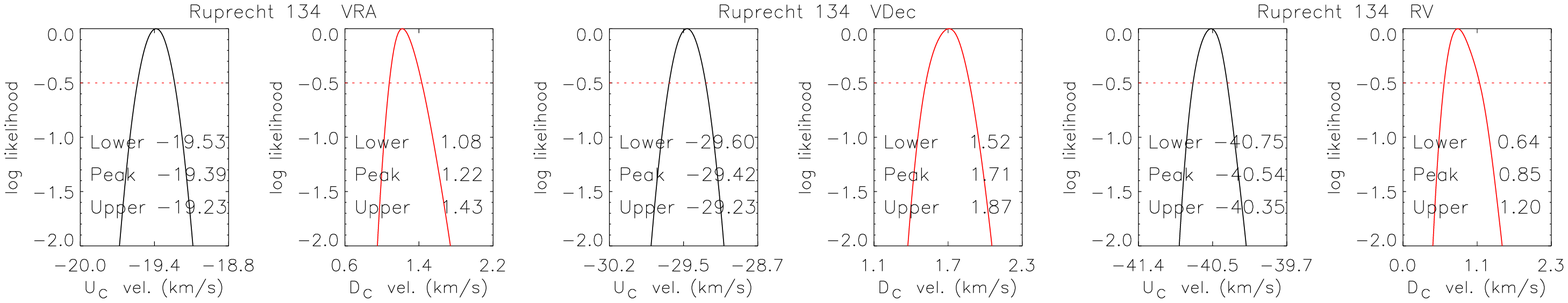}\\
\end{minipage}
\label{figB:37}
\end{figure*}
%%%%%%%%%%%%%%%%%%%%%%%%%%%%%%%%%%%%
\begin{figure*}
\begin{minipage}[t]{0.98\textwidth}
\centering
\includegraphics[width = 145mm]{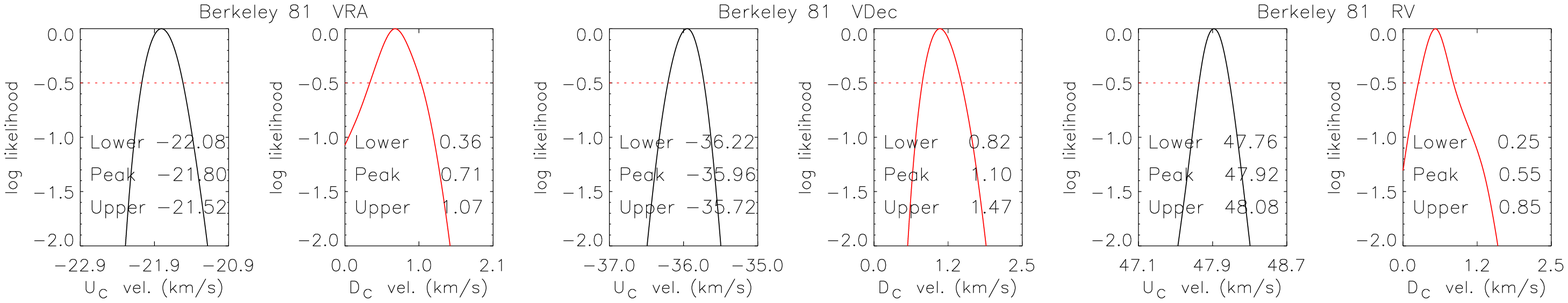}\\
\end{minipage}
\label{figB:38}
\end{figure*}
%%%%%%%%%%%%%%%%%%%%%%%%%%%%%%%%%%%%
\begin{figure*}
\begin{minipage}[t]{0.98\textwidth}
\centering
\includegraphics[width = 145mm]{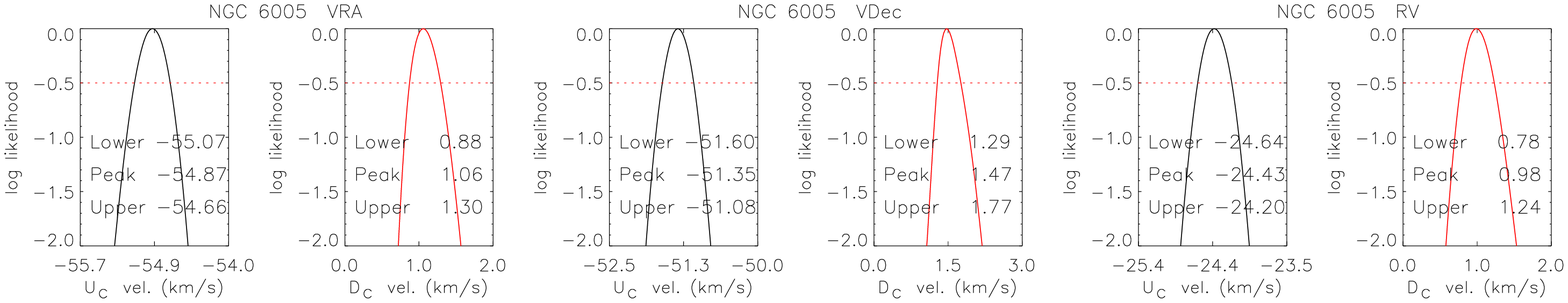}\\
\end{minipage}
\label{figB:39}
\end{figure*}
%%%%%%%%%%%%%%%%%%%%%%%%%%%%%%%%%%%%
\begin{figure*}
\begin{minipage}[t]{0.98\textwidth}
\centering
\includegraphics[width = 145mm]{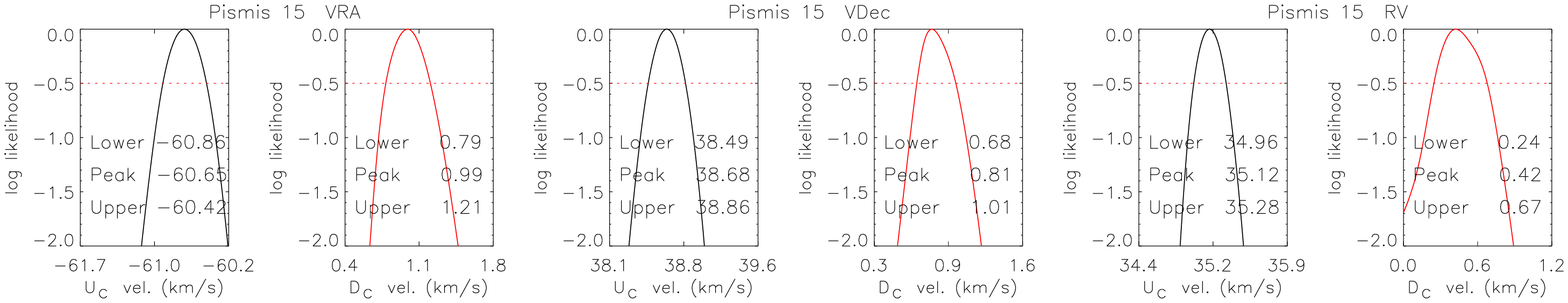}\\
\end{minipage}
\label{figB:40}
\end{figure*}
%%%%%%%%%%%%%%%%%%%%%%%%%%%%%%%%%%%%
\begin{figure*}
\begin{minipage}[t]{0.98\textwidth}
\centering
\includegraphics[width = 145mm]{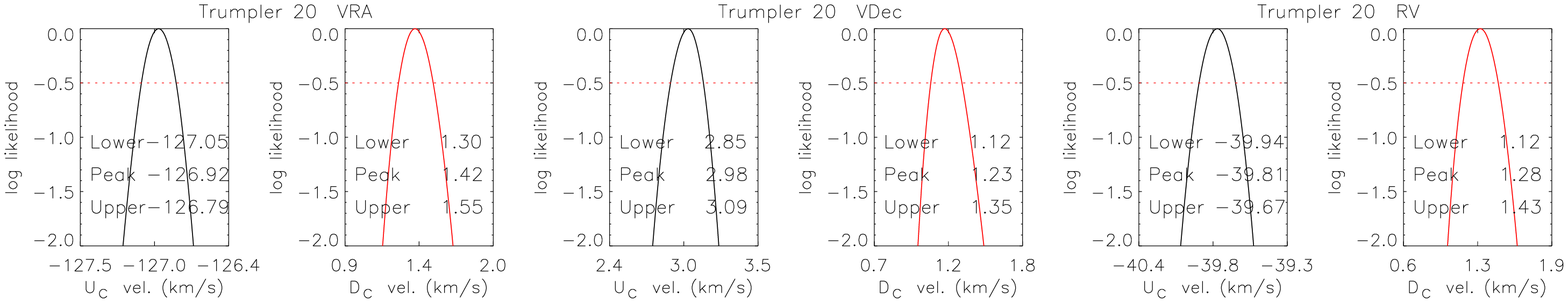}\\
\end{minipage}
\label{figB:41}
\end{figure*}
%%%%%%%%%%%%%%%%%%%%%%%%%%%%%%%%%%%%
\begin{figure*}
\begin{minipage}[t]{0.98\textwidth}
\centering
\includegraphics[width = 145mm]{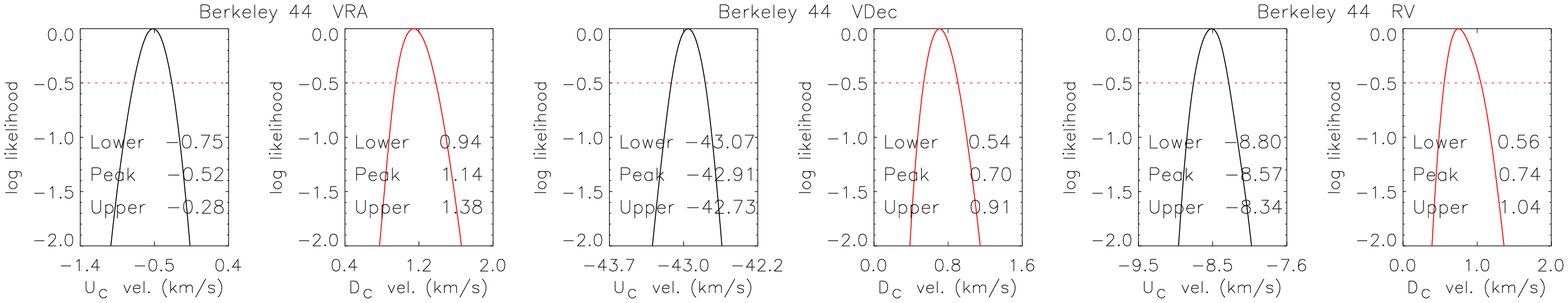}\\
\end{minipage}
\label{figB:42}
\end{figure*}
%%%%%%%%%%%%%%%%%%%%%%%%%%%%%%%%%%%%
\begin{figure*}
\begin{minipage}[t]{0.98\textwidth}
\centering
\includegraphics[width = 145mm]{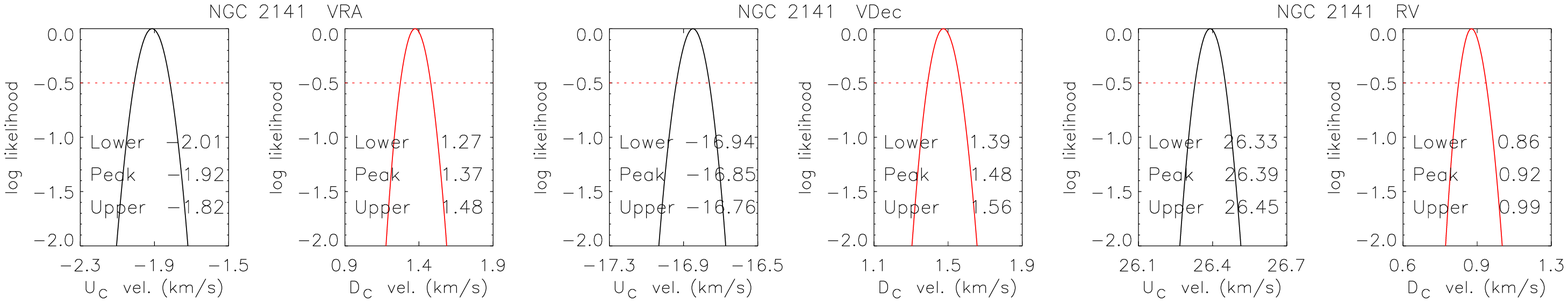}\\
\end{minipage}
\label{figB:43}
\end{figure*}
%%%%%%%%%%%%%%%%%%%%%%%%%%%%%%%%%%%%
\begin{figure*}
\begin{minipage}[t]{0.98\textwidth}
\centering
\includegraphics[width = 145mm]{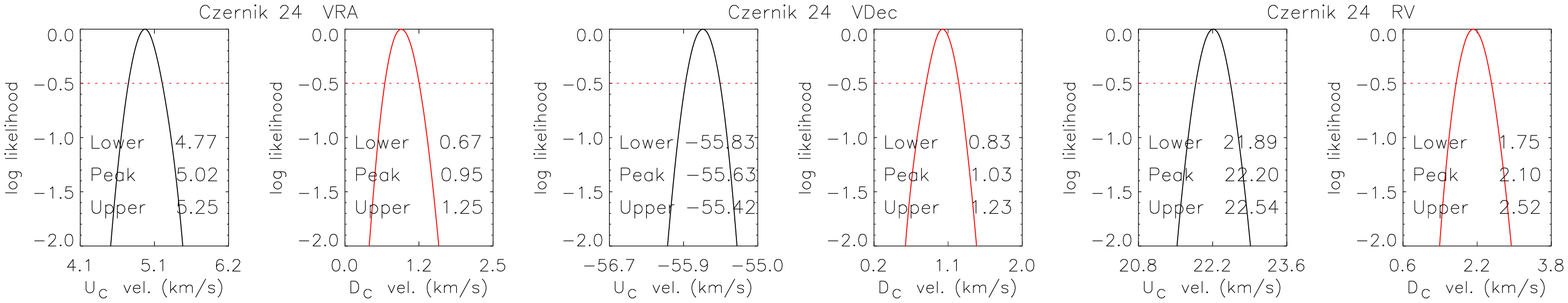}\\
\end{minipage}
\label{figB:44}
\end{figure*}
%%%%%%%%%%%%%%%%%%%%%%%%%%%%%%%%%%%%
\begin{figure*}
\begin{minipage}[t]{0.98\textwidth}
\centering
\includegraphics[width = 145mm]{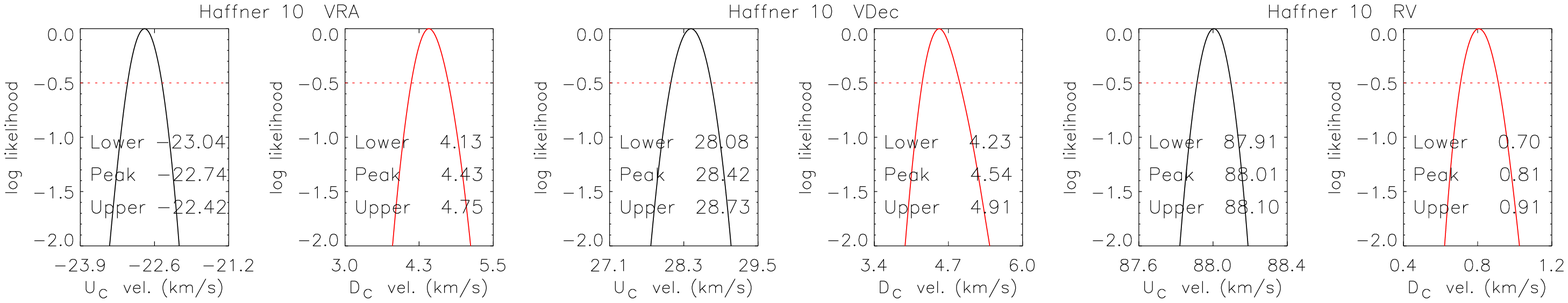}\\
\end{minipage}
\label{figB:45}
\end{figure*}
%%%%%%%%%%%%%%%%%%%%%%%%%%%%%%%%%%%%
\begin{figure*}
\begin{minipage}[t]{0.98\textwidth}
\centering
\includegraphics[width = 145mm]{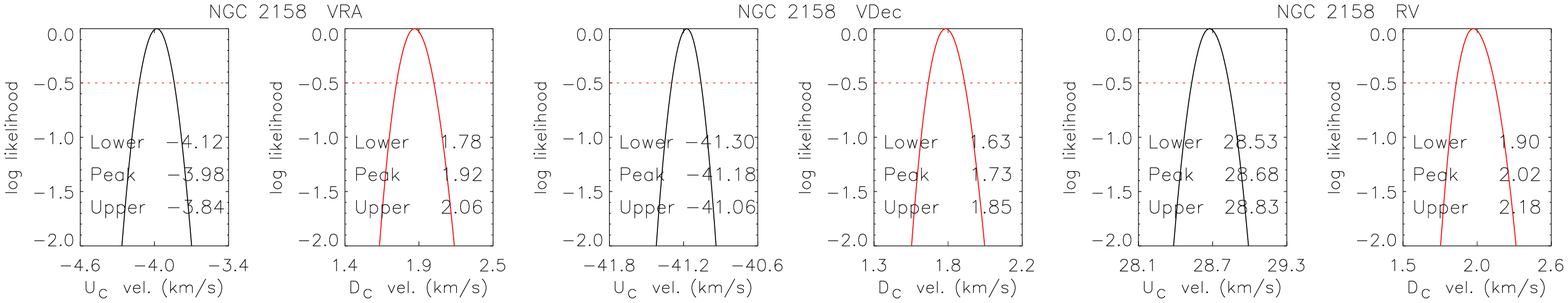}\\
\end{minipage}
\label{figB:46}
\end{figure*}
%%%%%%%%%%%%%%%%%%%%%%%%%%%%%%%%%%%%
\begin{figure*}
\begin{minipage}[t]{0.98\textwidth}
\centering
\includegraphics[width = 145mm]{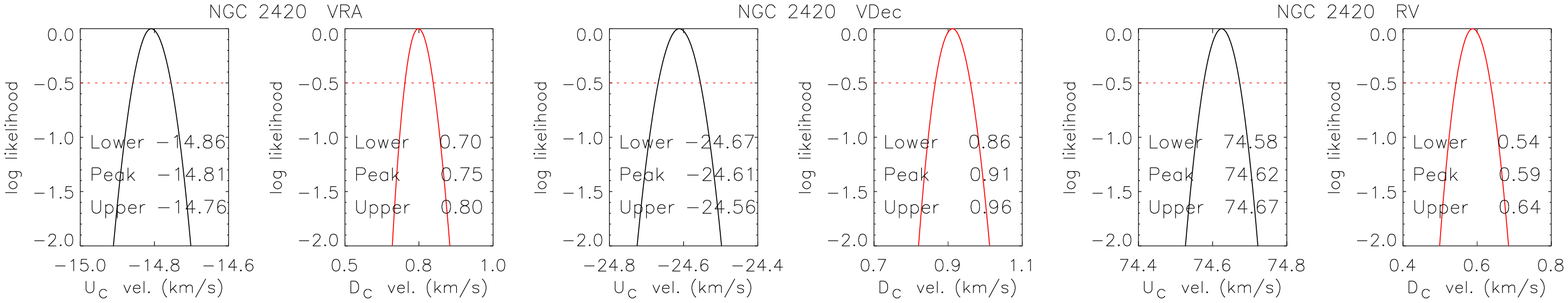}\\
\end{minipage}
\label{figB:47}
\end{figure*}
%%%%%%%%%%%%%%%%%%%%%%%%%%%%%%%%%%%%
\begin{figure*}
\begin{minipage}[t]{0.98\textwidth}
\centering
\includegraphics[width = 145mm]{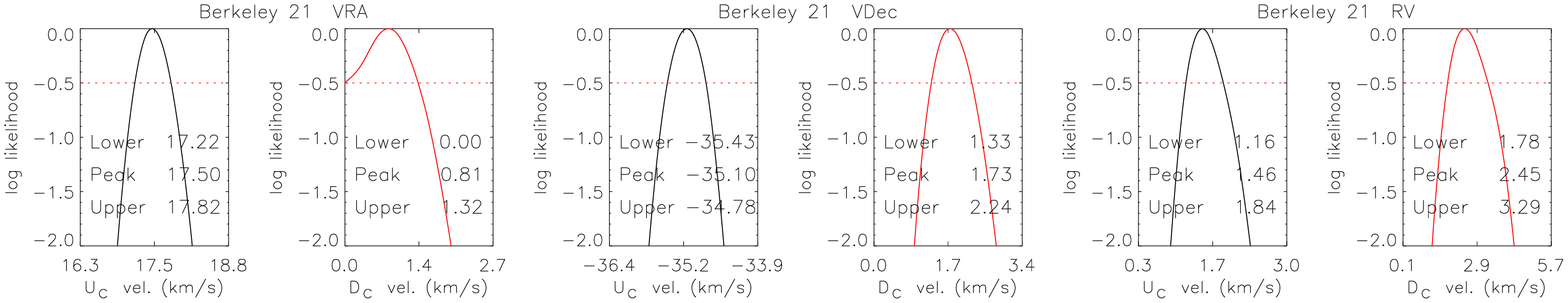}\\
\end{minipage}
\label{figB:48}
\end{figure*}
%%%%%%%%%%%%%%%%%%%%%%%%%%%%%%%%%%%%
\clearpage
\newpage
%%%%%%%%%%%%%%%%%%%%%%%%%%%%%%%%%%%%
\begin{figure*}
\begin{minipage}[t]{0.98\textwidth}
\centering
\includegraphics[width = 145mm]{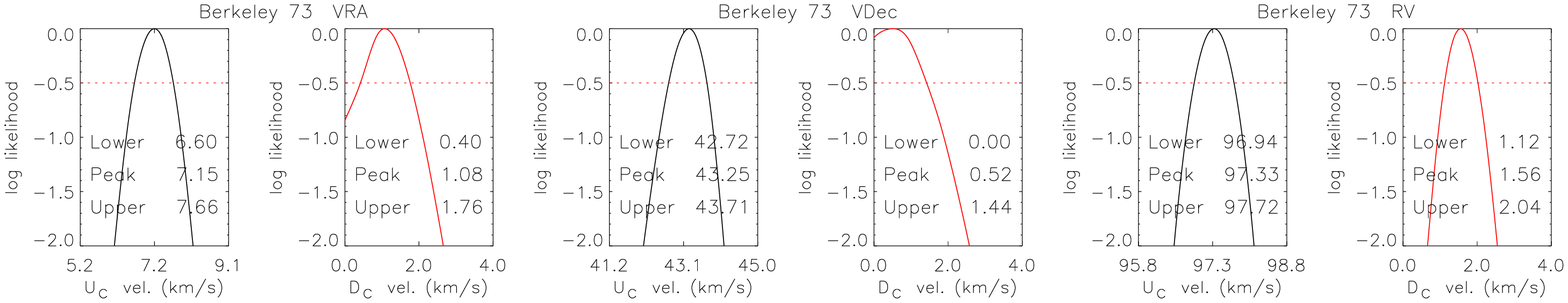}\\
\end{minipage}
\label{figB:49}
\end{figure*}
%%%%%%%%%%%%%%%%%%%%%%%%%%%%%%%%%%%%
\begin{figure*}
\begin{minipage}[t]{0.98\textwidth}
\centering
\includegraphics[width = 145mm]{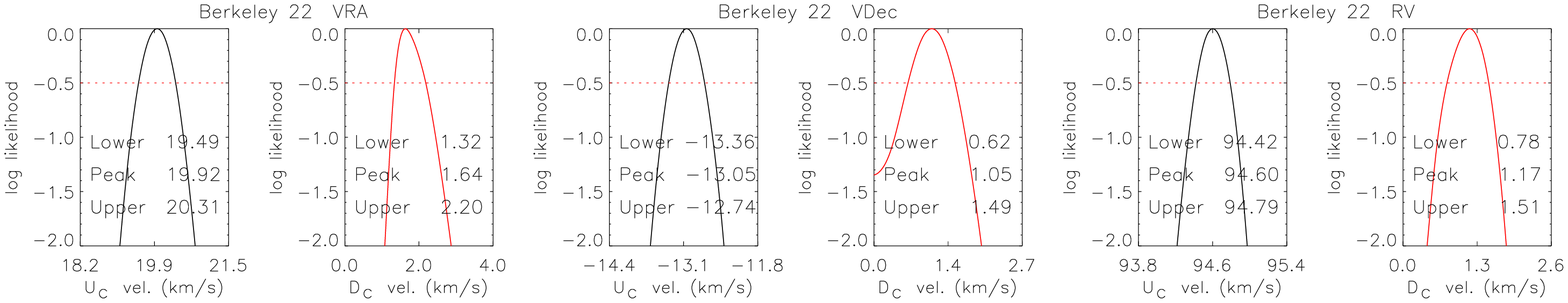}\\
\end{minipage}
\label{figB:50}
\end{figure*}
%%%%%%%%%%%%%%%%%%%%%%%%%%%%%%%%%%%%
\begin{figure*}
\begin{minipage}[t]{0.98\textwidth}
\centering
\includegraphics[width = 145mm]{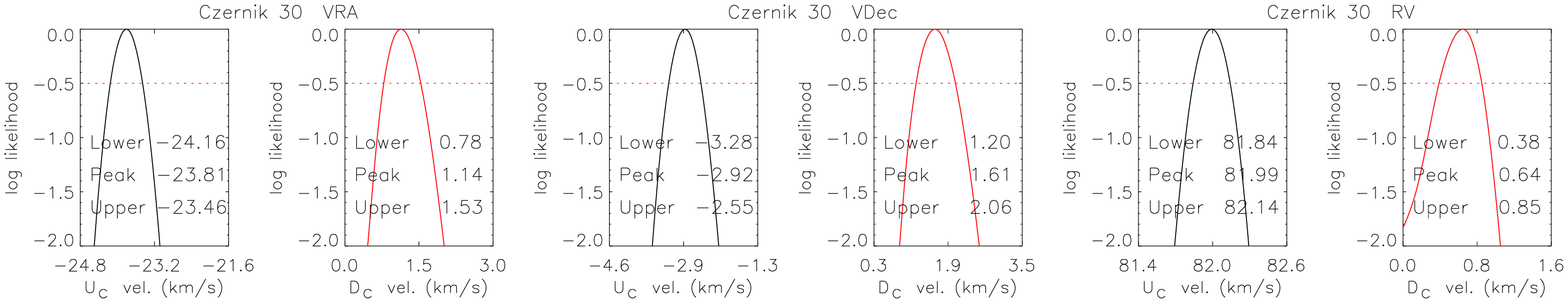}\\
\end{minipage}
\label{figB:51}
\end{figure*}
%%%%%%%%%%%%%%%%%%%%%%%%%%%%%%%%%%%%
\begin{figure*}
\begin{minipage}[t]{0.98\textwidth}
\centering
\includegraphics[width = 145mm]{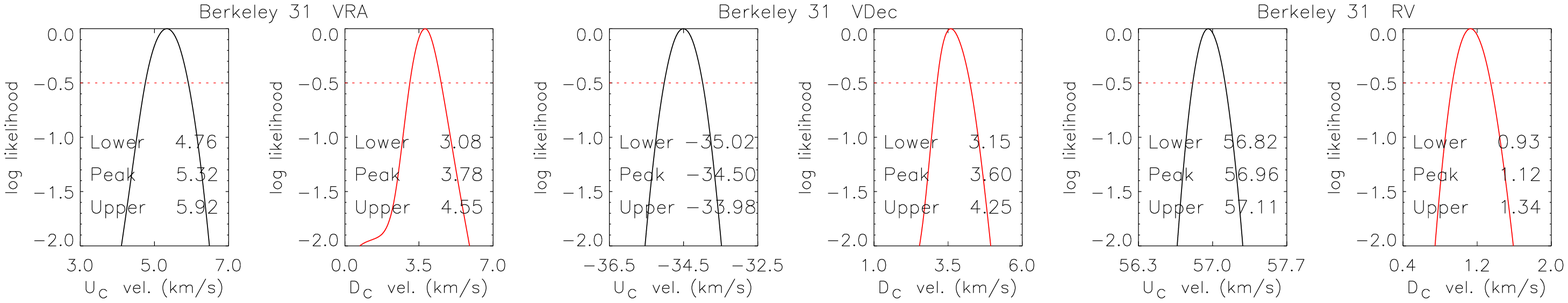}\\
\end{minipage}
\label{figB:52}
\end{figure*}
%%%%%%%%%%%%%%%%%%%%%%%%%%%%%%%%%%%%
\begin{figure*}
\begin{minipage}[t]{0.98\textwidth}
\centering
\includegraphics[width = 145mm]{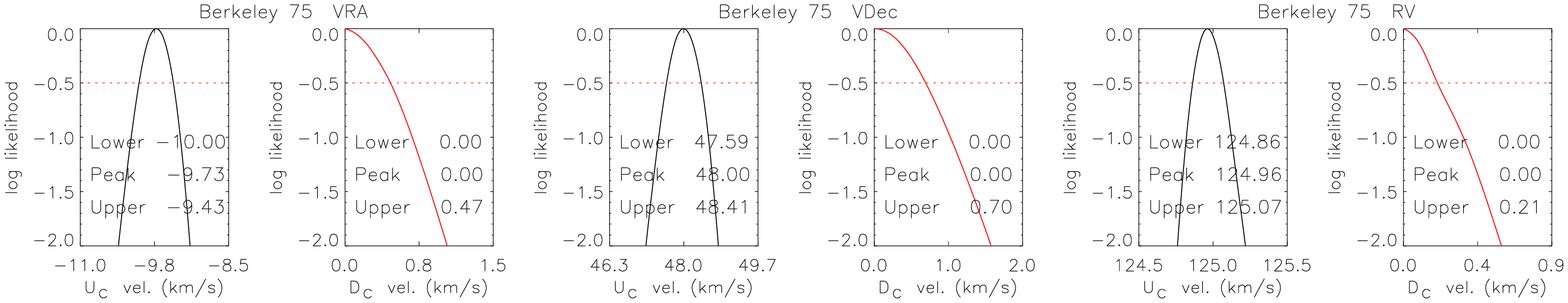}\\
\end{minipage}
\label{figB:53}
\end{figure*}
%%%%%%%%%%%%%%%%%%%%%%%%%%%%%%%%%%%%
\begin{figure*}
\begin{minipage}[t]{0.98\textwidth}
\centering
\includegraphics[width = 145mm]{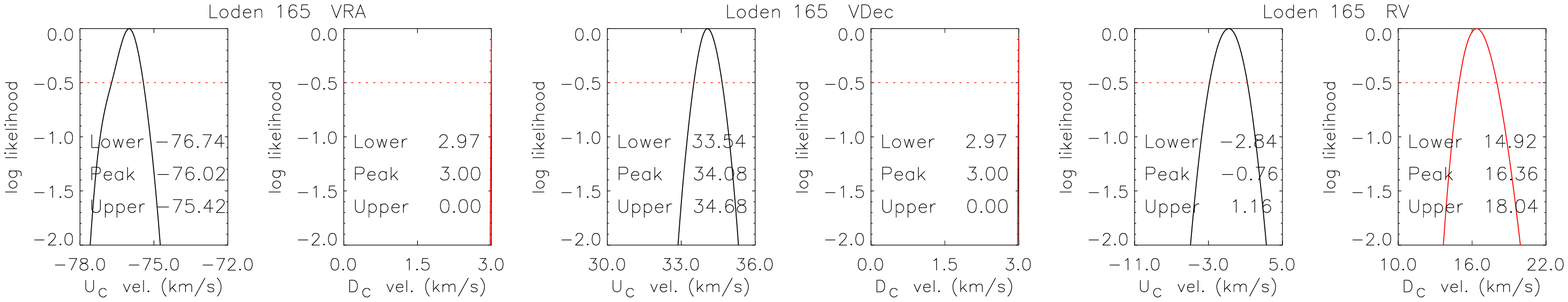}\\
\end{minipage}
\label{figB:54}
\end{figure*}
%%%%%%%%%%%%%%%%%%%%%%%%%%%%%%%%%%%%
\clearpage
\newpage
%%%%%%%%%%%%%%%%%%%%%%%%%%%%%%%%%%%%
\begin{figure*}
\begin{minipage}[t]{0.98\textwidth}
\centering
\includegraphics[width = 145mm]{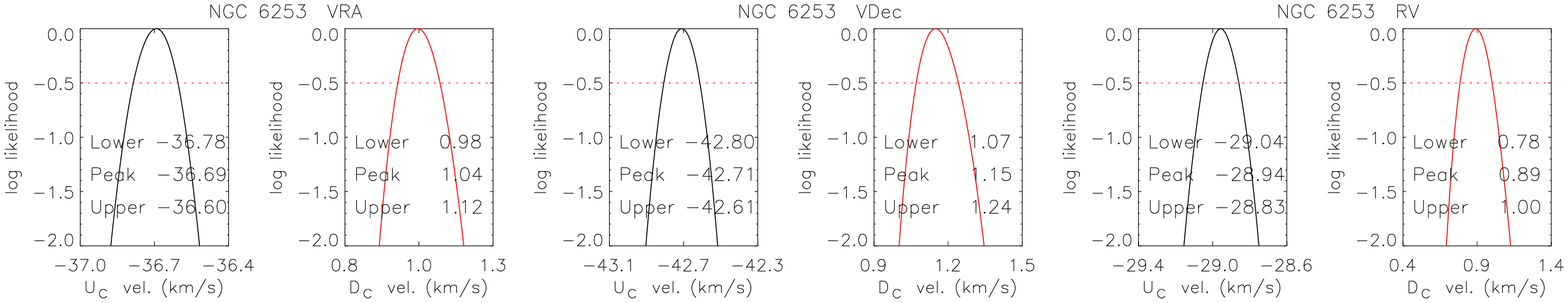}\\
\end{minipage}
\label{figB:55}
\end{figure*}
%%%%%%%%%%%%%%%%%%%%%%%%%%%%%%%%%%%%
\begin{figure*}
\begin{minipage}[t]{0.98\textwidth}
\centering
\includegraphics[width = 145mm]{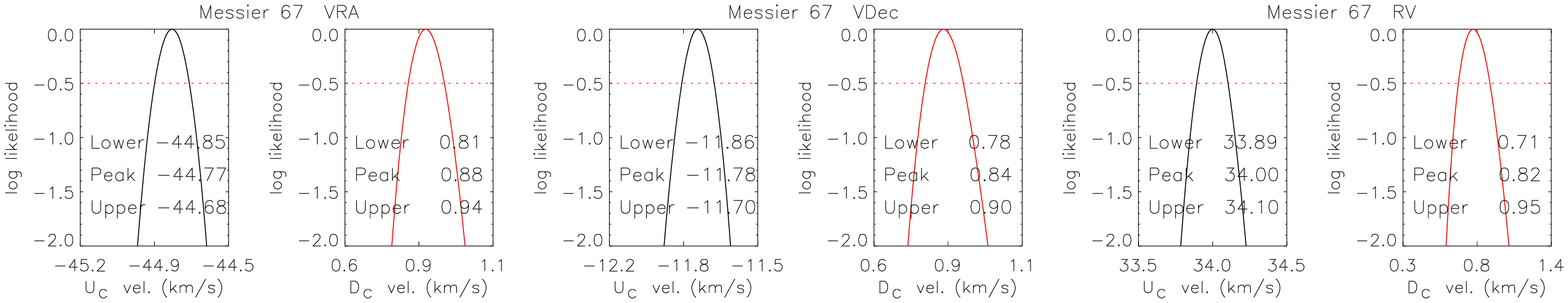}\\
\end{minipage}
\label{figB:56}
\end{figure*}
%%%%%%%%%%%%%%%%%%%%%%%%%%%%%%%%%%%%
\begin{figure*}
\begin{minipage}[t]{0.98\textwidth}
\centering
\includegraphics[width = 145mm]{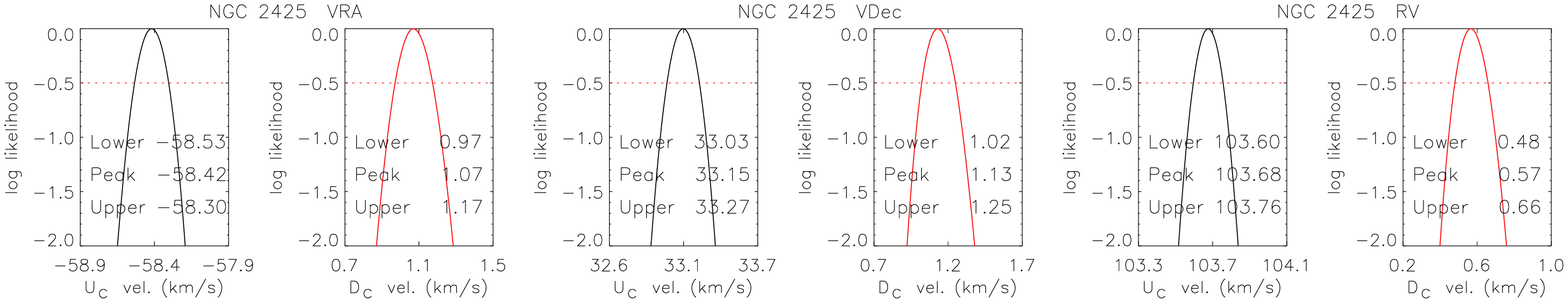}\\
\end{minipage}
\label{figB:57}
\end{figure*}
%%%%%%%%%%%%%%%%%%%%%%%%%%%%%%%%%%%%
\begin{figure*}
\begin{minipage}[t]{0.98\textwidth}
\centering
\includegraphics[width = 145mm]{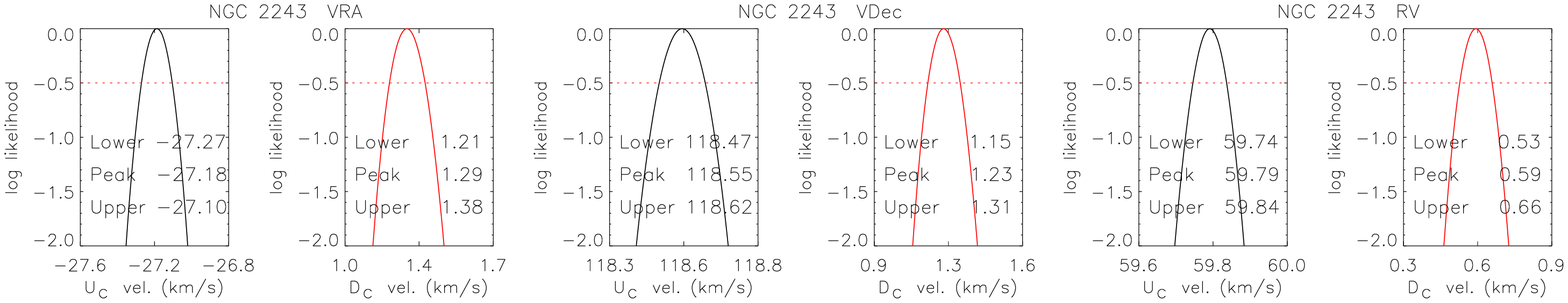}\\
\end{minipage}
\label{figB:58}
\end{figure*}
%%%%%%%%%%%%%%%%%%%%%%%%%%%%%%%%%%%%
\begin{figure*}
\begin{minipage}[t]{0.98\textwidth}
\centering
\includegraphics[width = 145mm]{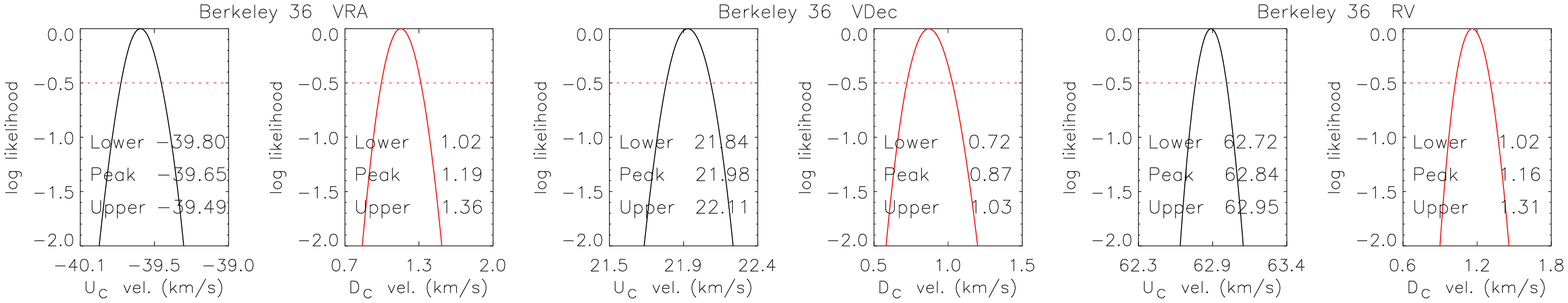}\\
\end{minipage}
\label{figB:59}
\end{figure*}
%%%%%%%%%%%%%%%%%%%%%%%%%%%%%%%%%%%%
\begin{figure*}
\begin{minipage}[t]{0.98\textwidth}
\centering
\includegraphics[width = 145mm]{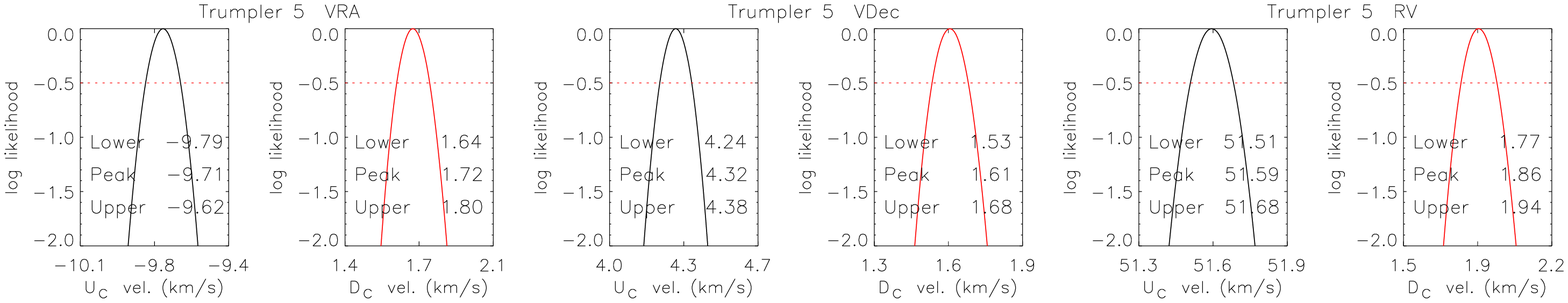}\\
\end{minipage}
\label{figB:60}
\end{figure*}
%%%%%%%%%%%%%%%%%%%%%%%%%%%%%%%%%%%%
\clearpage
\newpage
%%%%%%%%%%%%%%%%%%%%%%%%%%%%%%%%%%%%
\begin{figure*}
\begin{minipage}[t]{0.98\textwidth}
\centering
\includegraphics[width = 145mm]{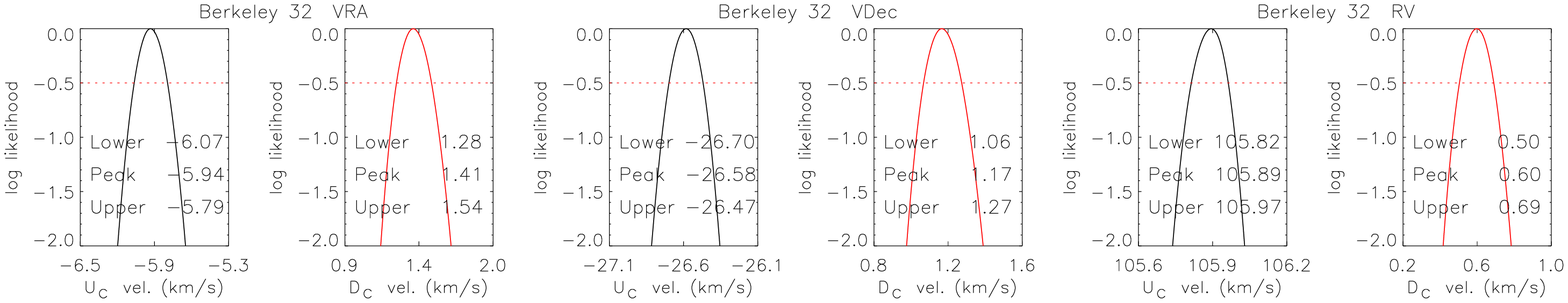}\\
\end{minipage}
\label{figB:61}
\end{figure*}
%%%%%%%%%%%%%%%%%%%%%%%%%%%%%%%%%%%%
\begin{figure*}
\begin{minipage}[t]{0.98\textwidth}
\centering
\includegraphics[width = 145mm]{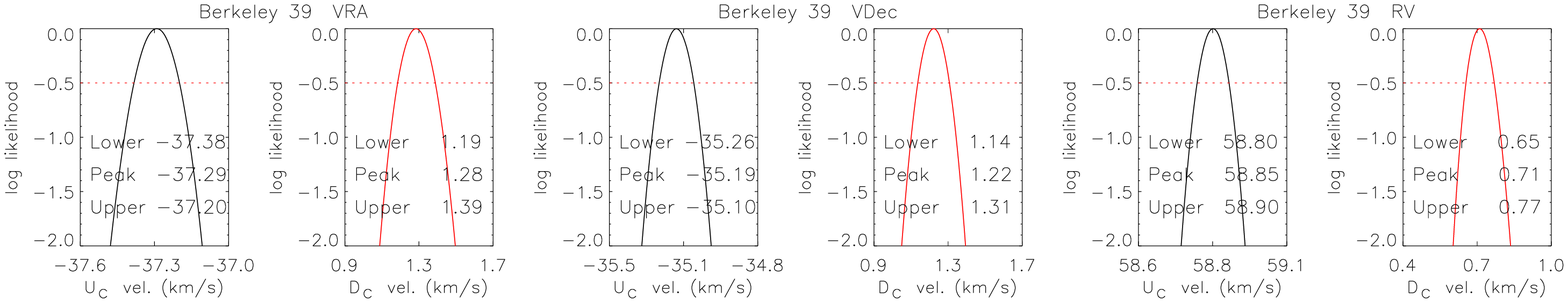}\\
\end{minipage}
\label{figB:62}
\end{figure*}
%%%%%%%%%%%%%%%%%%%%%%%%%%%%%%%%%%%%
\begin{figure*}
\begin{minipage}[t]{0.98\textwidth}
\centering
\includegraphics[width = 145mm]{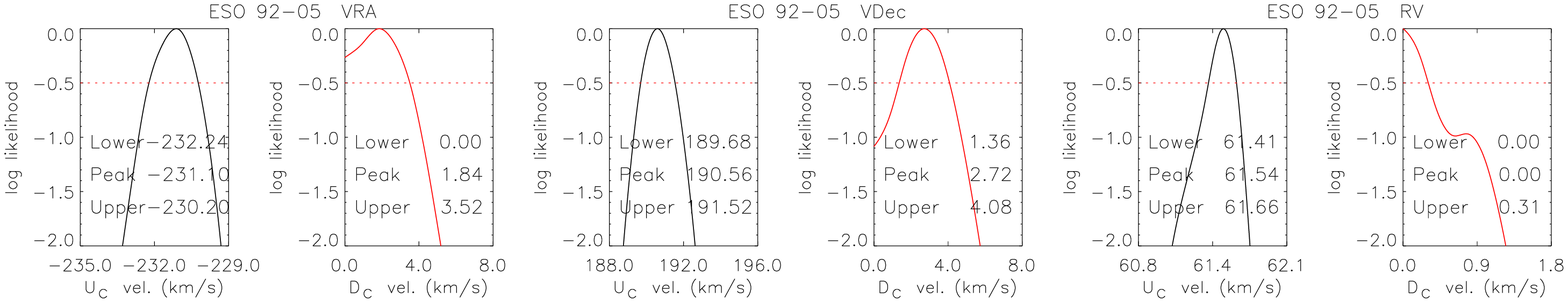}\\
\end{minipage}
\label{figB:63}
\end{figure*}
%%%%%%%%%%%%%%%%%%%%%%%%%%%%%%%%%%%%
\begin{figure*}
\begin{minipage}[t]{0.98\textwidth}
\centering
\includegraphics[width = 145mm]{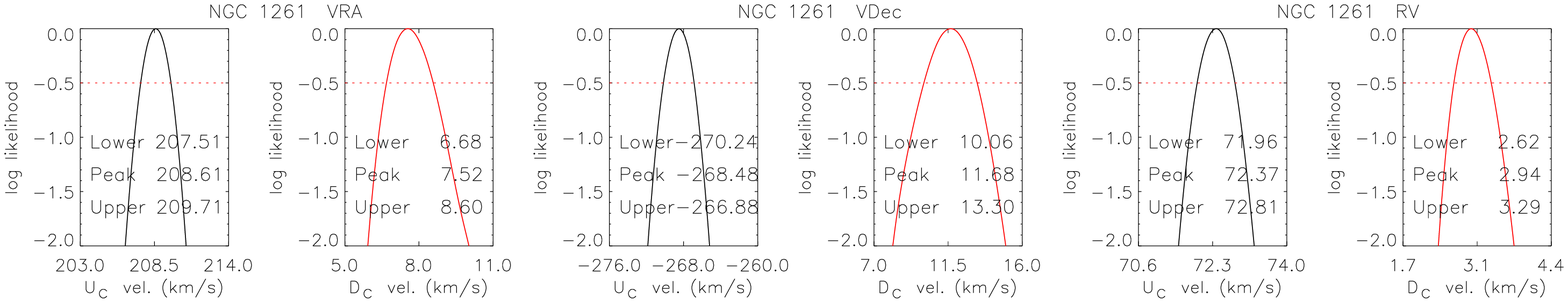}\\
\end{minipage}
\label{figB:64}
\end{figure*}
%%%%%%%%%%%%%%%%%%%%%%%%%%%%%%%%%%%%
\begin{figure*}
\begin{minipage}[t]{0.98\textwidth}
\centering
\includegraphics[width = 145mm]{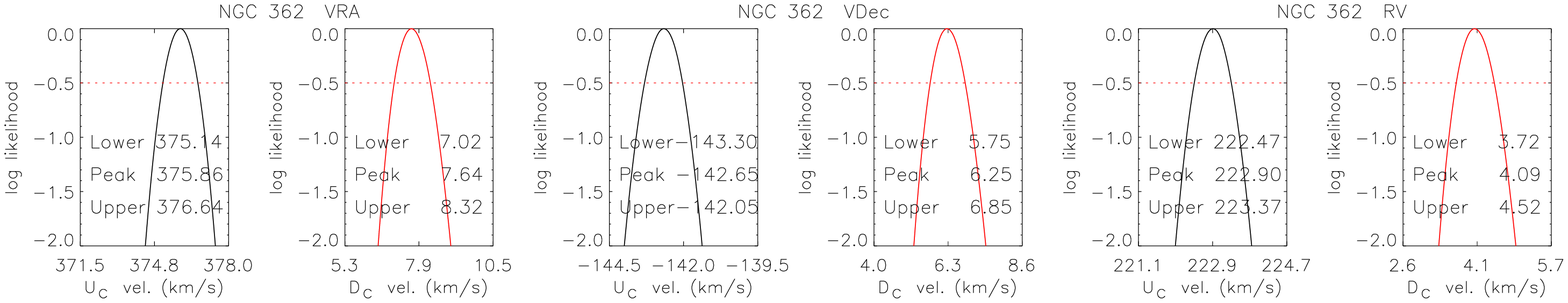}\\
\end{minipage}
\label{figB:65}
\end{figure*}
%%%%%%%%%%%%%%%%%%%%%%%%%%%%%%%%%%%%
\begin{figure*}
\begin{minipage}[t]{0.98\textwidth}
\centering
\includegraphics[width = 145mm]{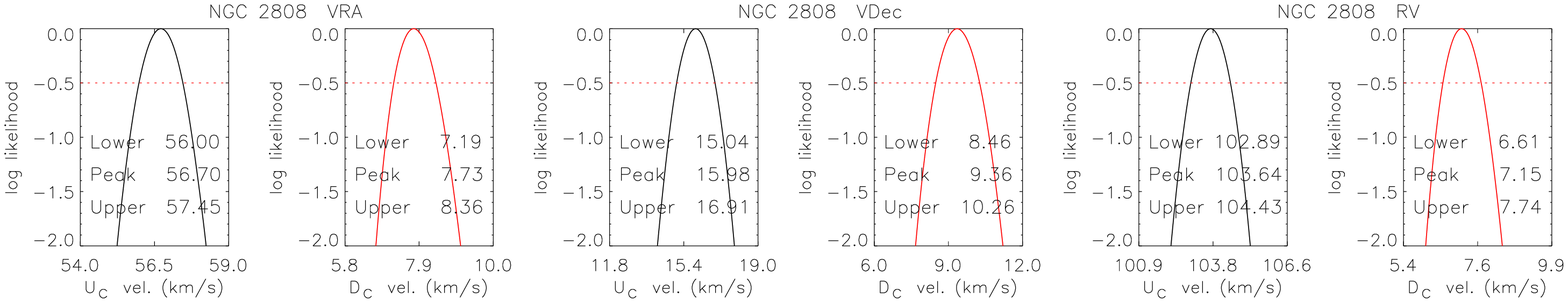}\\
\end{minipage}
\label{figB:66}
\end{figure*}
%%%%%%%%%%%%%%%%%%%%%%%%%%%%%%%%%%%%
\clearpage
\newpage
%%%%%%%%%%%%%%%%%%%%%%%%%%%%%%%%%%%%
\begin{figure*}
\begin{minipage}[t]{0.98\textwidth}
\centering
\includegraphics[width = 145mm]{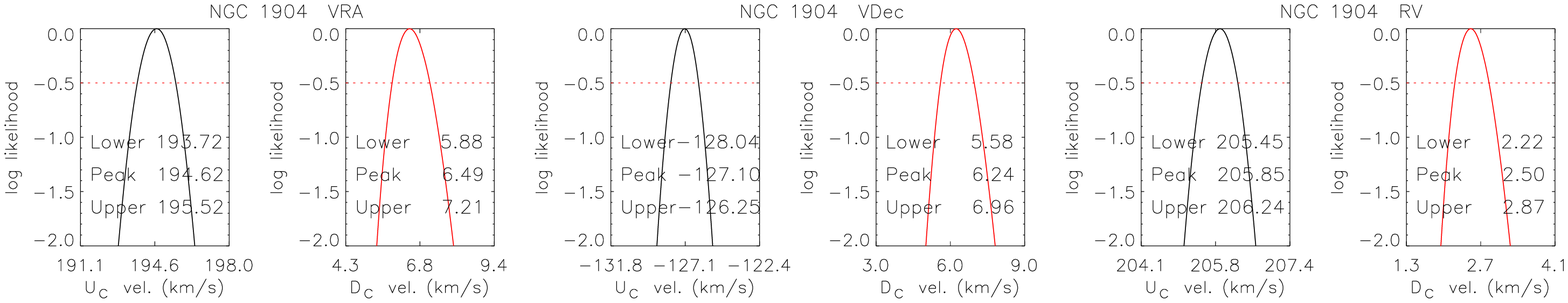}\\
\end{minipage}
\label{figB:67}
\end{figure*}
%%%%%%%%%%%%%%%%%%%%%%%%%%%%%%%%%%%%
\begin{figure*}
\begin{minipage}[t]{0.98\textwidth}
\centering
\includegraphics[width = 145mm]{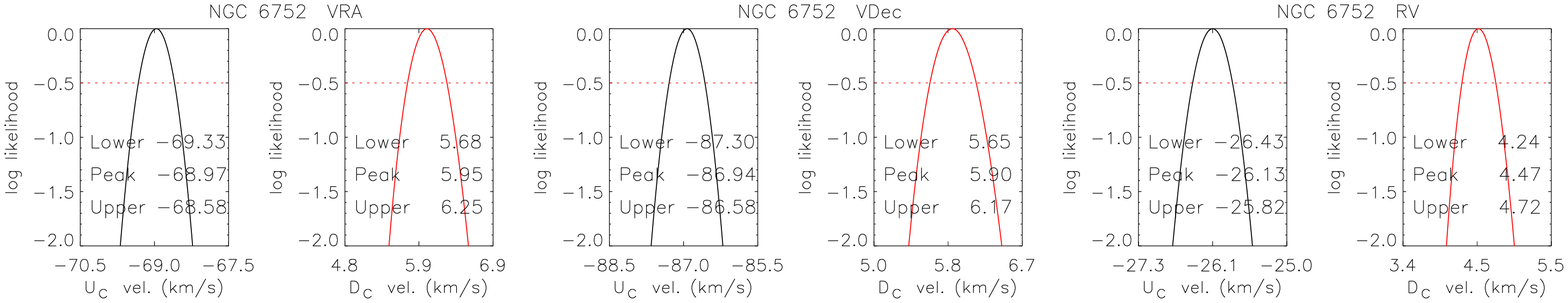}\\
\end{minipage}
\label{figB:68}
\end{figure*}
%%%%%%%%%%%%%%%%%%%%%%%%%%%%%%%%%%%%
\begin{figure*}
\begin{minipage}[t]{0.98\textwidth}
\centering
\includegraphics[width = 145mm]{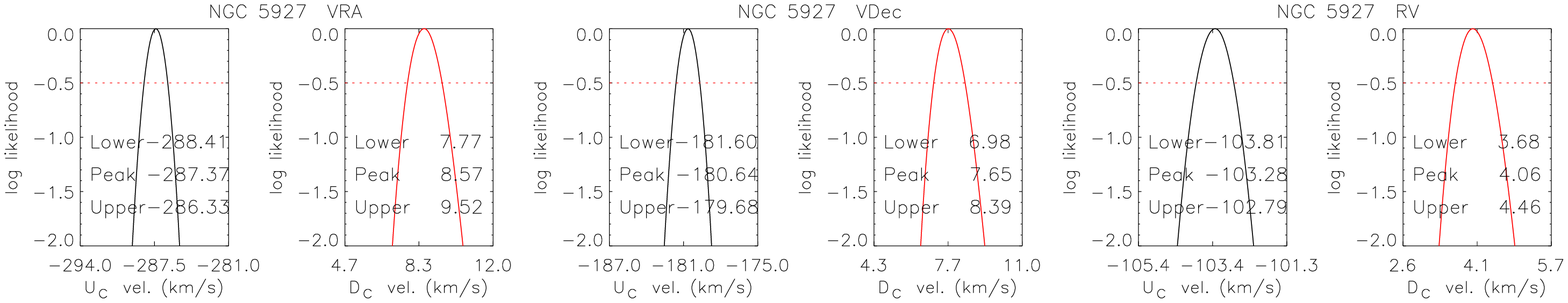}\\
\end{minipage}
\label{figB:69}
\end{figure*}
%%%%%%%%%%%%%%%%%%%%%%%%%%%%%%%%%%%%
\begin{figure*}
\begin{minipage}[t]{0.98\textwidth}
\centering
\includegraphics[width = 145mm]{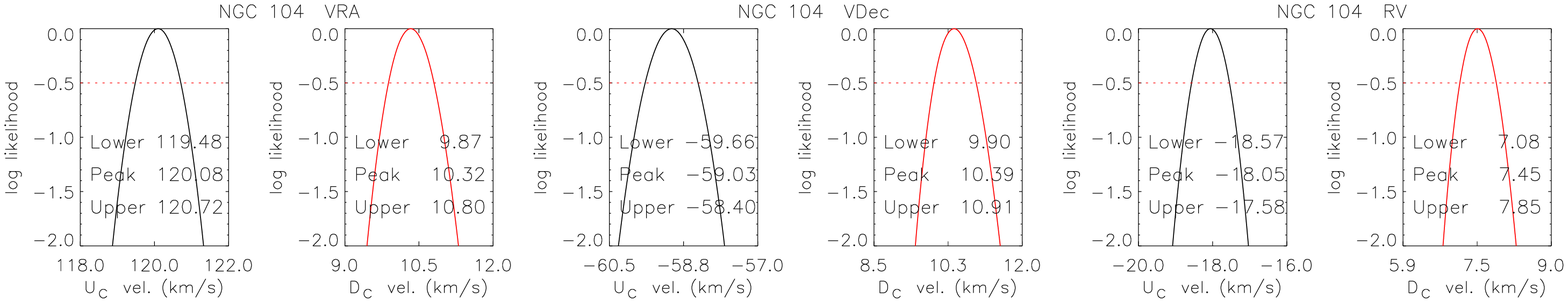}\\
\end{minipage}
\label{figB:70}
\end{figure*}
%%%%%%%%%%%%%%%%%%%%%%%%%%%%%%%%%%%%
%%%%%%%%%%%%%%%%%%%%%%%%%%%%%%%%%%%%
%%%%%%%%%%%%%%%%%%%%%%%%%%%%%%%%%%%%
%%%%%%%%%%%%%%%%%%%%%%%%%%%%%%%%%%%%
%%%%%%%%%%%%%%%%%%%%%%%%%%%%%%%%%%%%
%%%%%%%%%%%%%%%%%%%%%%%%%%%%%%%%%%%%
%%%%%%%%%%%%%%%%%%%%%%%%%%%%%%%%%%%%
\clearpage
\newpage
%%%%%%%%%%%%%%%%%%%%%%%%%%%%%%%%%%%%
\begin{figure*}
\begin{minipage}[t]{0.98\textwidth}
\centering
\includegraphics[width = 145mm]{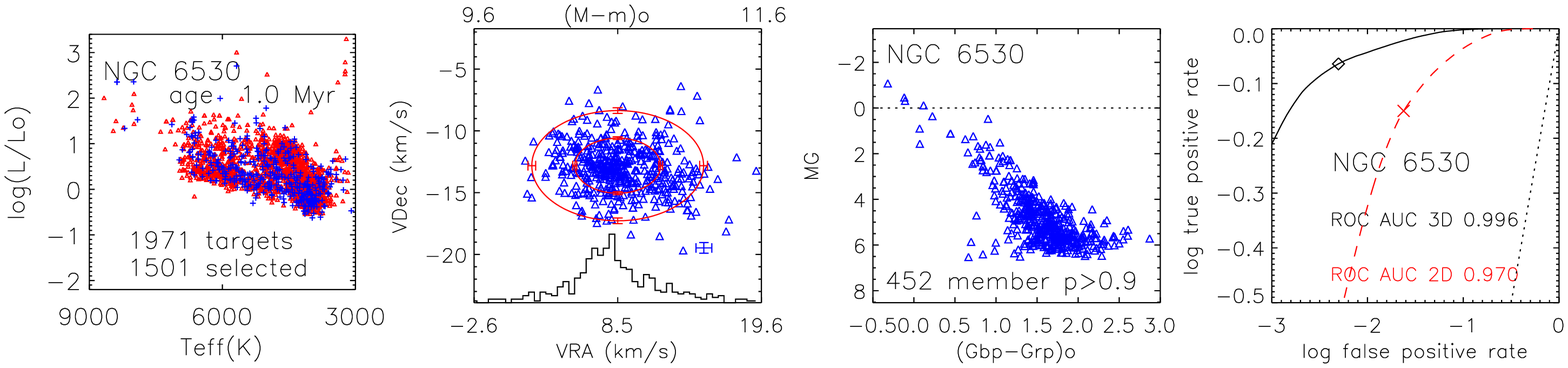}\\
\end{minipage}
\label{figB:1}
\end{figure*}
%%%%%%%%%%%%%%%%%%%%%%%%%%%%%%%%%%%%
\begin{figure*}
\begin{minipage}[t]{0.98\textwidth}
\centering
\includegraphics[width = 145mm]{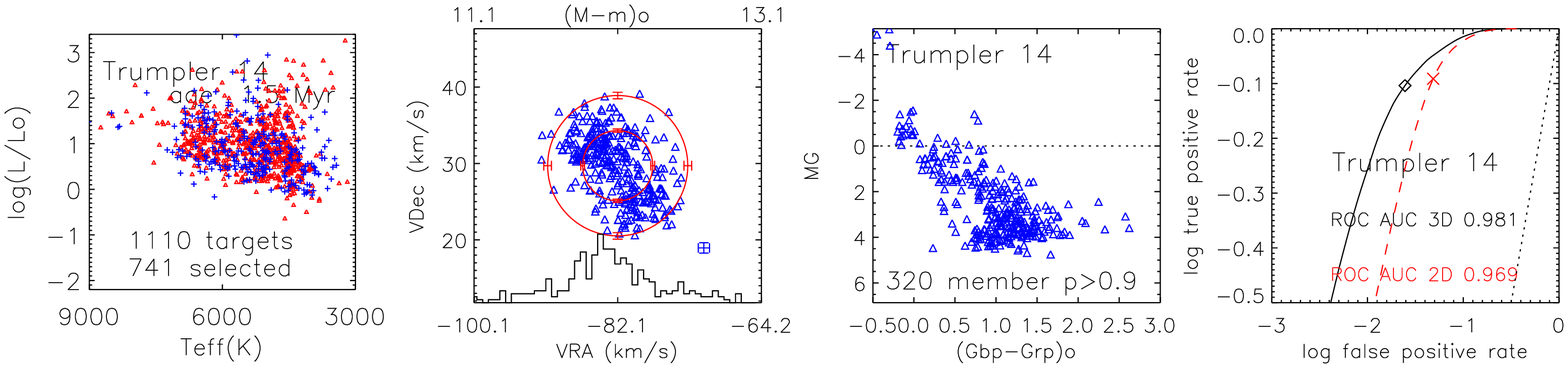}\\
\end{minipage}
\label{figB:2}
\end{figure*}
%%%%%%%%%%%%%%%%%%%%%%%%%%%%%%%%%%%%
\begin{figure*}
\begin{minipage}[t]{0.98\textwidth}
\centering
\includegraphics[width = 145mm]{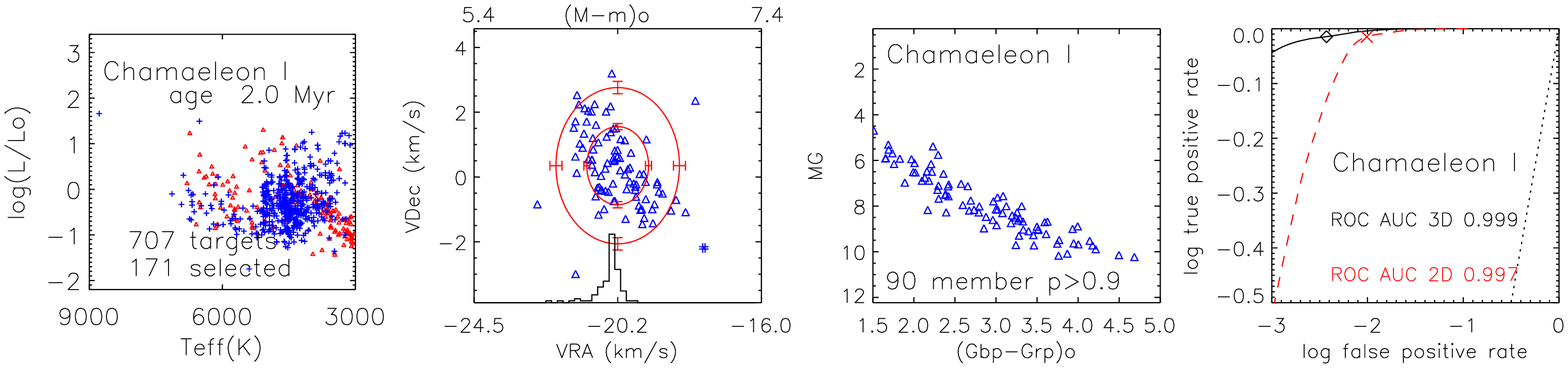}\\
\end{minipage}
\label{figB:3}
\end{figure*}
%%%%%%%%%%%%%%%%%%%%%%%%%%%%%%%%%%%%
\begin{figure*}
\begin{minipage}[t]{0.98\textwidth}
\centering
\includegraphics[width = 145mm]{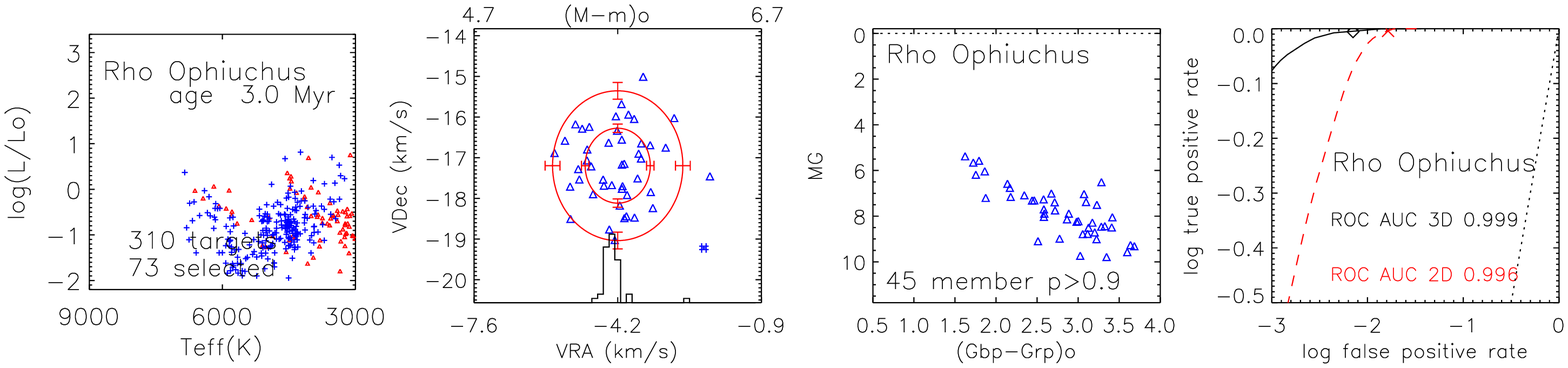}\\
\end{minipage}
\label{figB:4}
\end{figure*}
%%%%%%%%%%%%%%%%%%%%%%%%%%%%%%%%%%%%
\begin{figure*}
\begin{minipage}[t]{0.98\textwidth}
\centering
\includegraphics[width = 145mm]{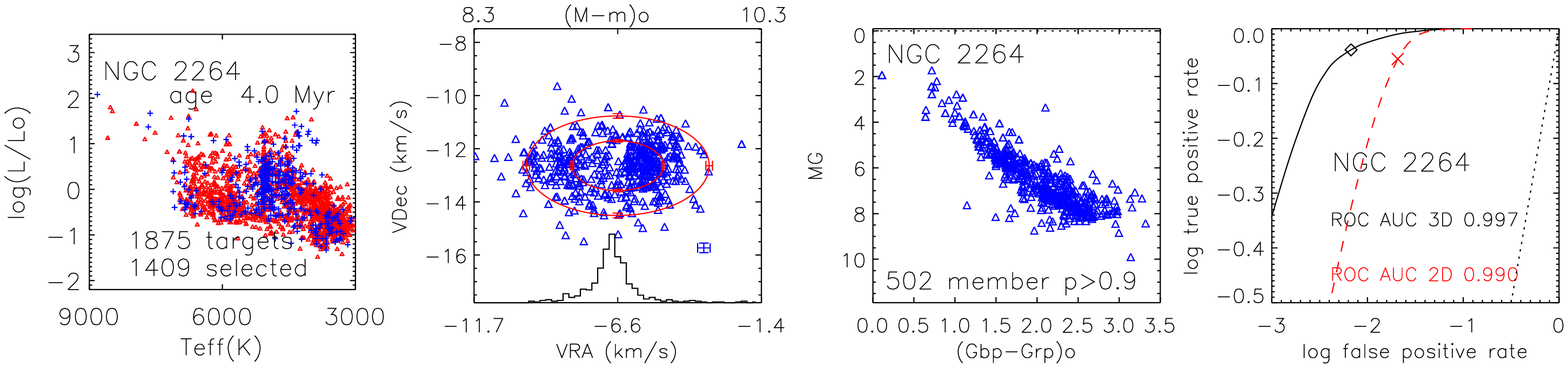}\\
\end{minipage}
\label{figB:5}
\end{figure*}
%%%%%%%%%%%%%%%%%%%%%%%%%%%%%%%%%%%%
\begin{figure*}
\begin{minipage}[t]{0.98\textwidth}
\centering
\includegraphics[width = 145mm]{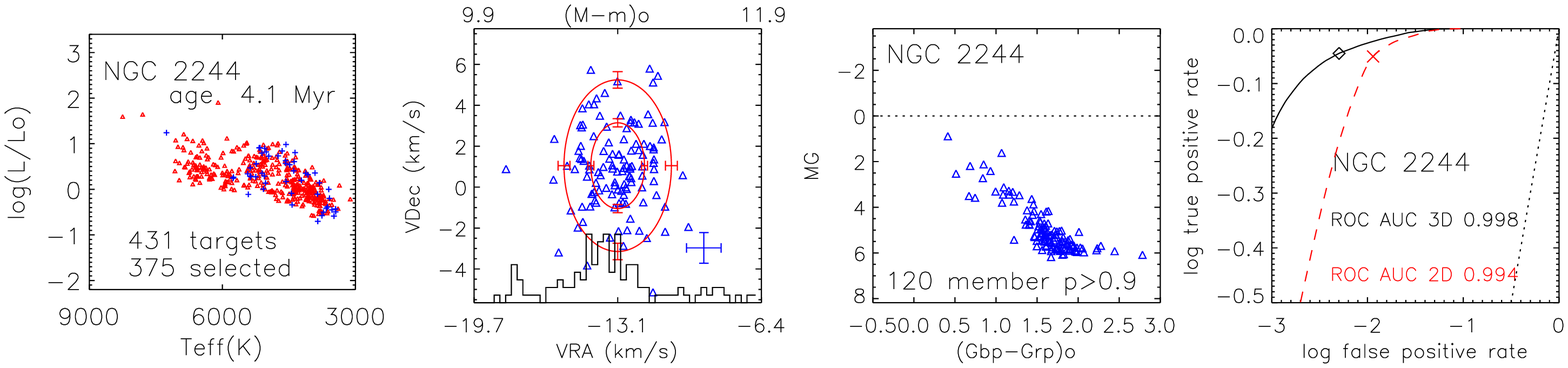}\\
\end{minipage}
\label{figB:6}
\end{figure*}
%%%%%%%%%%%%%%%%%%%%%%%%%%%%%%%%%%%%
\clearpage
\newpage
%%%%%%%%%%%%%%%%%%%%%%%%%%%%%%%%%%%%
\begin{figure*}
\begin{minipage}[t]{0.98\textwidth}
\centering
\includegraphics[width = 145mm]{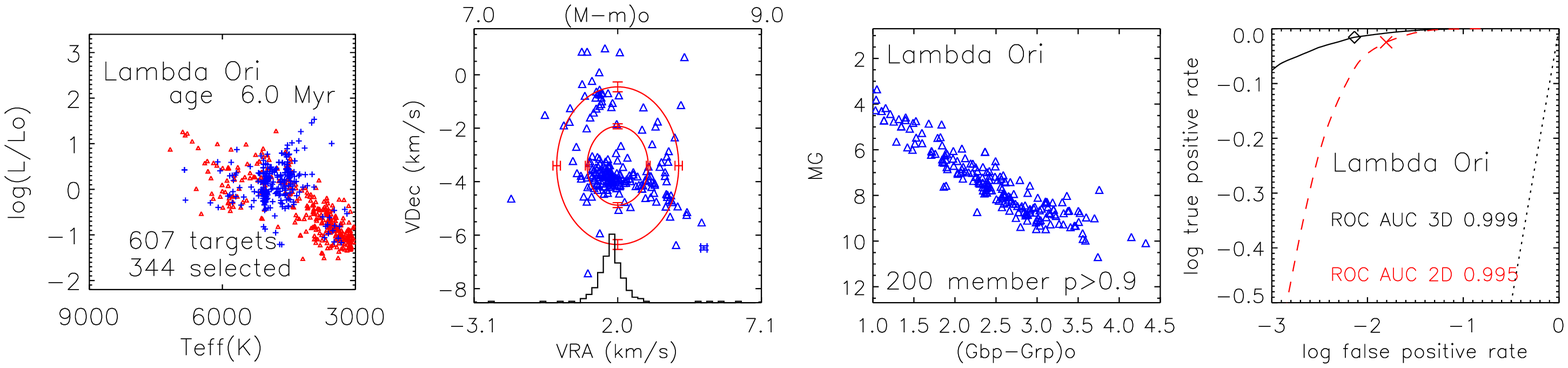}\\
\end{minipage}
\label{figB:7}
\end{figure*}
%%%%%%%%%%%%%%%%%%%%%%%%%%%%%%%%%%%%
\begin{figure*}
\begin{minipage}[t]{0.98\textwidth}
\centering
\includegraphics[width = 145mm]{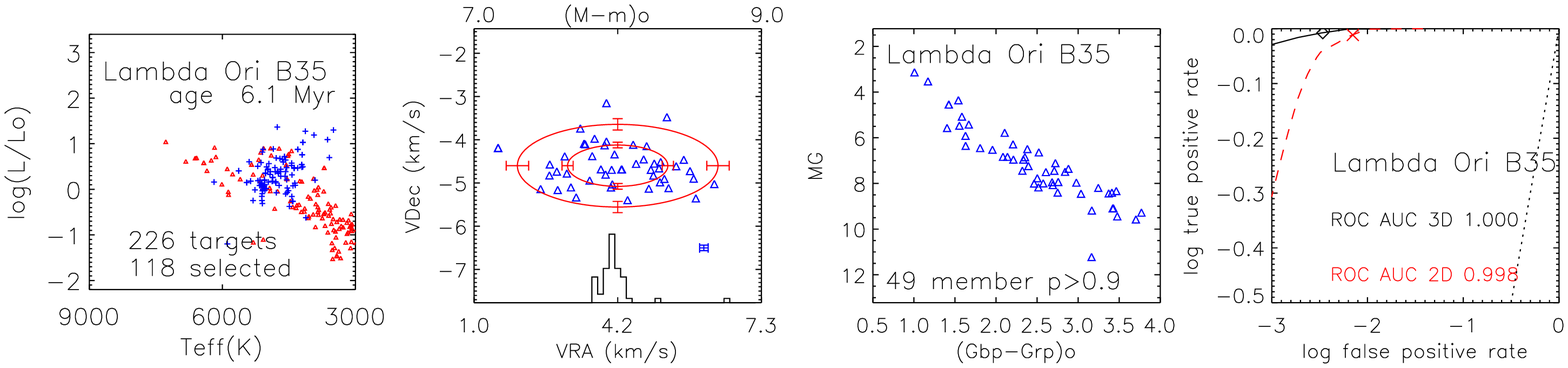}\\
\end{minipage}
\label{figB:8}
\end{figure*}
%%%%%%%%%%%%%%%%%%%%%%%%%%%%%%%%%%%%
\begin{figure*}
\begin{minipage}[t]{0.98\textwidth}
\centering
\includegraphics[width = 145mm]{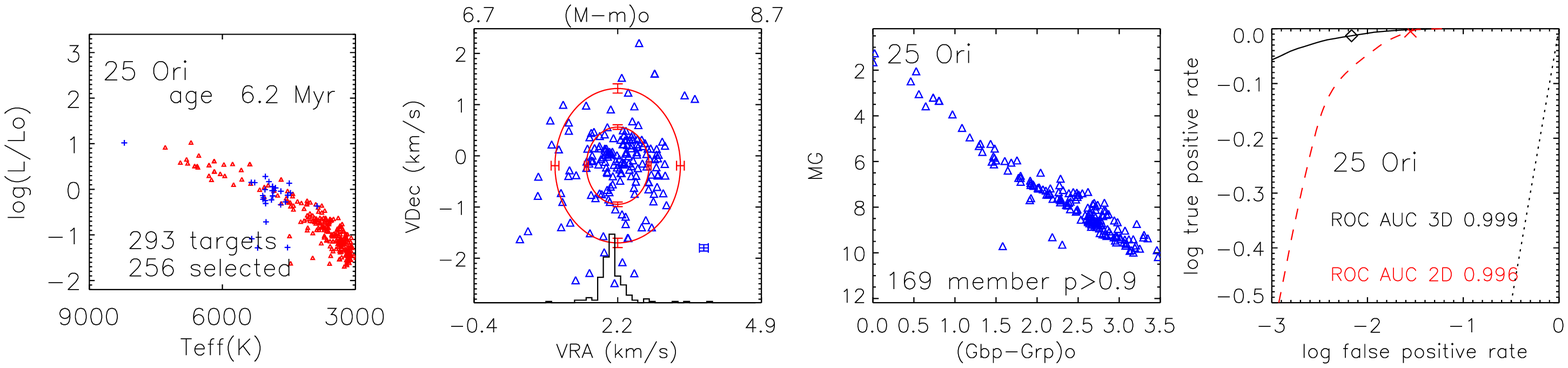}\\
\end{minipage}
\label{figB:9}
\end{figure*}
%%%%%%%%%%%%%%%%%%%%%%%%%%%%%%%%%%%%
\begin{figure*}
\begin{minipage}[t]{0.98\textwidth}
\centering
\includegraphics[width = 145mm]{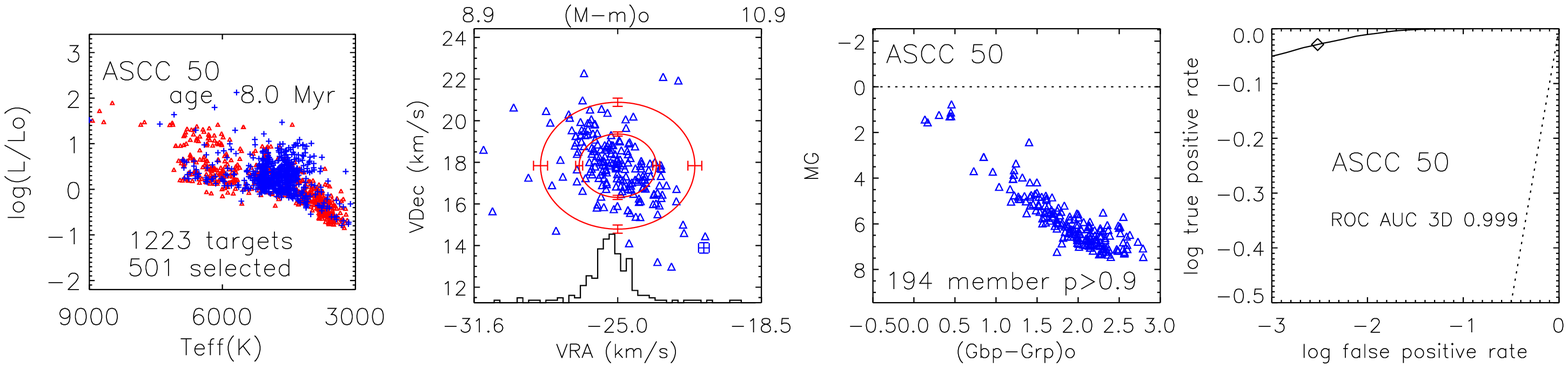}\\
\end{minipage}
\label{figB:10}
\end{figure*}
%%%%%%%%%%%%%%%%%%%%%%%%%%%%%%%%%%%%
\begin{figure*}
\begin{minipage}[t]{0.98\textwidth}
\centering
\includegraphics[width = 145mm]{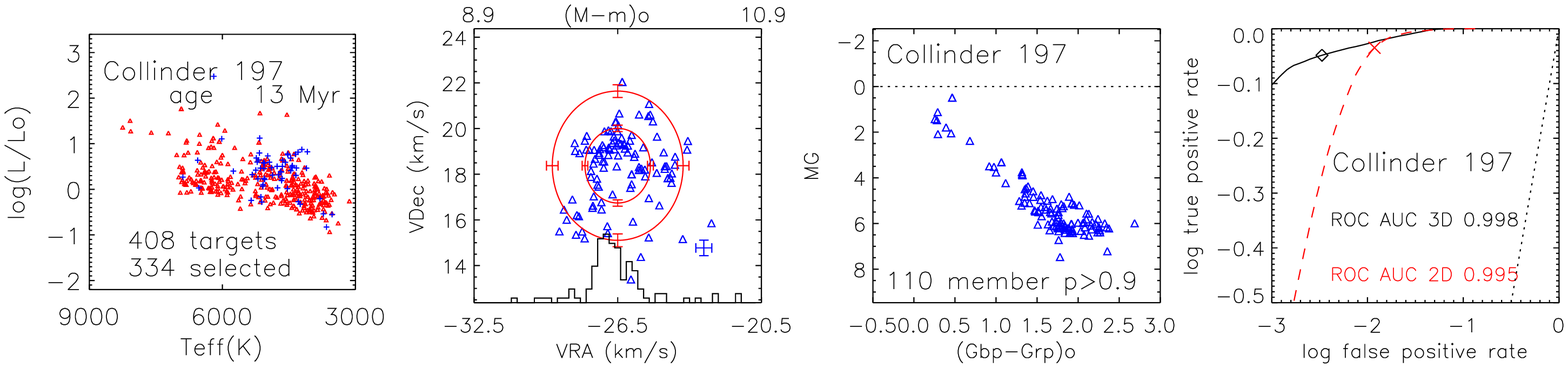}\\
\end{minipage}
\label{figB:11}
\end{figure*}
%%%%%%%%%%%%%%%%%%%%%%%%%%%%%%%%%%%%
\begin{figure*}
\begin{minipage}[t]{0.98\textwidth}
\centering
\includegraphics[width = 145mm]{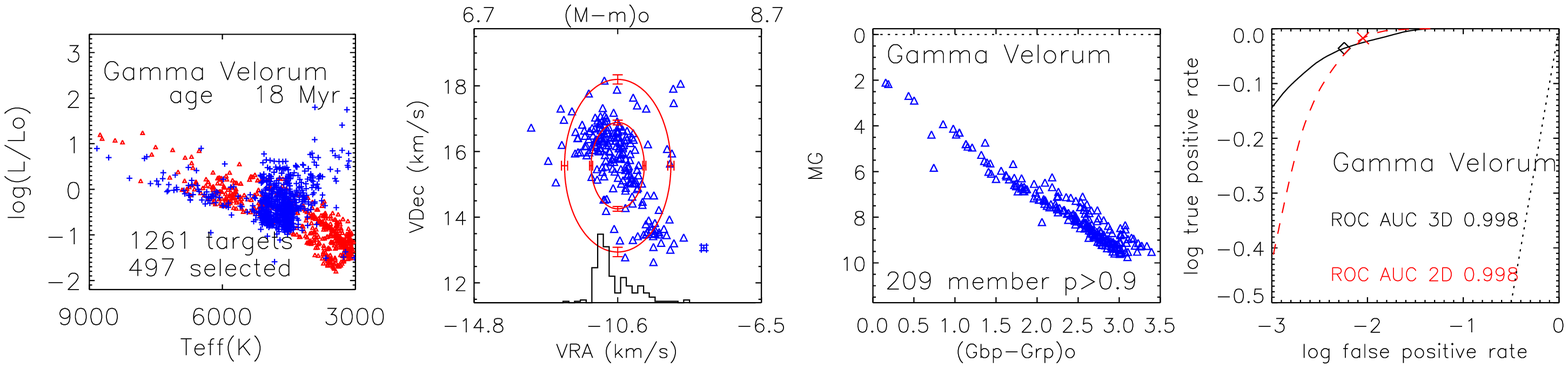}\\
\end{minipage}
\label{figB:12}
\end{figure*}
%%%%%%%%%%%%%%%%%%%%%%%%%%%%%%%%%%%%
\clearpage
\newpage
%%%%%%%%%%%%%%%%%%%%%%%%%%%%%%%%%%%%
\begin{figure*}
\begin{minipage}[t]{0.98\textwidth}
\centering
\includegraphics[width = 145mm]{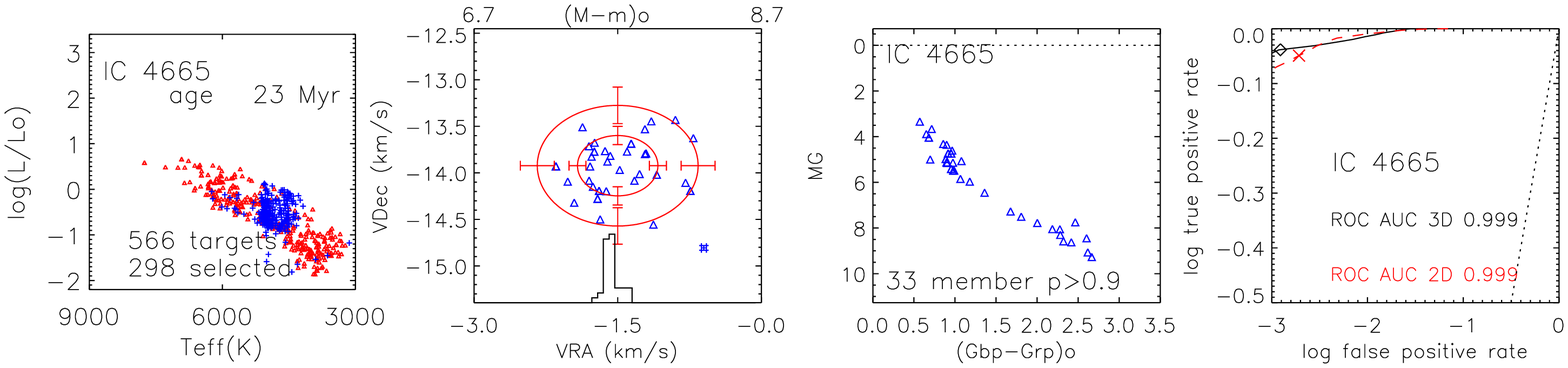}\\
\end{minipage}
\label{figB:13}
\end{figure*}
%%%%%%%%%%%%%%%%%%%%%%%%%%%%%%%%%%%%
\begin{figure*}
\begin{minipage}[t]{0.98\textwidth}
\centering
\includegraphics[width = 145mm]{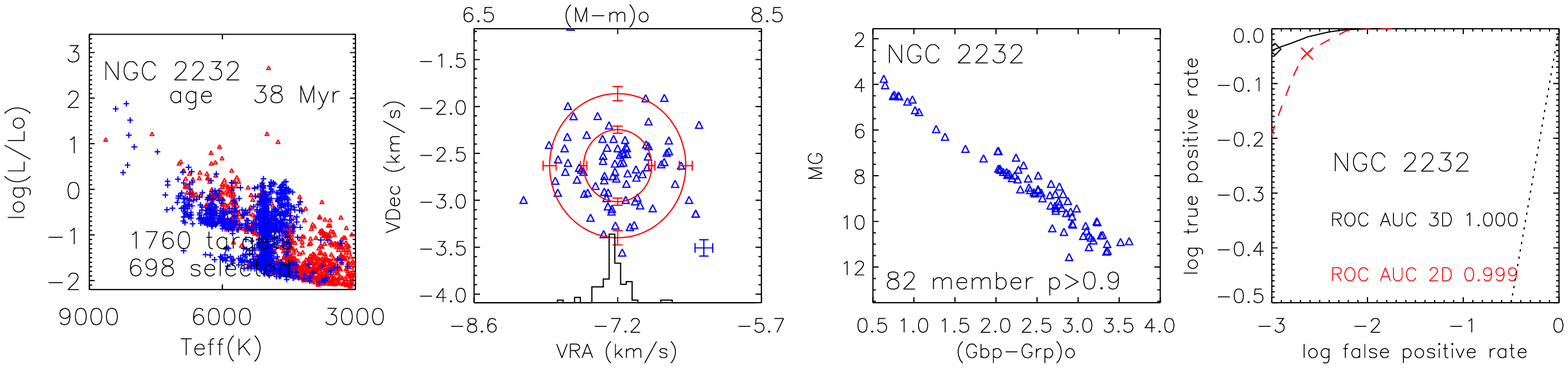}\\
\end{minipage}
\label{figB:14}
\end{figure*}
%%%%%%%%%%%%%%%%%%%%%%%%%%%%%%%%%%%%
\begin{figure*}
\begin{minipage}[t]{0.98\textwidth}
\centering
\includegraphics[width = 145mm]{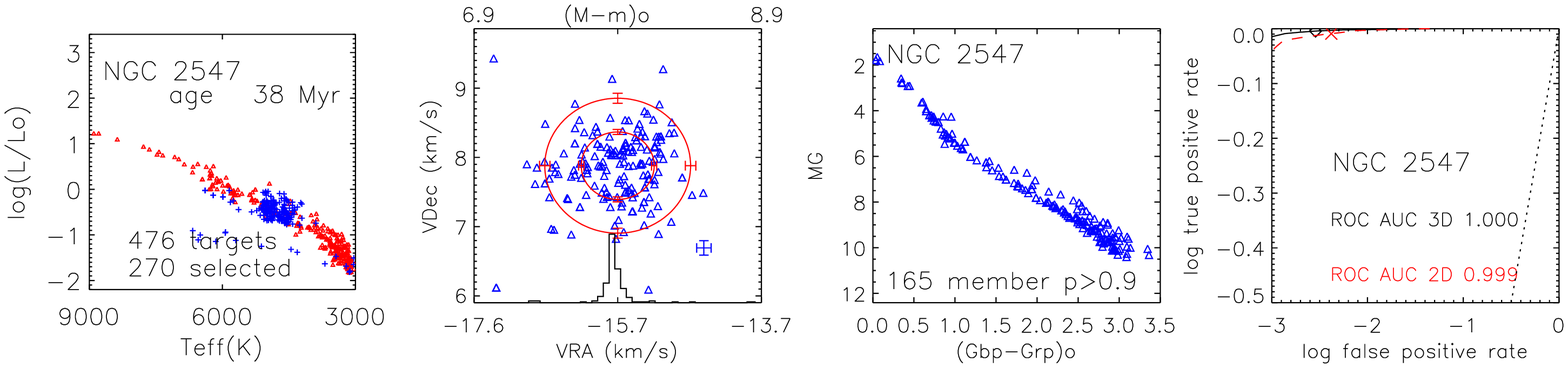}\\
\end{minipage}
\label{figB:15}
\end{figure*}
%%%%%%%%%%%%%%%%%%%%%%%%%%%%%%%%%%%%
\begin{figure*}
\begin{minipage}[t]{0.98\textwidth}
\centering
\includegraphics[width = 145mm]{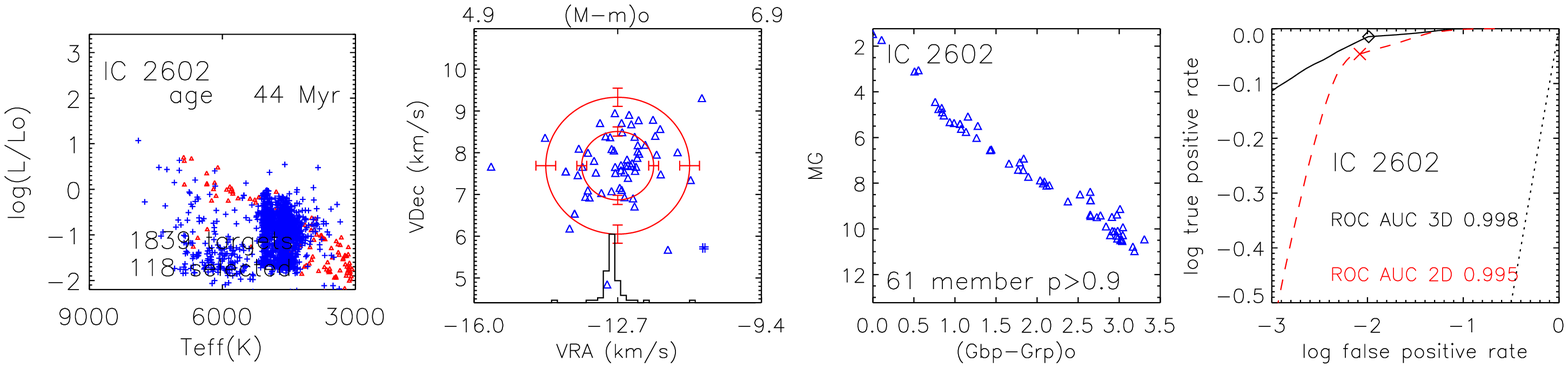}\\
\end{minipage}
\label{figB:16}
\end{figure*}
%%%%%%%%%%%%%%%%%%%%%%%%%%%%%%%%%%%%
\begin{figure*}
\begin{minipage}[t]{0.98\textwidth}
\centering
\includegraphics[width = 145mm]{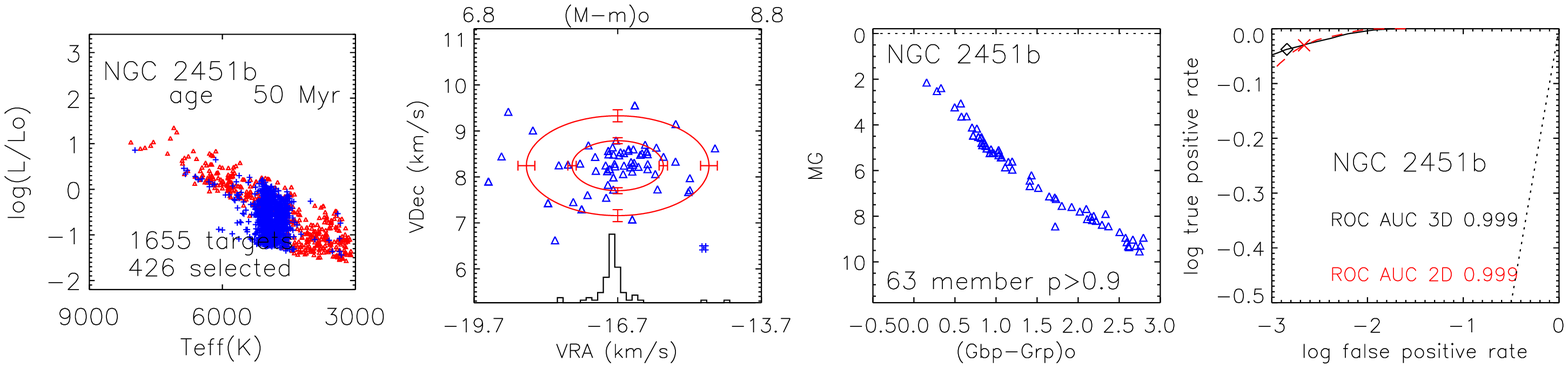}\\
\end{minipage}
\label{figB:17}
\end{figure*}
%%%%%%%%%%%%%%%%%%%%%%%%%%%%%%%%%%%%
\begin{figure*}
\begin{minipage}[t]{0.98\textwidth}
\centering
\includegraphics[width = 145mm]{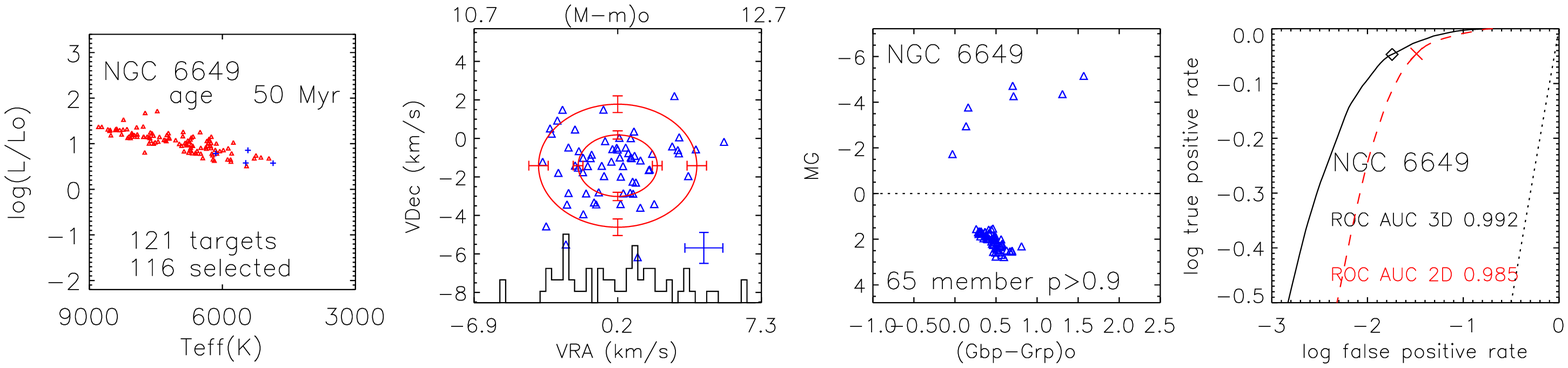}\\
\end{minipage}
\label{figB:18}
\end{figure*}
%%%%%%%%%%%%%%%%%%%%%%%%%%%%%%%%%%%%
\clearpage
\newpage
%%%%%%%%%%%%%%%%%%%%%%%%%%%%%%%%%%%%
\begin{figure*}
\begin{minipage}[t]{0.98\textwidth}
\centering
\includegraphics[width = 145mm]{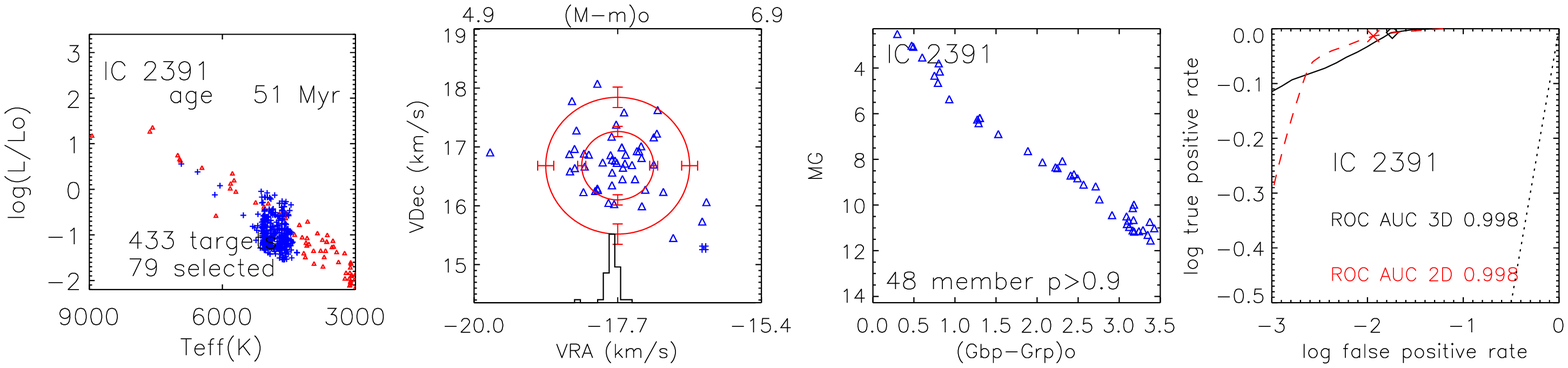}\\
\end{minipage}
\label{figB:19}
\end{figure*}
%%%%%%%%%%%%%%%%%%%%%%%%%%%%%%%%%%%%
\begin{figure*}
\begin{minipage}[t]{0.98\textwidth}
\centering
\includegraphics[width = 145mm]{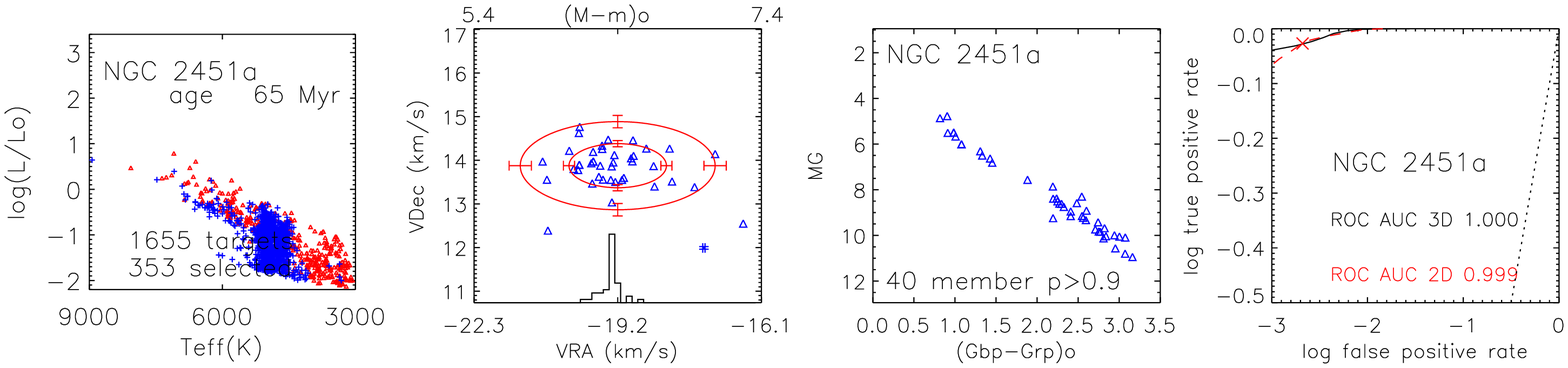}\\
\end{minipage}
\label{figB:20}
\end{figure*}
%%%%%%%%%%%%%%%%%%%%%%%%%%%%%%%%%%%%
\begin{figure*}
\begin{minipage}[t]{0.98\textwidth}
\centering
\includegraphics[width = 145mm]{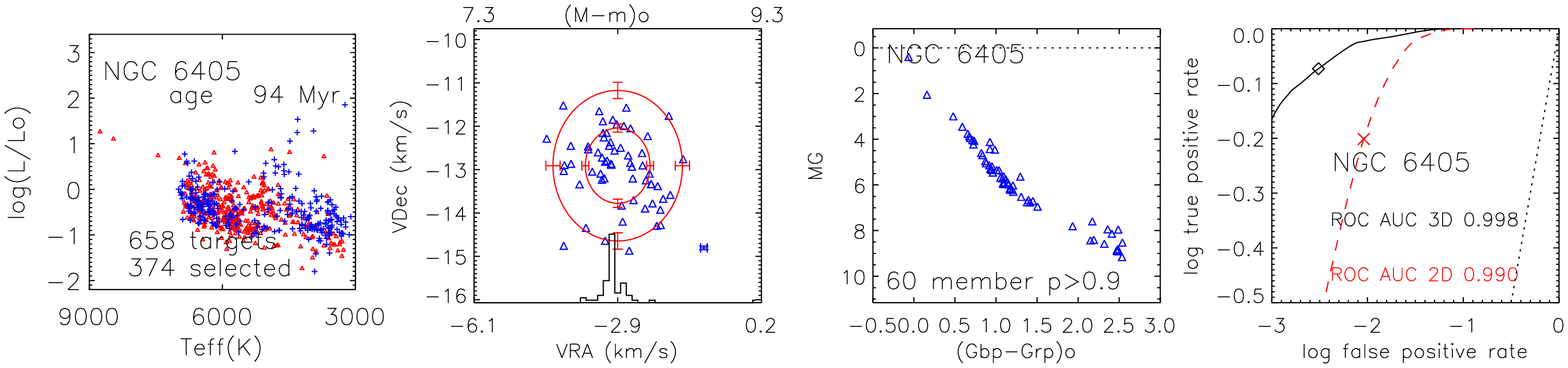}\\
\end{minipage}
\label{figB:21}
\end{figure*}
%%%%%%%%%%%%%%%%%%%%%%%%%%%%%%%%%%%%
\begin{figure*}
\begin{minipage}[t]{0.98\textwidth}
\centering
\includegraphics[width = 145mm]{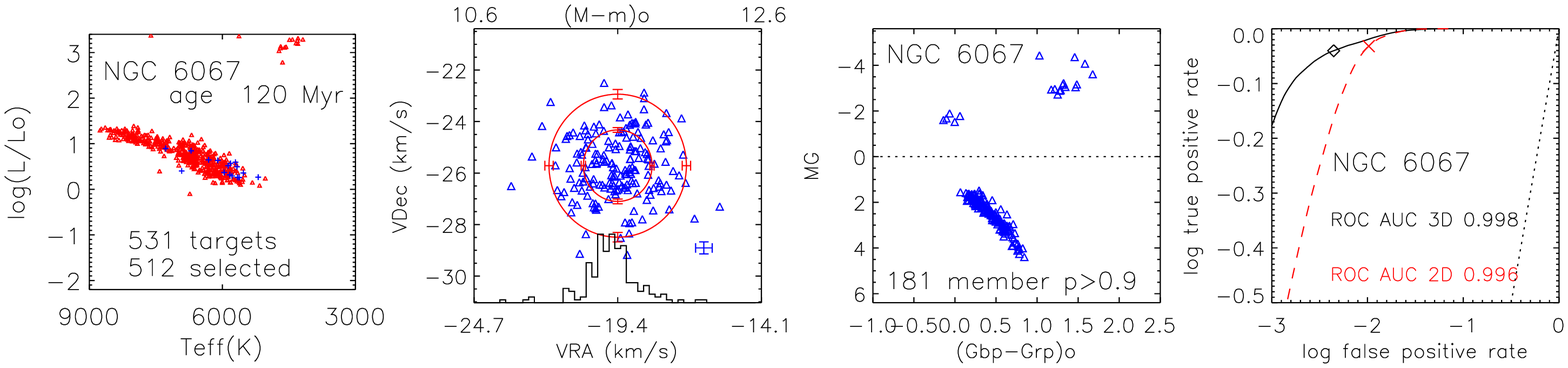}\\
\end{minipage}
\label{figB:22}
\end{figure*}
%%%%%%%%%%%%%%%%%%%%%%%%%%%%%%%%%%%%
\begin{figure*}
\begin{minipage}[t]{0.98\textwidth}
\centering
\includegraphics[width = 145mm]{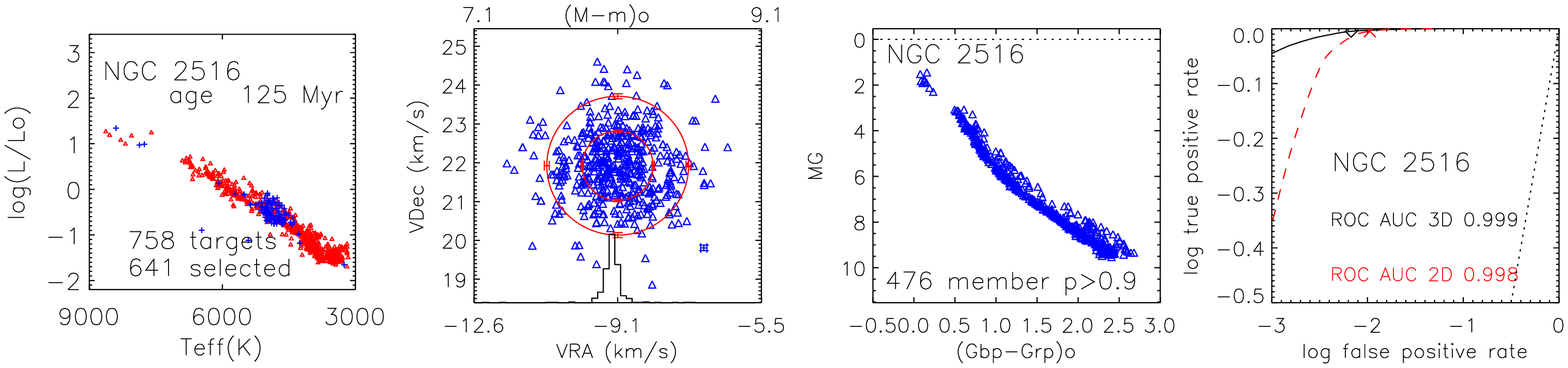}\\
\end{minipage}
\label{figB:23}
\end{figure*}
%%%%%%%%%%%%%%%%%%%%%%%%%%%%%%%%%%%%
\begin{figure*}
\begin{minipage}[t]{0.98\textwidth}
\centering
\includegraphics[width = 145mm]{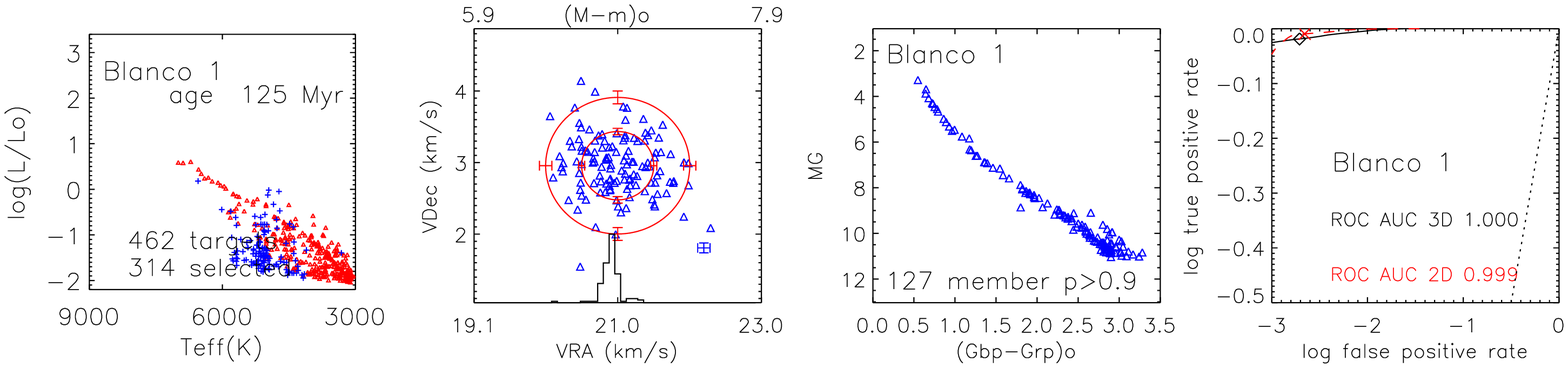}\\
\end{minipage}
\label{figB:24}
\end{figure*}
%%%%%%%%%%%%%%%%%%%%%%%%%%%%%%%%%%%%
\clearpage
\newpage
%%%%%%%%%%%%%%%%%%%%%%%%%%%%%%%%%%%%
\begin{figure*}
\begin{minipage}[t]{0.98\textwidth}
\centering
\includegraphics[width = 145mm]{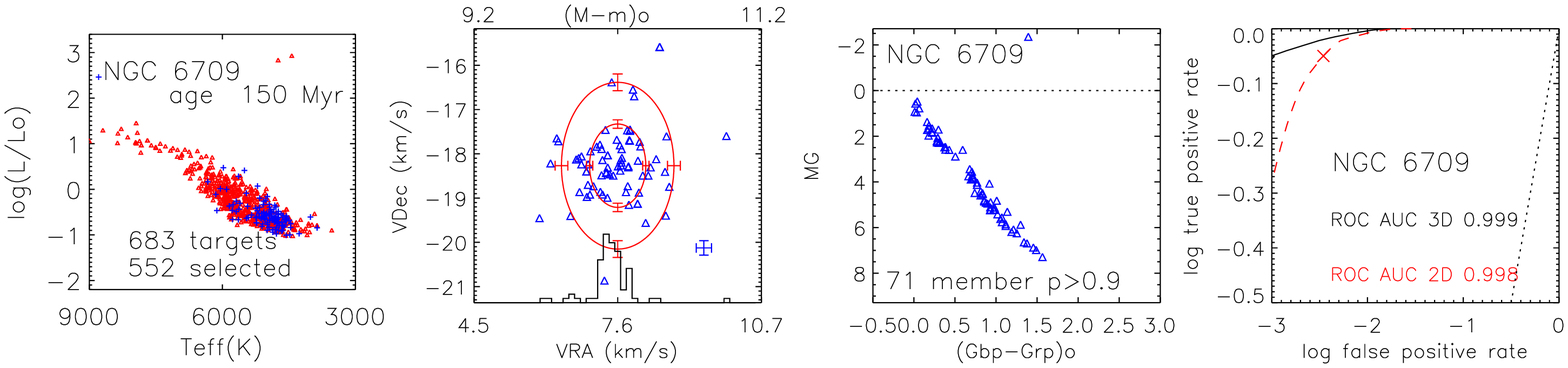}\\
\end{minipage}
\label{figB:25}
\end{figure*}
%%%%%%%%%%%%%%%%%%%%%%%%%%%%%%%%%%%%
\begin{figure*}
\begin{minipage}[t]{0.98\textwidth}
\centering
\includegraphics[width = 145mm]{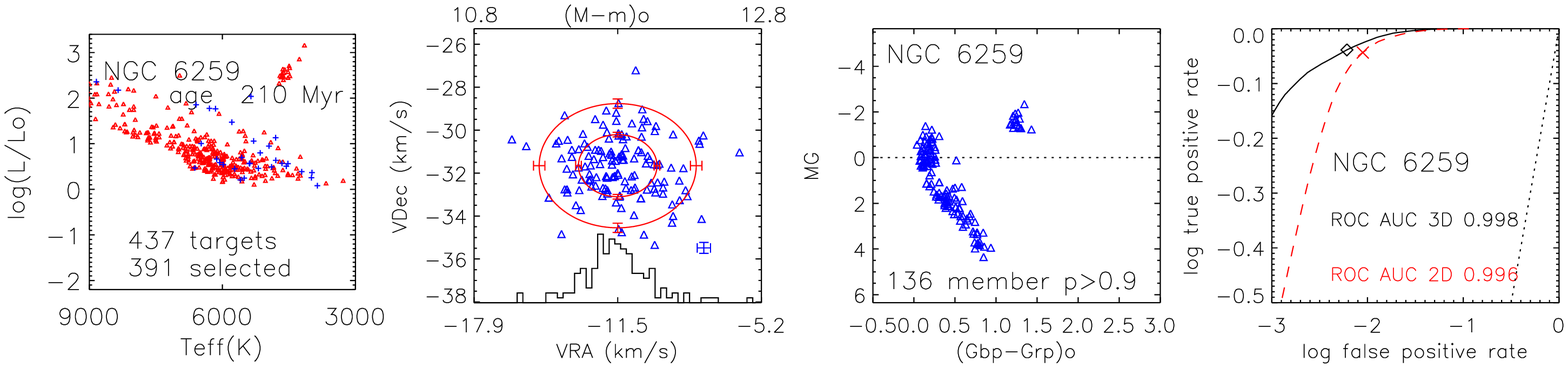}\\
\end{minipage}
\label{figB:26}
\end{figure*}
%%%%%%%%%%%%%%%%%%%%%%%%%%%%%%%%%%%%
\begin{figure*}
\begin{minipage}[t]{0.98\textwidth}
\centering
\includegraphics[width = 145mm]{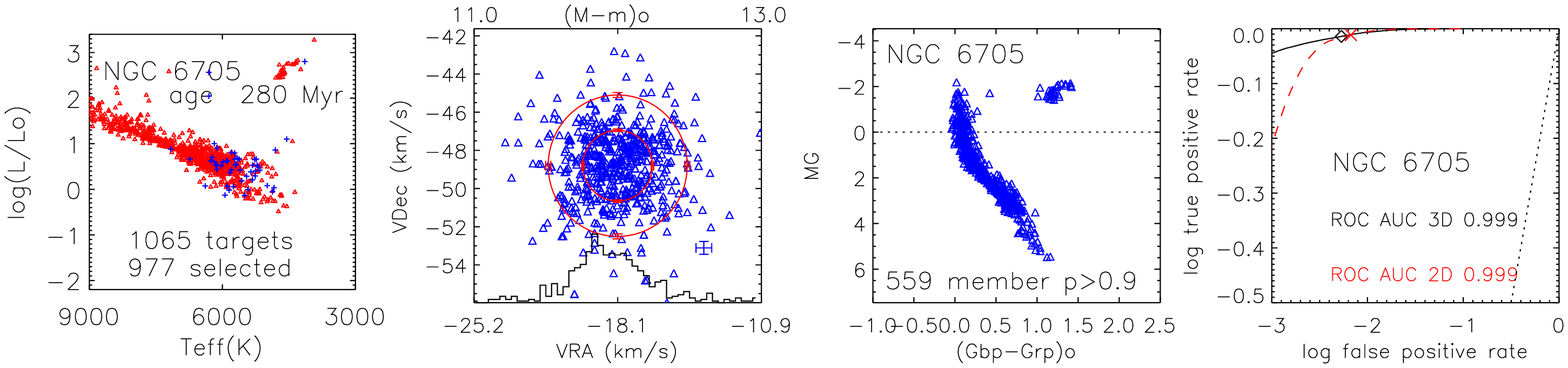}\\
\end{minipage}
\label{figB:27}
\end{figure*}
%%%%%%%%%%%%%%%%%%%%%%%%%%%%%%%%%%%%
\begin{figure*}
\begin{minipage}[t]{0.98\textwidth}
\centering
\includegraphics[width = 145mm]{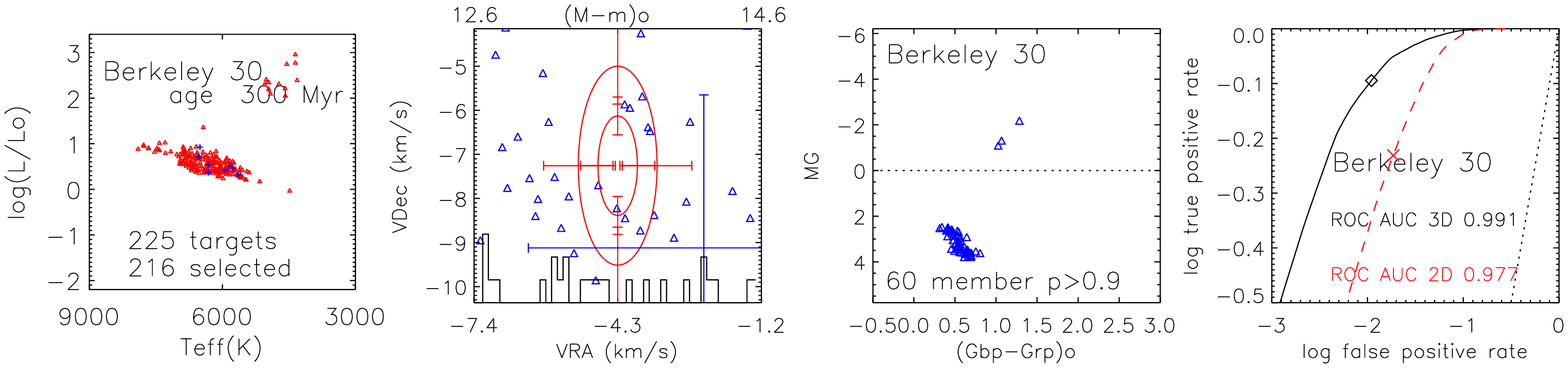}\\
\end{minipage}
\label{figB:28}
\end{figure*}
%%%%%%%%%%%%%%%%%%%%%%%%%%%%%%%%%%%%
\begin{figure*}
\begin{minipage}[t]{0.98\textwidth}
\centering
\includegraphics[width = 145mm]{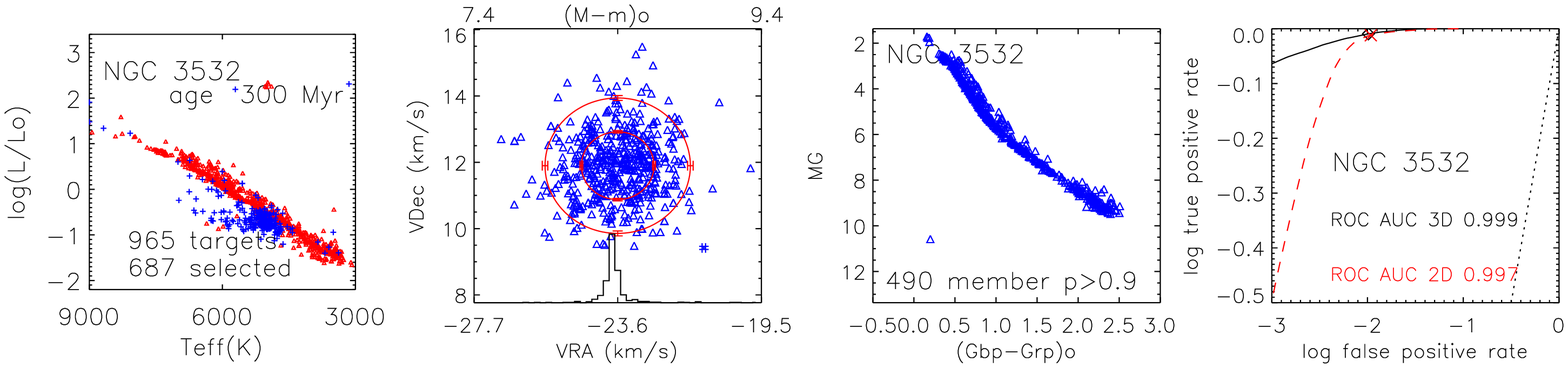}\\
\end{minipage}
\label{figB:29}
\end{figure*}
%%%%%%%%%%%%%%%%%%%%%%%%%%%%%%%%%%%%
\begin{figure*}
\begin{minipage}[t]{0.98\textwidth}
\centering
\includegraphics[width = 145mm]{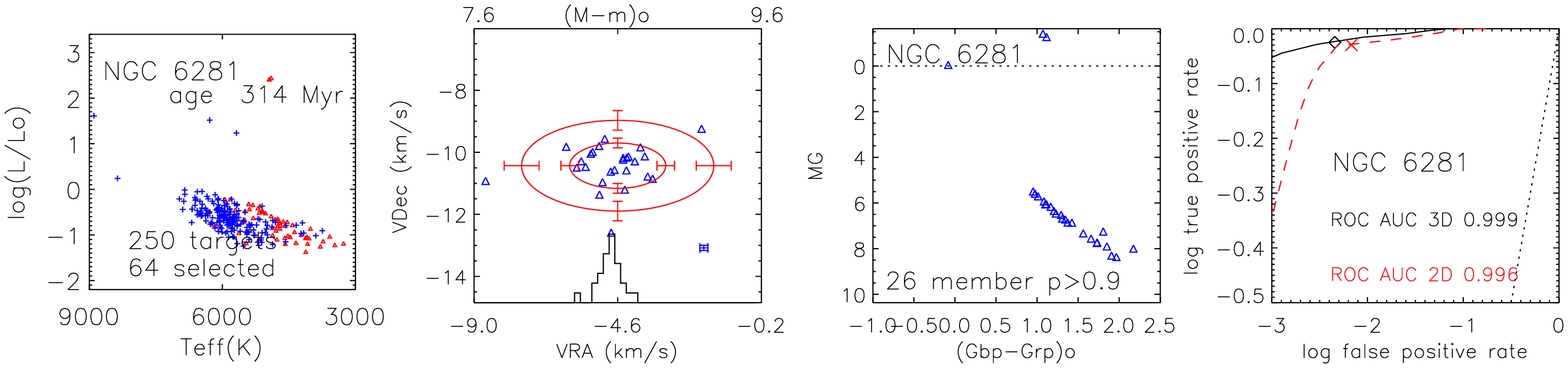}\\
\end{minipage}
\label{figB:30}
\end{figure*}
%%%%%%%%%%%%%%%%%%%%%%%%%%%%%%%%%%%%
\clearpage
\newpage
%%%%%%%%%%%%%%%%%%%%%%%%%%%%%%%%%%%%
\begin{figure*}
\begin{minipage}[t]{0.98\textwidth}
\centering
\includegraphics[width = 145mm]{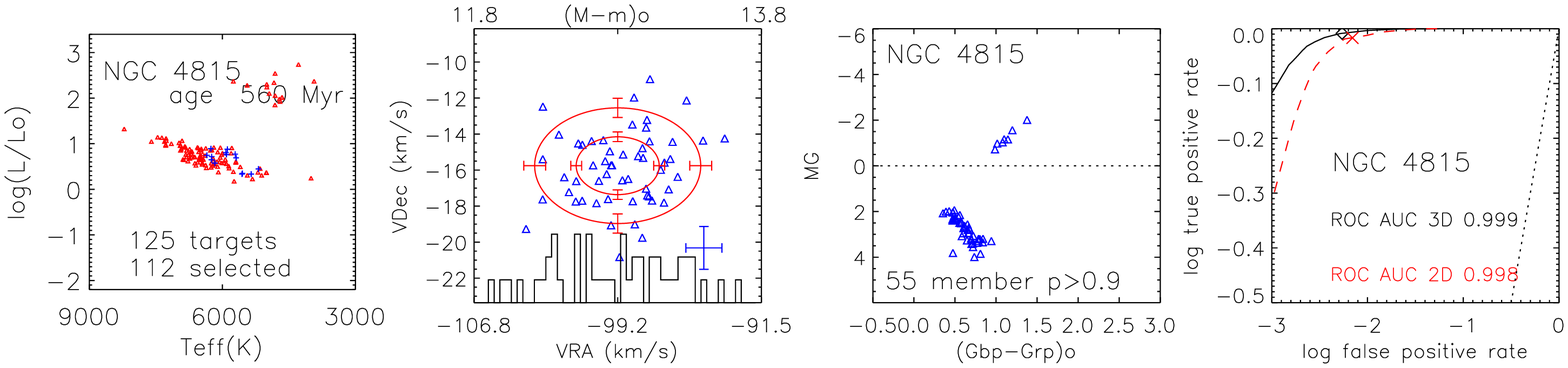}\\
\end{minipage}
\label{figB:31}
\end{figure*}
%%%%%%%%%%%%%%%%%%%%%%%%%%%%%%%%%%%%
\begin{figure*}
\begin{minipage}[t]{0.98\textwidth}
\centering
\includegraphics[width = 145mm]{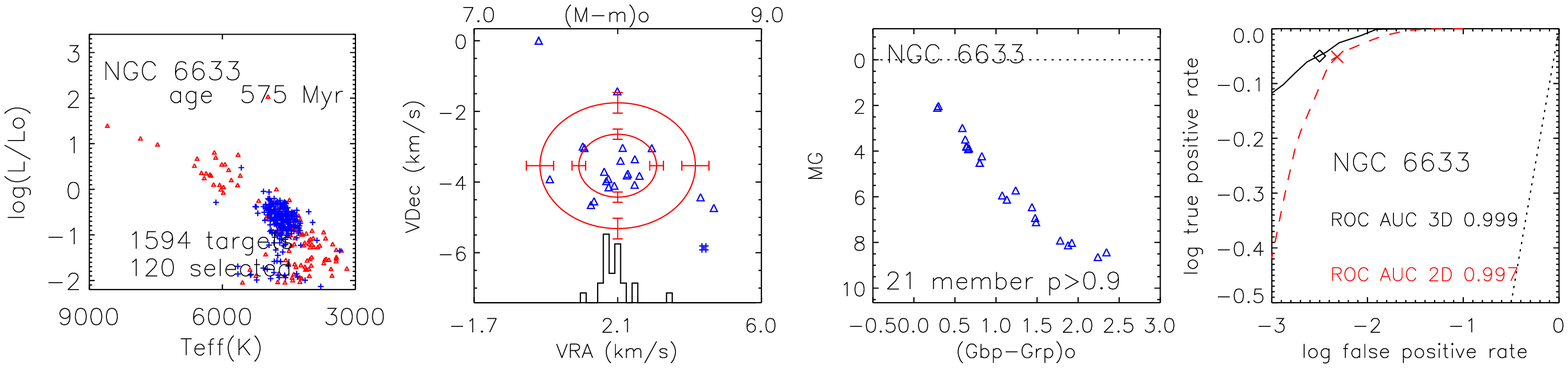}\\
\end{minipage}
\label{figB:32}
\end{figure*}
%%%%%%%%%%%%%%%%%%%%%%%%%%%%%%%%%%%%
\begin{figure*}
\begin{minipage}[t]{0.98\textwidth}
\centering
\includegraphics[width = 145mm]{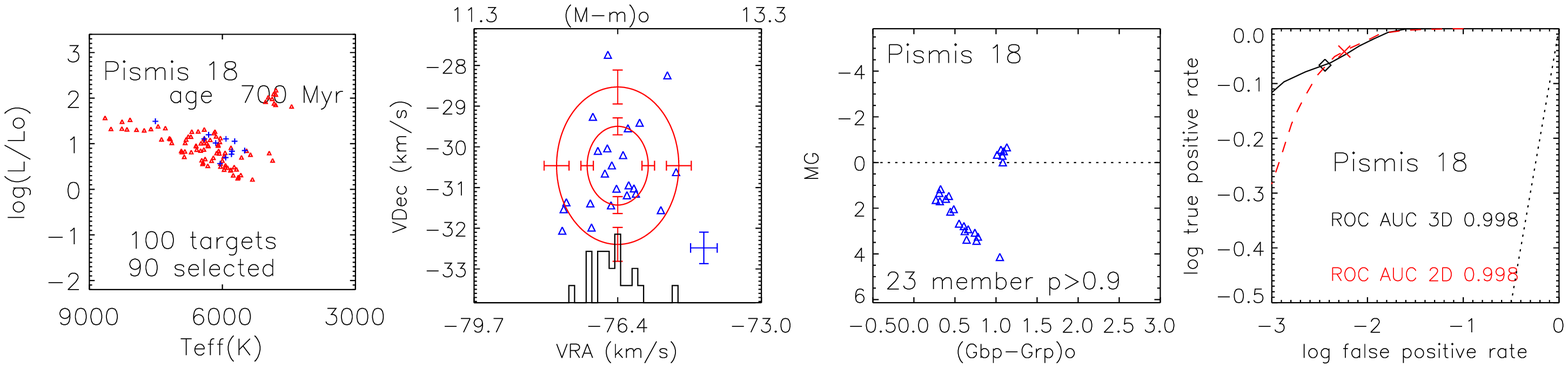}\\
\end{minipage}
\label{figB:33}
\end{figure*}
%%%%%%%%%%%%%%%%%%%%%%%%%%%%%%%%%%%%
\begin{figure*}
\begin{minipage}[t]{0.98\textwidth}
\centering
\includegraphics[width = 145mm]{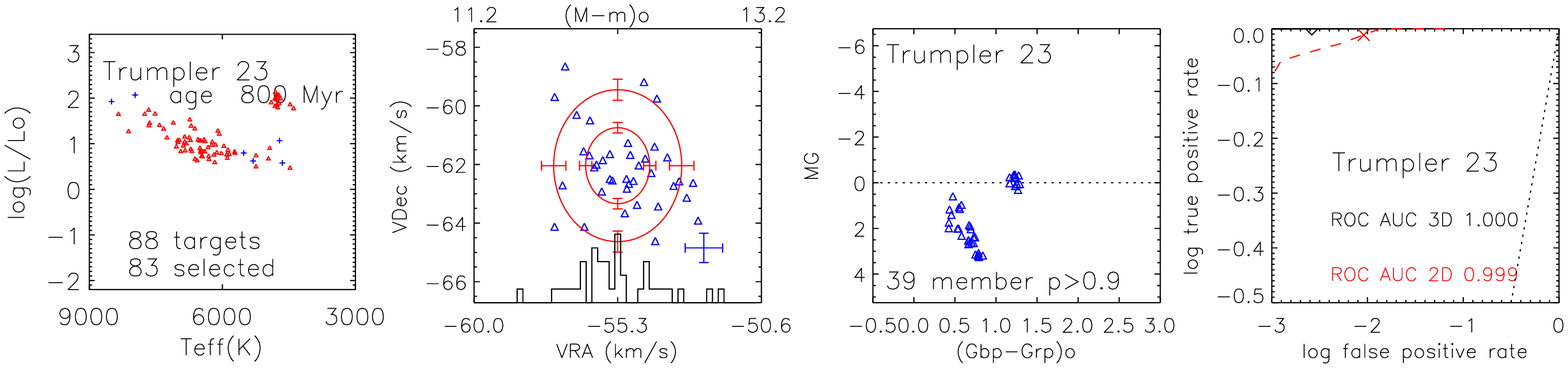}\\
\end{minipage}
\label{figB:34}
\end{figure*}
%%%%%%%%%%%%%%%%%%%%%%%%%%%%%%%%%%%%
\begin{figure*}
\begin{minipage}[t]{0.98\textwidth}
\centering
\includegraphics[width = 145mm]{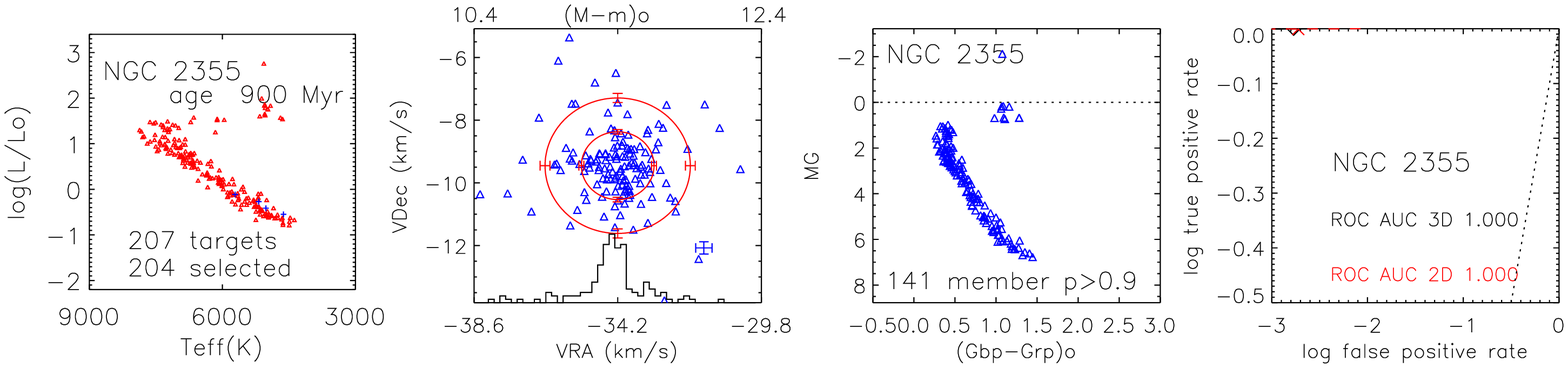}\\
\end{minipage}
\label{figB:35}
\end{figure*}
%%%%%%%%%%%%%%%%%%%%%%%%%%%%%%%%%%%%
\begin{figure*}
\begin{minipage}[t]{0.98\textwidth}
\centering
\includegraphics[width = 145mm]{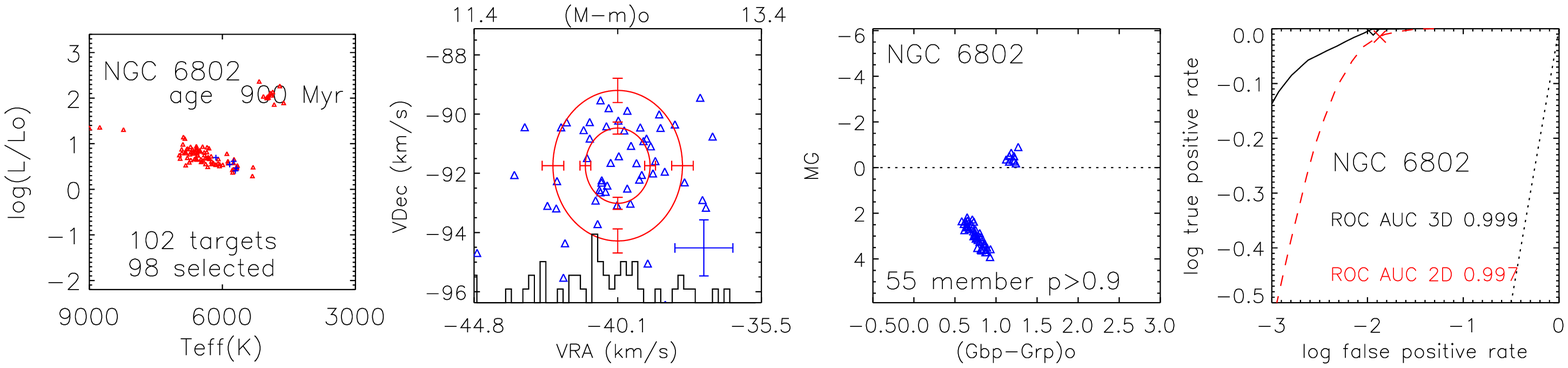}\\
\end{minipage}
\label{figB:36}
\end{figure*}
%%%%%%%%%%%%%%%%%%%%%%%%%%%%%%%%%%%%
\clearpage
\newpage
%%%%%%%%%%%%%%%%%%%%%%%%%%%%%%%%%%%%
\begin{figure*}
\begin{minipage}[t]{0.98\textwidth}
\centering
\includegraphics[width = 145mm]{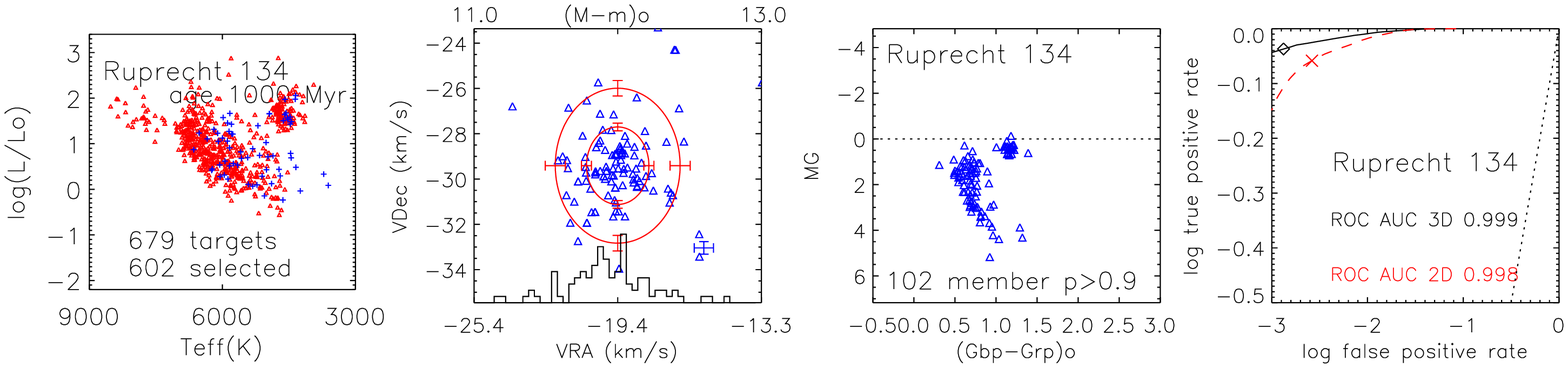}\\
\end{minipage}
\label{figB:37}
\end{figure*}
%%%%%%%%%%%%%%%%%%%%%%%%%%%%%%%%%%%%
\begin{figure*}
\begin{minipage}[t]{0.98\textwidth}
\centering
\includegraphics[width = 145mm]{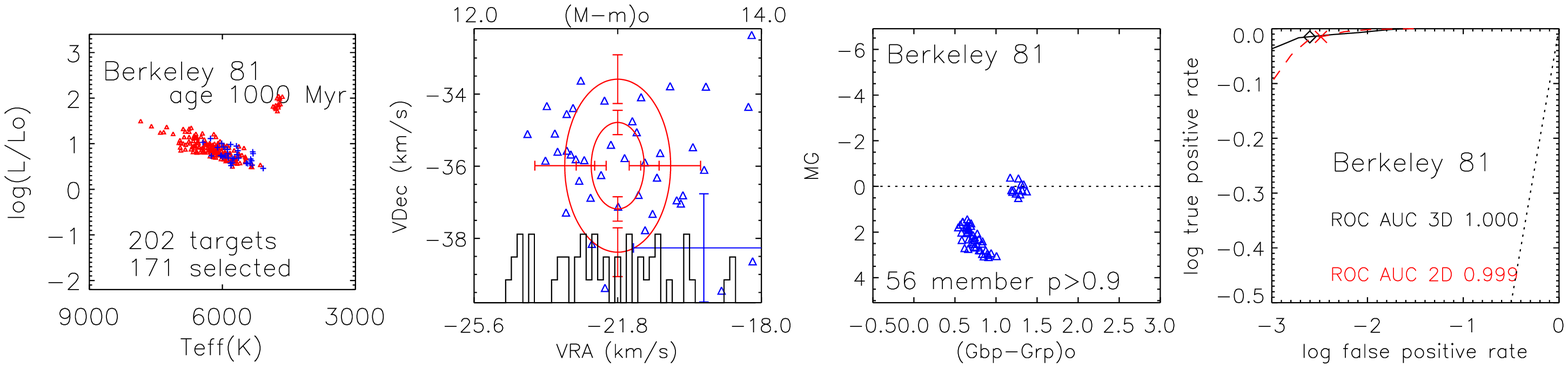}\\
\end{minipage}
\label{figB:38}
\end{figure*}
%%%%%%%%%%%%%%%%%%%%%%%%%%%%%%%%%%%%
\begin{figure*}
\begin{minipage}[t]{0.98\textwidth}
\centering
\includegraphics[width = 145mm]{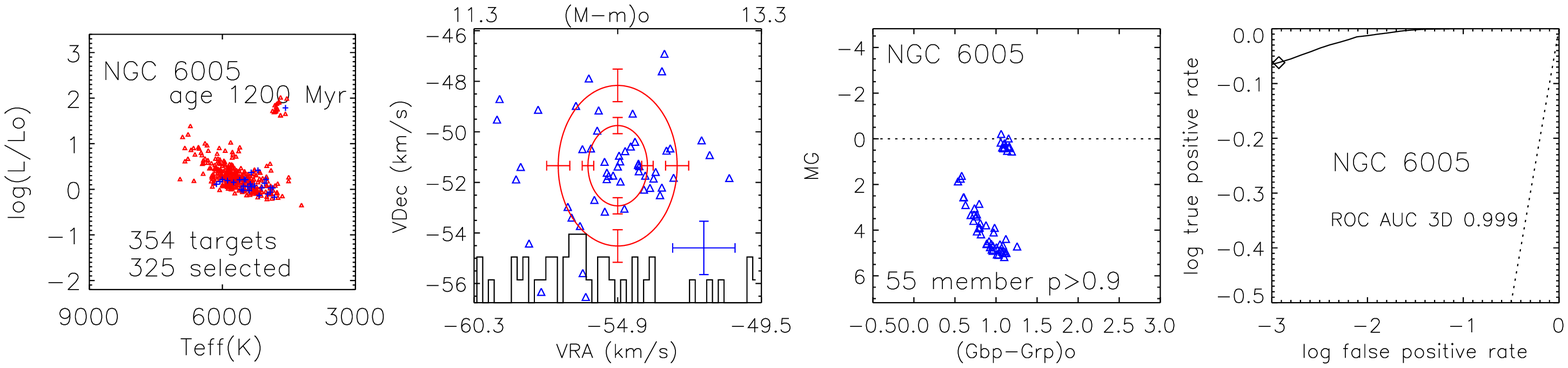}\\
\end{minipage}
\label{figB:39}
\end{figure*}
%%%%%%%%%%%%%%%%%%%%%%%%%%%%%%%%%%%%
\begin{figure*}
\begin{minipage}[t]{0.98\textwidth}
\centering
\includegraphics[width = 145mm]{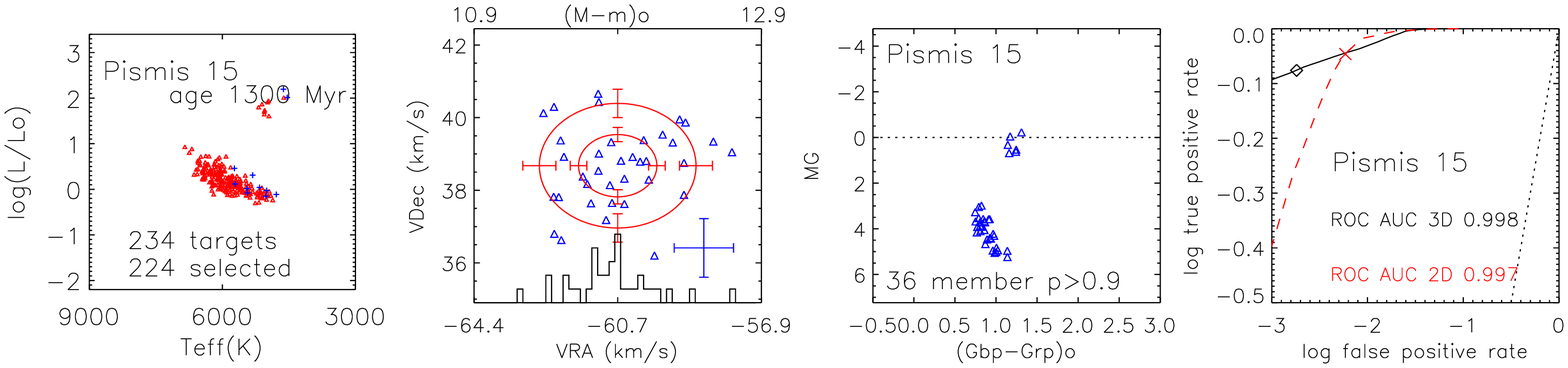}\\
\end{minipage}
\label{figB:40}
\end{figure*}
%%%%%%%%%%%%%%%%%%%%%%%%%%%%%%%%%%%%
\begin{figure*}
\begin{minipage}[t]{0.98\textwidth}
\centering
\includegraphics[width = 145mm]{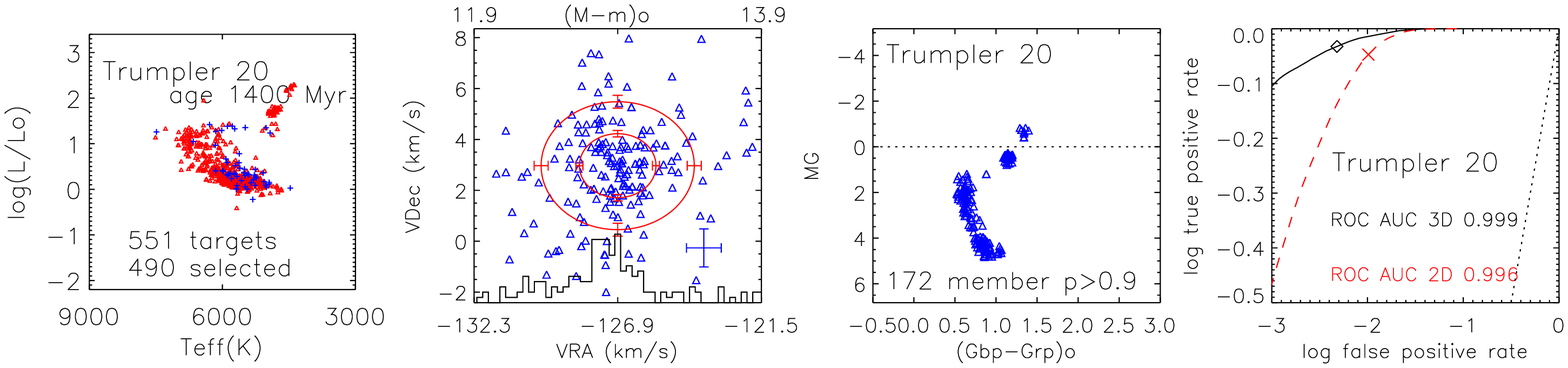}\\
\end{minipage}
\label{figB:41}
\end{figure*}
%%%%%%%%%%%%%%%%%%%%%%%%%%%%%%%%%%%%
\begin{figure*}
\begin{minipage}[t]{0.98\textwidth}
\centering
\includegraphics[width = 145mm]{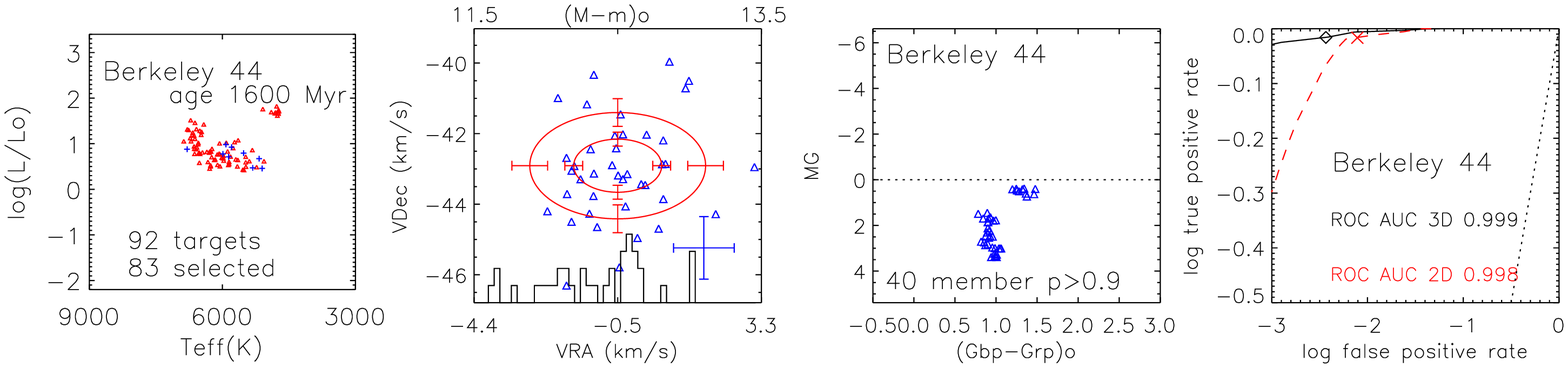}\\
\end{minipage}
\label{figB:42}
\end{figure*}
%%%%%%%%%%%%%%%%%%%%%%%%%%%%%%%%%%%%
\begin{figure*}
\begin{minipage}[t]{0.98\textwidth}
\centering
\includegraphics[width = 145mm]{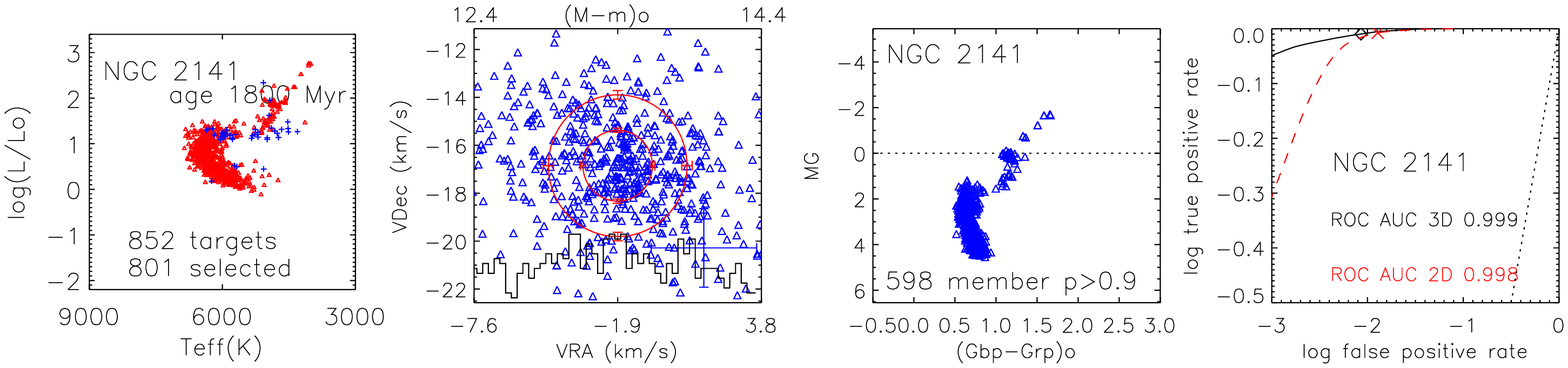}\\
\end{minipage}
\label{figB:43}
\end{figure*}
%%%%%%%%%%%%%%%%%%%%%%%%%%%%%%%%%%%%
\begin{figure*}
\begin{minipage}[t]{0.98\textwidth}
\centering
\includegraphics[width = 145mm]{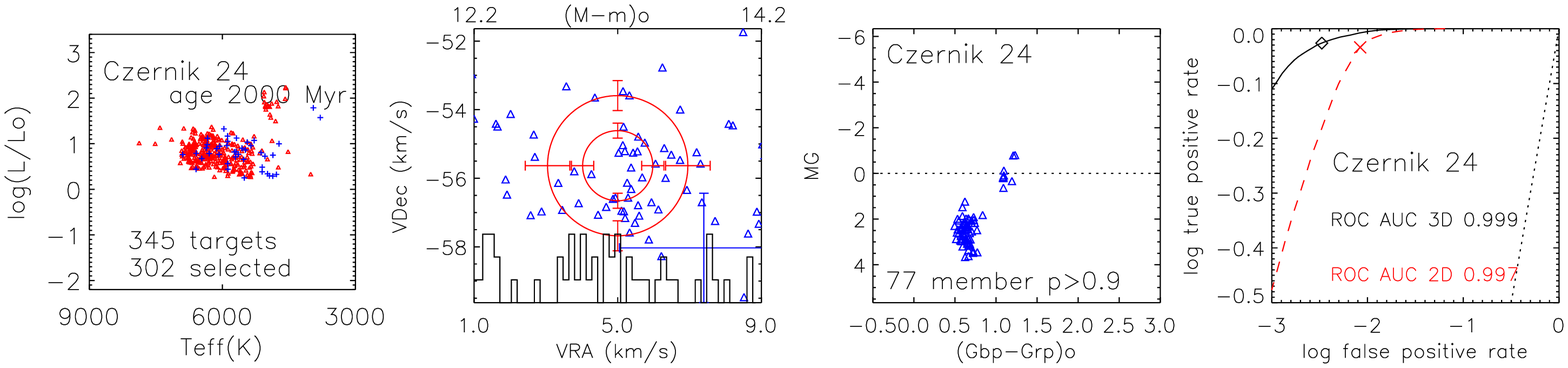}\\
\end{minipage}
\label{figB:44}
\end{figure*}
%%%%%%%%%%%%%%%%%%%%%%%%%%%%%%%%%%%%
\begin{figure*}
\begin{minipage}[t]{0.98\textwidth}
\centering
\includegraphics[width = 145mm]{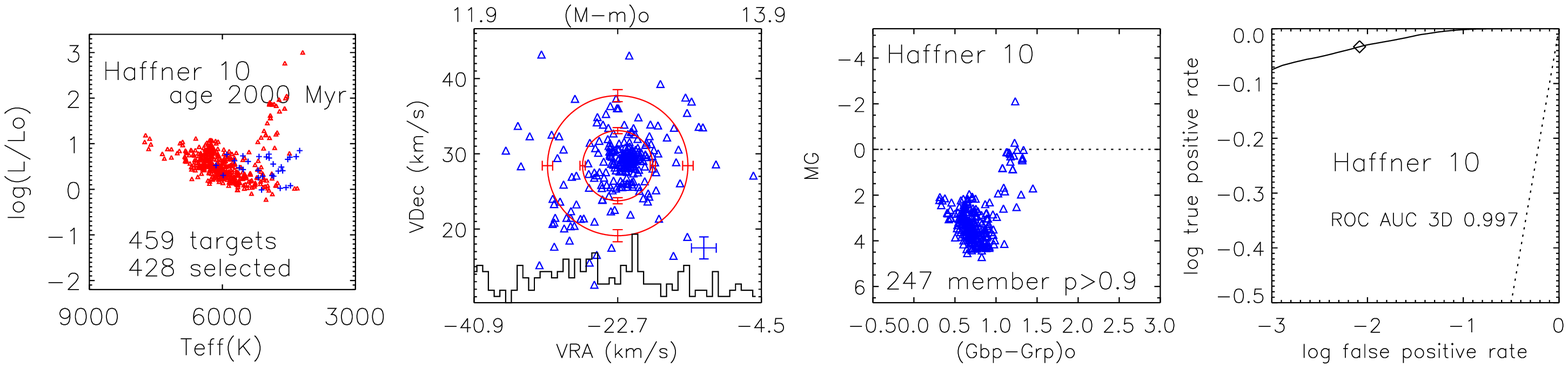}\\
\end{minipage}
\label{figB:45}
\end{figure*}
%%%%%%%%%%%%%%%%%%%%%%%%%%%%%%%%%%%%
\begin{figure*}
\begin{minipage}[t]{0.98\textwidth}
\centering
\includegraphics[width = 145mm]{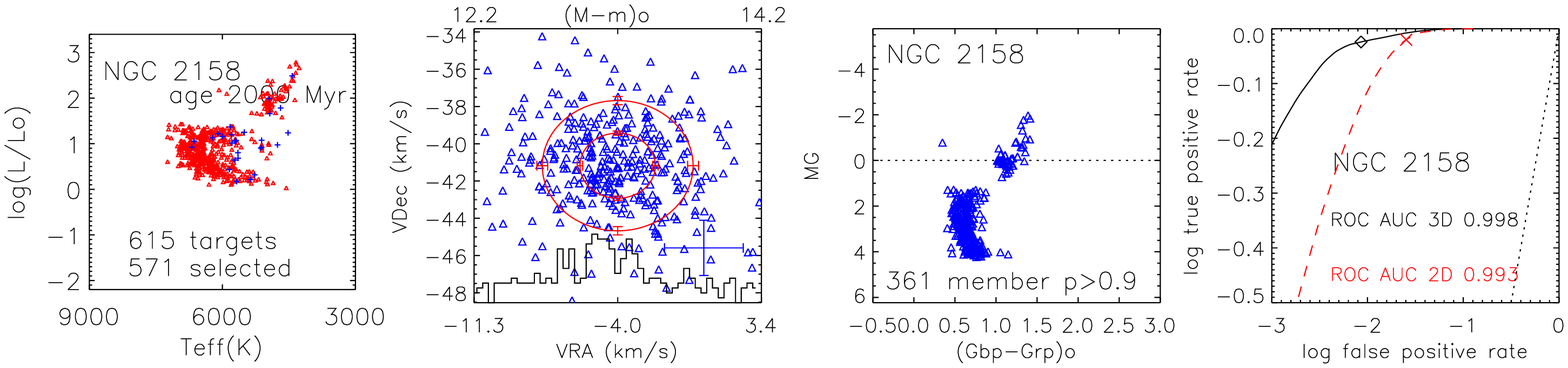}\\
\end{minipage}
\label{figB:46}
\end{figure*}
%%%%%%%%%%%%%%%%%%%%%%%%%%%%%%%%%%%%
\begin{figure*}
\begin{minipage}[t]{0.98\textwidth}
\centering
\includegraphics[width = 145mm]{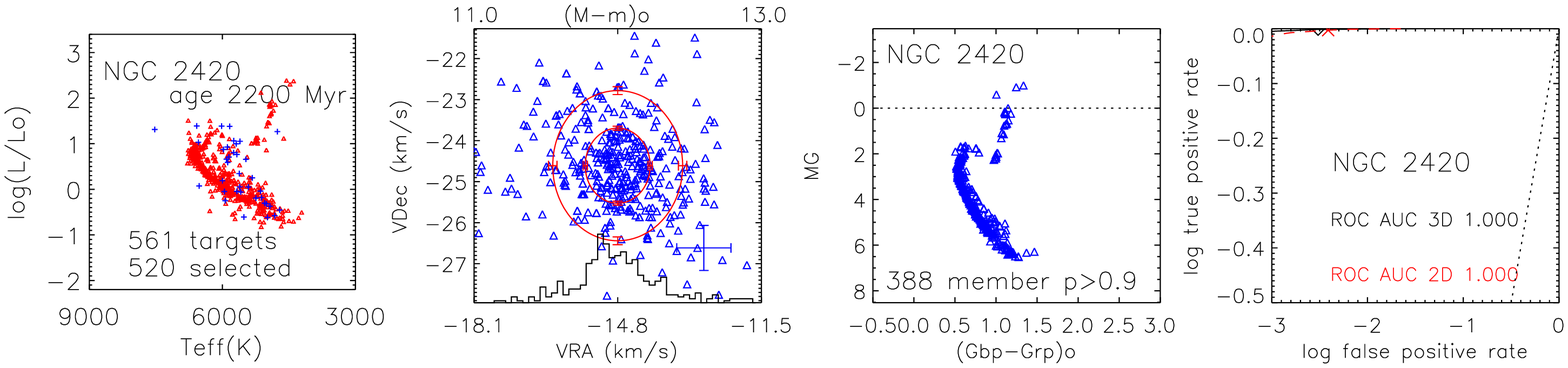}\\
\end{minipage}
\label{figB:47}
\end{figure*}
%%%%%%%%%%%%%%%%%%%%%%%%%%%%%%%%%%%%
\begin{figure*}
\begin{minipage}[t]{0.98\textwidth}
\centering
\includegraphics[width = 145mm]{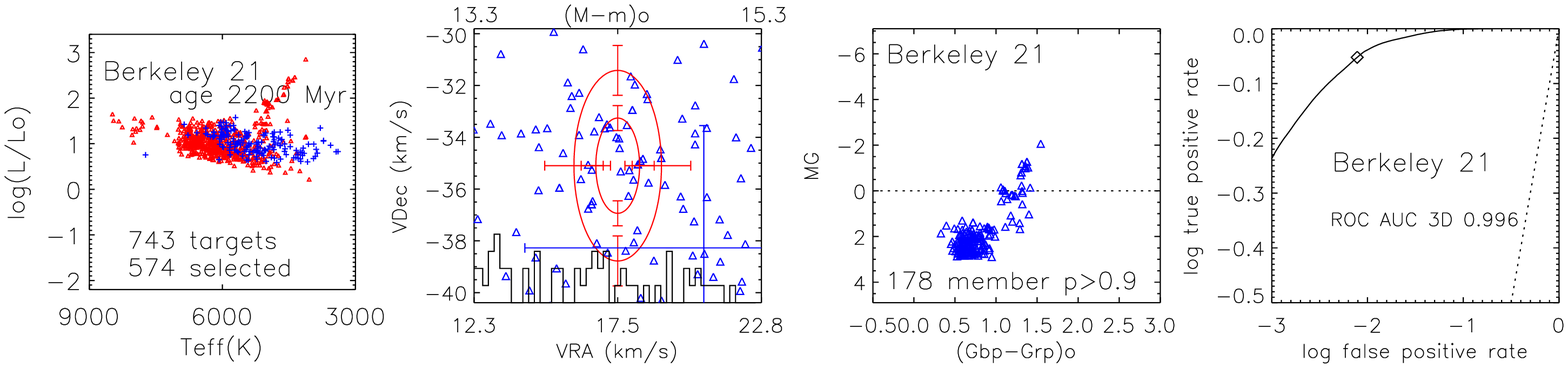}\\
\end{minipage}
\label{figB:48}
\end{figure*}
%%%%%%%%%%%%%%%%%%%%%%%%%%%%%%%%%%%%
\clearpage
\newpage
%%%%%%%%%%%%%%%%%%%%%%%%%%%%%%%%%%%%
\begin{figure*}
\begin{minipage}[t]{0.98\textwidth}
\centering
\includegraphics[width = 145mm]{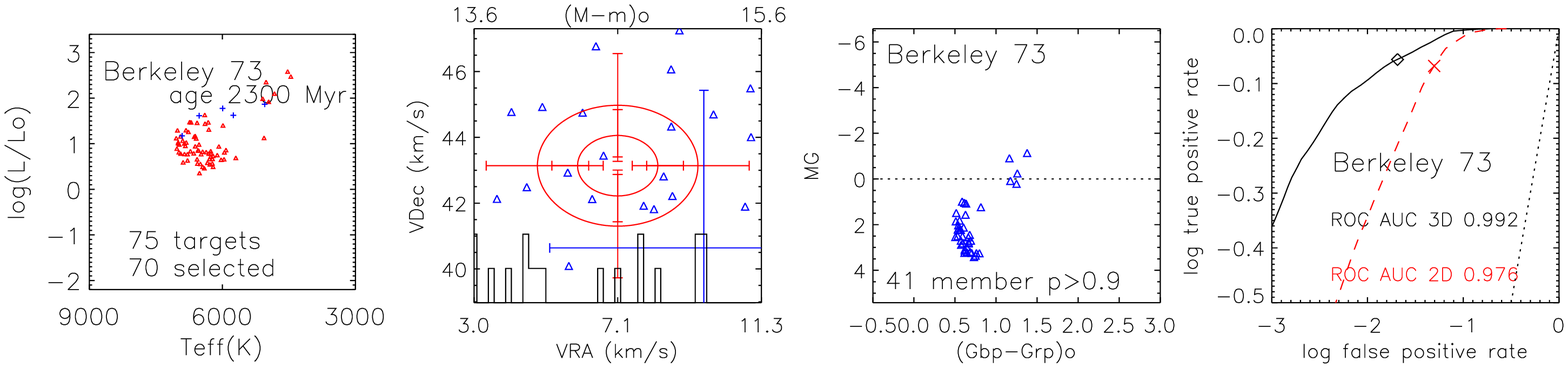}\\
\end{minipage}
\label{figB:49}
\end{figure*}
%%%%%%%%%%%%%%%%%%%%%%%%%%%%%%%%%%%%
\begin{figure*}
\begin{minipage}[t]{0.98\textwidth}
\centering
\includegraphics[width = 145mm]{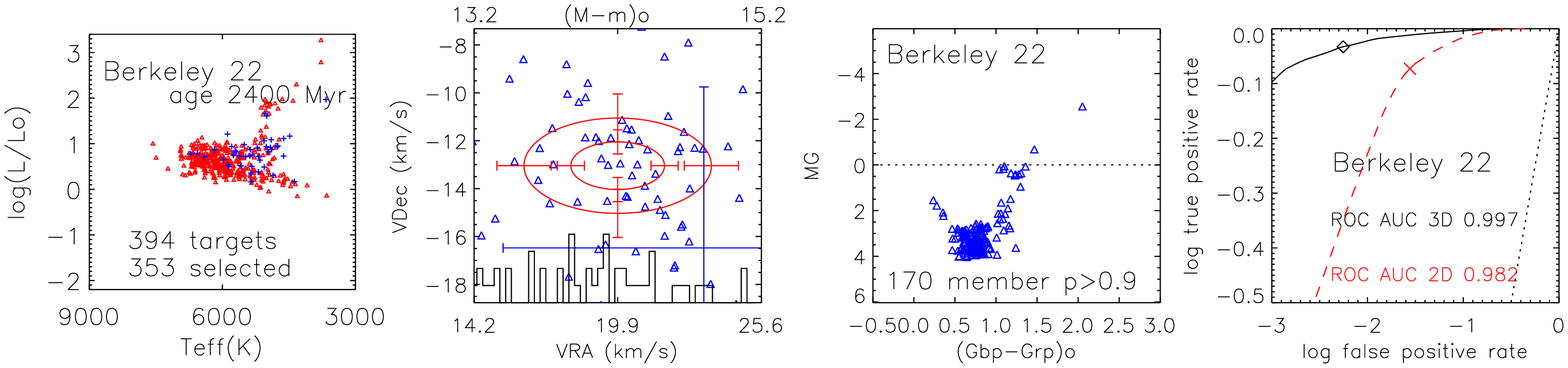}\\
\end{minipage}
\label{figB:50}
\end{figure*}
%%%%%%%%%%%%%%%%%%%%%%%%%%%%%%%%%%%%
\begin{figure*}
\begin{minipage}[t]{0.98\textwidth}
\centering
\includegraphics[width = 145mm]{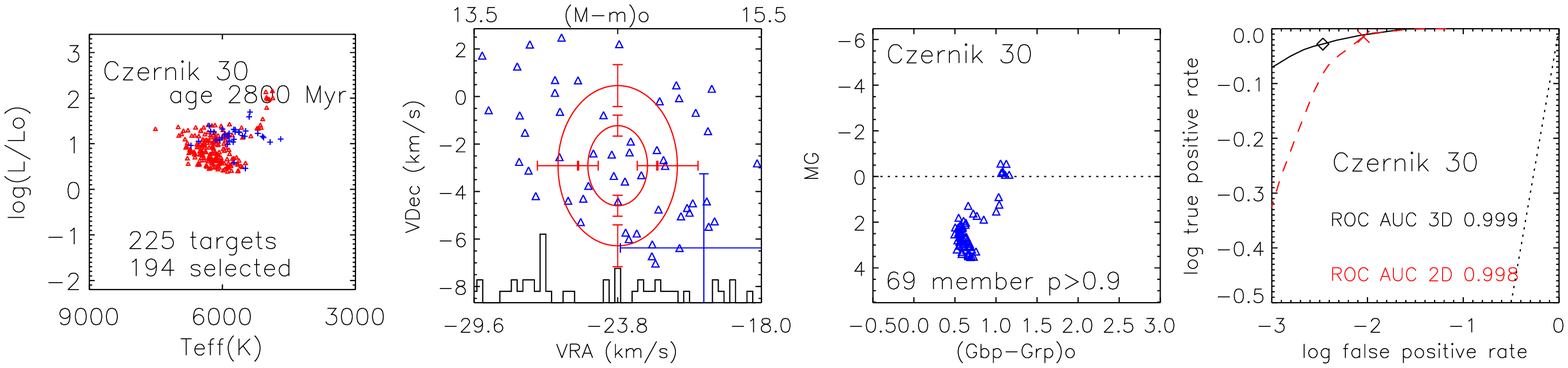}\\
\end{minipage}
\label{figB:51}
\end{figure*}
%%%%%%%%%%%%%%%%%%%%%%%%%%%%%%%%%%%%
\begin{figure*}
\begin{minipage}[t]{0.98\textwidth}
\centering
\includegraphics[width = 145mm]{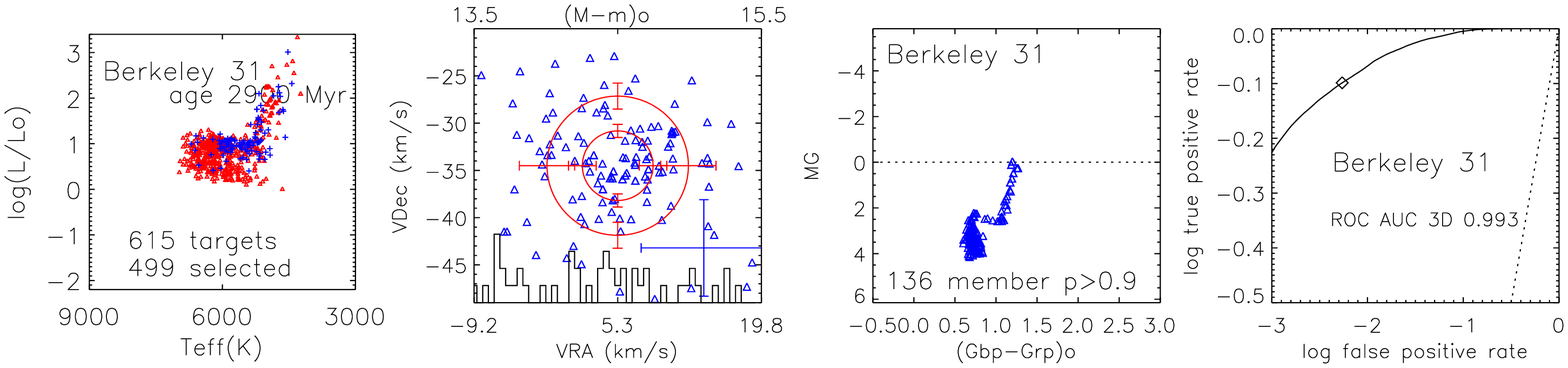}\\
\end{minipage}
\label{figB:52}
\end{figure*}
%%%%%%%%%%%%%%%%%%%%%%%%%%%%%%%%%%%%
\begin{figure*}
\begin{minipage}[t]{0.98\textwidth}
\centering
\includegraphics[width = 145mm]{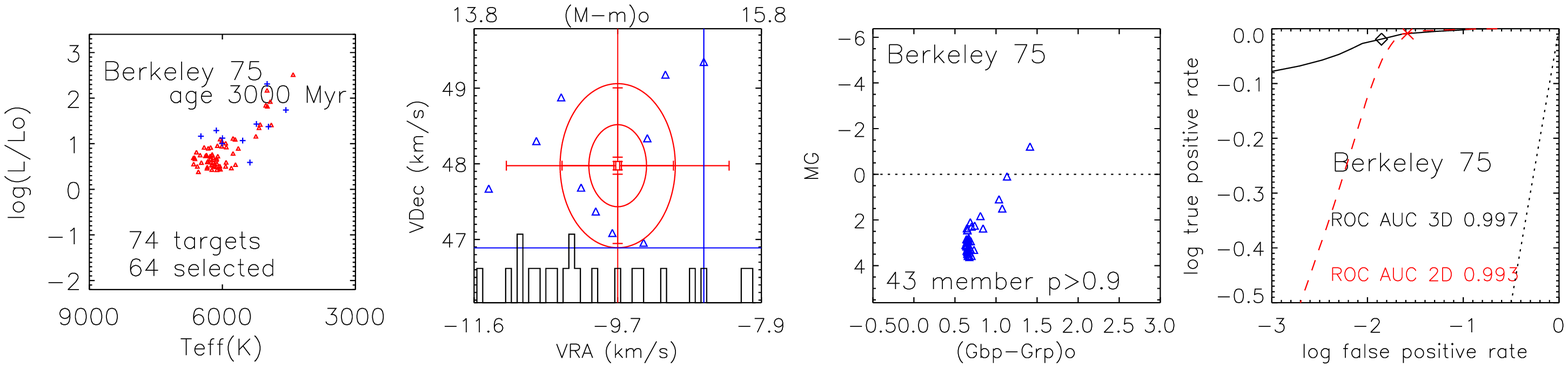}\\
\end{minipage}
\label{figB:53}
\end{figure*}
%%%%%%%%%%%%%%%%%%%%%%%%%%%%%%%%%%%%
\begin{figure*}
\begin{minipage}[t]{0.98\textwidth}
\centering
\includegraphics[width = 145mm]{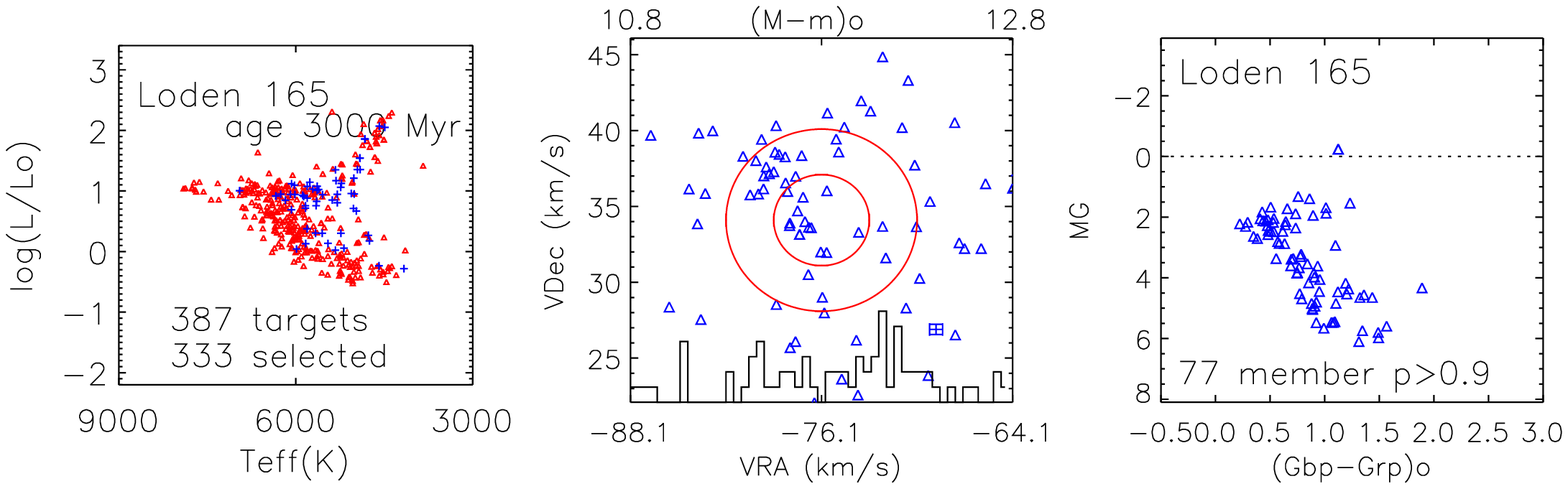}\\
\end{minipage}
\label{figB:54}
\end{figure*}
%%%%%%%%%%%%%%%%%%%%%%%%%%%%%%%%%%%%
\clearpage
\newpage
%%%%%%%%%%%%%%%%%%%%%%%%%%%%%%%%%%%%
\begin{figure*}
\begin{minipage}[t]{0.98\textwidth}
\centering
\includegraphics[width = 145mm]{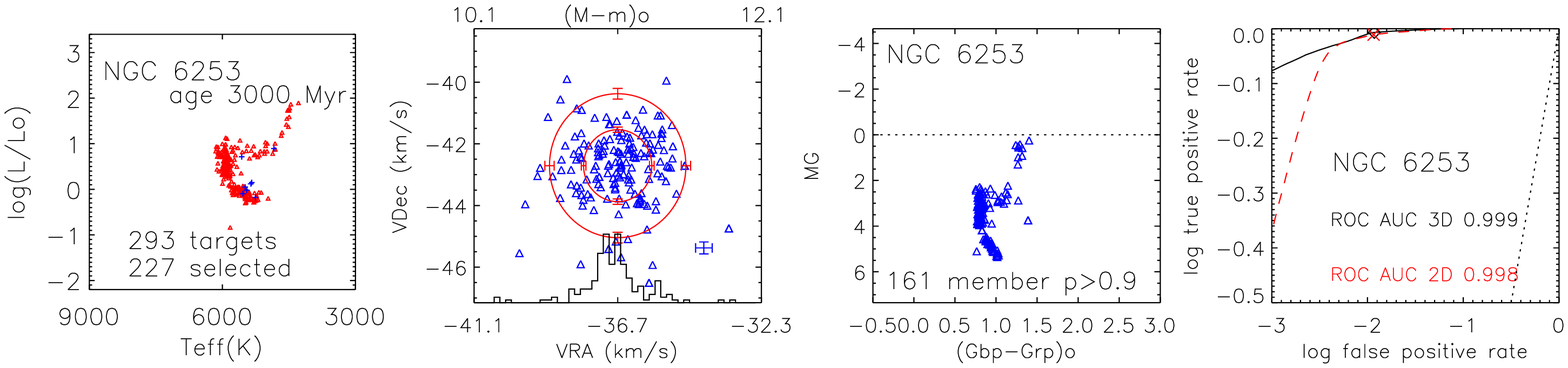}\\
\end{minipage}
\label{figB:55}
\end{figure*}
%%%%%%%%%%%%%%%%%%%%%%%%%%%%%%%%%%%%
\begin{figure*}
\begin{minipage}[t]{0.98\textwidth}
\centering
\includegraphics[width = 145mm]{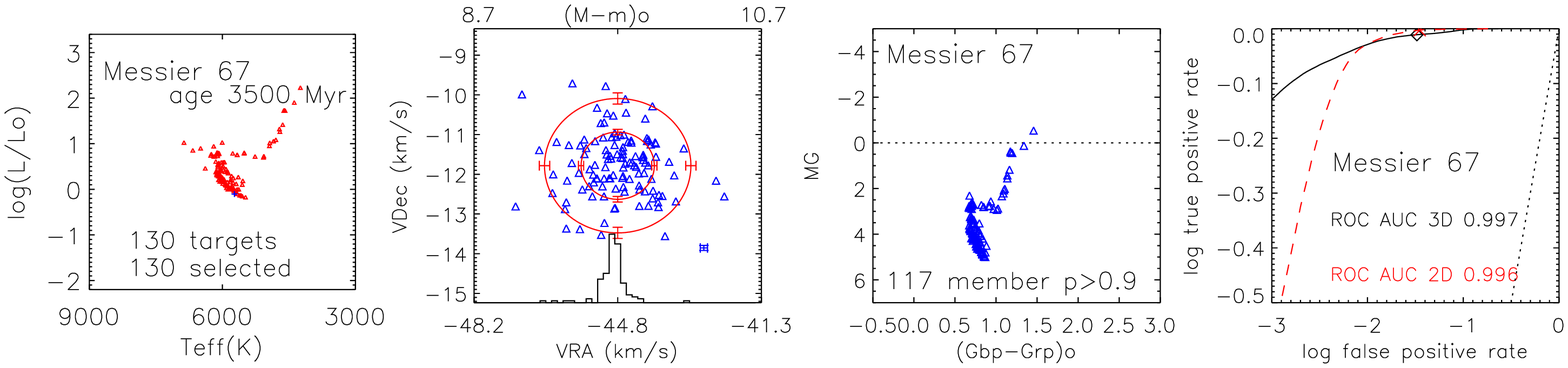}\\
\end{minipage}
\label{figB:56}
\end{figure*}
%%%%%%%%%%%%%%%%%%%%%%%%%%%%%%%%%%%%
\begin{figure*}
\begin{minipage}[t]{0.98\textwidth}
\centering
\includegraphics[width = 145mm]{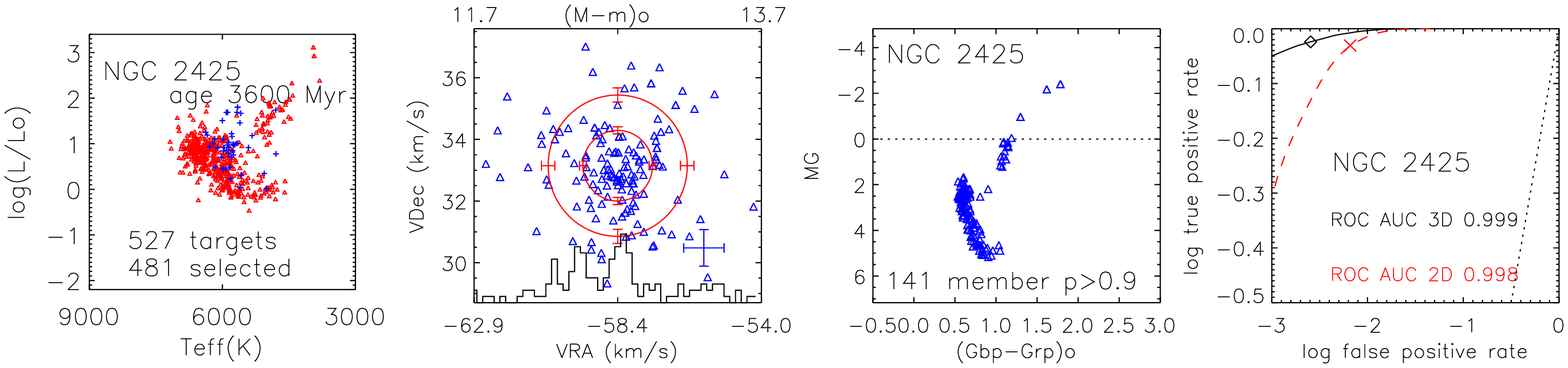}\\
\end{minipage}
\label{figB:57}
\end{figure*}
%%%%%%%%%%%%%%%%%%%%%%%%%%%%%%%%%%%%
\begin{figure*}
\begin{minipage}[t]{0.98\textwidth}
\centering
\includegraphics[width = 145mm]{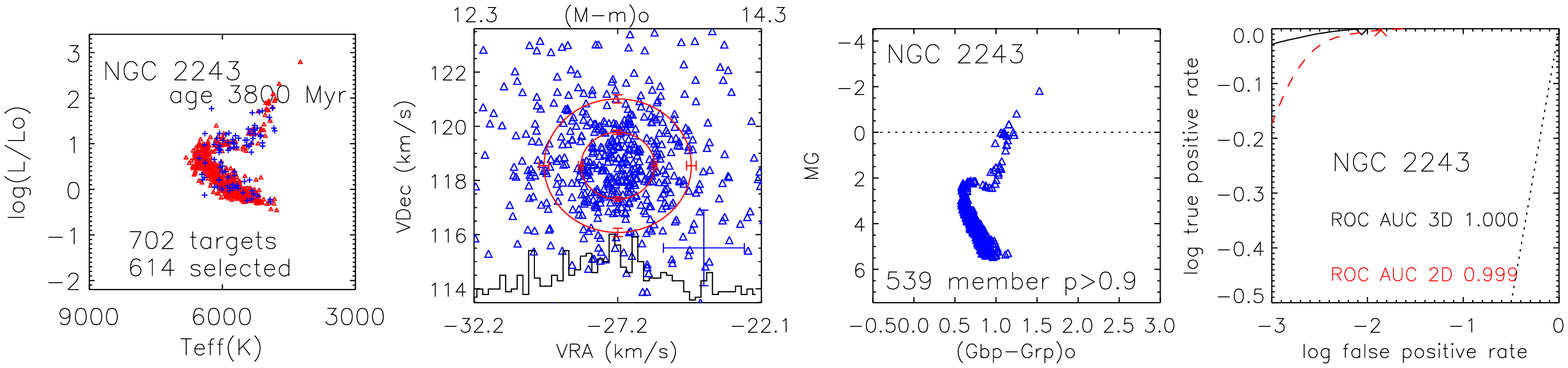}\\
\end{minipage}
\label{figB:58}
\end{figure*}
%%%%%%%%%%%%%%%%%%%%%%%%%%%%%%%%%%%%
\begin{figure*}
\begin{minipage}[t]{0.98\textwidth}
\centering
\includegraphics[width = 145mm]{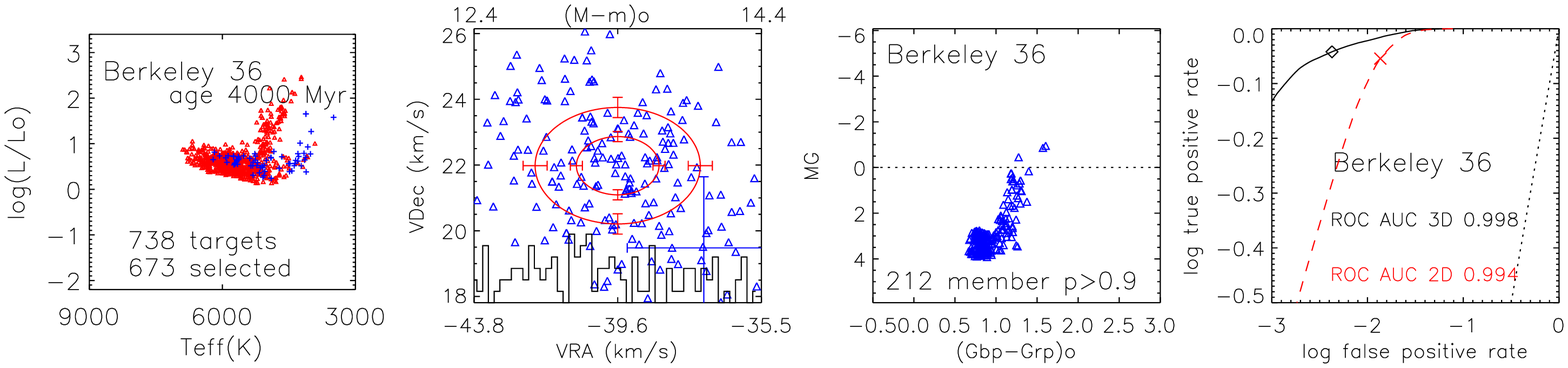}\\
\end{minipage}
\label{figB:59}
\end{figure*}
%%%%%%%%%%%%%%%%%%%%%%%%%%%%%%%%%%%%
\begin{figure*}
\begin{minipage}[t]{0.98\textwidth}
\centering
\includegraphics[width = 145mm]{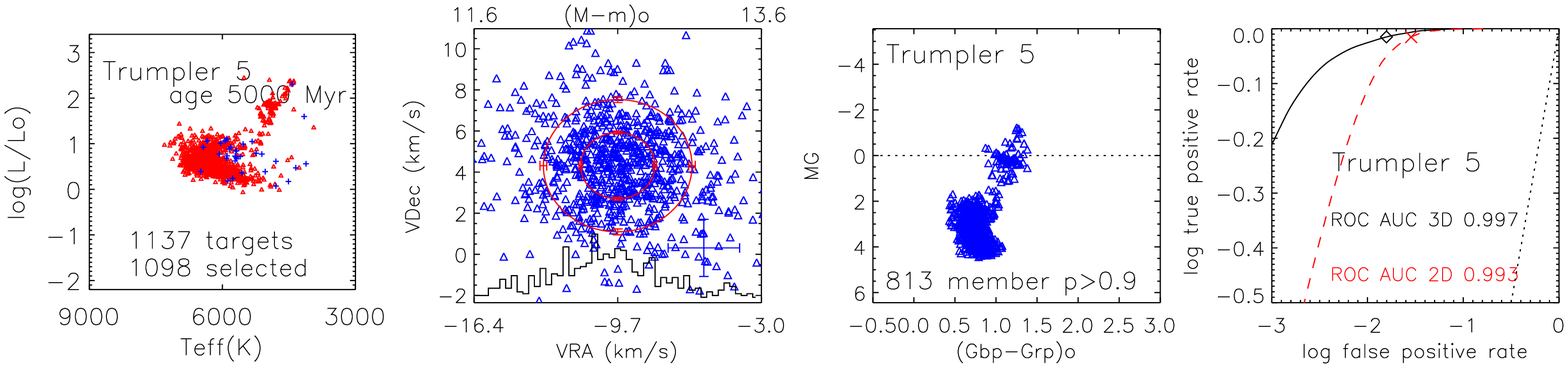}\\
\end{minipage}
\label{figB:60}
\end{figure*}
%%%%%%%%%%%%%%%%%%%%%%%%%%%%%%%%%%%%
\clearpage
\newpage
%%%%%%%%%%%%%%%%%%%%%%%%%%%%%%%%%%%%
\begin{figure*}
\begin{minipage}[t]{0.98\textwidth}
\centering
\includegraphics[width = 145mm]{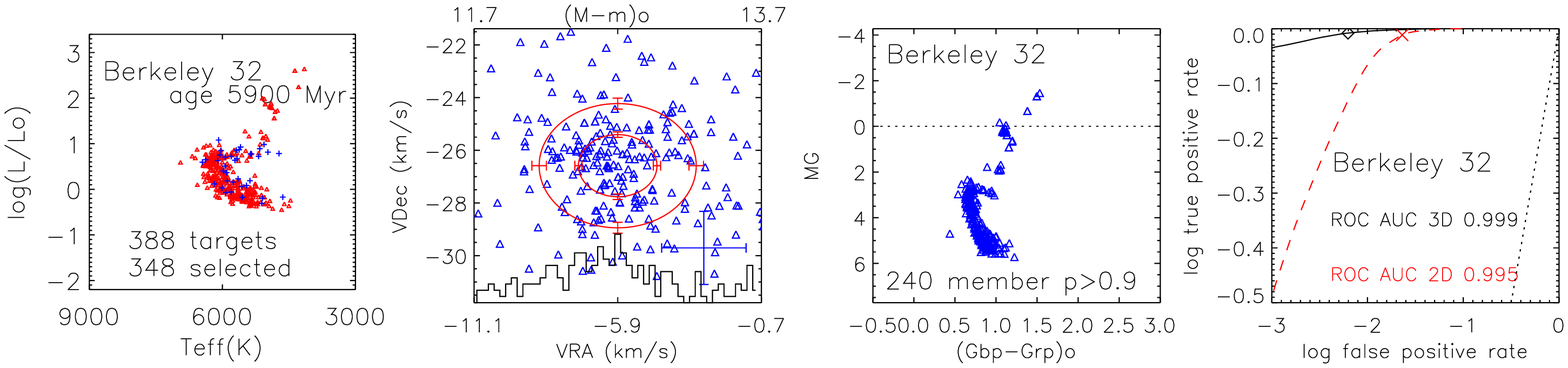}\\
\end{minipage}
\label{figB:61}
\end{figure*}
%%%%%%%%%%%%%%%%%%%%%%%%%%%%%%%%%%%%
\begin{figure*}
\begin{minipage}[t]{0.98\textwidth}
\centering
\includegraphics[width = 145mm]{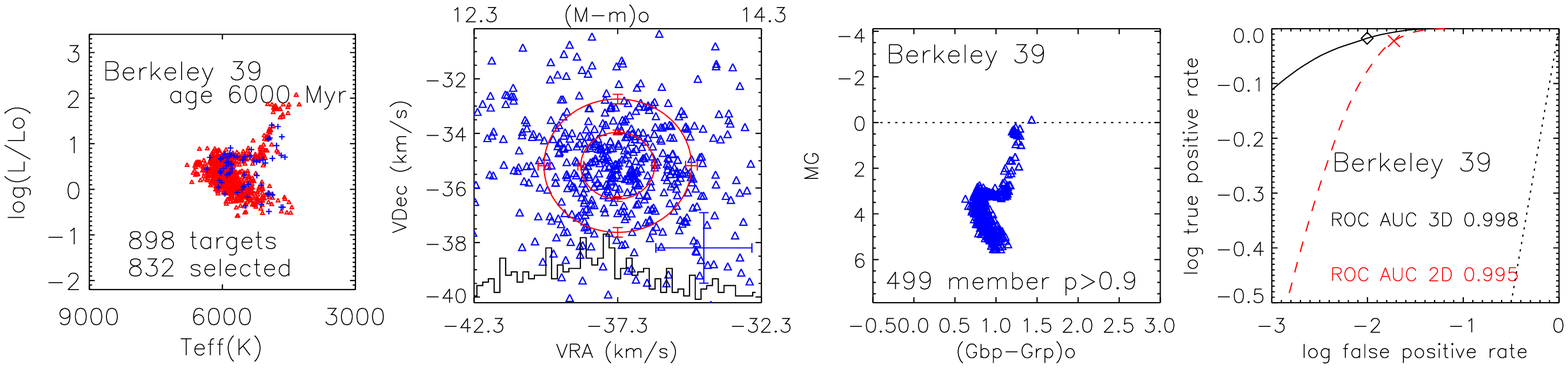}\\
\end{minipage}
\label{figB:62}
\end{figure*}
%%%%%%%%%%%%%%%%%%%%%%%%%%%%%%%%%%%%
\begin{figure*}
\begin{minipage}[t]{0.98\textwidth}
\centering
\includegraphics[width = 145mm]{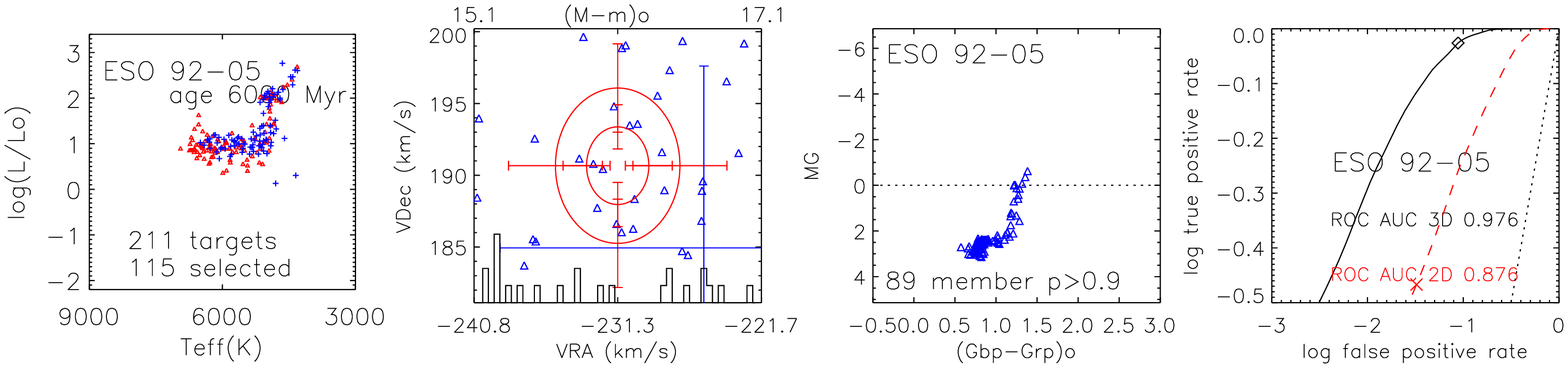}\\
\end{minipage}
\label{figB:63}
\end{figure*}
%%%%%%%%%%%%%%%%%%%%%%%%%%%%%%%%%%%%
\begin{figure*}
\begin{minipage}[t]{0.98\textwidth}
\centering
\includegraphics[width = 145mm]{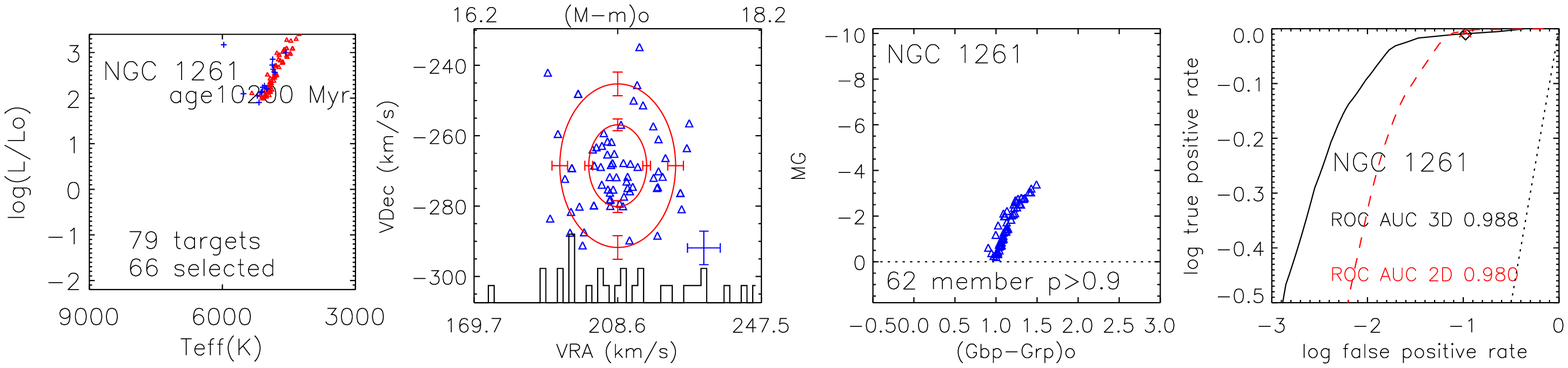}\\
\end{minipage}
\label{figB:64}
\end{figure*}
%%%%%%%%%%%%%%%%%%%%%%%%%%%%%%%%%%%%
\begin{figure*}
\begin{minipage}[t]{0.98\textwidth}
\centering
\includegraphics[width = 145mm]{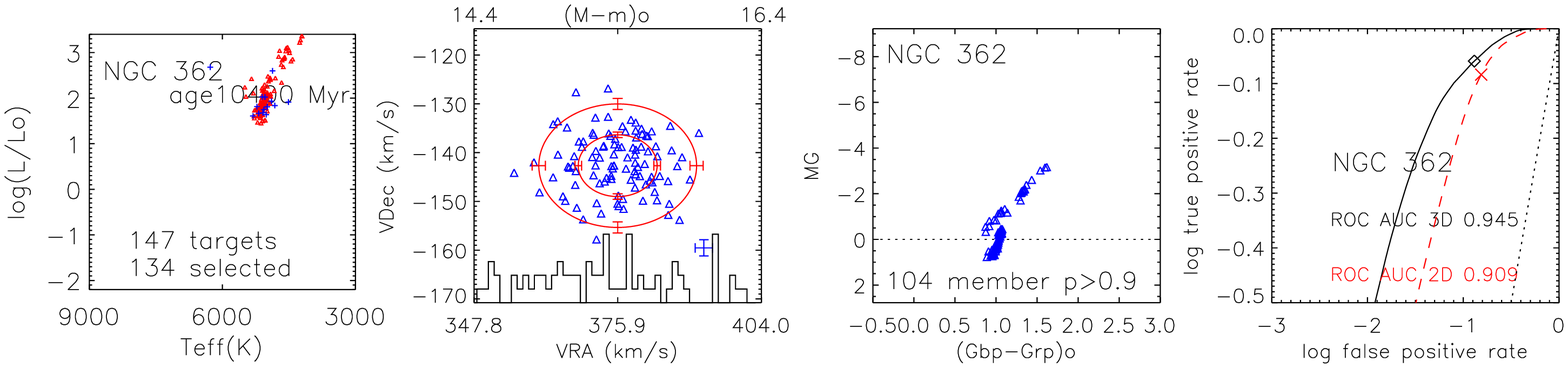}\\
\end{minipage}
\label{figB:65}
\end{figure*}
%%%%%%%%%%%%%%%%%%%%%%%%%%%%%%%%%%%%
\begin{figure*}
\begin{minipage}[t]{0.98\textwidth}
\centering
\includegraphics[width = 145mm]{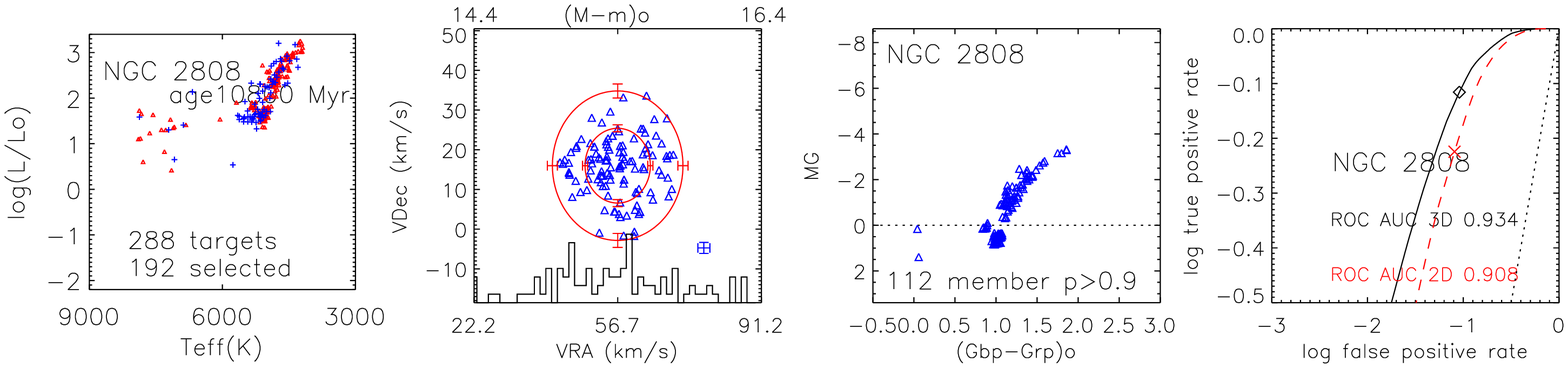}\\
\end{minipage}
\label{figB:66}
\end{figure*}
%%%%%%%%%%%%%%%%%%%%%%%%%%%%%%%%%%%%
\clearpage
\newpage
%%%%%%%%%%%%%%%%%%%%%%%%%%%%%%%%%%%%
\begin{figure*}
\begin{minipage}[t]{0.98\textwidth}
\centering
\includegraphics[width = 145mm]{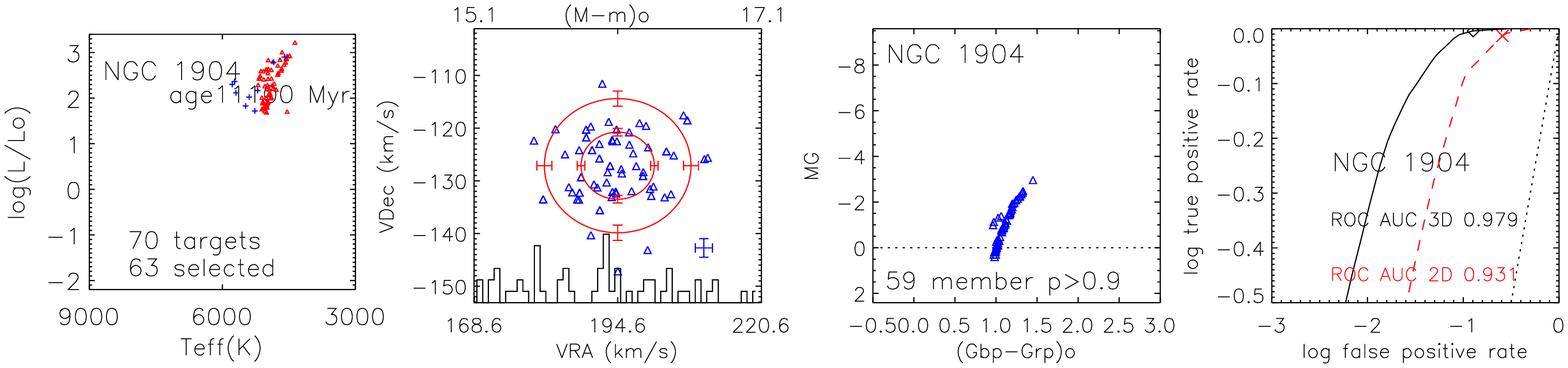}\\
\end{minipage}
\label{figB:67}
\end{figure*}
%%%%%%%%%%%%%%%%%%%%%%%%%%%%%%%%%%%%
\begin{figure*}
\begin{minipage}[t]{0.98\textwidth}
\centering
\includegraphics[width = 145mm]{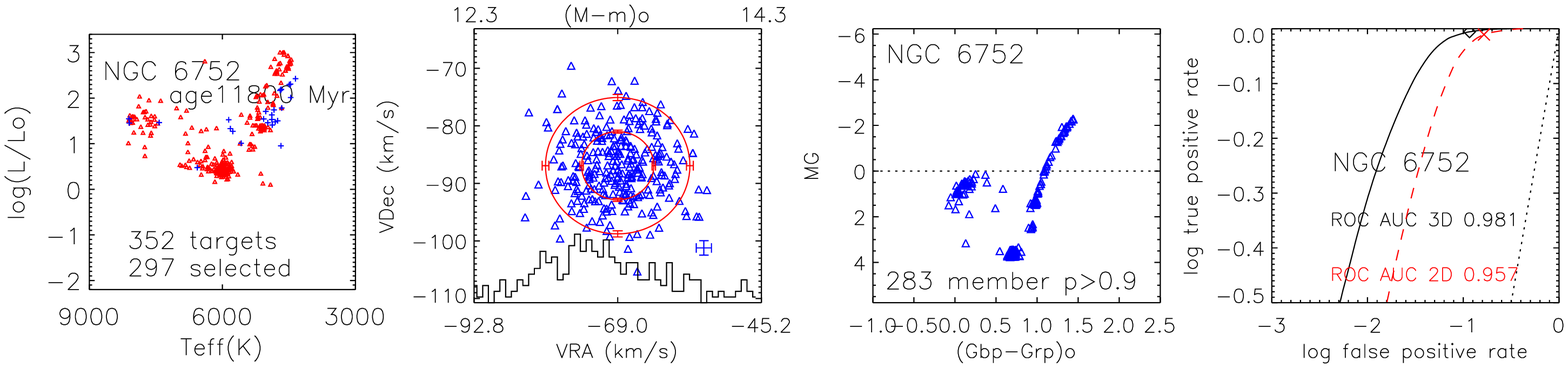}\\
\end{minipage}
\label{figB:68}
\end{figure*}
%%%%%%%%%%%%%%%%%%%%%%%%%%%%%%%%%%%%
\begin{figure*}
\begin{minipage}[t]{0.98\textwidth}
\centering
\includegraphics[width = 145mm]{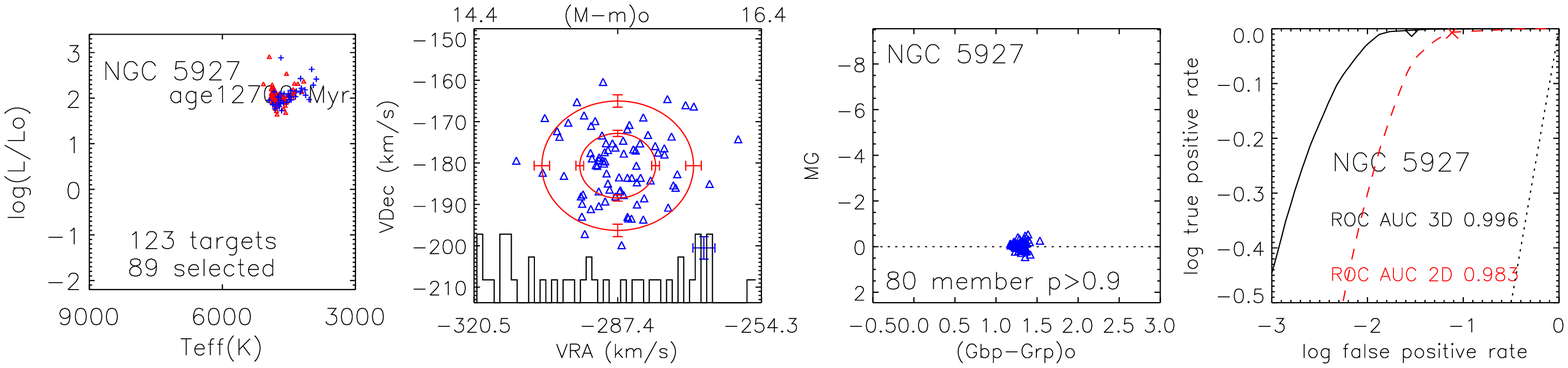}\\
\end{minipage}
\label{figB:69}
\end{figure*}
%%%%%%%%%%%%%%%%%%%%%%%%%%%%%%%%%%%%
\begin{figure*}
\begin{minipage}[t]{0.98\textwidth}
\centering
\includegraphics[width = 145mm]{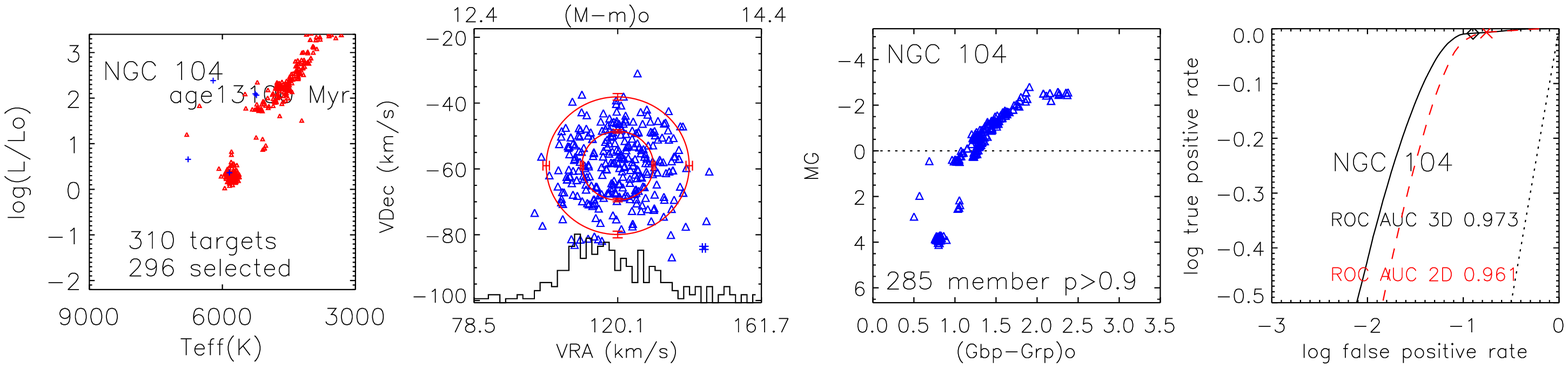}\\
\end{minipage}
\label{figB:70}
\end{figure*}
%%%%%%%%%%%%%%%%%%%%%%%%%%%%%%%%%%%%
\label{lastpage}
\end{document}